\documentclass[a4paper,fleqn]{cas-sc}

\usepackage[authoryear,longnamesfirst]{natbib}
\usepackage{dcolumn}                        
\newcolumntype{d}[1]{D{.}{.}{#1}}           
\newcommand\mc[1]{\multicolumn{1}{c}{#1}}   

\usepackage{changepage}
\usepackage[tableposition=top]{caption}
\usepackage{setspace}
\usepackage{tabularx}	
\usepackage{float}
\usepackage{bbm}
\usepackage{flafter}
\usepackage{float}
\floatstyle{plaintop}
\restylefloat{table}
\def\tsc#1{\csdef{#1}{\textsc{\lowercase{#1}}\xspace}}
\tsc{WGM}
\tsc{QE}
\tsc{EP}
\tsc{PMS}
\tsc{BEC}
\tsc{DE}

\ExplSyntaxOn
\cs_set:Npn \__first_footerline:
{
  \group_begin:
  \small
  \sffamily 
  { \rmfamily \itshape }
  \group_end:
}
\ExplSyntaxOff

\newtheorem{thm}{Theorem}
\numberwithin{thm}{section}

\newtheorem{prop}{Proposition}
\numberwithin{prop}{section}

\newtheorem{lem}{Lemma}
\numberwithin{lem}{section}

\newdefinition{rmk}{Remark}

\newdefinition{assumption}{Assumption}

\newcommand{\proofofref}{}
\newproof{zproofof}{\textit{Proof of \proofofref}}
\newenvironment{proofof}[1]
 {\renewcommand{\proofofref}{#1}\zproofof}
 \qed
 {\endzproofof}

\begin{document}
\let\WriteBookmarks\relax
\def\floatpagepagefraction{1}
\def\textpagefraction{.001}

\title{\bf 
Semiparametric inference for partially linear regressions with Box-Cox transformation}
\shorttitle{SmoothMD for semiparametric partially linear regressions with Box-Cox transformation}

\author[1]{Daniel Becker}[]

\address[1]{PhD-Student in Economics, University of Bonn, Germany, Adenauerallee 24-26, 53113 Bonn(e-mail: dbecker@uni-bonn.de)}

\author[2]{Alois Kneip}[]

\address[2]{Statistic Professor, University of Bonn, Germany, Adenauerallee 24-26, 53113 Bonn(e-mail: akneip@uni-bonn.de)}

\author[3]{Valentin Patilea}[]

\address[3]{Statistic Professor, CREST(Ensai), France, Campus de Ker-Lann, Rue Blaise Pascal - BP 37203 (e-mail: patilea@ensai.fr)}

\shortauthors{Becker et~al.}

\begin{abstract}
		In this paper, a semiparametric partially linear model in the spirit of Robinson (1988) with Box-Cox transformed dependent variable is studied.
	Transformation regression models are widely used in applied econometrics to avoid misspecification.
	In addition, a partially linear semiparametric model is an intermediate strategy that tries to balance advantages and disadvantages
	of a fully parametric model and nonparametric models.
	A combination of transformation and partially linear semiparametric model is, thus, a natural strategy.
	The model parameters are estimated by a semiparametric extension of the so called smooth minimum distance (SmoothMD) approach
	proposed by \citet{lavergne2013smooth}. SmoothMD is suitable for models defined by conditional moment conditions
	and allows the variance of the error terms to depend on the covariates. In addition, here we allow for infinite-dimension nuisance parameters.
	The asymptotic behavior of the new SmoothMD estimator is studied under general conditions and new inference methods are proposed.
	A simulation experiment illustrates the performance of the methods for finite samples.
\end{abstract}

\begin{keywords}
Semiparametric partially linear model, Nonparametric kernel estimators, Root N-consistent estimation, Conditional estimating equations, Hypothesis testing											
\end{keywords}

\maketitle

\section{Introduction} \label{intro}

The data consists of independent copies of a response variable $Y$ and a random covariate vector
$\left(\boldsymbol X^T,\boldsymbol Z^T \right)^T \in  \mathbb{R}^p \times \mathbb{R}^q$.\footnote{Herein, vectors are column matrices and for any matrix $\boldsymbol A$, $\boldsymbol A^T$ denotes its transpose.} We adopt a semiparametric approach to model the dependence of  $Y$ on $X$ and $Z$.
Starting with \citet{robinson1988root}, the use of partial linear models of the form $Y= \boldsymbol X^T\boldsymbol \beta + m(\boldsymbol Z) + \varepsilon$ for some unknown, nonparametric function $m$ has become popular  in this context. For an overview consider \citet{hardle2012partially} and \citet{li2007nonparametric}.

In many important economic applications, the response variable $Y$ is positive, i.e. $P(Y>0)=1$. It is then often questionable to assume (partial) linear regression models. A commonly used remedy is to apply a partial linear model to a suitable transformation of the response variable.
In econometric practice  the dependent variable is then frequently $\log$-transformed (see for example \citet{acemoglu2001colonial} and \citet{autor2013putting}). But usually no substantial knowledge exists ensuring that this specific transformation leads to the correct model. The transformation proposed by \citet{box1964analysis}
	\begin{align*}	\label{Box_Cox}
		T(Y,\lambda) =
		  \begin{cases}
		\frac{Y^\lambda -1}{\lambda}&, \lambda \neq 0 \\
		\log(Y) &, \lambda = 0
		\end{cases}
	\end{align*}
offers much more flexibility. Specifying the transformation up to a parameter and estimating the parameter together with the regression  coefficients, leads to more reliable results at the cost of having to estimate only one additional parameter. In addition, the common $\log$-transformation is nested in the Box-Cox transformation and, thus, can be confirmed by a statistical test.

This motivates the approach adopted in this paper. To model the relationship between a positive response variable and the covariate vectors, we consider a \emph{transformation partially linear} mean regression model given by
	\begin{equation}	\label{main_model}
		T(Y,\lambda) = \boldsymbol X^T\boldsymbol \beta + m(\boldsymbol Z) + \varepsilon,
	\end{equation}
where $m(\cdot)$ is an unknown function and

	\begin{align}	\label{mom_con}
		E[\varepsilon\mid\boldsymbol X,\boldsymbol Z] = 0.
	\end{align}
The true values $\lambda_0$ and $\boldsymbol \beta_0$ of the parameters $\lambda>0$ and  $\boldsymbol \beta\in \mathbb{R}^p$ are unknown and have to be estimated from an i.i.d.  sample $(Y_i,\boldsymbol X_i^T,\boldsymbol Z_i^T)$, $i=1,\dots,n$, of $(Y,\boldsymbol X^T,\boldsymbol Z^T)$. 
We impose no further assumption on the conditional distribution of $\varepsilon$. In particular, we allow for heteroscedasticity of unknown form. The vector $\boldsymbol Z$ contains only continuous variables, but the components of $\boldsymbol X$ need not be continuous.

 We wish to note that the Box-Cox transformation is widely used in applications, and is discussed in various textbooks, e.g. \citet{amemiya1985advanced, greene2003econometric,horowitz2012semiparametric,showalter1994monte,wooldridge1992some}. Furthermore, there exist several empirical studies that employ the Box-Cox transformation. See, for instance, \citet{berndt1993empirical,heckman1974empirical} or \citet{keane1988real}. For an overview of the Box-Cox transformation, consider \citet{horowitz2012semiparametric} and \citet{sakia1992box}. The semiparametric partially linear specification of the conditional mean of the response seems to be quite appealing as it allows a linear dependency on a subvector $\boldsymbol X$ of covariates, which could include discrete variables, and meanwhile allows a  nonparametric additive effect of the covariates $\boldsymbol Z$.  These features could help  practitioners faced with a large cross-sectional data set with independent observations including many candidate explanatory variables, who, on the basis of economic theory or past experience with similar data, feel able to parameterize only some of them.

Similar to other semiparametric models, the major challenge is to develop $\sqrt{n}$-consistent  estimators and corresponding inference procedures for the  parameters $(\lambda_0,\boldsymbol \beta_0^T)$.
 This requires a careful
methodological development, since to our knowledge, there is no established procedure  which can readily be applied under our general setup.
 Despite its popularity, even in a purely parametric framework, estimation and inference in a Box-Cox transformation model is a difficult statistical problem and is usually based on quite restrictive assumptions on the conditional law of the response. See for instance chapter 5 of \citet{horowitz2012semiparametric} for an illuminating discussion. In particular, ordinary least squares estimation of
  $(\lambda_0,\boldsymbol\beta_0^T)$ may lead to inconsistent results, and more sophisticated procedures, such as the nonlinear two-stage least squares (NL2SLS) estimator introduced by \citet{amemiya1981comparison}, have to be applied. The problem becomes even more complex in the case of the semiparametric regression (\ref{main_model}) where one only assumes the minimal identification condition (\ref{mom_con}).

In order to motivate our procedure, let us first consider the  special case that the true value $\lambda_0$ is known a priori.
  With $Y_0=T(Y,\lambda_0)$, we then arrive at a standard partial linear model
$Y_0=\boldsymbol X^T\boldsymbol \beta + m(\boldsymbol Z) + \varepsilon$. Consequently, $m(Z)=E(Y_0|Z)-E[\boldsymbol X\mid \boldsymbol Z])^T\boldsymbol \beta_0$ and
	\begin{align*}
0=	E[\varepsilon|\boldsymbol X,\boldsymbol Z] =E\bigg[	Y_0 - E[Y_0\mid \boldsymbol Z]- (\boldsymbol X-E[\boldsymbol X\mid \boldsymbol Z])^T\boldsymbol \beta_0\bigg| X,Z \bigg] .
	\end{align*}
The basic idea of \citet{robinson1988root} now consists in disentangling the nonparametric estimation of unknown functions and the parametric estimation of the coefficient vector $\boldsymbol\beta$. In a first step  Nadaraya-Watson kernel estimators are used for nonparametric estimation
of the functions $E[Y_0\mid \boldsymbol Z]$ and $E[\boldsymbol X\mid \boldsymbol Z]$. Plugging in these nonparametric function estimates,
an OLS regression of $Y_0- E[Y_0\mid \boldsymbol Z]$ on $\boldsymbol X-E[\boldsymbol X\mid \boldsymbol Z]$ then leads to an estimator
$\widehat{\boldsymbol\beta}$. If $q<4$ and standard nonparametric estimators based on second order kernels are applied, then, assuming homoscedasticity and suitable bandwidth sequences, Robinson showed that $\widehat{\boldsymbol\beta}$ is a $\sqrt{n}$-consistent, asymptotically normally distributed and efficient estimator of $\boldsymbol\beta_0$. If the dimensionality of $Z$ is larger, i.e. $q\ge 4$, then  $\sqrt{n}$-consistent estimators can only be achieved by using higher order kernels.

 The quite straightforward way to build efficient estimators made the  partially linear model quite a popular. Versions of this model have also been studied by \citet{engle1986semiparametric,heckman1986spline,shiller1984smoothness} and \citet{wahba1984partial}. In order to avoid the trimming introduced by \citet{robinson1988root} to ensure that the estimate of the density of $\boldsymbol Z$, $f_z(\boldsymbol Z)$, stays away from zero, \citet{li1996root}  considered as starting point the unfeasible OLS regression of  $(Y_0- E[Y_0\mid \boldsymbol Z])f_z(\boldsymbol Z) $ on $(\boldsymbol X-E[\boldsymbol X\mid \boldsymbol Z]) f_z(\boldsymbol Z) $. Premultiplying by the  density of $\boldsymbol Z$ does not break the consistency of the unfeasible OLS estimator since  $E[f_z(\boldsymbol Z) \varepsilon\mid \boldsymbol X,\boldsymbol Z] = f_z(\boldsymbol Z) E[\varepsilon\mid \boldsymbol X,\boldsymbol Z] = 0.$ Next, \citet{li1996root}  proposed to build OLS estimates using standard kernel estimators instead of the unfeasible response and covariates.  This new estimator is still $\sqrt{n}$-consistent and asymptotically normally distributed. Moreover, \citet{li1996root} relaxed the condition on the bandwidth with the consequence that the smoothing requires higher order kernels  only if the dimension of $ \boldsymbol Z$ is larger than 5, instead of larger than 3 as required in \citet{robinson1988root}.

 Let us now return to the general model \eqref{main_model} with unknown parameter $\lambda$. Adopting Li's idea of premultiplying with the density $f_z$ of $\boldsymbol Z$, the conditional moment condition $E[\varepsilon|\boldsymbol X,\boldsymbol Z] = 0$ leads to
	 \begin{equation} \label{CMC1}
		 E\left(\bigg(T(Y,\lambda) - E[T(Y,\lambda)\mid \boldsymbol Z]- (\boldsymbol X-E[\boldsymbol X\mid \boldsymbol Z])^T\boldsymbol \beta\bigg)f_z(\boldsymbol Z)\bigg| X,Z\right)=0\quad  \Longleftrightarrow 
		 \quad  \lambda=\lambda_0, \boldsymbol\beta=\boldsymbol\beta_0.
	 \end{equation}
 It is now immediately seen that the unknown, additional parameter $\lambda$ introduces a major complication. Unlike the standard partial linear model, there is no way to disentangle  nonparametric estimation of unknown functions and parametric estimation of the coefficient vector $(\lambda_0,\boldsymbol\beta_0^T)$. The reason is that $ E[T(Y,\lambda)\mid \boldsymbol Z]$ depends on $\lambda$. Indeed, there does not seem to exist a straightforward generalization of Robinson's approach which is able to cope with model \eqref{main_model}.

On the other hand, by \eqref{CMC1}, our  model  belongs to the large class of models identified by conditional moment restrictions. Methodologically however, we have to deal with
 the obvious facts that
a) the model is highly nonlinear in $\lambda$ and b) \eqref{CMC1} incorporates an infinite dimensional nuisance parameter $\boldsymbol \eta_\lambda$ consisting of the  functions
$f_z(z)$, $E[T(Y,\lambda)\mid \boldsymbol Z=z]$ and $E[\boldsymbol X\mid \boldsymbol Z=z]$.
 Even if $\boldsymbol\eta_\lambda$ were known a priori,  any use of  the generalized method of moments (GMM)  runs into the problem that the conditional moment restrictions identifying our model  imply an infinite number of unconditional moment restrictions, since the conditioning variables have a support with infinite cardinality. But GMM relies only on a finite number of instruments and, thus, in general, consistency of GMM requires regularization and   additional assumptions. See \citet{dominguez2004consistent}. This problem has already been pointed out for the Box-Cox transformation by \citet{foster2001estimation} and \citet{shin2008semiparametric} in the linear case. See also \citet{horowitz2012semiparametric}.

More recent work explicitly focuses on regularization techniques in order to account for complex conditional moment conditions.
Some methods rely on increasing the number of considered unconditional estimating equations (or instruments) with the sample size, such as the sieve minimum distance (SMD) approach of \citet{ai2003efficient}, or generalizations of GMM and empirical likelihood (EL) by \citet{donald2003empirical} and \citet{hjort2009extending}. \citet{carrasco2000generalization} use a regularization approach to generalize the GMM approach to a continuum of estimating equations. Other EL-type estimators use nonparametric smoothing to estimate conditional equations, such as \citet{antoine2007efficient}, \citet{kitamura2004empirical}, and \citet{smith2007efficient, smith2007local}. All these approaches share one common feature. The estimators’ sensitivity to the user-chosen parameter (number of estimating equations, regularization parameter, or smoothing parameter) remains largely unknown.

In this paper, we rely upon the SmoothMD approach proposed by \citet{lavergne2013smooth} to estimate $(\lambda_0,\boldsymbol\beta_0^T)$. Roughly speaking, SmoothMD can be seen as a new technique to translate conditional moment conditions into unconditional ones which can be approximated by sample averages. Although the method involves some tuning parameters, an attractive feature consists of the fact that a practical choice is quite uncritical, since asymptotic results can be established for a wide range of possible values of these tuning parameters (including values independent of the sample size). SmoothMD thus bridges a gap between Dominguez and Lobato’s method, which does not require a user-chosen parameter, and the competing SMD estimator and EL and GMM-type methods that rely on smoothing with restrictive conditions on the choice of smoothing parameters. Furthermore, although \citet{lavergne2013smooth} rely on a more standard setup. We will show that this technique can be well adapted to deal with complex functional nuisance parameters.

In order to explain the methodology, we  introduce some abbreviations in order to simplify the lengthy expressions in \eqref{CMC1}. Let $\boldsymbol W=\left(\boldsymbol X^T,\boldsymbol Z^T \right)^T \in   \mathbb{R}^p \times \mathbb{R}^q$ and $\boldsymbol U=\left(Y,\boldsymbol W^T \right)^T$.
Moreover, set $\boldsymbol\theta = \left(\lambda, \boldsymbol{\beta}^T\right)^T$, and for a real value $\gamma$ define
	\begin{equation}\label{def1}
		g(\boldsymbol U;  \boldsymbol \theta, 
		\boldsymbol \eta_\lambda) = \left(T(Y,\lambda) - E[T(Y,\lambda)\mid \boldsymbol Z]- (\boldsymbol X-E[\boldsymbol X\mid \boldsymbol Z])^T\boldsymbol \beta\right)f_z(\boldsymbol Z) . 
	\end{equation}
Recall that $\boldsymbol \eta_\lambda$  is an infinite-dimensional nuisance parameter defined by 
$$\boldsymbol \eta_\lambda=\boldsymbol\eta_\lambda(z)=(f_z(z), E[T(Y,\lambda)\mid \boldsymbol Z=z],E[\boldsymbol X\mid \boldsymbol Z=z]^T).$$ 
Condition \eqref{CMC1} is then equivalent to requiring
	\begin{equation} \label{CMC2}
		 E\left(g(\boldsymbol U; \boldsymbol \theta, 
		 \boldsymbol \eta_\lambda)\big| W\right)=0\quad  \Longleftrightarrow \quad 
		  \boldsymbol\theta = \left(\lambda, \boldsymbol{\beta}^T\right)^T=
		 \left(\lambda_0, \boldsymbol{\beta}_0^T\right)^T =:\boldsymbol\theta_0 . 
	 \end{equation}

SmoothMD is based on the following insight: Let $\boldsymbol U_1$ and $\boldsymbol U_2$ be two independent copies of $\boldsymbol U$ with corresponding subvectors $\boldsymbol  W_1$ and $\boldsymbol  W_2$. For any symmetric function  $\omega(\cdot)$  of $\boldsymbol W$ with positive Fourier transform, we than have that
	\begin{equation*}
		 E\left(g(\boldsymbol U; \boldsymbol \theta,
		 \boldsymbol \eta_\lambda)\big| W\right)=0\quad \text{if and only if}\quad Q(\boldsymbol \theta) = 
		 E[g(\boldsymbol U_1; \boldsymbol \theta, 
		 \boldsymbol \eta_{\lambda,1} )g(\boldsymbol U_2; \boldsymbol \theta, 
		 \boldsymbol \eta_{\lambda,2})
		 \omega ( \boldsymbol W_1 - \boldsymbol W_2) ] = 0.
	\end{equation*}
Here $\boldsymbol \eta_{\lambda,1} $ and $\boldsymbol \eta_{\lambda,2} $  are the  vector of nuisance functions corresponding to $\boldsymbol Z_1$ and $\boldsymbol Z_2$. 
Indeed, in Lemma  \ref{lem_ident} of Section \ref{identification}, it will be shown that
	\begin{align}	\label{MGA}
		Q(\boldsymbol \theta) = 
		  \begin{cases}
		 \hskip 0.35cm 0 &, \text{ if }  \boldsymbol\theta = \left(\lambda, \boldsymbol{\beta}^T\right)^T=
 	   	\left(\lambda_0, \boldsymbol{\beta_0}^T\right)^T  , \\
		>0 &, \text{ else}.
		\end{cases}
	\end{align}
\citet{lavergne2013smooth} list several possible ways to define $\omega(\cdot)$, but throughout this paper we will rely on the simple choice $\omega (\boldsymbol W) := \exp\left\{-\boldsymbol W^T \boldsymbol D\boldsymbol W\right\}$, where $\boldsymbol D$ is a diagonal matrix whose positive diagonal elements $d_1,\dots,d_{p+q}$ represent user selected tuning parameters. A sensible choice  consists of using the 
 standard deviations of the components of the vectors $(\boldsymbol X_i^T, \boldsymbol Z_i^T)^T$.

 If the functions which define $\boldsymbol \eta_\lambda$ were known, then  a data-based estimator
 of the unconditional moment $Q(\boldsymbol \theta)$ could be obtained by the sample averages
	\begin{equation} \label{MGA1}
	  	Q(\boldsymbol \theta) =  \frac{1}{n^2} \sum_{1\leq i,j \leq n}g(\boldsymbol U_i; \boldsymbol \theta, \boldsymbol \eta_{\lambda,i} )
	  	g(\boldsymbol U_j; \boldsymbol \theta, \boldsymbol \eta_{\lambda,j} ) \omega ( \boldsymbol W_i - \boldsymbol W_j),
	\end{equation}
where $\boldsymbol \eta_{\lambda,i} $ and $\boldsymbol \eta_{\lambda,j} $  are the nuisance parameter values corresponding to $\boldsymbol Z_i$ and $\boldsymbol Z_j$, respectively. 
The average scheme proposed by \citet{lavergne2013smooth} relies on leaving out diagonal elements with $i=j$. In our setup, inclusion of these diagonal terms provides more stable and reliable estimators. We will show that the resulting bias is asymptotically negligible. Note that by definition of $\omega(\cdot)$
	$$
		E\left[ Q_n\left(\boldsymbol \theta\right)\right]
		=
		Q(\boldsymbol \theta ) + \frac{1}{n} E\left( g(\boldsymbol U; \boldsymbol \theta,\boldsymbol \eta_\lambda)^2\right)
		=
		Q(\boldsymbol \theta)  + O(n^{-1}).
	$$

Under model \eqref{main_model}, the nuisance   parameter $\boldsymbol \eta_\lambda$ is unknown, and $Q_n(\cdot)$ cannot be directly computed. We therefore use kernel estimation to determine nonparametric estimators $\widehat{\boldsymbol \eta}_\lambda$. This then leads to a feasible version $\widehat{Q}_n(\cdot)$. More precisely, our estimation procedure can be described as follows. For each $\lambda$, we define the map
	$$
	  \boldsymbol \beta \mapsto \widehat Q_n   \left(\left( \lambda, \boldsymbol{\beta}^T\right)^T 
	  \right) 
	  = \frac{1}{n^2} \sum_{1\leq i,j \leq n}g(\boldsymbol U_i; \boldsymbol \theta, 
	  \widehat{\boldsymbol \eta}_{\lambda,i})g(\boldsymbol U_j; \boldsymbol \theta,
	  \widehat{\boldsymbol \eta}_{\lambda,j}) \omega ( \boldsymbol W_i - \boldsymbol W_j),
	$$
which is quadratic with an explicit unique minimum $\widehat {\boldsymbol \beta}(\lambda)$. Thus, we define a profile SmoothMD estimator of $\lambda_0$ as
	$$
	 \widehat \lambda = \arg\min_{\lambda } \widehat Q_n\left(\left(\lambda,\widehat {\boldsymbol \beta}(\lambda)^T\right)^T \right), 
	$$
and, with at hand the estimate $\widehat{\lambda}$, we eventually calculate  $\widehat{\boldsymbol \beta} (\widehat \lambda)$, the semiparametric SmoothMD estimate of $\boldsymbol \beta_0$. In a final step, the function $m(\cdot)$ in model \eqref{main_model} can be estimated from the residuals $\hat \epsilon_i=T(Y_i,\widehat{\lambda})-\boldsymbol X_i^T\widehat{\boldsymbol \beta}(\widehat{\lambda})$ by using any established smoothing procedure.

Details of the method are described in Section \ref{prelim}. Under mild regularity conditions it is then shown that our procedure leads to consistent estimators. Using standard kernel estimators based on second order kernels and suitable bandwidth sequences for nonparametric function estimation, we then establish $\sqrt{n}$-consistency and asymptotic normality, provided that the dimension of $\boldsymbol Z$ is  $q<4$. Corresponding test procedures are described in Section 4. All theoretical results are derived uniformly for all possible choices in a compact set for the   $d_1,\dots,d_{p+q}$  used in \eqref{MGA}. This provides theoretical grounds for a sample-based choice of $d_1,\dots,d_{p+q}$, such as the sample standard deviations of the components of the vectors $(\boldsymbol X_i^T, \boldsymbol Z_i^T)^T$.

We wish to note that the generality of our approach implies that the method may be used as a powerful tool to check parametric models. For example, in addition to verifying a log-transformation to the response variable, one may check linearity assumptions. The latter may be done by comparing the outputs of the parametric model with the results of a semiparametric analysis, where some of the regressors enter the model nonparametrically and define a corresponding vector $Z$. This is exemplified by our real data application in Section 4.

The remainder of the paper is organized as follows. In Section \ref{prelim}, we present our new estimation method and establish identification of the model parameters. Theoretical properties of the estimators are derived in Section  \ref{con_asy_norm}, while  in Section  \ref{Test}, we investigate a distance-metric procedure for testing restrictions on parameters. In Section \ref{small_sample_study}, we study the finite sample behavior by a simulation study and apply the estimator to a real data sample. Our estimator performs well in our experiments and our tests yield accurate levels and good power in moderate samples. Finally, in Section \ref{discussion} we formulate few conclusions and discuss the extension of our approach to higher-dimension vectors $\boldsymbol Z$ using higher-order kernels, as well as efficiency aspects. The proofs are left to the Appendix.

\section{The semiparametric SmoothMD approach} \label{prelim}

In this section we formally define our semiparametric estimator. First, we investigate two issues. On the one hand, we prove identification of the true value  $\boldsymbol \theta_0=(\lambda_0,\boldsymbol \beta_0^T)^T$ of the parameter of interest. Next, we discuss the recommendation appearing in the literature for normalizing the response variable. This issue is specific to the Box-Cox transformation, though similar problems occur with other families of transformations. Finally, we define our semiparametric SmoothMD estimator.

Before proceeding with this plan, let us slightly modify the definition of the conditional moment equation. The function  $g(\boldsymbol U; \boldsymbol \theta, \boldsymbol \eta_\lambda)$ defined in \eqref{def1} has zero-mean if $\boldsymbol \theta=\boldsymbol \theta_0$, but there is no reason to expect its sample version to be centered, as is the case when $\lambda_0$ is given and one uses least squares; see  \citet{li1996root}. We therefore propose to introduce an intercept and hereafter replace $g(\boldsymbol U; \boldsymbol \theta, \boldsymbol \eta_\lambda)$ with 
	\begin{equation}\label{def1_b}
		g(\boldsymbol U; \boldsymbol \theta, 
		\gamma, 
		\boldsymbol \eta_\lambda) = \left(T(Y,\lambda) - E[T(Y,\lambda)\mid \boldsymbol Z]- (\boldsymbol X-E[\boldsymbol X\mid \boldsymbol Z])^T\boldsymbol \beta\right) f_z(\boldsymbol Z) 
		- \gamma.
	\end{equation}
The true value of $\gamma$ is known to be $\gamma_0=0$, but this intercept slightly improves the results with finite samples, while it does not introduce any additional theoretical or computational complexity.

We use the following notation throughout the remainder of the paper. For $d_l, d_c\geq 1,$ let $\mathbb{R}^{d_l\times d_c}$ denote the set of  $d_l\times d_c-$ matrices with real elements. Let $\boldsymbol{1}_{d_l}$ (resp. $\boldsymbol{0}_{d_l}$) denote the vector with all components equal to 1 (resp. 0), $\boldsymbol{0}_{d_l \times d_c}$ the $d_l\times d_c-$null matrix and $\boldsymbol{I}_{d_l \times d_c}$ the identity matrix with dimension $d_l \times d_c$. For a matrix $\boldsymbol A$,  $\lVert \boldsymbol A\rVert$ is the Frobenius norm and $\lVert \boldsymbol A\rVert_{\rm{Sp}}$ the spectral norm. Below, $\boldsymbol D={\rm diag}(\boldsymbol d)$ is some positive definite diagonal matrix with $\boldsymbol d\in \mathcal{D} \subset \mathbb{R}^{p+q}_+$  being a diagonal vector with strictly positive components. Herein, $\mathcal{D}$ is a compact set and our asymptotic results are derived uniformly with respect to $\boldsymbol d\in \mathcal{D}$.

\subsection{Identification}  \label{identification}

Let $-\infty <\lambda_{\rm min}< \lambda_0 < \lambda_{\rm max} <\infty$, with $\lambda_{\min}< 0$ and $\lambda_{\max}> 0$. For any $\lambda\in 
[\lambda_{\rm min},\lambda_{\rm max}]$, let 
	\begin{equation}\label{beta_l}
		(\gamma(\lambda), \boldsymbol \beta (\lambda)^T)^T = \arg\min_{\gamma\in\mathbb{R},\boldsymbol\beta\in \mathbb{R} ^p} E[g(\boldsymbol U_1; \boldsymbol \theta,\gamma, \boldsymbol \eta_{\lambda,1})g(\boldsymbol U_2; \boldsymbol \theta,\gamma, \boldsymbol \eta_{\lambda,2})\omega ( \boldsymbol W_1 - \boldsymbol W_2) ], 
	\end{equation}
with $ g(\boldsymbol U_1; \boldsymbol \theta,\gamma, \boldsymbol \eta_{\lambda,1}) $ and $ g(\boldsymbol U_2; \boldsymbol \theta, \gamma , \boldsymbol \eta_{\lambda,2})$ being independent copies of $g(\boldsymbol U; \boldsymbol \theta,\gamma, \boldsymbol \eta_{\lambda}) $ defined in equation \eqref{def1_b} with $\boldsymbol U=\left(Y,\boldsymbol W^T \right)^T$, $\boldsymbol W=\left(\boldsymbol X^T,\boldsymbol Z^T \right)^T $,  $\boldsymbol \theta=(\lambda,\boldsymbol \beta^T)^T$ and $\boldsymbol \eta_{\lambda,k} =\boldsymbol \eta_{\lambda} (\boldsymbol Z_k)$, $k=1,2,$ where
	\begin{equation*}\label{eq_eta}
		\boldsymbol \eta_{\lambda} (\boldsymbol z)= (f_z(\boldsymbol z), E[T(Y,\lambda)\mid \boldsymbol Z= \boldsymbol z\;], E[\boldsymbol X\mid \boldsymbol Z=\boldsymbol z\;]^T)^T.
	\end{equation*}

\begin{assumption}\emph{Data Generating Process}
	\begin{enumerate} 
		\item The observations $\left(Y_i, \boldsymbol X_i^T, \boldsymbol Z_i^T \right)^T$ , $1 \leq i \leq n$, are i.i.d. copies of $\left(Y, \boldsymbol X^T,\boldsymbol Z^T \right)^T \in \mathbb{R} \times \mathbb{R}^p \times \mathbb{R}^q$. Moreover, there exists a constant $c>0$ such that $\mathbb{P}(Y > c)=1$.

	   \item The covariate vector $\boldsymbol Z$ admits a bounded density in $\mathbb{R}^q$. The covariate vector $\boldsymbol X$ is split into two subvectors $\boldsymbol X_c \in \mathbb{R}^{p_c}$ and $\boldsymbol X_d \in \mathbb{R}^{p_d} $ with $0\leq p_c, p_d\leq p$ and $p_c+p_d=p$. The subvector  $\boldsymbol X_c$   admits a bounded density in $\mathbb{R}^{p_c}$. The subvector  $\boldsymbol X_d$ takes values in a finite set. 
	   
  	   \item The  diagonal  of the matrix $\boldsymbol D$ belongs to the $(p+q)-$dimension cube $\mathcal{D}=  [d_L,d_U]^{p+q}$, with some fixed  $0<d_L<d_U<\infty$. 
	\end{enumerate}
	\label{ass_dgp}
\end{assumption}

The assumption that the discrete components of $\boldsymbol X$ take values in a finite set, is a technical condition that simplifies the proofs without significant restriction of the generality of the applications.

\begin{assumption}\emph{Identification}
	\begin{enumerate}
		
		\item $E\left[\| \boldsymbol X \|^2\right]<\infty$, $E\left[\| \boldsymbol Z \|^2\right]<\infty$, and $Var\left[\boldsymbol X - E[\boldsymbol X|\boldsymbol Z]\right]$ has full rank.

		\item The true value $\boldsymbol\beta_{0,c}\in \mathbb{R}^{p_c}$ of the subvector of coefficients corresponding to $\boldsymbol X_c$ is not equal to $\boldsymbol 0_{p_c}$. 
		
		\item The continuous random subvector  $\boldsymbol X_c$ is such that, for any $ \boldsymbol  b \in \mathbb{R}^{p_c}$, $\boldsymbol  b \neq \boldsymbol 0_{p_c}$, the variable 
$\boldsymbol X^T_c \boldsymbol b $ is  continuous with the support equal to the whole real line.
		
		\item Whenever $\lambda\neq \lambda_0$, for any $\boldsymbol z$ in the support of $\boldsymbol Z$  and $\boldsymbol x_d$ in the support of the discrete subvector $\boldsymbol X_d$, the set of values of the map 
		$ 
				\boldsymbol  x_c \mapsto E\left[ T(Y,\lambda) - T(Y,\lambda_0)\mid \boldsymbol X_c=\boldsymbol x_c, \boldsymbol X_d=\boldsymbol x_d,
				\boldsymbol Z=\boldsymbol z  
				\right] $,  
				 $\boldsymbol x_c \in \mathbb{R}^{p_c},
		$ 
		is unbounded. 
		 
		 \item $E\left[Y^{2C_\lambda}\right]<\infty$, where $C_\lambda=\max(|\lambda_{min}|,\lambda_{max})<\infty$.
		\end{enumerate}
	\label{ass_ident}
\end{assumption}

Note that $Var\left[(\boldsymbol{X}^T, \boldsymbol{Z}^T)^T\right]$ necessarily has full rank, by Assumption \ref{ass_ident}.1 and the fact that $\boldsymbol Z$ admits a density. 
The complete justification of this statement is given in the Appendix. 

With all this in hand, we can now state the following identification result. 

\begin{lem} Suppose  that Assumptions \ref{ass_dgp} and \ref{ass_ident} hold true.
		 Let
		$\gamma(\lambda)$ and $\boldsymbol \beta (\lambda)$ be defined as in 
		 \eqref{beta_l}. Then, $\gamma(\lambda_0)=0$ and $\boldsymbol {\beta} (\lambda_0) = \boldsymbol\beta_0$ and 
			\begin{align*}
				\mathbb{P} \left(E\left[  (T(Y,\lambda) - E[T(Y,\lambda)\mid \boldsymbol Z] ) f_z(\boldsymbol Z)   - \gamma - 
			                          (\boldsymbol X - E[\boldsymbol X\mid\boldsymbol Z])^T\boldsymbol\beta  f_z(\boldsymbol Z)  \mid \boldsymbol X,\boldsymbol Z  \right] = 0 \right) < 1,
			\end{align*}
		for all  $\gamma\in \mathbb{R}$ and $\boldsymbol \theta = (\lambda,\boldsymbol \beta^T)^T \in [\lambda_{\rm min},\lambda_{\rm max}]\times \mathbb{R}^p$ such that $(\gamma,\boldsymbol \theta^T)^T\neq (0,\boldsymbol {\theta}_0^T)^T$. Moreover, for any $\varepsilon >0$, 
			\begin{multline}
				\inf_{  
				 |\lambda-\lambda_0| \geq \varepsilon} \;  \inf_{\boldsymbol d \in\mathcal{D} } E\left[g\left(\boldsymbol U_1; (\lambda, \boldsymbol \beta (\lambda)^T)^T , \gamma (\lambda), \boldsymbol \eta_{\lambda,1}\right)g\left(\boldsymbol U_2; (\lambda, \boldsymbol \beta (\lambda)^T)^T , \gamma (\lambda), \boldsymbol \eta_{\lambda,2}\right)\right. \\ \times \left. \exp \left\{- (\boldsymbol W_1 - \boldsymbol W_2)^T \boldsymbol D (\boldsymbol W_1 - \boldsymbol W_2)\right\}  \right]  > 0.\label{well_sep_0}
			\end{multline}
		
		\label{lem_ident}
\end{lem}

\subsection{Box-Cox transformation and standardized responses} \label{stand_Y}

Let us note that 
$$
  \lim_{\lambda \uparrow \infty }\frac{y^\lambda -1}{\lambda} = 0 \quad \text{if $0< y<1$} \qquad and \qquad  \lim_{\lambda \downarrow -\infty }\frac{y^\lambda -1}{\lambda} = 0 \quad \text{if $y>1$}.
$$ 
In classical estimation approaches for parametric regression models with Box-Cox transformed response, this is likely to induce instability for the estimation of the parameter $\lambda$. See, e.g., \citet{khazzoom1989note}, \citet{powell1996rescaled} and \citet{showalter1994monte} for a discussion of this well-known issue. In order to avoid such problems, the common recommendation is to standardize the response by some constant, say $s$, such that 
$$
  \mathbb{P}\left( Y/s < 1  \right) >0 \qquad \text{and} \qquad  \mathbb{P}\left( Y/s > 1  \right) >0.
$$
The constant $s$ could be, for instance, the mean of $Y$ or the geometric mean of $Y$.\footnote{The geometric mean is defined as $G(Y)=\exp\{E(\log(Y)\}$ and the sample counterpart is $G_n = \prod\limits_{i=1}^{n} Y_i^{1/n}$.}
With finite samples, the practitioner would first estimate such a constant using the sample, and next would normalize the responses. The same type of problems might occur in our semiparametric extension of the Box-Cox transformation model.  For this reason, we will replace our function $g(\boldsymbol U; \boldsymbol \theta,\gamma,\boldsymbol \eta_\lambda) $ by a family of functions $s^{-\lambda} g(\boldsymbol U; \boldsymbol \theta,\gamma, \boldsymbol \eta_\lambda)$ also indexed by $s$ that we shall let depend on the sample. This change of the family of functions is equivalent to changing $Y$ to $Y/s$ in the definition \eqref{def1_b}, and a rescaling of the parameters $\boldsymbol \beta$ and  $\gamma$. By the profiling-based construction of our SmoothMD estimator, the replacement of the response $Y$ by $Y/s$ matters only for computing $\widehat \lambda$. Clearly, the identifiability property established in Lemma \ref{lem_ident} is preserved. In the remainder of the paper, we provide asymptotic results that are uniform with respect to $s$ in order to allow for a data-driven choice of $s$, such as for instance, the sample geometric mean of the response.

\subsection{The estimator} \label{estimator}

Given an independent sample $\left(Y_1,\boldsymbol X^T_1,\boldsymbol Z^T_1 \right)^T,\ldots, \left(Y_n,\boldsymbol X^T_n,\boldsymbol Z^T_n \right)^T$ from $\left(Y,\boldsymbol X^T,\boldsymbol Z^T \right)^T \in   \mathbb{R} \times \mathbb{R}^{p+q}$, let us define 
$$
  \widehat{\mathbb{Y}}_n(\lambda) = \left( (T(Y_1,\lambda) - \widehat{E}[T(Y_1,\lambda)\mid \boldsymbol Z_1] ) \widehat{ f} _z(\boldsymbol Z_1),\ldots, (T(Y_n,\lambda) - \widehat{E} [T(Y_n,\lambda)\mid \boldsymbol Z_n]) \widehat{f}_z(\boldsymbol Z_n)  \right)^T\in\mathbb{R}^n,
$$ 
  and 
$$
  \widehat{\mathbb{X}}_n  = \left( (\boldsymbol X_1-\widehat{E}[\boldsymbol X_1\mid \boldsymbol Z_1]) \widehat{f}_z(\boldsymbol Z_1),\ldots, (\boldsymbol X_n-\widehat{E}[\boldsymbol X_n\mid \boldsymbol Z_n]) \widehat{f}_z(\boldsymbol Z_n)\right)^T\in \mathbb{R}^{n\times p}.
$$ 
For  $1\leq i \leq n$, $\widehat{\boldsymbol \eta}_{\lambda,i} = (\widehat{f}_z(\boldsymbol Z_i) , \widehat{E}[T(Y_i,\lambda)\mid \boldsymbol Z_i], \widehat{E}[\boldsymbol X_i\mid \boldsymbol Z_i]^T)^T$ are nonparametric kernel estimates of $ \boldsymbol \eta_{\lambda,i}  =(f_z(\boldsymbol Z_i), E[T(Y_i,\lambda)\mid \boldsymbol Z_i], E[\boldsymbol X_i\mid\boldsymbol Z_i]^T)^T$. More precisely, 
$$
  \widehat{f}_z(\boldsymbol Z_i) = \frac{1}{nh^q}\sum_{j=1}^n K\left(  \frac{\boldsymbol Z_i-\boldsymbol Z_j}{h} \right),\quad 
\widehat{E}[T(Y_i,\lambda)\mid \boldsymbol Z_i]\widehat{f}_z(\boldsymbol Z_i)= \frac{1}{nh^q}\sum_{j=1}^nT(Y_j,\lambda) K\left(  \frac{\boldsymbol Z_i-\boldsymbol Z_j}{h} \right),
$$
  and 
$$
  \widehat{E}[\boldsymbol X_i\mid \boldsymbol Z_i]\widehat{f}_z(\boldsymbol Z_i)= \frac{1}{nh^q}\sum_{j=1}^n\boldsymbol X_j K\left(  \frac{\boldsymbol Z_i-\boldsymbol Z_j}{h} \right).
$$
  Here $K(\cdot)$ is a multivariate kernel function and $h$ is the bandwidth. Let $\boldsymbol \Omega_n$ be the $n\times n-$ symmetric matrix with elements
$$
  \boldsymbol{\Omega}_{n,ij} = \exp \{ - (\boldsymbol X_i^T-\boldsymbol X_j^T, \boldsymbol Z_i^T-\boldsymbol Z_j^T)\boldsymbol D (\boldsymbol X_i-\boldsymbol X_j, \boldsymbol Z_i-\boldsymbol Z_j)  \}, \qquad 1\leq i, j \leq n.
$$
Typically, the components of the vector $ \boldsymbol d$ defining the diagonal matrix $\boldsymbol D$, could be taken as proportional to the standard deviation of the components of the vectors $(\boldsymbol X_i^T, \boldsymbol Z_i^T)^T$. The definition of $\boldsymbol  \Omega_{n,i,j}$ allows also to take into account, discrete components of $\boldsymbol X$. For finite support discrete covariates, one could set some large value for the corresponding  diagonal element of $\boldsymbol  D$,  which in practice would be equivalent to an indicator of the event that the observations $i$ and $j$ have the same value for that covariate. 

We can now define, for any $\lambda$, the estimates of $(\gamma(\lambda), \boldsymbol \beta (\lambda)^T)^T\in \mathbb{R}^{1+p}$ introduced in
 \eqref{beta_l}. For any $s>0$, let 
$$
  \widehat Q_n\left(\left(\lambda, {\boldsymbol \beta}^T\right)^T, {\gamma}  ;s \right) = n^{-2}  s^{-2\lambda}\left(   \widehat{\mathbb{Y}}_n(\lambda) - \gamma\boldsymbol{1}_n - \widehat{\mathbb{X}}_n   \boldsymbol \beta \right)^T\boldsymbol  \Omega_n  \left( \widehat{\mathbb{Y}}_n(\lambda) - \gamma\boldsymbol{1}_n - \widehat{\mathbb{X}}_n \boldsymbol \beta\right).
$$
  For  fixed $s$ and $\lambda$, consider the generalized least-squares problem
	\begin{equation}\label{gls}
		\min_{\gamma,\boldsymbol \beta}  \widehat Q_n\left(\left(\lambda, {\boldsymbol \beta}^T\right)^T, {\gamma}  ;s \right). 
	\end{equation}
The solution of this problem does not depend on $ s^{-\lambda}$ and has the form of standard generalized least-squares estimators:
	\begin{equation*}\label{gamma_l_hat}
		\widehat \gamma(\lambda,{\boldsymbol \beta} (\lambda)) 
		= \frac{1}{\boldsymbol{1}_n^T \boldsymbol\Omega_n  \boldsymbol{1}_n} \boldsymbol{1}_n^T \boldsymbol\Omega_n \left(\widehat{\mathbb{Y}}_n(\lambda)  - \widehat{\mathbb{X}}_n {\boldsymbol \beta} (\lambda)  \right),
	\end{equation*}
and
	\begin{equation*}\label{beta_l_hat}
		\widehat{\boldsymbol \beta}  (\lambda) = \left(\widehat{\mathbb{X}}_n ^T{\mathbb{D}} _n\widehat{\mathbb{X}}_n   \right)^{-1} \widehat{\mathbb{X}}_n ^T {\mathbb{D}} _n \widehat{\mathbb{Y}}_n(\lambda), 
	\end{equation*}
with 
	\begin{equation} \label{d_n}
		{\mathbb{D}} _n = \boldsymbol\Omega_n   - \frac{1}{\boldsymbol{1}_n^T \boldsymbol\Omega_n  \boldsymbol{1}_n} \boldsymbol\Omega_n  \boldsymbol{1}_n  \boldsymbol{1}_n^T  \boldsymbol\Omega_n  \in\mathbb{R}^{n\times n}.
	\end{equation}

Next, plugging $(\widehat{\gamma}(\lambda,\widehat{\boldsymbol \beta} (\lambda)), \widehat{\boldsymbol \beta} (\lambda)^T)^T$ into the problem \eqref{gls}, for a given $s$, we define the SmoothMD estimator of $\lambda_0$ as
	\begin{equation}\label{lambda_l_hat}
		\widehat \lambda = \widehat \lambda (s) = \arg\min_{\lambda\in\Lambda }  s^{-\lambda} \widehat{\mathbb{Y}}_n(\lambda)^T \;  \widehat {\mathbb{B}} _n \; s^{-\lambda} \widehat{\mathbb{Y}}_n(\lambda), 
	\end{equation}
with
$$
  \widehat {\mathbb{B}}_n   = {\mathbb{D}} _n - {\mathbb{D}} _n \widehat{\mathbb{X}}_n   \left(\widehat{\mathbb{X}}_n ^T{\mathbb{D}} _n \widehat{\mathbb{X}}_n   \right)^{-1} \widehat{\mathbb{X}}_n ^T {\mathbb{D}} _n \in\mathbb{R}^{n\times n}.
$$
Note that, by construction, 
	\begin{equation*} 
		{\mathbb{D}} _n \boldsymbol{1}_n = \widehat {\mathbb{B}} _n \boldsymbol{1}_n = \boldsymbol 0_n  \quad \text{ and } \quad 
		\widehat {\mathbb{B}}_n \widehat{\mathbb{X}}_n = \boldsymbol {0}_{n\times p}.
	\end{equation*}
Finally, the SmoothMD estimator of $\boldsymbol \beta_0$ is $\widehat{\boldsymbol \beta}  (\widehat \lambda)$. We close this section by showing that our estimator is well-defined.

\begin{lem}\label{omega_mat}
If Assumptions \ref{ass_dgp}.3 and \ref{ass_ident} hold true, then, for each $n\geq 1$, 
	\begin{enumerate}
		\item the matrices $\boldsymbol  \Omega_n$ and $\widehat{\mathbb{X}}_n ^T\mathbb D_n \widehat{\mathbb{X}}_n $ are positive definite with probability 1. In particular, $\boldsymbol{1}_n^T \boldsymbol  \Omega_n  \boldsymbol{1}_n>0$ and $\widehat{\mathbb{X}}_n ^T\mathbb D_n \widehat{\mathbb{X}}_n $ is invertible with probability 1.
	
		\item the matrix $\widehat {\mathbb{B}} _n $ is positive semi-definite with probability 1.
	\end{enumerate}
\end{lem}

\begin{rmk}
The matrix ${\mathbb{D}}_n$ is defined in equation \eqref{d_n} and has dimension $n\times n$. Therefore, one might imagine that it becomes difficult to work with this matrix when the sample size is large. This is not the case because it is not necessary to estimate the matrix ${\mathbb{D}}_n$ itself. It suffices to compute $\widehat{\mathbb{X}}_n ^T{\mathbb{D}} _n$ and $\widehat{\mathbb{Y}}_n(\lambda)^T{\mathbb{D}} _n$ to be able to calculate $\widehat{\boldsymbol \beta}  (\lambda)$ and $\widehat \lambda$. In chapter \ref{small_sample_study} we shall show that the estimator can be easily applied even for $n > 100,000$.
\end{rmk}

\section{Consistency and asymptotic normality} \label{con_asy_norm}

The estimator $\widehat \lambda(s)$ depends on $s$, a value that  in practice could be calculated from the sample, such as for instance, the sample geometric mean. For this reason, our asymptotic results are stated uniformly with respect to $s$. 
Our asymptotic results are also stated uniformly with respect to the diagonal of the matrix $\boldsymbol D$. This ensures that we can use a data driven estimate of $\boldsymbol D$ proportional to the empirical standard deviation of $\boldsymbol X$ or $\boldsymbol Z$.   

Let's introduce some more notation: for each $\lambda \in\Lambda $, let  
$$
	{\mathbb{Y}}_n(\lambda) = \left( (T(Y_1,\lambda) -  {E}[T(Y_1,\lambda)\mid \boldsymbol Z_1] )  { f} _z(\boldsymbol Z_1),\ldots, (T(Y_n,\lambda) -  {E} [T(Y_n,\lambda)\mid \boldsymbol Z_n])  {f}_z(\boldsymbol Z_n)  \right)^T\in\mathbb{R}^n,
$$ 
and 
$$
	{\mathbb{X}}_n  = \left( (\boldsymbol X_1- {E}[\boldsymbol X_1\mid \boldsymbol Z_1]) {f}_z(\boldsymbol Z_1),\ldots, (\boldsymbol X_n- {E}[\boldsymbol X_n\mid \boldsymbol Z_n])  {f}_z(\boldsymbol Z_n)\right)^T\in \mathbb{R}^{n\times p}.
$$ 
Moreover, 
$$
	{\mathbb{B}} _n =  {\mathbb{D}} _n - {\mathbb{D}} _n  {\mathbb{X}}_n   \left( {\mathbb{X}}_n ^T{\mathbb{D}} _n  {\mathbb{X}}_n   \right)^{-1} {\mathbb{X}}_n ^T {\mathbb{D}} _n \in\mathbb{R}^{n\times n},
$$
with $   {\mathbb{D}}_n  $ defined in equation \eqref{d_n}. Again, by construction ${\mathbb{B}} _n \boldsymbol{1}_n  = \boldsymbol 0_n$ and 
${\mathbb{B}}_n  {\mathbb{X}}_n = \boldsymbol {0}_{n\times p}$.

\begin{assumption}\label{ass_con} \emph{Consistency}
	\begin{enumerate}
		\item The functions $f_z(\cdot)$, $(mf_z)(\cdot)$, $E[\| \boldsymbol X \|^2\mid \boldsymbol Z=\cdot\;]f_z(\cdot)$ and $\sup_{\lambda \in\Lambda}(\partial^2 /\partial \lambda^2) E[T(Y, \lambda)\mid \boldsymbol Z=\cdot\;]f_z(\cdot)$ have H\"older continuous partial derivatives of order four.
		
		\item The kernel $K(\cdot)$ is the product of $q$ univariate kernel functions $\widetilde K$ of bounded variation. Moreover, $\widetilde K$ is a symmetric function with integral equal to one and $\int t^2 \widetilde K(t) dt <\infty$. 
 
		\item The bandwidth $h$ belongs to a range $\mathcal{H}_{c,n}=[c_{min}n^{-\alpha}, c_{max}n^{-\alpha}],$ with  $0 < \alpha < 1/q$ and $c_{min}$, $c_{max}$ positive constants.
		
	\end{enumerate}
\end{assumption}

With all this in hand, we can now state the consistency of our estimator.

\begin{thm}[Consistency]\label{consist} 
Assume that Assumptions \ref{ass_dgp}, \ref{ass_ident} and \ref{ass_con} hold true. Let $s_0$ be some normalizing value such that 
$\mathbb{P}\left( Y/s_0 < 1  \right) >0$ and  $\mathbb{P}\left( Y/s_0 > 1  \right) >0$ and let $S_n$ be an arbitrary $o_{\mathbb{P}}(1)$ neighborhood of $s_0$. Then 
$$
	\sup_{h\in\mathcal{H}_{c,n}} \sup_{s\in S_n} \sup_{\boldsymbol d \in \mathcal{D} }\left|\widehat \lambda - \lambda_0 \right|= o_{\mathbb{P}}(1) \quad \text{ and } \quad \sup_{h\in\mathcal{H}_{c,n}} \sup_{s\in S_n} \sup_{\boldsymbol d \in \mathcal{D}} \left\| \widehat{\boldsymbol{\beta}} (\widehat \lambda) - \boldsymbol{\beta}_0 \right\| = o_{\mathbb{P}}(1).
$$ 
\end{thm}

In Theorem \ref{consist} we require that $h\in\mathcal{H}_{c,n}$.
This implies that $nh^q \rightarrow \infty$ and $h \rightarrow 0$ for $n \rightarrow \infty$ which is in line with \citet{robinson1988root} and \citet{li1996root}.

Next, we prove asymptotic normality for our estimator. For this purpose, we first derive the asymptotic  linear representation of $\widehat \lambda$ and and $\widehat {\boldsymbol \beta}(\widehat \lambda )$ from which the $\sqrt{n}-$asymptotic normality follows. In the following result, we show that $\widehat \lambda$ and $\widehat {\boldsymbol \beta}(\widehat \lambda )$ are asymptotically not equivalent to the infeasible  estimators of $\lambda_0$ and $\boldsymbol \beta_0$, one would  obtain when the infinite-dimensional parameter $\boldsymbol \eta_\lambda$ is given and the intercept $\gamma$ is equal to 0. This is in contrast to the results of \citet{li1996root} and \citet{robinson1988root}. The reason is that they can use the fact that $E[{\mathbb{X}}_{n,i}\mid \boldsymbol{Z}_i] = 0$ when controlling higher order terms. In our case, we weight the observations by $\boldsymbol{\Omega}_{n,ij}$ and, in general,  $E[{\mathbb{X}}_{n,i}\boldsymbol{\Omega}_{n,ij}\mid \boldsymbol{Z}_i] \neq 0$ for $i \neq j$. This is also the reason why we need to ask for $q < 4$ instead of $q < 6$ as in \citet{li1996root}. Therefore, we require that $h\in\mathcal{H}_{sc,n}$, where $\mathcal{H}_{sc,n}=[c_{min}n^{-\alpha}, c_{max}n^{-\alpha}]$, with  $\alpha \in (1/4,1/q)$. This is the small price we have to pay for getting what, to the best of our knowledge, is the first consistent estimation procedure for the semiparametric  transformation model we investigate. However, as discussed in section \ref{discussion}, one could still apply our approach with $q\geq 4$, provided one uses higher-order kernels.

\begin{assumption} 	\label{ass_asy_norm} \emph{Asymptotic Normality}
	\begin{enumerate}	
		\item $E\left[m(\boldsymbol{Z})^4\right] < \infty$ and $Var\left[\frac{\partial }{\partial \lambda} {T(Y,\lambda_0)}\right] > 0$.

		\item $E\left[\varepsilon^4\right] < \infty$ and $E\left[\varepsilon^2\mid\boldsymbol{X},\boldsymbol{Z} \right] = \sigma^2(\boldsymbol{X},\boldsymbol{Z})$ is in $L^1 \cap L^2$.		
		
		\item The bandwidth $h$ belongs to  $\mathcal{H}_{sc,n}=[c_{min}n^{-\alpha}, c_{max}n^{-\alpha}],$ with  $\alpha \in (1/4, 1/q)$ and $c_{min}$, $c_{max}>0$.
			\end{enumerate}

\end{assumption}

The results are again obtained uniformly with respect to the elements on the diagonal of the matrix $\boldsymbol D$ that determines $\boldsymbol{\Omega}_n$ and with respect to the scaling factor $s$ that could be used for numerical stability, as mentioned in Section \ref{stand_Y}. In addition, let $K_h (\cdot) = h^{-q}K(\cdot/h)$ and, for any $1\leq i, j \leq n$, let
	$$
		K_{h,ij}=K_h (\boldsymbol Z_i-\boldsymbol Z_j). 
	$$

\begin{prop}[Asymptotic representation]\label{AN_prop} 
Assume that the conditions of Theorem \ref{consist} and Assumption \ref{ass_asy_norm} hold true. Then, uniformly with respect to $h\in\mathcal{H}_{sc,n}$, $\boldsymbol d\in \mathcal{D}$ and $s\in S_n$,
	\begin{equation*}\label{lin_lambda}
		\widehat \lambda - \lambda_0  = - \left[  \frac{\partial }{\partial \lambda} {\mathbb{Y}}_n(\lambda_0)^T \;  {\mathbb{B}} _n \; \frac{\partial }{\partial \lambda}  {\mathbb{Y}}_n(\lambda_0)  \right]^{-1}    \;  \frac{\partial }{\partial \lambda} {\mathbb{Y}}_n(  \lambda_0)^T \;{\mathbb{B}} _n \left[(\boldsymbol{\varepsilon f_z})_n - \left(\boldsymbol{\widehat{\varepsilon}_{|z}\widehat{f}_z}\right)_n \right]+ o_{\mathbb{P}}(n^{-1/2}) = O_{\mathbb{P}}(n^{-1/2}),
	\end{equation*}
and
	\begin{equation*}\label{beta_l_hat_0}
		\widehat {\boldsymbol \beta}(\widehat \lambda ) - \boldsymbol {\beta}_0 = 
		\left( {\mathbb{X}}_n ^T {\mathbb{D}}_n  {\mathbb{X}}_n   \right)^{-1}  {\mathbb{X}}_n ^T {\mathbb{D}}_n  \left[ (\boldsymbol{\varepsilon f_z})_n  -\left(\boldsymbol{\widehat{\varepsilon}_{|z}\widehat{f}_z}\right)_n
		+  \frac{\partial }{\partial \lambda}  {\mathbb{Y}}_n(\lambda_0) \left( \widehat \lambda - \lambda_0 \right)\right] + o_{\mathbb{P}}(n^{-1/2})
		= O_{\mathbb{P}}(n^{-1/2}),
	\end{equation*}
where $(\boldsymbol{\varepsilon f_z})_n = (\varepsilon_1 {f}_z(\boldsymbol Z_1),\ldots,\varepsilon_n {f}_z(\boldsymbol Z_n))^T$
and $\left(\boldsymbol{\widehat{\varepsilon}_{|z}\widehat{f}_z}\right)_n = \left(\frac{1}{n}\sum\limits_{k=1, k\neq 1}^n \varepsilon_k  K_{h,1k}, \ldots, \frac{1}{n}\sum\limits_{k=1, k\neq n}^n \varepsilon_k  K_{h,nk} \right)^T$.
\end{prop}

Note that the asymptotic representation of $\widehat \lambda$ does not depend on $s_0$, i.e. the choice of $s_0$ does not influence the asymptotic behavior of $\widehat \lambda$. This result is in line with the result of \citet{powell1996rescaled}.

Next, we state the asymptotic normality of our estimator. We use the notation $\boldsymbol\Omega_{n,i,j}(\boldsymbol{d})= \boldsymbol\Omega_{n,i,j}$ and
	\begin{equation*}
		{\mathbb{D}}_n(\boldsymbol{d}) = \boldsymbol\Omega_n(\boldsymbol{d})   - \frac{1}{\boldsymbol{1}_n^T \boldsymbol\Omega_n(\boldsymbol{d})  \boldsymbol{1}_n} \boldsymbol\Omega_n(\boldsymbol{d})  \boldsymbol{1}_n  \boldsymbol{1}_n^T  \boldsymbol\Omega_n(\boldsymbol{d}),
	\end{equation*}
to make the dependence of $\boldsymbol\Omega_{n}$ on $\boldsymbol{d}$ explicit. Note that with
	\begin{align*}
		\boldsymbol \Omega^X_{n,ij} (\boldsymbol d) = \boldsymbol \Omega^X_{n,ij} &= \exp \{ - (\boldsymbol X_i-\boldsymbol X_j)^T{\rm diag}(d_1,\ldots,d_p)  (\boldsymbol X_i-\boldsymbol X_j)  \} \quad\qquad \text{and}\\
		\boldsymbol \Omega^Z_{n,ij}(\boldsymbol d) = \boldsymbol \Omega^Z_{n,ij} &= \exp \{ - (\boldsymbol Z_i-\boldsymbol Z_j)^T{\rm diag}(d_{p+1},\ldots,d_{p+q})  (\boldsymbol Z_i-\boldsymbol Z_j)  \}, \quad 1\leq i, j \leq n,
	\end{align*} 
$\boldsymbol \Omega_{n,ij}(\boldsymbol d) = \boldsymbol \Omega^X_{n,ij}(\boldsymbol d) \boldsymbol \Omega^Z_{n,ij}(\boldsymbol d)$. Furthermore, we define, for $1 \leq i \leq n$,
	\begin{align*}
	\!\!\!\!\!\!	\boldsymbol{\tau}_i(\boldsymbol{d})
		:= 
		\Bigg(\!
		\left(\!\frac{\partial}{\partial \lambda}\mathbb{Y}_{n,i} -\frac{1}{E\left[\boldsymbol{1}_n^T \boldsymbol\Omega_n(\boldsymbol{d})  \boldsymbol{1}_n\right]}E\left[\frac{\partial}{\partial \lambda}\mathbb{Y}_n ^T\boldsymbol\Omega_n(\boldsymbol{d}) \boldsymbol{1}_n \right]\right)	,
		-\left(\!\mathbb{X}_{n,i}^T -\frac{1}{E\left[\boldsymbol{1}_n^T \boldsymbol\Omega_n(\boldsymbol{d})  \boldsymbol{1}_n\right]}E\left[ \boldsymbol{1}_n^T\boldsymbol\Omega_n(\boldsymbol{d}) {\mathbb{X}}_n \right]\right)	
		\!\Bigg)^{\!T}\!,
	\end{align*}
where $\frac{\partial}{\partial \lambda}\mathbb{Y}_{n,i}(\lambda) = \left(\frac{\partial}{\partial \lambda}T(Y_i,\lambda) -  {E}[\frac{\partial}{\partial \lambda}T(Y_i,\lambda)\mid \boldsymbol Z_i]\right) { f} _z(\boldsymbol Z_i)$ and $\mathbb{X}_{n,i} = (\boldsymbol X_i- {E}[\boldsymbol X_i\mid \boldsymbol Z_i]) {f}_z(\boldsymbol Z_i)$. In addition, let
	\begin{align*}
		\boldsymbol \Phi^X_{n,ij}(\boldsymbol d) = \boldsymbol \Omega^X_{n,ij}(\boldsymbol d) - E\left[\boldsymbol \Omega^X_{n,ik}(\boldsymbol d)\mid \boldsymbol X_i\right].
	\end{align*}
Finally, for any vector $\boldsymbol a$, we denote $\boldsymbol a^{\otimes 2} = \boldsymbol a \boldsymbol a^T$. With all this in hand, we can state the following result.

\begin{thm}[Asymptotic normality]\label{AN}
Assume that the conditions of Proposition \ref{AN_prop} hold true. Then, uniformly with respect to $h\in\mathcal{H}_{sc,n}$,  $\boldsymbol d
\in \mathcal{D}$ and $s\in S_n$,   
	\begin{align*}
		\sqrt{n}\left((\widehat \lambda, \widehat {\boldsymbol \beta}(\widehat \lambda )^T)^T  - (\lambda_0, \boldsymbol {\beta}_0 ^T)^T\right) + \boldsymbol{V}(\boldsymbol{d})^{-1}\left(\frac{1}{\sqrt{n}}\sum_{j=1}^{n}\varepsilon_j{f}_z(\boldsymbol Z_j)E\left[\boldsymbol{\tau}_i(\boldsymbol{d})\,\boldsymbol\Omega_{n,ij}^Z(\boldsymbol{d}) \boldsymbol \Phi^X_{n,ij}(\boldsymbol d)\mid \boldsymbol X_j, \boldsymbol Z_j\right]\right)  = 
  		o_{\mathbb{P} }\left(1\right),
	\end{align*}

	where 
	$$
	\boldsymbol V(\boldsymbol{d}) = \underset{n \rightarrow \infty}{\lim}	
		\begin{pmatrix}
			E\left[n^{-2}\frac{\partial}{\partial \lambda}\mathbb{Y}_n(\lambda_0)^T{\mathbb{D}}_n(\boldsymbol{d})\frac{\partial}{\partial \lambda}\mathbb{Y}_n(\lambda_0) \right]& 
			- E\left[n^{-2}\frac{\partial}{\partial \lambda}\mathbb{Y}_n(\lambda_0)^T{\mathbb{D}}_n(\boldsymbol{d}) \mathbb{X}_n\right]\\
			- E\left[n^{-2}\mathbb{X}_n^T{\mathbb{D}}_n(\boldsymbol{d})\frac{\partial}{\partial \lambda}\mathbb{Y}_n(\lambda_0)\right] &E\left[n^{-2}\mathbb{X}_n^T{\mathbb{D}}_n(\boldsymbol{d})\mathbb{X}_n\right]
		\end{pmatrix}.
$$
As a consequence, $\sqrt{n}\left((\widehat \lambda, \widehat {\boldsymbol \beta}(\widehat \lambda )^T)^T  - (\lambda_0, \boldsymbol {\beta}_0 ^T)^T\right)$ converges in distribution to a $(p+1)-$dimension centered Gaussian vector with variance $\boldsymbol V(\boldsymbol{d})^{-1}\boldsymbol\Delta(\boldsymbol{d})\boldsymbol V(\boldsymbol{d})^{-1}$
where 
$$
	\boldsymbol\Delta(\boldsymbol{d}) =  E\left\{Var\left[\varepsilon_j\mid \boldsymbol X_j, \boldsymbol Z_j\right]{f}^2_z(\boldsymbol Z_j)
	\left( E  \left[\boldsymbol{\tau}_i(\boldsymbol{d})\,\boldsymbol\Omega_{n,ij}^Z(\boldsymbol{d}) \boldsymbol \Phi^X_{n,ij}(\boldsymbol d)\mid \boldsymbol X_j, \boldsymbol Z_j\right]\right)^{\otimes 2} 
	\right\}.
$$
\end{thm}

If $\boldsymbol{\eta}_\lambda$ were known, which corresponds to the case studied by \citet{lavergne2013smooth},  $\boldsymbol \Phi^X_{n,ij}(\boldsymbol d)$ should be replaced by $\boldsymbol \Omega^X_{n,ij}(\boldsymbol d)$ in the expression of $\boldsymbol\Delta(\boldsymbol{d})$.

We can estimate the covariance matrix by $\widehat{\boldsymbol V}(\boldsymbol{d})^{-1} \widehat{\boldsymbol\Delta}(\boldsymbol{d}) \widehat{\boldsymbol V}(\boldsymbol{d})^{-1}$, where 
	\begin{equation}\label{variance_est}
		\begin{aligned}
			\widehat{\boldsymbol V}(\boldsymbol{d}) &= 		
			\begin{pmatrix}
				n^{-2}\frac{\partial}{\partial \lambda}\widehat{\mathbb{Y}}_n(\widehat \lambda)^T{\mathbb{D}}_n(\boldsymbol{d})\frac{\partial}{\partial \lambda}\widehat{\mathbb{Y}}_n(\widehat\lambda) & - n^{-2}\frac{\partial}{\partial \lambda}\widehat{\mathbb{Y}}_n(\widehat\lambda)^T{\mathbb{D}}_n(\boldsymbol{d}) \widehat{\mathbb{X}}_n\\
				- n^{-2}\widehat{\mathbb{X}}_n^T{\mathbb{D}}_n(\boldsymbol{d})\frac{\partial}{\partial \lambda}\widehat{\mathbb{Y}}_n(\widehat\lambda) &n^{-2} \widehat{\mathbb{X}}_n^T{\mathbb{D}}_n(\boldsymbol{d})\widehat{\mathbb{X}}_n
			\end{pmatrix}
		\\ \text{and} \quad \\
		 \widehat{\boldsymbol\Delta}(\boldsymbol{d}) 
		& = n^{-3}\left(\frac{\partial}{\partial \lambda}\widehat{\mathbb{Y}}_n(\widehat \lambda), -\widehat{\mathbb{X}}_n\right)^T{\mathbb{D}}_{n,inf}(\boldsymbol{d})\widehat{\boldsymbol \Phi}_{n}(\boldsymbol{d})\widehat{\boldsymbol{\Sigma}}_n \widehat{\boldsymbol \Phi}^T_{n}(\boldsymbol{d}){\mathbb{D}}^T_{n,inf}(\boldsymbol{d})\left(\frac{\partial}{\partial \lambda}\widehat{\mathbb{Y}}_n(\widehat \lambda), -\widehat{\mathbb{X}}_n\right).	
		\end{aligned}
	\end{equation} 
Here, $\widehat{\boldsymbol \Phi}^X_{n}$ and $\widehat{\boldsymbol \Phi}_{n}$ are the $n\times n-$ symmetric matrices with elements
	\begin{align*}
		\widehat{\boldsymbol \Phi}^X_{n,ij}(\boldsymbol d) &= \boldsymbol \Omega^X_{n,ij}(\boldsymbol d) - \frac{1}{n}\sum\limits_{k=1}^{n} \boldsymbol \Omega^X_{n,ik}(\boldsymbol d), 
		\quad 1\leq i, j \leq n \\ 
		\widehat{\boldsymbol \Phi}_{n,ij}(\boldsymbol d) &= \widehat{\boldsymbol \Phi}^X_{n,ij}(\boldsymbol d)\boldsymbol \Omega^Z_{n,ij}(\boldsymbol d), 
		\hskip 1.74cm 1\leq i, j \leq n \\	
		and \quad	
		{\mathbb{D}}_{n,inf}(\boldsymbol{d}) &= \boldsymbol{I}_{n\times n}   - \frac{1}{\boldsymbol{1}_n^T \boldsymbol\Omega_n(\boldsymbol{d})  \boldsymbol{1}_n} \boldsymbol\Omega_n(\boldsymbol{d})  \boldsymbol{1}_n  \boldsymbol{1}_n^T.
	\end{align*}
$\widehat{\boldsymbol{\Sigma}}_n = $ \rm{diag}$\left(\widehat{Var}\left[\varepsilon_1 f_z(\boldsymbol Z_1)\mid \boldsymbol X_1, \boldsymbol Z_1\right], \ldots, \widehat{Var}\left[\varepsilon_n{f}_z(\boldsymbol Z_n)\mid \boldsymbol X_n, \boldsymbol Z_n\right]\right)$ is an estimator of 
$\rm{diag}\Big(Var\left[\varepsilon_1 f_z(\boldsymbol Z_1)\mid \boldsymbol X_1, \boldsymbol Z_1\right],\\ \ldots, Var\left[\varepsilon_n{f}_z(\boldsymbol Z_n)\mid \boldsymbol X_n, \boldsymbol Z_n\right]\Big)$. 
One can use a nonparametric estimator for the conditional variance or alternatively, use an estimate of the error terms to approximate the conditional variance in the spirit of the Eiker-White variance estimator. Consistency of the above estimators is straightforward to establish.

\section{Testing based on SmoothMD for parameter restrictions} \label{Test}

In Section \ref{con_asy_norm} we established consistency and asymptotic normality of our estimator. 
The asymptotic behavior of our estimator is not influenced by the standardization with $s$ but the asymptotic variance is affected by the estimation of $\boldsymbol{\eta}_\lambda$. 
In addition, the behavior of our estimator is, even asymptotically, influenced by the vectors $\boldsymbol{d}_1$ and $\boldsymbol{d}_2$. When developing a test theory, we should
take that influence into account in order to get reliable results. That's what we do in the following.

\subsection{Testing the transformation parameter}
When it comes to testing parameter restrictions in the semiparametric partially linear regression model with Box-Cox transformation we might be mainly interested in testing if $\lambda$ is zero or not and if the components of $\boldsymbol{\beta}$ are zero. However, we shall consider here a more general approach to allow for more complex hypotheses as well. We separate the discussion into two parts. In the first part, we consider only restrictions for $\lambda$ and in the second part we consider restrictions for $\boldsymbol\beta$ with and without restricting $\lambda$.	
	
Suppose we want to test the restriction for $\lambda$ given by 	 
	\begin{align}\label{test_lambda}
		H_0: \lambda_0 = \lambda_R.
	\end{align}
In order to test this restriction, we can use the distance metric statistic proposed by \citet{lavergne2013smooth}. Adapted to our case and for testing \eqref{test_lambda}, we consider the distance 	
	\begin{align*}
		DM_{\lambda} = 
		\frac{1}{n}\widehat{\mathbb{Y}}_n(\lambda_R)^T \;  \widehat {\mathbb{B}} _n \widehat{\mathbb{Y}}_n(\lambda_R)
		-
		\frac{1}{n}\widehat{\mathbb{Y}}_n(\widehat\lambda)^T \;  \widehat {\mathbb{B}} _n \widehat{\mathbb{Y}}_n(\widehat\lambda).		
	\end{align*}	

The distance metric is based on the object that is minimized to get the estimate for $\lambda$, see equation \eqref{lambda_l_hat}. However, the test statistic is not standardized by $s^{-\lambda}$ as we only need this for the estimation of $\lambda$.

Let
	\begin{align*}
			\boldsymbol{A}_n = 
			\begin{pmatrix}
			\frac{\partial}{\partial \lambda}\mathbb{Y}_n(\lambda_0)^T\\
			- \mathbb{X}_n^T
			\end{pmatrix}
			{\mathbb{D}}_n\left( (\boldsymbol{\varepsilon f_z})_n - \left(\boldsymbol{\widehat{\varepsilon}_{|z}\widehat{f}_z}\right)_n\right).
	\end{align*}
We can therefore now state the following Proposition.

\begin{prop}\label{prop_test_lambda}
Assume that the conditions of Proposition \ref{AN_prop} hold true. Then, uniformly with respect to $h\in\mathcal{H}_{sc,n}$, $\boldsymbol d\in \mathcal{D}$ and $s\in S_n$,   
$$
	DM_{\lambda} - 
	(1, \boldsymbol{0}_p^T)\boldsymbol V(\boldsymbol{d})^{-1} n^{-3/2}\boldsymbol{A}_n n^{-3/2}\boldsymbol{A}_n^T \boldsymbol V(\boldsymbol{d})^{-1}(1, \boldsymbol{0}_p^T)^T 
	E\left[  \frac{\partial }{\partial \lambda} {\mathbb{Y}}_n(\lambda_0) ^T   {\mathbb{B}} _n   \frac{\partial }{\partial \lambda}  {\mathbb{Y}}_n(\lambda_0)  \right]
	= o_{\mathbb{P}}(1),
$$
under $H_0$ and $\mathbb{P}(n^{-1}DM_{\lambda} > c) \rightarrow 1$ for any $c>0$ if $H_0$ does not hold.
\end{prop}
	
The process $(1, \boldsymbol{0}_p^T)\boldsymbol V(\boldsymbol{d})^{-1} n^{-3/2}\boldsymbol{A}_n n^{-3/2}\boldsymbol{A}_n^T \boldsymbol V(\boldsymbol{d})^{-1}(1, \boldsymbol{0}_p^T)^T E\left[  \frac{\partial }{\partial \lambda} {\mathbb{Y}}_n(\lambda_0) ^T   {\mathbb{B}} _n \frac{\partial }{\partial \lambda}  {\mathbb{Y}}_n(\lambda_0)  \right]$ is asymptotically tight and for each $\boldsymbol{d}$ behaves asymptotically as a chi-square times \\
$(1, \boldsymbol{0}_p^T)\boldsymbol V(\boldsymbol{d})^{-1}  \boldsymbol\Delta(\boldsymbol{d}) 
\boldsymbol V(\boldsymbol{d})^{-1}(1, \boldsymbol{0}_p^T)^T E\left[  \frac{\partial }{\partial \lambda} {\mathbb{Y}}_n(\lambda_0) ^T   {\mathbb{B}} _n  \frac{\partial }{\partial \lambda}  {\mathbb{Y}}_n(\lambda_0)  \right]$, see \citet{johnson1995}. The distribution of the distance metric statistic therefore is, in general, non-pivotal. Determining critical values requires the estimation of 
$(1, \boldsymbol{0}_p^T)\boldsymbol V(\boldsymbol{d})^{-1}\boldsymbol \Delta(\boldsymbol{d}) 
\boldsymbol V(\boldsymbol{d})^{-1}(1, \boldsymbol{0}_p^T)^T E\left[ \frac{\partial }{\partial \lambda} {\mathbb{Y}}_n(\lambda_0) ^T   {\mathbb{B}} _n \frac{\partial }{\partial \lambda}  {\mathbb{Y}}_n(\lambda_0)  \right]$, which can rely on the estimators stated in \eqref{variance_est}. 

\subsection{Testing the slope coefficients}

In the next part, we consider restrictions for $\boldsymbol\beta$. Suppose we want to test $r$ linear restrictions for $\boldsymbol\beta$ given by 	 
	\begin{align}\label{test_beta}
		H_0: \boldsymbol{R}\boldsymbol\beta_0 = \boldsymbol c,
	\end{align}
where $\boldsymbol{R}$ is a $r \times p-$ matrix of full rank and $\boldsymbol c \in \mathbb{R}^r$. In order to test the restrictions, we need to find the restricted estimators for $\boldsymbol{\beta}_0$, $\widehat{\boldsymbol{\beta}}_R(\lambda)$, and $\lambda_0$, $\widehat \lambda_R$. We minimize
	\begin{align*}
		n^{-2}  s^{-2\lambda}\left(   \widehat{\mathbb{Y}}_n(\lambda) - \widehat{\mathbb{X}}_n   \boldsymbol \beta \right)^T\boldsymbol 
		{\mathbb{D}}_n  \left( \widehat{\mathbb{Y}}_n(\lambda) - \widehat{\mathbb{X}}_n \boldsymbol \beta\right)
		\quad s.t. \quad 
		\boldsymbol{R}\boldsymbol{\beta} = \boldsymbol c,
	\end{align*}
with respect to $\boldsymbol \beta$ and get that 
	\begin{align*}
		\widehat{\boldsymbol{\beta}}_R(\lambda) = \widehat{\boldsymbol{\beta}}(\lambda) - \left(\widehat{\mathbb{X}}_n^T{\mathbb{D}}_n\widehat{\mathbb{X}}_n\right)^{-1} \boldsymbol{R}^T 
		\left(\boldsymbol{R} \left(\widehat{\mathbb{X}}_n^T{\mathbb{D}}_n\widehat{\mathbb{X}}_n\right)^{-1} \boldsymbol{R}^T \right)^{-1}
		\left(\boldsymbol{R}\widehat{\boldsymbol{\beta}}(\lambda) - \boldsymbol c\right).
	\end{align*}
The restricted estimator for $\lambda_0$ is then given by  
	\begin{equation}\label{lambda_l_hat_R}
		\widehat \lambda_R = \widehat \lambda_R (s) = \arg\min_{\lambda\in\Lambda }  
		s^{-\lambda}\left( \widehat{\mathbb{Y}}_n(\lambda)  - \widehat{\mathbb{X}}_n \widehat{\boldsymbol{\beta}}_R(\lambda)\right)^T {\mathbb{D}}_n \, 
        s^{-\lambda}\left( \widehat{\mathbb{Y}}_n(\lambda)  - \widehat{\mathbb{X}}_n \widehat{\boldsymbol{\beta}}_R(\lambda)\right).
	\end{equation}
With all the estimators in hand, we can now define our distance metric statistic for testing \eqref{test_beta}.
	\begin{align*}
		DM_{\boldsymbol\beta} = \frac{1}{n}
		\left( \widehat{\mathbb{Y}}_n(\widehat \lambda_R)  - \widehat{\mathbb{X}}_n \widehat{\boldsymbol{\beta}}_R(\widehat \lambda_R)\right)^T 
		{\mathbb{D}}_n \, 
	    \left( \widehat{\mathbb{Y}}_n(\widehat \lambda_R)  - \widehat{\mathbb{X}}_n \widehat{\boldsymbol{\beta}}_R(\widehat \lambda_R)\right)
		-
		\frac{1}{n}\widehat{\mathbb{Y}}_n(\widehat\lambda)^T \;  \widehat {\mathbb{B}} _n \widehat{\mathbb{Y}}_n(\widehat\lambda).	
	\end{align*}
The distance metric is based on the object that is minimized to get the restricted estimate for $\lambda$, see equation \eqref{lambda_l_hat_R}. Once again the test statistic is not standardized by $s^{-\lambda}$ as we only need this for the estimation of $\lambda$.
Let, 
	$$
		{\mathbb{B}}_{n,R} = {\mathbb{B}}_n + 
		 {\mathbb{D}}_n {\mathbb{X}}_n\left({\mathbb{X}}_n^T{\mathbb{D}}_n{\mathbb{X}}_n\right)^{-1} \boldsymbol{R}^T 
		 \left(\boldsymbol{R} \left({\mathbb{X}}_n^T{\mathbb{D}}_n{\mathbb{X}}_n\right)^{-1} \boldsymbol{R}^T \right)^{-1}
		 \boldsymbol{R} \left({\mathbb{X}}_n^T{\mathbb{D}}_n{\mathbb{X}}_n\right)^{-1}{\mathbb{X}}_n^T{\mathbb{D}}_n,
	$$
and 
	$$
		\boldsymbol{V}_R(\boldsymbol{d}) = 
		\left( 
		E\left[\frac{\partial }{\partial \lambda} {\mathbb{Y}}_n(\lambda_0)^T{\mathbb{B}}_{n,R}\frac{\partial }{\partial \lambda} {\mathbb{Y}}_n(\lambda_0)\right]^{-1},
		E\left[\frac{\partial }{\partial \lambda} {\mathbb{Y}}_n(\lambda_0)^T{\mathbb{B}}_{n,R}\frac{\partial }{\partial \lambda} {\mathbb{Y}}_n(\lambda_0)\right]^{-1}
		E\left[\frac{\partial }{\partial \lambda} {\mathbb{Y}}_n(\lambda_0)^T {\mathbb{D}}_n{\mathbb{X}}_n{\mathbb{B}}_n^+\right]
		\right),
	$$
where 
	$$
		{\mathbb{B}}_n^+ = \left({\mathbb{X}}_n^T{\mathbb{D}}_n{\mathbb{X}}_n\right)^{-1} 
		- \left({\mathbb{X}}_n^T{\mathbb{D}}_n{\mathbb{X}}_n\right)^{-1} \boldsymbol{R}^T 
		 \left(\boldsymbol{R} \left({\mathbb{X}}_n^T{\mathbb{D}}_n{\mathbb{X}}_n\right)^{-1} \boldsymbol{R}^T \right)^{-1}
		 \boldsymbol{R} \left({\mathbb{X}}_n^T{\mathbb{D}}_n{\mathbb{X}}_n\right)^{-1}.
	$$
We can therefore now state the following proposition.

\begin{prop} \label{prop_test_beta}
Assume that the conditions of Proposition \ref{AN_prop} hold true. Then, uniformly with respect to $h\in\mathcal{H}_{sc,n}$, $\boldsymbol d\in \mathcal{D}$ and $s\in S_n$,   
	\begin{align*}
			DM_{\boldsymbol\beta}&\\
			-
			&n^{-3/2}\boldsymbol{A}_n^T\Bigg(
			\left(\boldsymbol{0}_{p\times 1}, \boldsymbol{I}_{p\times p}\right)^T E\left[{\mathbb{X}}_n^T{\mathbb{D}}_n{\mathbb{X}}_n\right]^{-1} \boldsymbol{R}^T 
			\left(\boldsymbol{R} E\left[{\mathbb{X}}_n^T{\mathbb{D}}_n{\mathbb{X}}_n\right]^{-1} \boldsymbol{R}^T \right)^{-1}
			\boldsymbol{R} E\left[{\mathbb{X}}_n^T{\mathbb{D}}_n{\mathbb{X}}_n\right]^{-1}\left(\boldsymbol{0}_{p\times 1}, \boldsymbol{I}_{p\times p}\right)\\
			&\quad -
			\boldsymbol{V}_R(\boldsymbol{d})^T\boldsymbol{V}_R(\boldsymbol{d})
			E\left[\frac{\partial }{\partial \lambda} {\mathbb{Y}}_n(\lambda_0) ^T   {\mathbb{B}}_{n,R}  \frac{\partial }{\partial \lambda}  {\mathbb{Y}}_n(\lambda_0)  \right] \\
			&\quad +  \boldsymbol V(\boldsymbol{d})^{-1}(1, \boldsymbol{0}_p^T)^T (1, \boldsymbol{0}_p^T)\boldsymbol V(\boldsymbol{d})^{-1} 
			E\left[ \frac{\partial }{\partial \lambda} {\mathbb{Y}}_n(\lambda_0) ^T   {\mathbb{B}} _n \frac{\partial }{\partial \lambda}  {\mathbb{Y}}_n(\lambda_0)  \right]
			\Bigg)\boldsymbol{A}_nn^{-3/2} 	= o_{\mathbb{P}}(1),
	\end{align*}
under $H_0$ and $\mathbb{P}(n^{-1}DM_{\boldsymbol\beta} > c) \rightarrow 1$ for any $c>0$ if $H_0$ does not hold. 
\end{prop}

The process in Proposition \ref{prop_test_beta} is asymptotically tight and for each $\boldsymbol{d}$ behaves asymptotically as a weighted sum of $p +1 - r$ independent chi-squares, where the weights are the positive eigenvalues of 
	\begin{equation*}\label{dist_beta}
		\begin{aligned}
			&\left(\boldsymbol{0}_{p\times 1}, \boldsymbol{I}_{p\times p}\right)^T E\left[{\mathbb{X}}_n^T{\mathbb{D}}_n{\mathbb{X}}_n\right]^{-1} \boldsymbol{R}^T 
			\left(\boldsymbol{R} E\left[{\mathbb{X}}_n^T{\mathbb{D}}_n{\mathbb{X}}_n\right]^{-1} \boldsymbol{R}^T \right)^{-1}
			\boldsymbol{R} E\left[{\mathbb{X}}_n^T{\mathbb{D}}_n{\mathbb{X}}_n\right]^{-1}\left(\boldsymbol{0}_{p\times 1}, \boldsymbol{I}_{p\times p}\right)
			\boldsymbol\Delta(\boldsymbol{d})
			\\
			&\quad -
			\boldsymbol{V}_R(\boldsymbol{d})^T\boldsymbol{V}_R(\boldsymbol{d})\boldsymbol\Delta(\boldsymbol{d}, \boldsymbol{d})
			E\left[\frac{\partial }{\partial \lambda} {\mathbb{Y}}_n(\lambda_0) ^T   {\mathbb{B}}_{n,R}  \frac{\partial }{\partial \lambda}  {\mathbb{Y}}_n(\lambda_0)  \right] \\
			&\quad +  \boldsymbol V(\boldsymbol{d})^{-1}(1, \boldsymbol{0}_p^T)^T (1, \boldsymbol{0}_p^T)\boldsymbol V(\boldsymbol{d})^{-1} 
			\boldsymbol\Delta(\boldsymbol{d})
			E\left[ \frac{\partial }{\partial \lambda} {\mathbb{Y}}_n(\lambda_0) ^T   {\mathbb{B}} _n \frac{\partial }{\partial \lambda}  {\mathbb{Y}}_n(\lambda_0)  \right],
		\end{aligned}
    \end{equation*}
see \citet{johnson1995}. Determining critical values requires the estimation of the last expression. We can use the estimators stated in \eqref{variance_est} and for all other components, we simply replace the unknown expressions by their sample version, e.g. estimate ${\mathbb{B}}_{n,R}$ by 
	$$
		\widehat{\mathbb{B}}_n + {\mathbb{D}}_n \widehat{\mathbb{X}}_n\left(\widehat{\mathbb{X}}_n^T{\mathbb{D}}_n\widehat{\mathbb{X}}_n\right)^{-1} \boldsymbol{R}^T 
		 \left(\boldsymbol{R} \left(\widehat{\mathbb{X}}_n^T{\mathbb{D}}_n\widehat{\mathbb{X}}_n\right)^{-1} \boldsymbol{R}^T \right)^{-1}
		 \boldsymbol{R} \left(\widehat{\mathbb{X}}_n^T{\mathbb{D}}_n\widehat{\mathbb{X}}_n\right)^{-1}\widehat{\mathbb{X}}_n^T{\mathbb{D}}_n.
	$$

\subsection{Testing the transformation parameter and the slope coefficients}

Finally, we consider the combined restrictions for $\boldsymbol\beta$ and $\lambda$. Suppose we want to test 	 
	\begin{align*} 
		H_0: \boldsymbol{R}\boldsymbol\beta_0 = \boldsymbol c \quad and \quad \lambda_0 = \lambda_R.
	\end{align*}
In contrast to the hypothesis stated in \eqref{test_beta} we do not need to estimate $\widehat{\lambda}_R$. Therefore, the distance metric statistic is for this case given by
	\begin{align*}
		DM_{\boldsymbol\beta,\lambda} = \frac{1}{n}
		\left( \widehat{\mathbb{Y}}_n(\lambda_R)  - \widehat{\mathbb{X}}_n \widehat{\boldsymbol{\beta}}_R(\lambda_R)\right)^T 
		{\mathbb{D}}_n \, 
	    \left( \widehat{\mathbb{Y}}_n(\lambda_R)  - \widehat{\mathbb{X}}_n \widehat{\boldsymbol{\beta}}_R(\lambda_R)\right)	
		-
		\frac{1}{n}\widehat{\mathbb{Y}}_n(\widehat\lambda)^T \;  \widehat {\mathbb{B}} _n \widehat{\mathbb{Y}}_n(\widehat\lambda).	
	\end{align*}
In addition, $DM_{\boldsymbol\beta,\lambda}$ does not converge to the same expression as $DM_{\boldsymbol\beta}$ as $\lambda_R$ is fixed. Therefore, we state the following proposition.
\begin{prop} \label{prop_test_beta_lambda}
Assume that the conditions of Proposition \ref{AN_prop} hold true. Then, uniformly with respect to $h\in\mathcal{H}_{sc,n}$, $\boldsymbol d\in \mathcal{D}$ and $s\in S_n$,   
	\begin{align*}
			DM_{\boldsymbol\beta, \lambda}&\\
			-
			&n^{-3/2}\boldsymbol{A}_n^T\Bigg(
			\left(\boldsymbol{0}_{p\times 1}, \boldsymbol{I}_{p\times p}\right)^T E\left[{\mathbb{X}}_n^T{\mathbb{D}}_n{\mathbb{X}}_n\right]^{-1} \boldsymbol{R}^T 
			\left(\boldsymbol{R} E\left[{\mathbb{X}}_n^T{\mathbb{D}}_n{\mathbb{X}}_n\right]^{-1} \boldsymbol{R}^T \right)^{-1}
			\boldsymbol{R} E\left[{\mathbb{X}}_n^T{\mathbb{D}}_n{\mathbb{X}}_n\right]^{-1}\left(\boldsymbol{0}_{p\times 1}, \boldsymbol{I}_{p\times p}\right)\\
			&\quad +  \boldsymbol V(\boldsymbol{d})^{-1}(1, \boldsymbol{0}_p^T)^T (1, \boldsymbol{0}_p^T)\boldsymbol V(\boldsymbol{d})^{-1} 
			E\left[ \frac{\partial }{\partial \lambda} {\mathbb{Y}}_n(\lambda_0) ^T   {\mathbb{B}} _n \frac{\partial }{\partial \lambda}  {\mathbb{Y}}_n(\lambda_0)  \right]
			\Bigg)\boldsymbol{A}_nn^{-3/2} 	= o_{\mathbb{P}}(1),
	\end{align*}
under $H_0$ and $\mathbb{P}(n^{-1}DM_{\boldsymbol\beta, \lambda} > c) \rightarrow 1$ for any $c>0$ if $H_0$ does not hold. 
\end{prop}

The process in Proposition \ref{prop_test_beta_lambda} is asymptotically tight and for each $\boldsymbol{d}$ behaves asymptotically as a weighted sum of $p - r$ independent chi-squares, where the weights are the positive eigenvalues of 
	\begin{equation*}\label{dist_beta_lambda}
		\begin{aligned}
			&\left(\boldsymbol{0}_{p\times 1}, \boldsymbol{I}_{p\times p}\right)^T E\left[{\mathbb{X}}_n^T{\mathbb{D}}_n{\mathbb{X}}_n\right]^{-1} \boldsymbol{R}^T 
			\left(\boldsymbol{R} E\left[{\mathbb{X}}_n^T{\mathbb{D}}_n{\mathbb{X}}_n\right]^{-1} \boldsymbol{R}^T \right)^{-1}
			\boldsymbol{R} E\left[{\mathbb{X}}_n^T{\mathbb{D}}_n{\mathbb{X}}_n\right]^{-1}\left(\boldsymbol{0}_{p\times 1}, \boldsymbol{I}_{p\times p}\right)
		    \boldsymbol\Delta(\boldsymbol{d})		
			\\
			&\quad +  \boldsymbol V(\boldsymbol{d})^{-1}(1, \boldsymbol{0}_p^T)^T (1, \boldsymbol{0}_p^T)\boldsymbol V(\boldsymbol{d})^{-1} 
			\boldsymbol\Delta(\boldsymbol{d})
			E\left[ \frac{\partial }{\partial \lambda} {\mathbb{Y}}_n(\lambda_0) ^T   {\mathbb{B}} _n \frac{\partial }{\partial \lambda}  {\mathbb{Y}}_n(\lambda_0)  \right],
		\end{aligned}
	\end{equation*}
see \citet{johnson1995}. Determining critical values requires the estimation of the last expression. We can use the estimators stated in \eqref{variance_est} and for all other components, we simply replace the unknown expressions by their sample version.

\begin{rmk}

The Propositions of Section \ref{Test} are also valid if the unknown parameters $\lambda$ and $\boldsymbol{\beta}$ are estimated without the intercept nuisance parameter $\gamma$. In that case ${\mathbb{D}}_{n}$ is replaced by $\boldsymbol \Omega_{n}$ in the statements. Moreover, when estimating the unknown variance ${\mathbb{D}}_{n,inf}$ has to be replaced by $\boldsymbol{I}_{n\times n}$.

\end{rmk}

\section{Small sample study and real data application}   \label{small_sample_study}
In this section we consider the small sample behavior of our estimator. We conduct several simulation experiments to consider bias and standard deviation for the estimated parameters. In addition, we conduct hypothesis tests as discussed in Section \ref{Test}.

In the proof of Theorem \ref{AN} it was established that
	\begin{align*}
		\left((\widehat \lambda, \widehat {\boldsymbol \beta}(\widehat \lambda )^T)^T  - (\lambda_0, \boldsymbol {\beta}_0 ^T)^T\right) &= - \boldsymbol{V}(\boldsymbol{d})^{-1}\Bigg(\frac{1}{n}\sum_{j=1}^{n}\varepsilon_j{f}_z(\boldsymbol Z_j)  E\left[\boldsymbol{\tau}_i(\boldsymbol{d})\,\boldsymbol\Omega_{n,ij}(\boldsymbol{d})\mid \boldsymbol X_j, \boldsymbol Z_j\right]\\
		&\quad -
		\frac{1}{n}	\sum_{k=1}^{n}\varepsilon_k f_z(\boldsymbol Z_k) E\left[\boldsymbol{\tau}_i(\boldsymbol{d})  \boldsymbol\Omega_{n,ik}^Z \boldsymbol\Omega_{n,ij}^X\mid \boldsymbol Z_k\right]
		\Bigg)  + 
		  o_{\mathbb{P} }\left(n^{-1/2}\right).
	\end{align*}	  
The second sum in this asymptotic representation is due to the estimation of $\boldsymbol \eta_\lambda$. In order to propose a simpler procedure, in our simulation experiments we also investigated what happens when one neglects  the second part in the asymptotic representation, that is abusively consider
$
E\left[\boldsymbol{\tau}_i(\boldsymbol{d})  \boldsymbol\Omega_{n,ik}^Z \boldsymbol\Omega_{n,ij}^X\mid \boldsymbol Z_k\right]= \boldsymbol 0_{p+1}	$. The estimator is labeled SmoothMD* in this section. The reason for this investigation is that $E\left[\boldsymbol{\tau}_i(\boldsymbol{d})  \boldsymbol\Omega_{n,ij}^Z \boldsymbol\Omega_{n,ij}^X\mid \boldsymbol Z_k\right]$ is indeed null. It appears  that considering an intercept $\gamma$ compensates for this \emph{ad-hoc} simplification and allows reasonably accurate results to be obtained.\footnote{Note that when estimating the model without constant one has to replace $\boldsymbol{\tau}_i(\boldsymbol{d})$ by 
$
		\widetilde{\boldsymbol{\tau}}_i(\boldsymbol{d})
		:= 
		\left(\frac{\partial}{\partial \lambda}\mathbb{Y}_{n,i},-\mathbb{X}_{n,i}^T \right)^{T},
$
but $E\left[\widetilde{\boldsymbol{\tau}}_i(\boldsymbol{d})  \boldsymbol\Omega_{n,ij}^Z \boldsymbol\Omega_{n,ij}^X\mid \boldsymbol Z_k\right] \neq \boldsymbol 0_{p+1}	$.
}

\subsection{Simulation setup} \label{sim}

During the simulation, we consider four different models. The models are given by 
	\begin{enumerate}
		\item[Model 1:] $T(Y,\lambda_0) = X\beta_0 + m(Z) + \varepsilon$,
						$m(Z) = \frac{\exp\{Z\}}{1 + \exp\{Z\}} + \frac{1}{3}$ with $Z \sim N(1,1)$, $\lambda_0 = 0$ and $\beta_0 = 1$, 
						$X = -\frac{2}{3} Z + u$ with $u \sim N(0,1)$ and $\varepsilon = \sqrt{\frac{1 + X^2}{2}} \; \widetilde u$ with  $\widetilde u \sim N\left(0,\frac{1}{13}\right)$. 
						
		\item[Model 2:] $T(Y,\lambda_0) = X\beta_0 + m(Z) + \varepsilon$, $m(Z) = \frac{\exp\{Z\}}{1 + \exp\{Z\}} + 3$ with $Z \sim N(1,1)$, $\lambda_0 = 0.5$
						and $\beta_0 = 1$, $X = -\frac{2}{3} Z + u$ with $u \sim N(0,1)$ and $\varepsilon  \sim N\left(0,\frac{1}{9}\right)$.
						
		\item[Model 3:] $T(Y,\lambda_0) = X\beta_0 + m(Z) + \varepsilon$, $m(Z) = \frac{\exp\{Z\}}{1 + \exp\{Z\}} - 1$ with $Z \sim U(-3,-1)$, $\lambda_0 = -1$
						and $\beta_0 = 1$, $X = \frac{2}{3} Z + u$ with $u \sim U(-1,1)$ and $\varepsilon  \sim U\left(-\sqrt{1/9},\sqrt{1/9}\right)$.
		
		\item[Model 4:] $T(Y,\lambda_0) = X_1\beta_{10} + \boldsymbol X_2\boldsymbol \beta_{20} + m(Z_1, Z_2) +  \varepsilon$, 
						$m(Z_1, Z_2) = \frac{1}{3} + Z_1 + Z_2 + Z_1Z_2$ with $Z_1, Z_2 \sim N(0,1)$, $\lambda_0 = 0$, $\beta_{10} = 1$, 
						$X_1 = -\frac{1}{3} \left( Z_1 + Z_2\right) + u$ with $u \sim N(0,1)$, $X_{2,l} \overset{i.i.d.}{\sim} Ber(0.2)$ and $\beta_{2,l} \overset{i.i.d.}{\sim} U(-1,1)$ for $l=1,\ldots,30$ , $\varepsilon  \sim N\left(0,\frac{1}{9}\right)$.	 				
	\end{enumerate}
	
The main difference in the models is the transformation parameter $\lambda$. Model 1 and Model 4 have $\lambda_0 = 0$, whereas Model 2 has $\lambda_0 = 0.5$ and Model 3 $\lambda_0 = -1$. To ensure that $Y > 0$ in Model 3 we draw the random variables from uniform distributions. In all other models positivity of $Y$ is ensured as well. Model 1 has heteroskedastic error terms which are captured by the developed theory. Model 4 contains 30 dummy variables, $\boldsymbol X_2$, which take the value 1 with probability $20\%$. 
In addition, Model 4 has the same structure as the model of the application we consider in section \ref{real_data}.

The estimators are computed by employing a normal kernel for $K(\cdot)$. $\boldsymbol Z$ is standardized componentwise by the corresponding standard deviations and $h \propto n^{-1/3.5}$. This bandwidth choice satisfies the assumptions of Theorem \ref{AN}.   
The components of $\boldsymbol d$ defining the diagonal matrix $\boldsymbol{D}$ in $\boldsymbol{\Omega}_n$, are set equal to the componentwise standard deviations of $X$ and $\boldsymbol Z$ when $X$ is continuous. In the case of the dummy variables $\boldsymbol{X}_2$, an indicator of the event that the observations have the same value, is employed. For Model 4, we ensure in the simulations that for every observation there exists at least 4 observations with the same dummy variable combination.

In the estimation, we define a grid for values of $\lambda$ that are considered during the optimization. 
This optimization grid for $\lambda$ is given in our simulation by the grid $[\lambda_0 -0.8, \lambda_0 + 0.8]$ with step size $0.001$. We minimize $G_n^{-\lambda} \widehat{\mathbb{Y}}_n(\lambda)^T \;  \widehat {\mathbb{B}} _n \; G_n^{-\lambda} \widehat{\mathbb{Y}}_n(\lambda) $ over the defined grid to get $\widehat \lambda$ and $\boldsymbol{\widehat\beta}(\widehat \lambda)$, where $G_n = \prod\limits_{i=1}^{n} Y_i^{1/n}$ is the sample geometric mean.

In the simulation, we compare the proposed estimator where $\gamma$ is employed with the estimator that does not use $\gamma$. Both estimators converge asymptotically to a normal distribution. The only difference is that we have to replace  $\boldsymbol{\tau}_i(\boldsymbol{d})$ by 
$
		\widetilde{\boldsymbol{\tau}}_i(\boldsymbol{d})
		:= 
		\left(\frac{\partial}{\partial \lambda}\mathbb{Y}_{n,i},-\mathbb{X}_{n,i}^T \right)^{T}
$
in the case of the estimator without $\gamma$. It is therefore interesting to compare both estimators.

We consider the bias and standard deviation of the estimators as well as the power and size of the distance metric statistics proposed in Section \ref{Test}. In addition, we test by a simple Z-Test, if the estimated parameters are significantly different from the true value. Therefore, we employ the variance estimator stated in equation \eqref{variance_est}, and the necessary adjustments for the estimator without $\gamma$ are replacing $\mathbb{D}_{n,inf}$ by $\boldsymbol I_{n\times n}$ and $\mathbb{D}_n$ by $\boldsymbol \Omega_n$.
To estimate the error variance, we employ the Eiker-White variance estimator. In order to see the influence of the estimated $\boldsymbol \eta_\lambda$ on the variance, we consider all tests also without taking the estimation error of $\boldsymbol \eta_\lambda$ into account. Therefore, we replace $\widehat{\boldsymbol \Phi}_{n}$ by $\boldsymbol{\Omega}_n$ in the variance estimator. As mentioned before, we label this estimator SmoothMD*.

In addition, the Nonlinear two-stage Least Squares (NL2SLS) estimator for the Box-Cox model introduced by \citet{amemiya1981comparison} is considered as a competitor. In order to be able to employ this estimator, it is assumed that the function $m(\cdot)$ is known and, thus, 
$m(\boldsymbol Z)$ can be added as additional regressor. The instruments are given therefore by $\boldsymbol V_i = (1,\boldsymbol X_i,\boldsymbol X_i^2,m(\boldsymbol Z_i),m(\boldsymbol Z_i)^2)$. We consider the Z-Test for the NL2SLS estimator as well where we employ the Eiker-White variance estimator again.

\subsection{Simulation results}
Table \ref{Bias_Std_Model1} states the results for the bias and standard deviation for $\lambda$ and $\beta$ in Model 1. All three estimators have comparable results for the bias and the bias decreases with sample size for $\beta$, whereas it is the lowest for $n=500$ for the SmoothMD estimators in the case of $\lambda$. Surprisingly, the standard deviation is also comparable for all three estimators even though $m(\cdot)$ is given for the NL2SLS estimator. 
		\begin{table}[H]
			\caption{\textit{Bias and Standard Deviation of the estimators for $\lambda$ and $\beta$ in Model 1. } }
			\begin{tabular}{@{}lcd{3.5}d{3.5}d{3.5}d{3.5}d{3.5}d{3.5}}
    \toprule\midrule
    & $s$   & \multicolumn{3}{c}{Bias} &\multicolumn{3}{c}{St. dev.}\\\cmidrule(lr){3-5} \cmidrule(lr){6-8} $n$ &  & \mc{250} & \mc{500} & \mc{1000} & \mc{250} & \mc{500} & \mc{1000} \\ 
   \midrule
$\lambda$ estimator &  &  &  &  &  &  &  \\ 
   \midrule
  SmoothMD with $\gamma$ & $G_n$    & 0.003  & 0.0001 & 0.001  & 0.042 & 0.03  & 0.021 \\ 
  SmoothMD without $\gamma$ & $G_n$ & 0.002  & 0.0001 & 0.001  & 0.041 & 0.029 & 0.02 \\ 
  NL2SLS                    & $G_n$ & -0.003 & -0.001 & 0.0001 & 0.042 & 0.029 & 0.02 \\ 
   \midrule
$\beta$ estimator &  &  &  &  &  &  &  \\ 
   \midrule
  SmoothMD with $\gamma$ & $G_n$    & -0.001 & -0.001  & 0.0004 & 0.036 & 0.025 & 0.017\\ 
  SmoothMD without $\gamma$ & $G_n$ & -0.001 & -0.001  & 0.0004 & 0.035 & 0.024 & 0.017 \\ 
  NL2SLS                    & $G_n$ & -0.002 & -0.001  & 0.0002 & 0.035 & 0.024 & 0.016 \\  
  \midrule\bottomrule
\end{tabular}
			\vskip 0.1cm
			\textit{Notes: For the SmoothMD estimators, $h \propto n^{-1/3.5}$. The components of $\boldsymbol d$ are set equal to the componentwise standard deviations for all variables. The grid for $\lambda$ is $[\lambda_0-0.8,\lambda_0+0.8]$. 2000 Monte Carlo samples were used for all simulations. }
			\label{Bias_Std_Model1}
		\end{table}

Table \ref{Bias_Std_Model4} states the results for the bias and standard deviation for $\lambda$, $\beta_1$ and $\beta_2$, one representative parameter out of the 30 parameters in $\boldsymbol{\beta}_2$, in Model 4. All three estimators have comparable results for the bias and the bias decreases with sample size for all three parameters. In contrast to the results for Model 1, the standard deviation is smaller in the case of the NL2SLS estimator for $\beta_1$ and $\beta_2$. This result should be expected as $m(\cdot)$ is given for the NL2SLS estimator. The standard deviations for the SmoothMD estimators with and without $\gamma$ are as in Model 1, nearly the same.

		\begin{table}[H]
			\caption{\textit{Bias and Standard Deviation of the estimators for $\lambda$, $\beta_1$ and $\beta_2$ in Model 4. } }
			\begin{tabular}{@{}lcd{3.5}d{3.5}d{3.5}d{3.5}d{3.5}d{3.5}}
    \toprule\midrule
    & $s$   & \multicolumn{3}{c}{Bias} &\multicolumn{3}{c}{St. dev.}\\\cmidrule(lr){3-5} \cmidrule(lr){6-8} $n$ &  & \mc{250} & \mc{500} & \mc{1000} & \mc{250} & \mc{500} & \mc{1000} \\ 
   \midrule
$\lambda$ estimator &  &  &  &  &  &  &  \\ 
   \midrule
  SmoothMD with $\gamma$ & $G_n$    & 0.0001   & -0.0002 & -0.0001   & 0.015 & 0.01 & 0.008 \\ 
  SmoothMD without $\gamma$ & $G_n$ & 0.0001   & -0.0002 & -0.0001   & 0.015 & 0.01 & 0.008 \\ 
  NL2SLS                    & $G_n$ & -0.0004  & -0.0002 & -0.0002   & 0.014 & 0.01 & 0.005 \\ 
   \midrule
$\beta_1$ estimator &  &  &  &  &  &  &  \\ 
   \midrule
  SmoothMD with $\gamma$ & $G_n$    & -0.002 & -0.002   &  -0.001  & 0.036 & 0.023 & 0.017\\ 
  SmoothMD without $\gamma$ & $G_n$ & -0.002 & -0.002   &  -0.001  & 0.036 & 0.023 & 0.017 \\ 
  NL2SLS                    & $G_n$ & 0.0004 & -0.0001  &  -0.0001 & 0.025 & 0.015 & 0.011 \\ 
   \midrule  
$\beta_2$ estimator &  &  &  &  &  &  &  \\ 
   \midrule
  SmoothMD with $\gamma$ & $G_n$    & 0.004  & -0.001  & -0.0002  & 0.133 & 0.065 & 0.042\\ 
  SmoothMD without $\gamma$ & $G_n$ & 0.004  & -0.001  & -0.0002  & 0.133 & 0.065 & 0.042 \\ 
  NL2SLS                    & $G_n$ & -0.001 & 0.002   & -0.0006  & 0.091 & 0.046 & 0.028 \\     
   \midrule\bottomrule
\end{tabular}
			\vskip 0.1cm
			\textit{Notes: For the SmoothMD estimators, $h \propto n^{-1/3.5}$. The components of $\boldsymbol d$ are set equal to the componentwise standard deviations for all continuous variables and for the dummy variables an indicator of the event that the observations have the same value is employed. The grid for $\lambda$ is $[\lambda_0-0.8,\lambda_0+0.8]$. 2000 Monte Carlo samples were used for all simulations. }
			\label{Bias_Std_Model4}
		\end{table}

 		\begin{table}[H]
 			\caption{\textit{Empirical Level for distance metric statistics of the estimators for $\lambda$ and $\beta$ in Model 2. } }
\begin{tabular}{@{}lcd{3.5}d{3.5}d{3.5}d{3.5}d{3.5}d{3.5}}
    \toprule\midrule& $s$   & \multicolumn{3}{c}{5\% level} &\multicolumn{3}{c}{10\% level}\\\cmidrule(lr){3-5} \cmidrule(lr){6-8} $n$ &  & \mc{250} & \mc{500} & \mc{1000} & \mc{250} & \mc{500} & \mc{1000} \\ 
   \midrule
Test for $\lambda$  &  &  &  &  &  &  &  \\ 
   \midrule
SmoothMD with $\gamma$     & $G_n$ & 10.3  & 8.15 & 7.8  & 12.75 & 10.65 & 11.45 \\ 
SmoothMD* with $\gamma$    & $G_n$ & 10.15 & 7.95 & 7.7  & 12.75 & 10.6  & 11.45 \\ 
SmoothMD without $\gamma$  & $G_n$ & 9.55  & 7.0  & 6.0  & 12.45 & 10.85 & 10.35 \\ 
   \midrule
Test for $\beta$ &  &  &  &  &  &  &  \\ 
   \midrule
SmoothMD with $\gamma$     & $G_n$ & 10.4 & 8.25 & 7.95 & 12.55 & 11.1 & 11.9 \\ 
SmoothMD* with $\gamma$    & $G_n$ & 10.2 & 8.3  & 7.7  & 12.55 & 11.3 & 11.1 \\ 
SmoothMD without $\gamma$  & $G_n$ & 9.7  & 6.85 & 6.7  & 12.55 & 10.2 & 11.2 \\ 
\midrule \bottomrule
\end{tabular}

 			\vskip 0.1cm
 			\textit{Notes: For the SmoothMD estimators, $h \propto n^{-1/3.5}$. The components of $\boldsymbol d$ are set equal to the componentwise standard deviations for all variables. The variances are estimated by the Eiker-White variance estimator. For SmoothMD* the additional variance part due to the estimation of $\boldsymbol{\eta_\lambda}$ is not taken into account. For SmoothMD the additional variance part is taken into account. 2000 Monte Carlo samples were used for all simulations.}
 			\label{Emp_DistM_Model2}
 		\end{table}

Table \ref{Emp_DistM_Model2} states the empirical level for distance metric statistics of the estimators for $\lambda$ and $\beta$ in Model 2. Here we state the results for the SmoothMD estimator with correctly estimated variance as well as with the variance estimate that does not account for the estimation of $\boldsymbol \eta_\lambda$. For all three estimators, the empirical levels converge to the nominal levels if the sample size increases and $\beta$ seems to need a larger sample size than $\lambda$ to get close to the nominal level. However, in this setup, the results of the SmoothMD estimators with and without $\gamma$ differ. In addition, it can be seen that the estimator SmoothMD* leads to almost the same results as SmoothMD.

Table \ref{Emp_LevZ_Model4} states the empirical level for the Z-Tests for $\lambda$, $\beta_1$ and $\beta_2$ in Model 4. The fact that we do not consider the estimation error has almost no influence on the results. In addition, both SmoothMD versions lead to similar results. However, in order to get close to the nominal level, the sample size needs to be large as only for $n = 1000$ do the SmoothMD estimators get close to the nominal level. The NL2SLS estimator gives more convincing results for smaller sample sizes. Note that the dummy variable coefficient $\beta_2$ seems to require a larger sample size than the other two parameters to get close to the nominal level when employing the SmoothMD estimators. 	

		\begin{table}[H]
			\caption{\textit{Empirical Level for Z-Tests of the estimators for $\lambda$, $\beta_1$ and $\beta_2$ in Model 4. } }
\begin{tabular}{@{}lcd{3.5}d{3.5}d{3.5}d{3.5}d{3.5}d{3.5}}
    \toprule\midrule& $s$   & \multicolumn{3}{c}{5\% level} &\multicolumn{3}{c}{10\% level}\\\cmidrule(lr){3-5} \cmidrule(lr){6-8} $n$ &  & \mc{250} & \mc{500} & \mc{1000} & \mc{250} & \mc{500} & \mc{1000} \\ 
   \midrule
Test for $\lambda$  &  &  &  &  &  &  &  \\ 
   \midrule
SmoothMD with $\gamma$     & $G_n$ & 8.4  & 7.1  & 5.5  & 15.45 & 13.5  & 10.55 \\ 
SmoothMD* with $\gamma$    & $G_n$ & 8.5  & 7.1  & 5.45 & 15.3  & 13.45 & 10.55 \\ 
SmoothMD without $\gamma$  & $G_n$ & 9.6  & 8.15 & 5.2  & 16.45 & 12.6  & 11.65  \\ 
  NL2SLS                   & $G_n$ & 9.6  & 7.6  & 6.0  & 16.8  & 12.5  & 11.1  \\ 
   \midrule
Test for $\beta_1$ &  &  &  &  &  &  &  \\ 
   \midrule
SmoothMD with $\gamma$     & $G_n$ & 11.8   & 8.55 & 6.4  & 18.25 & 14.05 & 11.75 \\ 
SmoothMD* with $\gamma$    & $G_n$ & 11.8   & 8.55 & 6.45 & 18.25 & 14.05 & 11.8  \\ 
SmoothMD without $\gamma$  & $G_n$ & 11.75  & 8.9  & 6.4  & 18.35 & 14.8  & 12.1  \\ 
  NL2SLS                   & $G_n$ & 8.25   & 4.95 & 5.1  & 13.95 & 10.4  & 10.35 \\ 
   \midrule
Test for $\beta_2$ &  &  &  &  &  &  &  \\ 
   \midrule
SmoothMD with $\gamma$     & $G_n$ & 13.6  & 8.25 & 7.05 & 20.7  & 15.65 & 12.7  \\ 
SmoothMD* with $\gamma$    & $G_n$ & 13.7  & 8.25 & 7.05 & 20.75 & 15.55 & 12.75 \\ 
SmoothMD without $\gamma$  & $G_n$ & 13.75 & 8.35 & 6.65 & 20.6  & 14.55 & 12.25  \\ 
  NL2SLS                   & $G_n$ &  7.65 & 6.55 & 4.7  & 13.45 & 13.05 & 9.55  \\ 
 \midrule \bottomrule 
\end{tabular}

			\vskip 0.1cm
			\textit{Notes: For the SmoothMD estimators, $h \propto n^{-1/3.5}$. The components of $\boldsymbol d$ are set equal to the componentwise standard deviations for all continuous variables and for the dummy variables, an indicator of the event that the observations have the same value is employed. The variances are estimated by the Eiker-White variance estimator. For SmoothMD* the additional variance part due to the estimation of $\boldsymbol{\eta_\lambda}$, is not taken into account. For SmoothMD, the additional variance part is taken into account. $\beta_2$ is one representative parameter out of the 30 parameters in $\boldsymbol{\beta}_2$. 2000 Monte Carlo samples were used for all simulations.}
			\label{Emp_LevZ_Model4}
		\end{table}

	\begin{figure}[h]
		\begin{minipage}{0.4\textwidth} 
				\caption{\textit{Power function of the distance metric \\ statistic 
				 for $\lambda$ of Model 3 with $n=250$.}}
					\includegraphics[scale=0.4]{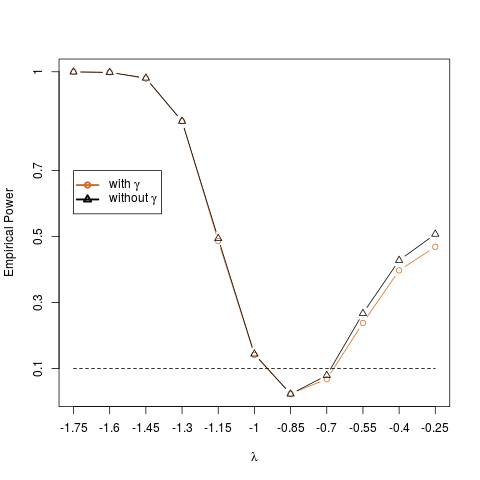}
				\label{fig_power_4.1}
		\end{minipage}
		\hfill
		\begin{minipage}{0.4\textwidth}
				\caption{\textit{Power function of the distance metric \\ statistic 
				for $\lambda$ of Model 3 with $n=1000$.}}
				\includegraphics[scale=0.4]{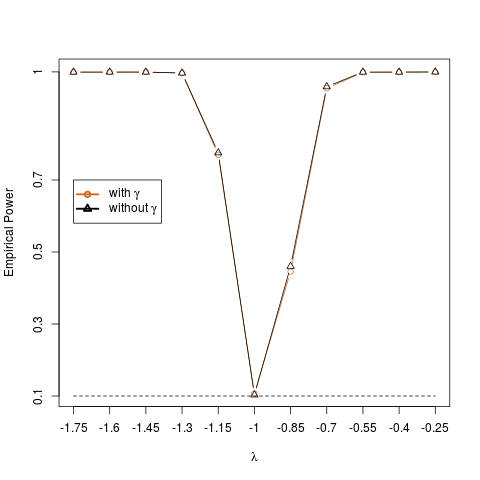}
				\label{fig_power_4.2}
		\end{minipage}
	\begin{spacing}{0.8}
		\flushleft{\textit{\footnotesize Notes: For the SmoothMD estimators, $h \propto n^{-1/3.5}$. The components of $\boldsymbol d$ are set equal to the componentwise standard deviations for all variables. The variances are estimated by the Eiker-White variance estimator. Only the SmoothMD estimators that take the additional variance part due to the estimation of $\boldsymbol{\eta_\lambda}$ into account are considered. 2000 Monte Carlo samples were used for all simulations. The nominal level is $ 10\%$.}}		
	\end{spacing}	
	\end{figure}

Figures \ref{fig_power_4.1} and \ref{fig_power_4.2} state the power functions of the distance metric statistic for $\lambda$ in Model 3 with $n=250$ and $n = 1000$. In the case of $n=250$, the power function is skewed and the power for values larger than $-1$ is small. In addition, the power function is smaller than the nominal value at $-0.85$ and $-0.7$.  	
For the SmoothMD estimator without $\gamma$, the power function is larger than for the SmoothMD estimator with $\gamma$ at values larger than $-1$. These issues disappear for the larger sample size $n=1000$.

\begin{figure}[h!]
		\begin{minipage}{0.4\textwidth} 
				\caption{\textit{Power function of the distance metric \\ statistic 
				for $\beta$ of Model 2 with $n=250$.}}
				\includegraphics[scale=0.4]{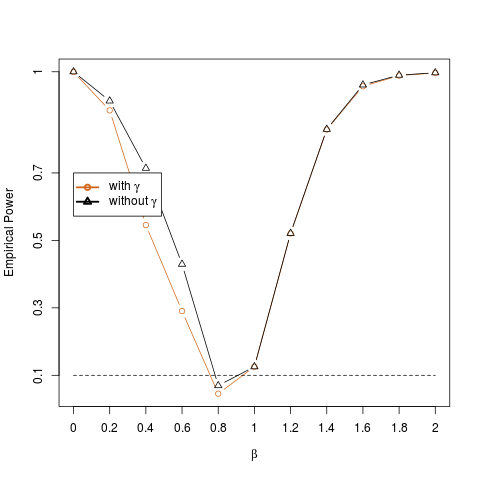}
				\label{fig_power_2.1}
		\end{minipage}
		\hfill
		\begin{minipage}{0.4\textwidth}
				\caption{\textit{Power function of the distance metric \\ statistic 
				for $\beta$ of Model 2 with $n=500$.}}
				\includegraphics[scale=0.4]{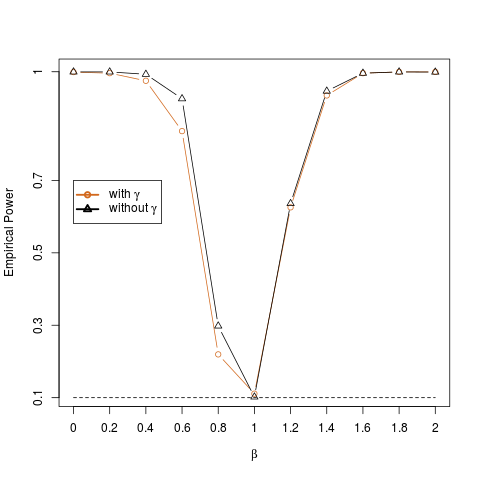}
				\label{fig_power_2.2}
		\end{minipage}
	\begin{spacing}{0.8}
		\flushleft{\textit{\footnotesize Notes: For the SmoothMD estimators, $h \propto n^{-1/3.5}$. The components of $\boldsymbol d$ are set equal to the componentwise standard deviations for all variables. The variances are estimated by the Eiker-White variance estimator. Only the SmoothMD estimators that take the additional variance part due to the estimation of $\boldsymbol{\eta_\lambda}$ into account are considered. 2000 Monte Carlo samples were used for all simulations. The nominal level is $ 10\%$.}}		
	\end{spacing}
	\end{figure}

Figures \ref{fig_power_2.1} and \ref{fig_power_2.2} state the power functions of the distance metric statistic for $\beta$ in Model 2 with $n=250$ and $n = 500$.
As in Figure \ref{fig_power_4.1} the power function for $n=250$ is skewed but the effect is less distinct. However, the power function is smaller than the nominal value at $0.8$.	For the SmoothMD estimator without $\gamma$, the power function is larger than for the SmoothMD estimator with $\gamma$ at values smaller than $1$ for both sample sizes. For $n=500$, the skewness is less pronounced and the power function has no values lower than the nominal value. The main conclusion from both power functions is that the samples size should not be too small so that the tests have a reasonable power.

\begin{figure}[h!]
		\begin{minipage}{0.4\textwidth} 
				\caption{\textit{Estimated $m(Z)$ for Model 1 with $n=250$.}}
				\includegraphics[scale=0.4]{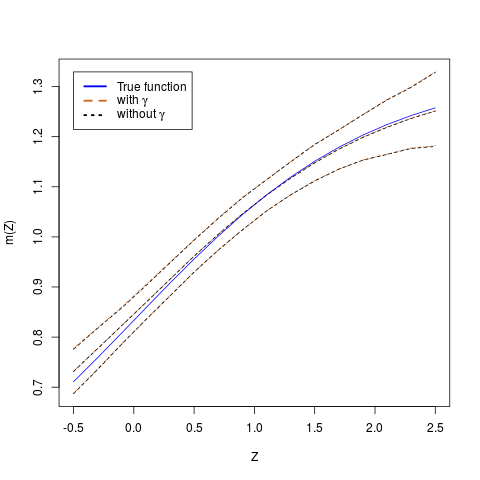}
				\label{fig_m_1.1}
		\end{minipage}
		\hfill
		\begin{minipage}{0.4\textwidth}
				\caption{\textit{Estimated $m(Z)$ for Model 3\\  with $n=500$.}}
				\includegraphics[scale=0.4]{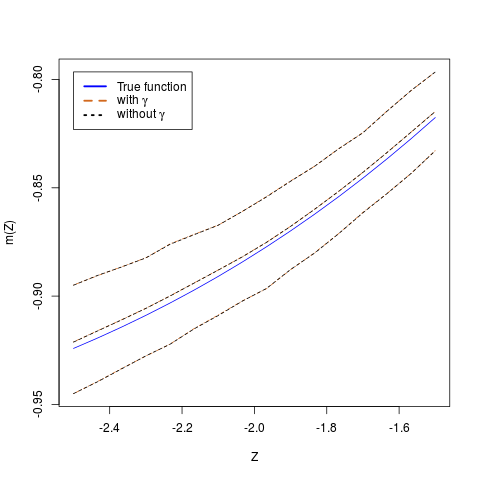}
				\label{fig_m_3.1}	
		\end{minipage}
	\begin{spacing}{0.8}
		\flushleft{\textit{\footnotesize Notes: For the estimation the NW estimator with normal kernel and $h \propto n^{-1/3.5}$ is employed. The $25\%$ and $75\%$ quantiles as well as the mean are reported. 2000 Monte Carlo samples were used for all simulations.}}		
	\end{spacing}	
	\end{figure}

Before we consider a real data application, we close the discussion with Figures \ref{fig_m_1.1} and \ref{fig_m_3.1}. The figures state the estimated $m(Z)$ for Model 1 with $n=250$ and for Model 3 with $n=500$. For the estimation, the NW estimator with same kernel and bandwidth as for the SmoothMD estimator was used. No matter if the SmoothMD estimator with or without $\gamma$ is employed, the results are very accurate. In practice one can of course use cross validation to choose the bandwidth or employ the local linear estimator instead of the NW estimator.
		
\subsection{Real data application} \label{real_data}

We consider in this section an application of our estimator to investigate the returns of social and cognitive skills in the labor market. For this purpose, we apply the proposed transformation partially linear estimator to a dataset studied in \citet{deming2017growing}. In particular, we consider the regressions (4) and (5) in TABLE I of \citet{deming2017growing} that are based on the National Longitudinal Survey of Youth 1979 (NLSY79). NLSY79 is a nationally representative sample taken in the US, of young people aged from 14 to 22. The survey was conducted yearly from 1979 to 1993 and biannually from 1994 through 2012. 
\citet{deming2017growing} estimates the model 
	\begin{equation}
		\begin{aligned}
			\log(wage_{ijt}) &= \alpha + \beta_1 \cdot COG_i + \beta_2 \cdot SS_i + \beta_3 \cdot COG_i \times SS_i + \beta_4 \cdot NCOG_i \\
							 &  \quad + \boldsymbol{C}_{ijt}^T \boldsymbol{\rho} + \delta_j + \zeta_t  + \varepsilon_{ijt},
		\end{aligned}
		\label{reg_dem}	
		\end{equation}
where $COG$, $SS$ and $NCOG$ denote measures of cognitive, social and noncognitive skills. The model includes controls $\boldsymbol{C}$ for race-by-gender indicators, indicators for region and urbanicity as well as age (indexed by $j$) and year (indexed by $t$) fixed effects.

In his paper, \citet{deming2017growing} develops a theoretical model that is written in levels instead of logs as in equation \eqref{reg_dem}. Nevertheless, he estimates the log-linearized model in his paper to follow standard practice in the literature, as he argues. Results for the model in levels are stated in an online appendix. Therefore, it makes sense to use the Box-Cox transformation for $wage$ and estimate the transformation parameter $\lambda$ together with the remaining model parameters to decide whether the model in logs or in levels is more appropriate.   

Furthermore, we consider an unknown functional form for cognitive and social skills to see if the linear form, $\beta_1 \cdot COG + \beta_2 \cdot SS + \beta_3 \cdot COG \times SS$ used by \citet{deming2017growing}, is reasonable. The transformation partially linear model is, thus, given by
	\begin{align}
		T(wage_{ijt},\lambda) =  m(COG_i, SS_i) + \beta \cdot NCOG_i + \boldsymbol{C}_{ijt}^T \boldsymbol{\rho} + \delta_j + \zeta_t + \varepsilon_{ijt},
		\label{reg_smooth}
	\end{align}
where $m(\cdot)$ is an unknown function. The model stated in \eqref{reg_smooth} is closely related to Model 4 of the simulations in section \ref{sim}, where $Y = wage$, $\boldsymbol Z = (COG, SS)^T$, $X_1 = NCOG$ and $X_2 = (\boldsymbol{C}^T,1,1)^T$.

As proxy for cognitive skills, the Armed Forces Qualifying Test (AFQT) was taken. \citet{deming2017growing} uses raw scores from \citet{altonji2012changes}
and normalizes them to have mean 0 and standard deviation 1. The social skill measure is constructed from the following four variables of the NLSY79:
	\begin{enumerate}
		\item Self-reported sociability in 1981 (extremely shy, somewhat shy, somewhat outgoing, extremely outgoing)
		
		\item Self-reported sociability in 1981 at age 6 (retrospective)
		
		\item The number of clubs in which the respondent participated in high school
		
		\item Participation in high school sports (yes/no).
	\end{enumerate}

Each variable is normalized to have mean 0 and standard deviation 1. The social skill measure is the average of these four normalized variables (also normalized to standard deviation 1). In addition to social and cognitive skill measures \citet{deming2017growing} includes a noncognitive skill measure in his regression. He uses the Rotter Locus of Control and the Rosenberg Self-Esteem Scale as also used by \citet{heckman2006effects}. In the following discussion we use these variables to estimate the models stated in \eqref{reg_dem} and \eqref{reg_smooth}.    

\citet{deming2017growing} used a weighted log-linearized OLS estimator to estimate the returns of cognitive and social skills on wage and excluded respondents under the age of $23$ or who were enrolled in school. The weighting was necessary as in each survey year of the NLSY79 a set of sampling weights was constructed. These weights provided the researcher with an estimate of how many individuals in the United States each respondent's answers represented. We also employ these weights in our analysis.

Table \ref{base_reg} shows the regression results. The first column, (4), provides the results of \citet{deming2017growing} estimating equation \eqref{reg_dem}. The second and third column state the transformation partially linear estimator of equation \eqref{reg_smooth} with and without employing $\gamma$. The fourth and fifth column state the transformation partially linear estimator of equation \eqref{reg_smooth} with and without employing $\gamma$  imposing $\lambda=0$. This is a standard partially linear model as studied by \citet{robinson1988root} and \citet{li1996root}. 

For the inner smoothing of the estimations in column 2-5, we use a normal kernel with $h \propto n^{-1/3.5}$, which is in line with the developed theory and usual bandwidth choices for bivariate smoothing with the Nadaraya-Watson estimator. The components of $\boldsymbol d$ defining the diagonal matrix $\boldsymbol{D}$ in $\boldsymbol{\Omega}_n$ are set equal to the componentwise standard deviations for all continuous variables. In the case of the controls and fixed effects, an indicator of the event that the observations have the same value is employed.

The optimization grid for $\lambda$ is given by $[-0.1, 0.1]$ with step size $0.001$.\footnote{We evaluated subsamples of the dataset before we made the final estimation. The estimated $\lambda$'s in the subsamples are contained in the used grid.} We minimize 
$G_n^{-\lambda} \widehat{\mathbb{Y}}_n(\lambda)^T \;  \widehat {\mathbb{B}} _n \; G_n^{-\lambda} \widehat{\mathbb{Y}}_n(\lambda) $ over the defined grid to get $\widehat \lambda$ and the estimates of the remaining coefficients, where $G_n$
is the sample geometric mean.

The results in the first column of Table \ref{base_reg} show that all \citet{deming2017growing} estimated coefficients are significantly different from $0$. In the remaining four columns, we cannot state parameter estimates for cognitive and social skills and the interaction of both as these variables are contained in $m(\cdot)$. However, the parameter estimates for noncognitive skills are comparable to the estimate from the first column. In addition, the estimates for $\lambda$ with and without $\gamma$ are equal and close to zero which would imply that a log-transformation of the dependent variable is appropriate. The estimated coefficient for noncognitive skills is significantly different from $0$ in all SmoothMD estimations whereas both estimates for $\lambda$ are not significantly different from $0$.

In order to check if the linear specification for cognitive and social skills employed by \citet{deming2017growing} is reasonable, we proceed as follows. We estimate the parameters of the  transformation partially linear model as stated in \eqref{reg_smooth} to obtain the residuals 
$$
	\widehat\varepsilon_{ijt} = T(wage_{ijt},\widehat\lambda) - \widehat\beta \cdot NCOG_i - \boldsymbol{C}_{ijt}^T \widehat{\boldsymbol{\rho}}- \widehat \delta_j - \widehat \zeta_t.
$$
We now estimate the unknown function $m(\cdot)$ by smoothing $\widehat\varepsilon$ with the NW estimator. In addition, we also regress $\widehat\varepsilon$ on $COG$, $SS$ and $COG \times SS$. To see if the linear specification is appropriate, we compare the MSE of the linear and nonlinear estimates. We employ a normal density kernel for the NW estimator and let $h \propto n^{-1/6}$.	

Table \ref{base_OLS} states the results where OLS indicates that we used the linear model to fit the residuals. The MSE of the linear and nonlinear estimates are identical no matter if we use the SmoothMD estimator with or without $\gamma$ to estimate the unknown model parameters. The same holds true for the SmoothMD estimator with or without $\gamma$ where $\lambda = 0$ is imposed. All results show that the linear representation of \citet{deming2017growing} seems to be reasonable.
\begin{table}[H]		
	\begin{adjustwidth}{+0.0cm}{+0.0cm}    
		\caption{\textit{Labor Market Returns to Cognitive and Social Skills in the NLSY79}}
			\begin{tabular}{@{}lccccc}
		    \toprule
		    \midrule
			 Outcome: (log) hourly wage    &(4)  & SmoothMD      & SmoothMD         & SmoothMD                    & SmoothMD  \\ 
			\hskip 1.6cm(in 2012 dollars)  &     & with $\gamma$ & without $\gamma$ & with $\gamma$, $\lambda = 0$ & without $\gamma$, $\lambda = 0$ \\ 			 
		    \midrule
		    $\lambda$                      & -        & \multicolumn{1}{l}{\;-0.007}    & \multicolumn{1}{l}{\;-0.007} & -           & -  \\ 
		    		    		           &          & [0.005]                         & [0.005]                      &             &    \\
		    Cognitive skills               & 0.189*** & -                               & -                            & -           & -  \\ 
		    		                       & [0.007]  &                                 &                              &             &    \\
		    Social skills                  & 0.043*** & -                               & -        					   & -           & -  \\ 
		    		    		           & [0.006]  &                                 &          					   &             &    \\
		    Cognitive $\times$ Social      & 0.019*** & -                               & -        					   & -           & -  \\
		    		    		           & [0.006]  &                                 &          					   &             &     \\ 
		    Noncognitive skills            & 0.048*** & 0.047***                        & 0.047*** 					   & 0.048***    & 0.048***  \\ 
		    		    		           & [0.006]  & [0.004]                         & [0.004]  					   & [0.004]     & [0.004]\\
		    Demographics and age/          & X        & X                               & X        					   & X           & X  \\ 
		    year fixed effects             &          &                                 &          					   &             &    \\ 
		    Number of Observations         & 126191   & 126191                          & 126191   					   & 126191      & 126191 \\
		    \midrule
		    \bottomrule
	\end{tabular}
		\vskip 0.1cm		
		\textit{Notes: The data source is the National Longitudinal Survey of Youth 1979 cohort (NLSY79). (4) denotes the OLS regression proposed by \citet{deming2017growing}. In all SmoothMD estimations, $h \propto n^{-1/3.5}$. The components of $\boldsymbol d$ are set equal to the componentwise standard deviations for all continuous variables and for controls and fixed effects, an indicator of the event that the observations have the same value is employed. The grid for $\lambda$ is $[-0.1,0.1]$ and $s = G_n$. Cognitive skills are measured by each NLSY79 respondent's score on the Armed Forces Qualifiying Test (AFQT) and are normalized to have mean 0 and standard deviation 1. The AFQT score crosswalk of \citet{altonji2012changes} is used. Social skill is a standardized composite of four variables, (i) sociability in childhood, (ii) sociability in adulthood, (iii) participation in high school clubs and (iv) participation in team sports; see the text and \citet{deming2017growing} for details on the construction of the social skills measure. The noncognitive skills measure is the normalized average of the Rotter and Rosenberg scores in the NLSY. The regressions also control for race-by-gender indicator variables, age, year, census region and urbanicity. Standard errors are in brackets and are clustered at the individual level for (4). The remaining standard errors are estimated by the Eiker-White variance estimator. ***$p < .01$, **$p < .05$, *$p < .1$}
		\label{base_reg}
     \end{adjustwidth}   
\end{table}

\begin{table}[H]
		\caption{\textit{MSE of estimated nonlinear part in the transformation partially linear model}}
		\begin{tabular}{@{}lcccccccc}
	  \toprule \midrule
	   & \multicolumn{2}{c}{SmoothMD} &\multicolumn{2}{c}{SmoothMD} & \multicolumn{2}{c}{SmoothMD} &\multicolumn{2}{c}{SmoothMD}\\
	   & \multicolumn{2}{c}{with $\gamma$} &\multicolumn{2}{c}{without $\gamma$} & \multicolumn{2}{c}{with $\gamma$, $\lambda = 0$} &\multicolumn{2}{c}{without $\gamma$, $\lambda = 0$}\\	   
	  \cmidrule(lr){2-3} \cmidrule(lr){4-5}  \cmidrule(lr){6-7} \cmidrule(lr){8-9} 
                              & OLS & NW & OLS& LL & OLS & NW & OLS& LL\\ 
	 \midrule
	 MSE                          & 0.282  & 0.277  & 0.282  & 0.277   & 0.293  & 0.288  & 0.293  & 0.288   \\
	 Number of Observations       & 126191 & 126191 & 126191 & 126191  & 126191 & 126191 & 126191 & 126191  \\  
	 \midrule
	 \bottomrule
\end{tabular}
		\vskip 0.1cm		
	\textit{Notes: For the NW estimator a normal kernel with $h \propto n^{-1/6}$ is employed. OLS indicates that the linear model is used to fit the residuals.}
		 \label{base_OLS}
\end{table}

In a second step, we include \textit{years of completed education} as an additional explanatory variable in the regression models. In one of his estimations \citet{deming2017growing} controls for years of education as well. Table \ref{edu_reg} states the regression results for all considered  models. 
The first column, (5), provides the results of \citet{deming2017growing} estimating equation \eqref{reg_dem} with \textit{years of completed education} as control. The results show that all  \citet{deming2017growing} estimated coefficients are significantly different from $0$. However, the coefficients become smaller compared to the first specification. In addition, the coefficient of the interactive effect is only significant at the $10\%$ level. In the remaining four columns, the parameter estimates for noncognitive skills are comparable to the estimate from the first column. In addition, the estimates for $\lambda$ with and without $\gamma$ are equal and close to zero which would imply that a log-transformation of the dependent variable is appropriate. The estimated coefficient for noncognitive skills is significantly different from $0$ in all SmoothMD estimations whereas both estimates for $\lambda$ are not significantly different from $0$. 

Table \ref{edu_OLS} states the MSE of the estimated nonlinear part in the transformation partially linear models. The MSE of the linear and nonlinear estimates are identical no matter if we use the SmoothMD estimator with or without $\gamma$ to estimate the unknown model parameters. The same holds true for the SmoothMD estimator with or without $\gamma$ where $\lambda = 0$ is imposed. All results show that the linear representation of \citet{deming2017growing} seems to be reasonable.

	\begin{table}[H]		
		\begin{adjustwidth}{+0.0cm}{+0.0cm}    
			\caption{\textit{Labor Market Returns to Cognitive and Social Skills in the NLSY79 controlling for education}}
				\begin{tabular}{@{}lccccc}
		    \toprule
		    \midrule
			 Outcome: (log) hourly wage    &(5)  & SmoothMD      & SmoothMD         & SmoothMD                    & SmoothMD  \\ 
			\hskip 1.6cm(in 2012 dollars)  &     & with $\gamma$ & without $\gamma$ & with $\gamma$, $\lambda = 0$ & without $\gamma$, $\lambda = 0$ \\ 			 
		    \midrule
		    $\lambda$                      & -        & 0.002~~~~~   & 0.002~~~~~   & -           & -  \\ 
		    		    		           &          & [0.005]      & [0.005]      &             &    \\
		    Cognitive skills               & 0.126*** & -            & -            & -           & -  \\ 
		    		                       & [0.008]  &              &              &             &    \\
		    Social skills                  & 0.029*** & -            & -            & -           & -  \\ 
		    		    		           & [0.006]  &              &              &             &    \\
		    Cognitive $\times$ Social      & 0.011*~~~& -            & -            & -           & -  \\
		    		    		           & [0.006]  &              &              &             &     \\ 
		    Noncognitive skills            & 0.040*** & 0.037***     & 0.037***     & 0.037***    & 0.037***  \\ 
		    		    		           & [0.006]  & [0.004]      & [0.004]      & [0.004]     & [0.004]\\
		    Demographics and age/          & X        & X            & X            & X           & X      \\ 
		    year fixed effects             &          &              &              &             &        \\ 
		    Years of completed education   & X        & X            & X            & X           & X      \\  
		    Number of Observations         & 126191   & 126191       & 126191       & 126191      & 126191 \\
		    \midrule
		    \bottomrule
	\end{tabular}
			\vskip 0.1cm		
			\textit{Notes: The data source is the National Longitudinal Survey of Youth 1979 cohort (NLSY79). (5) denotes the OLS regression proposed by \citet{deming2017growing}. In all SmoothMD estimations, $h \propto n^{-1/3.5}$. The components of $\boldsymbol d$ are set equal to the componentwise standard deviations for all continuous variables and for controls and fixed effects, an indicator of the event that the observations have the same value is employed. The grid for $\lambda$ is $[-0.1,0.1]$ and $s = G_n$. Cognitive skills are measured by each NLSY79 respondent's score on the Armed Forces Qualifiying Test (AFQT) and are normalized to have mean 0 and standard deviation 1. The AFQT score crosswalk of \citet{altonji2012changes} is used. Social skill is a standardized composite of four variables (i) sociability in childhood, (ii) sociability in adulthood, (iii) participation in high school clubs and (iv) participation in team sports; see the text and \citet{deming2017growing} for details on the construction of the social skills measure. The noncognitive skills measure is the normalized average of the Rotter and Rosenberg scores in the NLSY. The regressions also control for race-by-gender indicator variables, age, year, census region, urbanicity and years of completed education. Standard errors are in brackets and are clustered at the individual level for (5). The remaining standard errors are estimated by the Eiker-White variance estimator. ***$p < .01$, **$p < .05$, *$p < .1$}
			\label{edu_reg}
	     \end{adjustwidth}   
	\end{table}

\begin{figure}[h]
			\begin{minipage}{0.4\textwidth} 
				\caption{\textit{Estimated Labor Market Returns to\\ Cognitive  
				and Social Skills in the NLSY79.}}
				\includegraphics[scale=0.8]{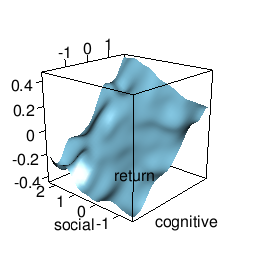}
				\label{fig_app_1.1}
			\end{minipage}
			\hfill
			\begin{minipage}{0.4\textwidth}
				\caption{\textit{Estimated Labor Market Returns to \\ Cognitive 
				and Social Skills in the NLSY79 \\ controlling  for education.}}
				\includegraphics[scale=0.8]{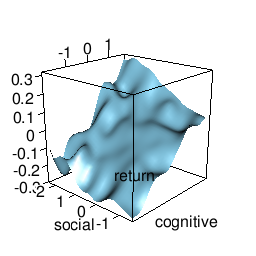}
				\label{fig_app_2.2}	
			\end{minipage}
		\begin{spacing}{0.8}
			\flushleft{\textit{\footnotesize Notes: The coefficients are estimated by the SmoothMD estimator with $\gamma$. For the NW estimator a normal kernel with $h \propto n^{-1/6}$ is employed.}}		
		\end{spacing}	
		\end{figure}

Before we close the section, we plot the estimated labor market returns to cognitive and social skills of model \eqref{reg_smooth} with and without controlling for years of completed education. The returns are estimated with the NW estimator employing a normal kernel with $h \propto n^{-1/6}$. Figures \ref{fig_app_1.1} and \ref{fig_app_2.2} present the results. Note that the mean was subtracted. The return increases no matter if the social or cognitive indicator is increased. However, the cognitive effect seems to be stronger. In addition, if we control for education, it seems that being too social might sometimes lower the wage somewhat. Nevertheless, both plots confirm that the linear model used by \citet{deming2017growing} is reasonable.

	\begin{table}[H]
			\caption{\textit{MSE of estimated nonlinear part in the transformation partially linear model controlling for education}}
			\begin{tabular}{@{}lcccccccc}
	  \toprule \midrule
	   & \multicolumn{2}{c}{SmoothMD} &\multicolumn{2}{c}{SmoothMD} & \multicolumn{2}{c}{SmoothMD} &\multicolumn{2}{c}{SmoothMD}\\
	   & \multicolumn{2}{c}{with $\gamma$} &\multicolumn{2}{c}{without $\gamma$} & \multicolumn{2}{c}{with $\gamma$, $\lambda = 0$} &\multicolumn{2}{c}{without $\gamma$, $\lambda = 0$}\\	   
	  \cmidrule(lr){2-3} \cmidrule(lr){4-5}  \cmidrule(lr){6-7} \cmidrule(lr){8-9} 
                              & OLS & NW & OLS& LL & OLS & NW & OLS& LL\\ 
	 \midrule
	 MSE                          & 0.288  & 0.283  & 0.288  & 0.283   & 0.284  & 0.280  & 0.284  & 0.280   \\
	 Number of Observations       & 126191 & 126191 & 126191 & 126191  & 126191 & 126191 & 126191 & 126191  \\  
	 \midrule
	 \bottomrule
\end{tabular}
			\vskip 0.1cm		
			\textit{Notes: For the NW estimator a normal kernel with $h \propto n^{-1/6}$ is employed. OLS indicates that the linear model is used to fit the residuals.}
			\label{edu_OLS}
	\end{table}

\section{Discussion} \label{discussion}

In this paper, we study the semiparametric partially linear model with Box-Cox transformed dependent variable. We follow the SmoothMD approach introduced by \citet{lavergne2013smooth}, which is based on conditional moment restrictions, and we extend it to the case with an infinite-dimensional nuisance parameter. Our results are new both for the transformation regression models and the semiparametric partially linear model. We establish model identification,  consistency as well as $\sqrt{n}$-asymptotic normality. In addition, we proposed a distance metric statistic to test the model parameters. A Monte Carlo experiment showed the usefulness of the proposed estimator in finite samples. An application to a large real data sample already studied in Labor Economics is also reported. 

The SmoothMD approach is a convenient approach for nonlinear regression models. It could be interpreted as a generalized least-squares method where the weights are given by a suitable positive-definite matrix with each entry representing a measure of discrepancy between a pair of covariate vectors.  It is unnecessary to  localize the measure of discrepancy between the observed covariate vectors, and this represents a significant advantage for the practitioner who thus avoids the choice of an additional tuning parameter. One price to pay is on the semiparametric efficiency for the estimators of the finite dimensional parameters. However, \citet{lavergne2013smooth} showed that a two-step procedure, where the first step allows nonparametrically suitable weights to be estimated for building an asymptotically optimal measure of discrepancy, allows to achieve semiparametric efficiency. An efficient estimator would also induce a distance metric test statistic with a usual asymptotic chi-square distribution under the null hypothesis. We expect that the same results extend to the present framework. However, with our semiparametric model, the two-step procedure would likely result in a  numerically unstable, complex to calibrate, inference procedure. We  believe this theoretical, and quite technical, refinement to be of little use for the applications, and we therefore do not consider it. 

Another slight drawback for using SmoothMD with a fixed weighting matrix, is the condition that $\boldsymbol Z$ should be of dimension $q$ less than or equal to 3. This restriction would not be binding in most applications. However, if necessary, one could use higher-order kernels to diminish the bias  induced by the nonparametric estimation of the nuisance parameter, and thus allow for larger $q$. It is noticeable that SmoothMD does not involve any denominator and therefore the higher-order kernels would not induce the numerical problems, due to division by zero, usually encountered in semiparametric methods.

\addcontentsline{toc}{section}{Acknowledgments}
\section*{Acknowledgments}

This research was supported by the DFG through KN 567/5-1 and the Hausdorff Center for Mathematics. We furthermore thank the Regional Computing Center of the 
University of Cologne (RRZK) for providing computing time on the DFG-funded High Performance Computing (HPC) system CHEOPS as well as support. This work was started while the first author was visiting CREST-ENSAI. V. Patilea acknowledges support from ‘Models and mathematical processing of very large data’, a Joint Research Initiative under the aegis of Risk Foundation, with partnership of MEDIAMETRIE and GENES, and from the Romanian Minister of Education and Research, CNCS – UEFISCDI, project number PN-III-P4-ID-PCE-2020-1112, within PNCDI III.

\clearpage
\newpage

\begin{appendix}
\numberwithin{equation}{section}
\section*{Appendix} 
\addcontentsline{toc}{section}{Appendix}
\setcounter{section}{1} 
\setcounter{equation}{0}

\subsection*{Appendix A: Proofs}
\addcontentsline{toc}{subsection}{Appendix A: Proofs}

\begin{proofof}{ the statement that $Var\left[(\boldsymbol{X}^T, \boldsymbol{Z}^T)^T\right]$ has  full rank}\quad\\
Indeed, for any $\boldsymbol u \in\mathbb{R}^p$ and $\boldsymbol v \in\mathbb{R}^q$ such that  $(\boldsymbol u^T, \boldsymbol v^T)^T\neq  \boldsymbol 0_{p+q},$ we can write 
	$$
		Var\left[ \boldsymbol  u ^T\boldsymbol{X} + \boldsymbol v ^T  \boldsymbol{Z} \right] = 
		E \left[ Var\left[ \boldsymbol  u ^T(\boldsymbol{X} -E\left[\boldsymbol{X}\mid \boldsymbol{Z} \right] )\mid  \boldsymbol{Z}\right] \right]  + Var\left[ \boldsymbol  u ^TE\left[\boldsymbol{X}\mid \boldsymbol{Z} \right] + \boldsymbol v ^T  \boldsymbol{Z} \right] .
	$$
If $\boldsymbol u\neq  \boldsymbol 0_{p}$,
	\begin{multline*}
		Var\left[ \boldsymbol  u ^T\boldsymbol{X} + \boldsymbol v ^T  \boldsymbol{Z} \right] \geq 
		E \left[ Var\left[ \boldsymbol  u ^T(\boldsymbol{X} -E\left[\boldsymbol{X}\mid \boldsymbol{Z} \right] )\mid  \boldsymbol{Z}\right] \right] \\ =\boldsymbol  u ^T E \left[ Var\left[ \boldsymbol{X} -E\left[\boldsymbol{X}\mid \boldsymbol{Z} \right]\mid  \boldsymbol{Z}\right] \right]\boldsymbol  u = \boldsymbol  u ^T Var\left[ \boldsymbol{X} -E\left[\boldsymbol{X}\mid \boldsymbol{Z} \right]\right]\boldsymbol  u>0,
	\end{multline*}
where the last inequality is guaranteed by Assumption \ref{ass_ident}.1. When $\boldsymbol u =  \boldsymbol 0_{p}$, we obtain
	$$
		Var\left[ \boldsymbol  u ^T\boldsymbol{X} + \boldsymbol v ^T  \boldsymbol{Z} \right]  = Var\left[ \boldsymbol v ^T  \boldsymbol{Z} \right] = \boldsymbol v ^T Var\left[ \boldsymbol{Z} \right]\boldsymbol v >0,
	$$
where the last inequality holds because  $\boldsymbol v \neq \boldsymbol 0_{q}$ and $Var\left[\boldsymbol{Z}\right]$ has necessarily full rank provided $ \boldsymbol{Z}$ admits a density. \\
\end{proofof}

\begin{proofof}{Lemma \ref{lem_ident}}\quad\\
By  construction, $\gamma(\lambda_0)=0$ and $\boldsymbol {\beta} (\lambda_0) = \boldsymbol\beta_0$. Following the lines of \citet{shin2008semiparametric}, for any $(\gamma, \lambda,\boldsymbol \beta^T)^T $ we note that  
	\begin{multline*}
		\mathbb{P} \bigg(  E\left[ \left(T(Y,\lambda) - E[T(Y,\lambda)\mid \boldsymbol Z] \right) f_z(\boldsymbol Z)  - \gamma - (\boldsymbol X - E[\boldsymbol X\mid\boldsymbol Z])^T\boldsymbol \beta  f_z(\boldsymbol Z)  \mid\boldsymbol X,\boldsymbol Z  \right] = 0 \bigg)\\		
		=
		\mathbb{P} \bigg(  E\left[ T(Y,\lambda) - T(Y,\lambda_0)|\boldsymbol X,\boldsymbol Z  \right] - \boldsymbol X^T(\boldsymbol \beta-\boldsymbol \beta_{0})  =  E[ T(Y,\lambda)\mid\boldsymbol Z] - E[T(Y,\lambda_0)|\boldsymbol Z]\\ 
	                     +   \gamma f^{-1}_z(\boldsymbol Z) - E[\boldsymbol X\mid\boldsymbol Z]^T  (\boldsymbol\beta -\boldsymbol \beta_{0})  \bigg). 
	\end{multline*}
Hence, it suffices to prove that the last probability could not be equal to 1 when $(\gamma,\boldsymbol \theta^T)^T\neq (0,\boldsymbol {\theta}_0^T)^T$. 
Note that 
	\begin{align*}
		E[ T(Y,\lambda)\mid\boldsymbol Z] - E[T(Y,\lambda_0)\mid\boldsymbol Z]  + \gamma f^{-1}_z(\boldsymbol Z) - E[\boldsymbol X\mid\boldsymbol Z]^T  (\boldsymbol\beta -\boldsymbol \beta_{0}) ,
	\end{align*}
does not depend on $\boldsymbol X$ anymore but only on $\boldsymbol Z$.

If $\lambda = \lambda_0$ the result follows immediately from the full rank condition in Assumption \ref{ass_ident}.1. Indeed, by the variance decomposition formula and Assumption \ref{ass_ident}.1, for any $\boldsymbol a\in\mathbb{R}^p$, $\boldsymbol a\neq \boldsymbol 0_p$, 
$$
  \boldsymbol a^T Var(\boldsymbol X - E[\boldsymbol X|\boldsymbol Z]) \boldsymbol a = E\left[  Var(\boldsymbol a^T(\boldsymbol X - E[\boldsymbol X|\boldsymbol Z] ) \mid \boldsymbol Z) \right] >0.
$$
  This implies 
	\begin{multline*}
		\boldsymbol a^T Var(f_z(\boldsymbol Z) \left(\boldsymbol X - E[\boldsymbol X|\boldsymbol Z]\right) ) \boldsymbol a = E\left[ f^2_z(\boldsymbol Z)  \boldsymbol a^T Var(\left(\boldsymbol X - E[\boldsymbol X|\boldsymbol Z]\right) \mid \boldsymbol Z) \boldsymbol a^T \right] \\ = E\left[  f^2_z(\boldsymbol Z)   Var(\boldsymbol a^T\left(\boldsymbol X - E[\boldsymbol X|\boldsymbol Z] \right) \mid \boldsymbol Z) \right] >0.
	\end{multline*}
Thus, $f_z(\boldsymbol Z) (\boldsymbol X - E[\boldsymbol X\mid\boldsymbol Z]^T)^T  (\boldsymbol\beta -\boldsymbol \beta_{0})$ cannot be equal to a constant almost surely, as is necessarily the case when $\lambda=\lambda_0$. 

Next, consider  the case $\lambda \neq \lambda_0$. Without loss of generality, we could  assume that $\lambda > \lambda_0$.
\begin{enumerate}
	\item Consider the case  $\boldsymbol \beta \neq \boldsymbol \beta_{0} $ and let introduce the event   
		\begin{multline*}
			\mathcal{E}= 	\left\{   E\left[ T(Y,\lambda) - T(Y,\lambda_0)\mid \boldsymbol X, 
		                                \boldsymbol Z 
		                                \right] - \boldsymbol X^T(\boldsymbol \beta-\boldsymbol \beta_{0}) \right.\\\left.
			  = E[ T(Y,\lambda)\mid \boldsymbol Z]  - E[T(Y,\lambda_0)|\boldsymbol Z]  +  \gamma f^{-1}_z(\boldsymbol Z) - E[\boldsymbol X\mid\boldsymbol Z]^T  (\boldsymbol\beta -\boldsymbol \beta_{0})\right\}.
		\end{multline*}				
	
	Taking conditional expectation given $\boldsymbol Z$ on both sides, we deduce that on $\mathcal{E}$ we necessarily have $\gamma=0$. Thus it suffices to investigate the probability of the larger event 
		\begin{multline*}
			\mathcal{E}^\prime  =  \left\{ 		  E\left[ T(Y,\lambda) - T(Y,\lambda_0)\mid \boldsymbol X, 
		                                           \boldsymbol Z 
		                                           \right] - \boldsymbol X^T(\boldsymbol \beta-\boldsymbol \beta_{0}) \right.\\\left.
			  =  E[ T(Y,\lambda)\mid \boldsymbol Z
		        ]  - E[T(Y,\lambda_0)|\boldsymbol Z
		               ]   - E[\boldsymbol X\mid\boldsymbol Z]^T  (\boldsymbol\beta -\boldsymbol \beta_{0})\right\}.
		\end{multline*}
	Note that the right hand side equality does not depend on  $\boldsymbol X$. We distinguish two sub-cases. First, the case where the components of $(\boldsymbol \beta-\boldsymbol \beta_{0})$ corresponding to $\boldsymbol X_c$ are equal to zero. Thus  the linear combination $\boldsymbol X^T(\boldsymbol \beta-\boldsymbol \beta_{0})$ does not include any of the continuous components of $\boldsymbol X$. In this case, for any value of $\boldsymbol Z$, the support of $\boldsymbol X^T(\boldsymbol \beta-\boldsymbol \beta_{0})$ is finite and independent of the value of $\boldsymbol X_c$. Then Assumption \ref{ass_ident}.4 guarantees that the probability of the event $\mathcal{E}^\prime $ could not be equal to 1.  Next, consider the case where   $\boldsymbol X^T(\boldsymbol \beta-\boldsymbol \beta_{0})$  includes  continuous components of $\boldsymbol X$. In this case,  by  Assumption \ref{ass_ident}.3, the support of the variable $\boldsymbol X^T(\boldsymbol \beta-\boldsymbol \beta_{0})$ is the whole real line and, by the monotonicity  of $\lambda\mapsto T(y;\lambda)$ for each value $y>0$, $E\left[ T(Y,\lambda) - T(Y,\lambda_0)|\boldsymbol X,\boldsymbol Z \right] \geq 0$ almost surely, the statement follows again.
	
	\item Consider the case $\gamma\neq 0$ and $\boldsymbol \beta = \boldsymbol \beta_{0} $. In this case
	$$
	  f_z(\boldsymbol Z) \left(E\left[ T(Y,\lambda) - T(Y,\lambda_0)\mid \boldsymbol X,    \boldsymbol Z   \right] - E\left[ T(Y,\lambda) - T(Y,\lambda_0)\mid     \boldsymbol Z   \right]  \right) = \gamma  .
	$$	  
	  Taking expectation on both sides, we deduce that necessarily $\gamma=0$.

	\item Consider the case $(\gamma,\boldsymbol \beta^T)^T = (0, \boldsymbol \beta_{0}^T)^T $. Then necessarily 
		\begin{align*}
			E\bigg[ T(Y,\lambda) - T(Y,\lambda_0)|\boldsymbol X,\boldsymbol Z \bigg]  
			=  E[ T(Y,\lambda)|\boldsymbol Z] - E[T(Y,\lambda_0)|\boldsymbol Z] \quad \text{almost surely}.
		\end{align*}
	
	Once again the right hand side does not depend on $\boldsymbol X$, and thus the probability of the event $\mathcal{E} $ could not be equal to 1 because of Assumption  \ref{ass_ident}.3. 					 
\end{enumerate}
Therefore, the first statement follows. Consider now the second statement of the Lemma \ref{lem_ident}. First, note that the maps $\lambda\mapsto \gamma(\lambda)$, $\lambda\mapsto \beta(\lambda)$ and 
	\begin{equation}\label{third_map}
	  \lambda \mapsto E\left[g\left(\boldsymbol U_1; 
	  \boldsymbol \theta (\lambda),
	  \gamma (\lambda), \boldsymbol \eta_{\lambda,1} \right)g\left(\boldsymbol U_2; (\lambda, \boldsymbol \beta (\lambda)^T)^T \!, \gamma (\lambda), \boldsymbol \eta_{\lambda,2}\right) \exp \left\{- (\boldsymbol W_1 - \boldsymbol W_2)^T \boldsymbol D (\boldsymbol W_1 - \boldsymbol W_2)\right\}  \right] , 
	\end{equation}
with   $\lambda \in\Lambda : = [\lambda_{\rm min},\lambda_{\rm max}]$ and $\boldsymbol \theta (\lambda) = (\lambda,\boldsymbol \beta(\lambda)^T)^T$, are continuous. Indeed, it is quite clear that $\inf_{\boldsymbol d \in \mathcal D} E[\omega ( \boldsymbol W_1 - \boldsymbol W_2)] >0$. Next, by Lebesgue's Dominated Convergence Theorem we have 
$$
  \gamma (\lambda) = \frac{1}{E[\omega ( \boldsymbol W_1 - \boldsymbol W_2)]}
	E\left[ \left(T(Y_1,\lambda) - E[T(Y_1,\lambda)\mid \boldsymbol Z_1]- (\boldsymbol X_1-E[\boldsymbol X_1\mid \boldsymbol Z_1])^T\boldsymbol \beta\right)f_z(\boldsymbol Z_1) \omega ( \boldsymbol W_1 - \boldsymbol W_2) \right],
$$
  and 
	\begin{multline*}
		\boldsymbol\beta (\lambda) =E\left[ (\boldsymbol X_1-E[\boldsymbol X_1\mid \boldsymbol Z_1])(\boldsymbol X_2-E[\boldsymbol X_2\mid \boldsymbol Z_2])^Tf_z(\boldsymbol Z_1) f_z(\boldsymbol Z_2)\omega ( \boldsymbol W_1 - \boldsymbol W_2) \right]^{-1} \\
		E\left[(\boldsymbol X_1-E[\boldsymbol X_1\mid \boldsymbol Z_1])f_z(\boldsymbol Z_1) \left(\left(T(Y_2,\lambda) - E[T(Y_2,\lambda)\mid \boldsymbol Z_2]\right)f_z(\boldsymbol Z_2) - \gamma (\lambda)\right)
	       \omega ( \boldsymbol W_1 - \boldsymbol W_2) \right],
	\end{multline*}
which are clearly continuous. Again, by dominated convergence argument,  the continuity of the  map \eqref{third_map} follows. Finally, by the same inverse Fourier Transform argument used by \citet{lavergne2013smooth} we get that
	\begin{multline*}
		E\left[g\left(\boldsymbol U_1; \boldsymbol \theta,\gamma, \boldsymbol \eta_{\lambda,1} \right)g\left(\boldsymbol U_2; \boldsymbol \theta,\gamma, \boldsymbol \eta_{\lambda,2}\right)\omega ( \boldsymbol W_1 - \boldsymbol W_2) \right] = \frac{\pi^{-(p+q)/2}}{\sqrt{d_1\cdots d_{p+q}}} \\ \times  \int_{\mathbb{R}^{p+q}} \left|E\left[E [g\left(\boldsymbol U; \boldsymbol \theta,\gamma, \boldsymbol \eta_\lambda\right) \mid \boldsymbol X, \boldsymbol Z ] \exp\left\{2 i \pi \boldsymbol{w}^T \left( \boldsymbol X^T,\boldsymbol Z^T \right)^T\right\} \right] \right|^2 \exp\left\{- \boldsymbol{w}^T \boldsymbol D^{-1}  \boldsymbol{w}\right\}d\boldsymbol{w}.
	\end{multline*} 
For any $\boldsymbol x, \boldsymbol z$, the map $\lambda\mapsto E [g\left(\boldsymbol U; \boldsymbol \theta(\lambda),\gamma (\lambda),\boldsymbol \eta_\lambda \right) \mid \boldsymbol X=\boldsymbol x, \boldsymbol Z = \boldsymbol z ]$ is continuous. By Lebesgue Dominated Convergence Theorem, the map
$$
  \lambda \mapsto  \int_{\mathbb{R}^{p+q}} \left|E\left[E [g\left(\boldsymbol U; \boldsymbol \theta (\lambda),\gamma (\lambda), \boldsymbol \eta_\lambda \right) \mid \boldsymbol X, \boldsymbol Z ] \exp\left\{2 i \pi \boldsymbol{w}^T \left( \boldsymbol X^T,\boldsymbol Z^T \right)^T\right\} \right] \right|^2 \exp\left\{- \boldsymbol{w}^T {\rm diag}(d_L,\ldots,d_L)^{-1}  \boldsymbol{w}\right\}d\boldsymbol{w},
$$
  is continuous, and thus attains its minimum on the  compact set $\Lambda\setminus [\lambda_0 -\varepsilon,\lambda_0+ \varepsilon]$. The minimum value is necessarily positive. Since $(d_1\cdots d_{p+q})^{-1/2}\exp\left\{- \boldsymbol{w}^T \boldsymbol D  \boldsymbol{w}\right\} \geq d_U^{-(p+q)/2}\exp\left\{- \boldsymbol{w}^T  {\rm diag}(d_L,\ldots,d_L)^{-1} \boldsymbol{w} \right\}$, the last statement in the Lemma follows. 
\end{proofof}

\medskip

\begin{proofof}{Lemma \ref{omega_mat}}~~\\
\begin{enumerate}
\item[1.] First, we note that 
	\begin{equation}\label{sq1}
		\mathbb{P}\left(  \left\{ \boldsymbol u^T \boldsymbol{X} + \boldsymbol v^T \boldsymbol{Z} = 0 , \;\;\forall (\boldsymbol u^T, \boldsymbol v^T)^T \text{ with norm equal to } 1 \right\} \right) =  0. 
	\end{equation}
This is a consequence of the fact that $\boldsymbol{W}$ is not degenerated. Given a sample $\left( \boldsymbol X^T_1,\boldsymbol Z^T_1 \right)^T,\ldots, \left( \boldsymbol X^T_n,\boldsymbol Z^T_n \right)^T$, and a vector $\boldsymbol{a}= (a_1,\ldots,a_n)\in\mathbb{R}^{n}$, using the inverse Fourier Transform, we could write 
$$
  \boldsymbol{a}^T \boldsymbol\Omega_n \boldsymbol{a} = \frac{\pi^{-(p+q)/2}}{\sqrt{d_1\cdots d_{p+q}}}  \int_{\mathbb{R}^{p+q}} \left|\sum_{j=1}^n a_j \exp\left\{2 i \boldsymbol{w}^T \left( \boldsymbol X^T_j,\boldsymbol Z^T_j \right)^T\right\}  \right|^2 \exp\left\{- \boldsymbol{w}^T \boldsymbol D^{-1}  \boldsymbol{w}\right\}d\boldsymbol{w},
$$
  where $\boldsymbol D = {\rm diag}(d_1,\ldots,d_{p+q})$ with $d_1,\ldots,d_{p+q}\in [d_L,d_U]$; see Assumption \ref{ass_dgp}.3. Then, necessarily
	\begin{equation}\label{sq2}
	 	\boldsymbol{a}^T \boldsymbol\Omega_n \boldsymbol{a} = 0 \Longrightarrow  	\sum_{j=1}^n a_j \exp\left\{2 i  \pi \boldsymbol{w}^T \left( \boldsymbol X^T_j,\boldsymbol Z^T_j \right)^T\right\} = 0, \qquad \forall \boldsymbol{w}\in \mathbb{R}^{p+q}.
	\end{equation}
Equation \eqref{sq1} indicates that, with probability 1, the unique vector $\boldsymbol{a}$  which satisfies the right-hand side of \eqref{sq2} is $\boldsymbol{a}=\boldsymbol{0}_{n}$ .
This means that, with probability 1, the matrix $\boldsymbol\Omega_n$ is positive definite. 
																																																				
Next, we use the following Cauchy-Schwarz\footnote{A similar so-called Cauchy-Schwarz inequality was proposed by \citet{CSLavergne}. To justify the statement, it suffices to notice that 
  $
    \boldsymbol{B}^T\boldsymbol{B} - \boldsymbol{B}^T\boldsymbol{A} (\boldsymbol{A}^T\boldsymbol{A})^{-1} \boldsymbol{A}^T\boldsymbol{B} = \boldsymbol \Gamma ^T \boldsymbol \Gamma
    $
      with
    $
      \boldsymbol \Gamma = \boldsymbol{B} - \boldsymbol{A} (\boldsymbol{A}^T\boldsymbol{A})^{-1} \boldsymbol{A}^T\boldsymbol{B}.
      $
} inequality for matrices: 
  let $\boldsymbol{A} \in \mathbb{R}^{n \times p_1}$ such that $\boldsymbol{A}^T\boldsymbol{A}$ is invertible  and let  $\boldsymbol{B}\in \mathbb{R}^{n \times p_2}$. 
Then 
$$
  \boldsymbol{B}^T\boldsymbol{B} - \boldsymbol{B}^T\boldsymbol{A} (\boldsymbol{A}^T\boldsymbol{A})^{-1} \boldsymbol{A}^T\boldsymbol{B}\; \text{ is positive semi-definite. }
$$
  Moreover, the equality 
$
  \boldsymbol{B}^T\boldsymbol{B} = \boldsymbol{B}^T\boldsymbol{A} (\boldsymbol{A}^T\boldsymbol{A})^{-1} \boldsymbol{A}^T\boldsymbol{B}
$
  is equivalent to the relationship\\
$
  \boldsymbol{B} = \boldsymbol{A} (\boldsymbol{A}^T  \boldsymbol{A} )^{-1} \boldsymbol{A}^T \boldsymbol{B}.
$
  For any non null vector  $\boldsymbol u \in\mathbb{R}^p$, taking 
$$
  \boldsymbol{B} = \boldsymbol\Omega_n^{1/2}   \quad \text{ and  } \boldsymbol{A} = \boldsymbol\Omega_n^{1/2} \boldsymbol{1}_n,  
$$
  we deduce that $ \mathbb{D}_n$ is positive semi-definite and thus $\boldsymbol u^T \widehat{\mathbb{X}}_n^T  \mathbb{D}_n \widehat{\mathbb{X}}_n \boldsymbol u \geq 0 $. (Herein, $\boldsymbol\Omega_n^{1/2} $ is the positive definite square root of $\boldsymbol\Omega_n $.) Meanwhile, by elementary matrix algebra, we deduce that, for any  $\boldsymbol{a}\in \mathbb{R}^n$, 
$$
  \boldsymbol{a}^T \mathbb{D}_n \boldsymbol{a} = 0 \quad \Leftrightarrow \quad \left( \boldsymbol{a}^T \boldsymbol \Omega_n \boldsymbol{a}^T \right)
\left( \boldsymbol{1}_n^T \boldsymbol\Omega_n \boldsymbol{1}_n \right)= \left( \boldsymbol{a}^T \boldsymbol\Omega_n \boldsymbol{1}_n \right)^2. 
$$
  Then, the Cauchy-Schwarz inequality indicates that $\boldsymbol{a}^T \mathbb{D}_n \boldsymbol{a} = 0$ if and only if 
$\boldsymbol{a} = a  \boldsymbol{1}_n $ for some scalar $a\neq 0$. Thus, $\boldsymbol u^T \widehat{\mathbb{X}}_n^T  \mathbb{D}_n \widehat{\mathbb{X}}_n \boldsymbol u = 0 $ if and only if $\widehat{\mathbb{X}}_n \boldsymbol u= a  \boldsymbol{1}_n $ for some $a\neq 0$. By construction, $\boldsymbol{1}_n^T \widehat{\mathbb{X}}_n = \boldsymbol{0}_n $, and thus necessarily $a=0$, which leads to a contradiction. Thus, $\widehat{\mathbb{X}}_n^T  \mathbb{D}_n \widehat{\mathbb{X}}_n$ is almost surely invertible. Note that we could also write
$$
  \mathbb{D}_n = \left[ \boldsymbol{I}_{n\times n}  - \frac{1}{\boldsymbol{1}_n^T \boldsymbol\Omega_n  \boldsymbol{1}_n}  \boldsymbol{1}_n  \boldsymbol{1}_n^T  \boldsymbol\Omega_n  \right]^T \boldsymbol\Omega_n \left[ \boldsymbol{I}_{n\times n} - \frac{1}{\boldsymbol{1}_n^T \boldsymbol\Omega_n  \boldsymbol{1}_n}    \boldsymbol{1}_n  \boldsymbol{1}_n^T  \boldsymbol\Omega_n  \right]
$$
  and deduce the positive semi-definiteness of  $\mathbb{D}_n$ from the positive definiteness of $\boldsymbol\Omega_n$. 

\item[2.] We could rewrite $\widehat {\mathbb{B}}_n $ under the form
$$
  \widehat {\mathbb{B}}_n = \left[ \boldsymbol{I}_{n\times n}  -   \widehat{\mathbb{X}}_n   \left(\widehat{\mathbb{X}}_n ^T{\mathbb{D}} _n \widehat{\mathbb{X}}_n   \right)^{-1} \widehat{\mathbb{X}}_n ^T {\mathbb{D}} _n   \right]^T \mathbb{D}_n 
\left[ \boldsymbol{I}_{n\times n} -     \widehat{\mathbb{X}}_n   \left(\widehat{\mathbb{X}}_n ^T{\mathbb{D}} _n \widehat{\mathbb{X}}_n   \right)^{-1} \widehat{\mathbb{X}}_n ^T {\mathbb{D}} _n   \right]
$$
  and deduce its positive semi-definiteness  from the positive definiteness of $\mathbb{D}_n$.
\end{enumerate}
\end{proofof}

\begin{proofof}{Theorem \ref{consist}}~~\\
Recall that $\Lambda : = 
[\lambda_{\rm min},\lambda_{\rm max}]$ and let 
$$
	\widehat M_n (\lambda) =  n^{-2} \widehat{\mathbb{Y}}_n(\lambda)^T \;  \widehat {\mathbb{B}} _n \; \widehat{\mathbb{Y}}_n(\lambda) 
$$
such that 
$$
	s^{-2\lambda} \widehat M_n (\lambda) = n^{-2}s^{-\lambda} \widehat{\mathbb{Y}}_n(\lambda)^T \;  \widehat {\mathbb{B}} _n  \; s^{-\lambda} \widehat{\mathbb{Y}}_n(\lambda),
$$
and, thus, $\widehat \lambda  = \arg\min\limits_{\lambda \in\Lambda}s^{-2\lambda}  \widehat M_n (\lambda)$. Next, let 
$$
	M_n (\lambda) = n^{-2} {\mathbb{Y}}_n(\lambda)^T \;  {\mathbb{B}} _n  \; {\mathbb{Y}}_n(\lambda).
$$
By construction,
$$
	M_n (\lambda) = Q_n((\lambda, {\boldsymbol \beta}_n(\lambda) ^T)^T, {\gamma}_n(\lambda) ), 
$$
where 
$$
	Q_n((\lambda, {\boldsymbol \beta}  ^T)^T, {\gamma}  ) =   n^{-2} \left(    {\mathbb{Y}}_n(\lambda) - \gamma\boldsymbol{1}_n -  {\mathbb{X}}_n   \boldsymbol \beta \right)^T \boldsymbol  \Omega_n  \left(    {\mathbb{Y}}_n(\lambda) - \gamma\boldsymbol{1}_n -  {\mathbb{X}}_n   \boldsymbol \beta \right)
$$
and 
	\begin{equation*} 
		\gamma_n(\lambda)
		= \frac{1}{\boldsymbol{1}_n^T \boldsymbol\Omega_n  \boldsymbol{1}_n} \boldsymbol{1}_n^T \boldsymbol\Omega_n \left( {\mathbb{Y}}_n(\lambda)  -  {\mathbb{X}}_n {\boldsymbol \beta} (\lambda)  \right)
\qquad \text{ and } \qquad
		{\boldsymbol \beta}_n  (\lambda) = \left( {\mathbb{X}}_n ^T{\mathbb{D}} _n {\mathbb{X}}_n   \right)^{-1}  {\mathbb{X}}_n ^T {\mathbb{D}} _n {\mathbb{Y}}_n(\lambda) .
	\end{equation*}
Let 
$$
Q((\lambda, {\boldsymbol \beta}  ^T)^T, {\gamma}  )  = E\left[Q_n((\lambda, {\boldsymbol \beta}  ^T)^T, {\gamma}  ) \right], \qquad \lambda \in\Lambda, \boldsymbol \beta\in\mathbb{R}^p, \gamma\in\mathbb{R}. 
$$
Next, let $c>0$ be a lower bound of the support of $Y$. Then, necessarily $c<s_0$ and we could work on the event $c\leq \inf S_n$, that is $s$ stays away from zero. In order to prove the uniform consistency it will suffice to prove 
	\begin{equation}\label{unif_cvg1}
		\sup_{h\in\mathcal{H}^{c}_n} \sup_{\boldsymbol d \in \mathcal{D}  }
		\sup_{\lambda \in\Lambda} 
		\left| \widehat M_n (\lambda) -  M_n (\lambda)\right| = o_{\mathbb{P}}(1),
	\end{equation}
	\begin{equation}\label{unif_cvg1b}
		\sup_{h\in\mathcal{H}^{c}_n}\sup_{\boldsymbol d \in \mathcal{D} } \sup_{\lambda\in\Lambda}\left|Q_n (( \lambda,  {\boldsymbol \beta}_n( \lambda) ^T)^T,  {\gamma}_n( \lambda) ) - Q_n (( \lambda,  \widehat{\boldsymbol \beta}( \lambda) ^T)^T,  \widehat{\gamma}( \lambda) ) \right| = o_{\mathbb{P}}(1),
	\end{equation}
	\begin{equation}\label{unif_cvg2}
		\sup_{\boldsymbol d \in \mathcal{D}  }
		\sup_{\lambda \in\Lambda } 
		\sup_{ \boldsymbol \beta \in\mathbb{R}^p}\sup_{\gamma \in\mathbb{R} } 
		\left|  Q_n((\lambda, {\boldsymbol \beta}  ^T)^T, {\gamma}  )  - Q((\lambda, {\boldsymbol \beta}  ^T)^T, {\gamma}  )  \right| = o_{\mathbb{P}}(1),
	\end{equation}
and to show that $\lambda_0$ is a uniformly well-separated minimum value of $\lambda\mapsto  s^{-2\lambda}_0 Q((\lambda, {\boldsymbol \beta}(\lambda)  ^T)^T, {\gamma}(\lambda)  )  $, that is  for any $\varepsilon>0$,
	\begin{equation}\label{opt}
		\inf_{\lambda  \in\Lambda, |\lambda -\lambda_0|\geq \varepsilon }   \inf_{\boldsymbol d \in \mathcal{D} } s^{-2\lambda}_0 Q((\lambda, {\boldsymbol \beta}(\lambda)  ^T)^T, {\gamma}(\lambda)  )  >0,
	\end{equation}
with $(\gamma(\lambda), \boldsymbol \beta (\lambda)^T)^T$ defined in equation \eqref{beta_l}. 

For the uniform convergence \eqref{unif_cvg1}, we first decompose 
	\begin{align*}
		\left| \widehat M_n (\lambda) -  M_n (\lambda) \right| 
		&\leq   
		\left|   n^{-1}{\mathbb{Y}}_n(\lambda)	^T\left( \widehat {\mathbb{B}} _n - {\mathbb{B}} _n  \right) n^{-1}{\mathbb{Y}}_n(\lambda) \right|
		+ 2 \left| n^{-1} \left[ \widehat{\mathbb{Y}}_n(\lambda) -  {\mathbb{Y}}_n(\lambda) \right]^T \;  \widehat {\mathbb{B}} _n n^{-1}{\mathbb{Y}}_n(\lambda)\right| \\
		&\quad + n^{-1}    \left[ \widehat{\mathbb{Y}}_n(\lambda) -  {\mathbb{Y}}_n(\lambda) \right]^T \;  \widehat {\mathbb{B}} _n  n^{-1}    \left[ \widehat{\mathbb{Y}}_n(\lambda) -  {\mathbb{Y}}_n(\lambda) \right] \\
		&\leq 
		\left\|\widehat {\mathbb{B}} _n \right\|_{\rm{Sp}} \left(2 \left\|n^{-1}    \left[ \widehat{\mathbb{Y}}_n(\lambda) -  {\mathbb{Y}}_n(\lambda) \right] \right\| \; \left\| n^{-1}  {\mathbb{Y}}_n(\lambda) \right\| + \left\| n^{-1}     \left[\widehat{\mathbb{Y}}_n(\lambda) -  {\mathbb{Y}}_n(\lambda) \right] \right\|  ^2 \right)\\
		& \quad + \left\| n^{-1}  {\mathbb{Y}}_n(\lambda) \right\|\left\| \widehat {\mathbb{B}} _n -{\mathbb{B}} _n  \right\|_{\rm{Sp}} \left\| n^{-1}  {\mathbb{Y}}_n(\lambda) \right\|.
	\end{align*} 
Herein, For a matrix $\boldsymbol A$, $\left\| A  \right\|_{\rm{Sp}}$ denotes its spectral norms, that is the square root of the largest eigenvalue of $ \boldsymbol A^T\boldsymbol A$. 
Next, from Lemma \ref{spectral_B_hat} and \ref{spectral_B_hat2} we obtain that
$$
	\sup_{h\in\mathcal{H}^{c}_n} \sup_{\boldsymbol d \in \mathcal D} \left\| \widehat {\mathbb{B}} _n \right\|_{\rm{Sp}}= O_{\mathbb{P}}(n) \qquad \text{and} \qquad \sup_{h\in\mathcal{H}^{c}_n} \sup_{\boldsymbol d \in  \mathcal D} \left\| \widehat {\mathbb{B}} _n -{\mathbb{B}} _n  \right\|_{\rm{Sp}}= o_{\mathbb{P}}(n).
$$
Moreover, by Lemma \ref{yn_2}
$$
	\sup_{\lambda \in\Lambda}  \left\| n^{-1}  {\mathbb{Y}}_n(\lambda) \right\| = O_{\mathbb{P}}(n^{-1/2}),
$$
and by Lemma \ref{yn_hat2}
$$
	\sup_{h\in\mathcal{H}^{c}_n}   \sup_{\lambda \in\Lambda}  \left\| n^{-1}  \left[\widehat{\mathbb{Y}}_n(\lambda) -  {\mathbb{Y}}_n(\lambda) \right] 
	\right\| = o_{\mathbb{P}}(n^{-1/2}).
$$
Therefore, the uniform convergence \eqref{unif_cvg1} follows. Similarly, by a suitable decomposition and elementary matrix algebra calculations
	\begin{align*}
		Q_n (( \lambda,  {\boldsymbol \beta}_n( \lambda) ^T)^T,  {\gamma}_n( \lambda) ) &- Q_n (( \lambda,  \widehat{\boldsymbol \beta}( \lambda) ^T)^T,  \widehat{\gamma}( \lambda) )\\ 
		&= \left(\gamma^2_n  (\lambda) - \widehat \gamma^2 (\lambda) \right) n^{-2}\boldsymbol{1}_n^T\boldsymbol \Omega_n \boldsymbol{1}_n \\
		&\quad + n^{-2} \left( {\boldsymbol \beta}_n (\lambda) ^T \mathbb{X}_n^T 
		\boldsymbol \Omega_n \mathbb{X}_n {\boldsymbol \beta}_n (\lambda) - \widehat{\boldsymbol \beta}(\lambda)  ^T \mathbb{X}_n^T 
		\boldsymbol \Omega_n \mathbb{X}_n \widehat {\boldsymbol \beta} (\lambda)  \right) \\ 
		&\quad + 2 n^{-2} \left(\gamma_n (\lambda) \boldsymbol{1}_n^T\boldsymbol \Omega_n \mathbb{X}_n  {\boldsymbol \beta}_n(\lambda) -  \widehat \gamma (\lambda) \boldsymbol{1}_n^T\boldsymbol \Omega_n \mathbb{X}_n \widehat{\boldsymbol \beta}(\lambda)   \right)\\
		&\quad - 2 n^{-2} \left({\gamma }_n (\lambda) \boldsymbol{1}_n^T\boldsymbol \Omega_n \mathbb{Y}_n (\lambda) -  \widehat {\gamma } (\lambda) \boldsymbol{1}_n^T\boldsymbol \Omega_n \mathbb{Y}_n (\lambda)  \right)  \\
		&\quad - 2 n^{-2} \left(\mathbb{Y}_n (\lambda)^T 
		\boldsymbol \Omega_n \mathbb{X}_n  {\boldsymbol \beta}_n (\lambda) -  \mathbb{Y}_n (\lambda)^T 
		\boldsymbol \Omega_n \mathbb{X}_n \widehat {\boldsymbol \beta} (\lambda)  \right) \\
		&=\; O_{\mathbb{P}}(1) \times 
		\sup_{h\in\mathcal{H}^{c}_n}\sup_{\boldsymbol d \in \mathcal{D} } \sup_{\lambda\in\Lambda}\left( \left\| \widehat {\boldsymbol \beta}( \lambda) - {\boldsymbol \beta}_n( \lambda) \right\|+ \left| \widehat {\gamma}( \lambda) - {\gamma}_n( \lambda) \right| \right).
	\end{align*}
By the results of \citet{sherman1994maximal}, the rate $O_{\mathbb{P}}(1)$ is uniform with respect to $\boldsymbol d$ and $\lambda$.  See also below for an example of application of the results in \citet{sherman1994maximal}. The uniform convergence of $\left\| \widehat {\boldsymbol \beta}( \lambda) - {\boldsymbol \beta}_n( \lambda) \right\|$ and $  \widehat {\gamma}( \lambda) - {\gamma}_n( \lambda) $ follows by the same type of matrix algebra calculations and uniform rates of convergence for $U-$processes. Thus, the uniform convergence \eqref{unif_cvg1b} holds true. 

Next, by the properties of Euclidean families, see \citet{nolan1987u} and \citet{sherman1994maximal}, the families of functions 
$$
	\{ g\left(\boldsymbol u_1 ; \boldsymbol \theta , \gamma, \boldsymbol \eta_{\lambda,1} 
	\right)g\left(\boldsymbol u_2; \boldsymbol \theta, \gamma , \boldsymbol \eta_{\lambda,2}  \right) \exp \left\{- (\boldsymbol w_1 - \boldsymbol w_2)^T \boldsymbol D (\boldsymbol w_1 - \boldsymbol w_2)\right\} : \boldsymbol \theta = (\lambda,\boldsymbol \beta^T)^T \in  \Lambda\times \mathbb{R}^p, \gamma \in \mathbb{R},\boldsymbol d \in \mathcal{D} \}
$$ 
and $\{ g^2\left(\boldsymbol u ; \boldsymbol \theta , \gamma,   \boldsymbol \eta_{\lambda}\right): \boldsymbol \theta \in  \Lambda\times \mathbb{R}^p, \gamma\in\mathbb{R} \}$ are Euclidean for a squared envelope. Thus, decomposing $Q_n((\lambda, {\boldsymbol \beta}  ^T)^T, {\gamma}  ) $ is a $U-$process plus the sum of the diagonal terms, and using Corollary 4 of \citet{sherman1994maximal},  the uniform convergence \eqref{unif_cvg2} holds true.

By construction, condition \eqref{well_sep_0} in Lemma \ref{lem_ident} is equivalent with 
$$
	\inf_{\lambda  \in\Lambda, |\lambda -\lambda_0|\geq \varepsilon }   \inf_{\boldsymbol d \in \mathcal{D} }s^{-2\lambda}_0 \left(Q((\lambda, {\boldsymbol \beta}(\lambda)  ^T)^T, {\gamma}(\lambda)  ) - n^{-1} E\left[g^2\left(\boldsymbol U; (\lambda, \boldsymbol \beta (\lambda)^T)^T , \gamma (\lambda), \boldsymbol \eta_{\lambda}\right)\right]\right) >0 .
$$
Since the family $\{g^2\left(\boldsymbol u; \boldsymbol \theta  , \gamma ,   \boldsymbol \eta_{\lambda}\right): \boldsymbol \theta \in  \Lambda\times \mathbb{R}^p, \gamma\in\mathbb{R} \}$ has an integrable envelope, the expectation in the last display is finite. Thus, we deduce \eqref{opt} and $\lambda_0$ is a uniformly well-separated minimum.

Finally, to derive the uniform consistency of $\widehat \lambda$, we  adapt the steps in the proof of Theorem 5.7 of \citet{van2000asymptotic}.
First, for any sequence $s_n\in S_n$, $n\geq 1$, and $\widehat \lambda = \widehat \lambda (s_n)$ defined as in equation \eqref{lambda_l_hat}, 
	\begin{equation} \label{eqa1}
		\begin{aligned}
		0\leq s_n^{-2\widehat \lambda} \widehat M_n (\widehat \lambda) &\leq s_n^{-2\lambda_0} \widehat M_n (\lambda_0)  \\
		& = s_0^{-2\lambda_0}  M_n (\lambda_0)  
		+  s_n^{-2\lambda_0}\left( \widehat M_n (\lambda_0)  -   M_n (\lambda_0) \right) +  \left(s_n^{-2\lambda_0} - s_0^{-2\lambda_0}\right) M_n (\lambda_0)  \\
		& = s_0^{-2\lambda_0}  Q_n((\lambda_0, {\boldsymbol \beta}_n(\lambda_0) ^T)^T, {\gamma}_n(\lambda_0) )   
		+ o_{\mathbb{P}}(1) \\
		&\leq  s_0^{-2\lambda_0}  Q_n((\lambda_0, {\boldsymbol \beta}(\lambda_0) ^T)^T, {\gamma}(\lambda_0) )  
		+ o_{\mathbb{P}}(1) \\
		&= s_0^{-2\lambda_0}  Q((\lambda_0, {\boldsymbol \beta}(\lambda_0) ^T)^T, {\gamma}(\lambda_0) )  
		+ o_{\mathbb{P}}(1) = o_{\mathbb{P}}(1),
		\end{aligned}
	\end{equation}	
uniformly with respect to $h$ and  $\boldsymbol d $. (Note that $\widehat \lambda$ depends on $s_n$, but also on $\boldsymbol d$ and $h$.) For the last inequality in the last display we use the fact that, by definition, ${\boldsymbol \beta}_n(\lambda) $ and ${\gamma}_n(\lambda)$ minimize $Q_n((\lambda, {\boldsymbol \beta}^T)^T, {\gamma} )$ with respect to $\boldsymbol \beta$ and $\gamma$ given $\lambda$.

Meanwhile, from \eqref{unif_cvg1}, \eqref{unif_cvg1b} and \eqref{unif_cvg2} and the fact that $S_n$ is a $o_{\mathbb{P}}(1)$ neighborhood of $s_0$ that is contained in the support of $Y$, for any $s_n$, 
	\begin{equation}\label{eqa2}
		\begin{aligned}
		\Big| s_n^{-2\widehat \lambda} \widehat M_n (\widehat \lambda)& -  s_n^{-2\widehat \lambda}  Q ((\widehat \lambda,\widehat  {\boldsymbol \beta}(\widehat \lambda) ^T)^T, \widehat {\gamma}(\widehat \lambda) )
		\Big| \\
		\leq& \left| s_n^{-2\widehat \lambda} \widehat M_n (\widehat \lambda) - s_n^{-2\widehat \lambda}  Q_n ((\widehat \lambda,  \boldsymbol \beta_n(\widehat \lambda) ^T)^T,  {\gamma}_n(\widehat \lambda) )
		\right|\\
		&+   s_n^{-2\widehat \lambda}  \left|   Q_n ((\widehat \lambda,  {\boldsymbol \beta}_n(\widehat \lambda) ^T)^T,  {\gamma}_n(\widehat \lambda) ) -  Q_n ((\widehat \lambda,\widehat  {\boldsymbol \beta}(\widehat \lambda) ^T)^T, \widehat {\gamma}(\widehat \lambda) )
		\right| \\
		&+  s_n^{-2\widehat \lambda} \left|   Q_n ((\widehat \lambda,  \widehat{\boldsymbol \beta}(\widehat \lambda) ^T)^T,  \widehat{\gamma}(\widehat \lambda) ) - Q ((\widehat \lambda,\widehat  {\boldsymbol \beta}(\widehat \lambda) ^T)^T, \widehat {\gamma}(\widehat \lambda) )
		\right|\\
		\leq& \sup_{s\in S_n}  \sup_{\lambda\in\Lambda} s^{-2 \lambda} \times\sup_{h\in\mathcal{H}^{c}_n}  \sup_{\boldsymbol d \in \mathcal{D} } \sup_{\lambda\in\Lambda} \left|
		\widehat M_n ( \lambda) -   M_n (\lambda)\right|\\
		&+  \sup_{s\in S_n}  \sup_{\lambda\in\Lambda} s^{-2 \lambda} \times \sup_{h\in\mathcal{H}^{c}_n}\sup_{\boldsymbol d \in \mathcal{D} } \sup_{\lambda\in\Lambda} \left|Q_n (( \lambda,  {\boldsymbol \beta}_n( \lambda) ^T)^T,  {\gamma}_n( \lambda) ) - Q_n (( \lambda,  \widehat{\boldsymbol \beta}( \lambda) ^T)^T,  \widehat{\gamma}( \lambda) ) \right| \\
		&+  \sup_{s\in S_n}  \sup_{\lambda\in\Lambda} s^{-2 \lambda} \times  \sup_{\boldsymbol d \in \mathcal{D} } \sup_{\lambda\in\Lambda}\sup_{ \boldsymbol \beta \in\mathbb{R}^p}\sup_{\gamma \in\mathbb{R} }   \left|  Q_n ((\lambda, {\boldsymbol \beta} ^T)^T,  {\gamma})  - Q ((\lambda, {\boldsymbol \beta} ^T)^T,  {\gamma}) \right|  \\
		=&\; o_{\mathbb{P}}(1).
		\end{aligned}
	\end{equation}	
Next, by property \eqref{opt}, for any $\varepsilon >0$ there exists $\zeta >0$ (depending on $\varepsilon$, but also on the  endpoints of the sets $S_n$ and $\Lambda$) such that the probability of the event 
$$
	E_n = \left\{\inf_{\lambda\in\Lambda, |\lambda - \lambda_0|\geq \varepsilon } \inf_{\boldsymbol d \in \mathcal{D}  }\inf_{s\in S_n }s^{-2\lambda} Q((\lambda, {\boldsymbol \beta}(\lambda) ^T)^T, {\gamma}(\lambda) )    > s_0^{-2\lambda_0} Q((\lambda_0, {\boldsymbol \beta}(\lambda_0) ^T)^T, {\gamma}(\lambda_0) )   + \zeta  = \zeta\right\}
$$
tends to 1. Moreover,  the event
$$
	\left\{ \sup_{h\in\mathcal{H}^{c}_n} \sup_{s\in S_n} \sup_{\boldsymbol d \in \mathcal{D} }\left|\widehat \lambda(s) - \lambda_0 \right| \geq \varepsilon \right\} 
$$
is contained in the event
$$
	\left\{ \inf_{h\in\mathcal{H}^{c}_n} \inf_{s\in S_n } \inf_{\boldsymbol d \in \mathcal{D}  }s^{-2\widehat \lambda (s)} Q((\widehat \lambda(s), {\boldsymbol \beta}(\widehat \lambda(s)) ^T)^T, {\gamma}(\widehat \lambda(s)) ) > \zeta \right\} \cap E_n.
$$ 
By \eqref{eqa1} and \eqref{eqa2}, the probability of the intersection event tends to zero. Now the proof for $\widehat \lambda$ is complete.

Consider now the convergence of $\widehat{\boldsymbol \beta}  (\widehat \lambda)$. Given the uniform convergence of $\left\| \widehat {\boldsymbol \beta}( \lambda) - {\boldsymbol \beta}_n( \lambda) \right\|$ and the continuity of $\lambda \mapsto \boldsymbol \beta(\lambda)$, it suffices to obtain the convergence of  $\left\| {\boldsymbol \beta}_n( \lambda) -{\boldsymbol \beta}( \lambda) \right\|$ uniformly over $o_{\mathbb{P}}(1)$ neighborhoods of $\lambda_0$. By construction,  
	\begin{equation}\label{eqa1c}
		\begin{aligned}
			0\leq s_n^{-2\lambda}  M_n (\lambda) &= s_n^{-2\lambda}  Q_n((\lambda, {\boldsymbol \beta}_n(\lambda) ^T)^T, {\gamma}_n(\lambda) ) \\ 
			&\leq  s_n^{-2\lambda}  Q_n((\lambda, {\boldsymbol \beta}(\lambda) ^T)^T, {\gamma}(\lambda) )   
			= s_n^{-2\lambda}  Q((\lambda, {\boldsymbol \beta}(\lambda) ^T)^T, {\gamma}(\lambda) )     
			+ o_{\mathbb{P}}(1) , 
		\end{aligned}
	\end{equation}
uniformly with respect to $s_n$, $\lambda$, $h$ and  $\boldsymbol d $. Moreover, since $Q((\lambda_0, {\boldsymbol \beta}(\lambda_0) ^T)^T, {\gamma}(\lambda_0) )=0$, we have 
	\begin{equation}\label{eqa1c2}
		Q((\lambda, {\boldsymbol \beta}(\lambda) ^T)^T, {\gamma}(\lambda) )  = o_{\mathbb{P}}(1)
	\end{equation}
uniformly over $o_{\mathbb{P}}(1)$ neighborhoods of $\lambda_0$. Meanwhile, by \eqref{unif_cvg2}, for any $s_n$ and any $\lambda$, 
	\begin{equation}\label{eqa2c}
		\begin{aligned}
			\Big| s_n^{-2 \lambda}   M_n ( \lambda) - &s_n^{-2 \lambda}  Q ((  \lambda,  {\boldsymbol \beta}_n(  \lambda) ^T)^T,   {\gamma}_n(  \lambda) )
			\Big| \\
			&=s_n^{-2 \lambda}  \left|   Q_n (( \lambda,  {\boldsymbol \beta}_n( \lambda) ^T)^T,  {\gamma}_n(\lambda) ) -  Q (( \lambda,  {\boldsymbol \beta}_n( \lambda) ^T)^T,  {\gamma}_n(\lambda) )
			\right| \\
			&\leq   \sup_{s\in S_n}  \sup_{\lambda\in\Lambda} s^{-2 \lambda} \times  \sup_{\boldsymbol d \in \mathcal{D} } \sup_{\lambda\in\Lambda}\sup_{ \boldsymbol \beta \in\mathbb{R}^p}\sup_{\gamma \in\mathbb{R} }   \left|  Q_n ((\lambda, {\boldsymbol \beta} ^T)^T,  {\gamma})  - Q ((\lambda, {\boldsymbol \beta} ^T)^T,  {\gamma}) \right|  \\
			& = o_{\mathbb{P}}(1).
		\end{aligned}
	\end{equation}
Next, by the proof of property \eqref{opt} and continuity arguments, for any $\varepsilon >0$ there exists $\upsilon >0$ such that the probability of the event 
$$
	F_n = \left\{\inf_{\lambda\in\Lambda, |\lambda - \lambda_0| =o_{\mathbb{P}}(1) } \inf_{\|\boldsymbol \beta - \boldsymbol \beta _0\|> \varepsilon } \inf_{\gamma \in \mathbb{R}} \inf_{\boldsymbol d \in \mathcal{D}  }\inf_{s\in S_n }s^{-2\lambda} Q((\lambda, {\boldsymbol \beta} ^T)^T, {\gamma})    >  \upsilon \right\}
$$
tends to 1. Finally, note that the event
$$
	\left\{ \sup_{\lambda\in\Lambda, |\lambda - \lambda_0| =o_{\mathbb{P}}(1) } \sup_{s\in S_n} \sup_{\boldsymbol d \in \mathcal{D} }\left\| {\boldsymbol \beta}_n( \lambda) -{\boldsymbol \beta}( \lambda) \right\| \geq \varepsilon \right\} 
$$
is contained in the intersection 
$$
	\left\{ \inf_{\lambda\in\Lambda, |\lambda - \lambda_0| =o_{\mathbb{P}}(1)} \inf_{s\in S_n } \inf_{\boldsymbol d \in \mathcal{D}  }s^{-2 \lambda } Q(( \lambda, {\boldsymbol \beta}_n( \lambda) ^T)^T, {\gamma}_n( \lambda) )   >   \upsilon \right\} \cap F_n,
$$
which, by \eqref{eqa1c}, \eqref{eqa1c2} and \eqref{eqa2c}, has a probability tending to zero. Now the proof is complete. 
\end{proofof} 

\medskip

\begin{proofof}{Proposition \ref{AN_prop}}~~\\
As $\widehat \lambda - \lambda_0=o_{\mathbb{P}}(1)$ uniformly with respect to $h\in\mathcal{H}_{sc,n}$, $\boldsymbol d\in \mathcal{D}$ and $s\in S_n$, we get that
	\begin{align}
		0 =  
		n^{-1}s^{- \widehat \lambda} \widehat{\mathbb{Y}}_n(\widehat \lambda)^T \;  \widehat {\mathbb{B}} _n \frac{\partial }{\partial \lambda} &\left\{n^{-1}s^{- \widehat \lambda} \widehat{\mathbb{Y}}_n(\widehat \lambda) \right\} 
		=   
		n^{-1}s^{- \lambda_0} \widehat{\mathbb{Y}}_n(\lambda_0)^T \;  \widehat {\mathbb{B}} _n \frac{\partial }{\partial \lambda} \left\{n^{-1}s^{- \lambda_0} \widehat{\mathbb{Y}}_n(\lambda_0) \right\} \notag\\ 
		&+  
		\Bigg[  \frac{\partial }{\partial \lambda} \left\{ n^{-1}s^{-\lambda_0}  \widehat{\mathbb{Y}}_n(\lambda_0) \right\}^T  \widehat {\mathbb{B}} _n  \frac{\partial }{\partial \lambda} \left\{ n^{-1} s^{-\lambda_0}  \widehat{\mathbb{Y}}_n(\lambda_0) \right\}  \notag \\ 
		  &+
		     n^{-1}s^{-\lambda_0}  \widehat{\mathbb{Y}}_n(\lambda_0)^T\;  \widehat {\mathbb{B}} _n  \frac{\partial ^2}{\partial \lambda^2} \left\{ n^{-1}s^{-\lambda_0}  \widehat{\mathbb{Y}}_n(\lambda_0) \right\} + R_{1,n}(\widetilde \lambda,\lambda_0;s)\Bigg] \left(\widehat \lambda - \lambda_0\right),\label{maybe_we_need1}
	\end{align}
where $\widetilde \lambda = c \widehat \lambda + (1-c)\lambda_0$ for some $c\in(0,1)$. We have  $\sup_{h\in \mathcal H_{sc,n} }\sup_{\boldsymbol d \in \mathcal D} \sup_{s\in S_n}|  R_{1,n}(\widetilde\lambda,\lambda_0;s)| = o_{\mathbb{P}}(1)$, see Lemma \ref{remainder}. 

Note that $$ \frac{\partial }{\partial \lambda} \{n^{-1}s^{- \lambda_0} \widehat{\mathbb{Y}}_n(\lambda_0) \}  = s^{- \lambda_0}\frac{\partial }{\partial \lambda} \{n^{-1} \widehat{\mathbb{Y}}_n(\lambda_0) \} - \log(s)s^{- \lambda_0} n^{-1} \widehat{\mathbb{Y}}_n(\lambda_0).$$
First, we show that 
	\begin{equation}\label{neg1}
		\begin{aligned}
			n^{-1}s^{- \lambda_0}\widehat{\mathbb{Y}}_n(\lambda_0) ^T \;  \widehat {\mathbb{B}} _n \frac{\partial }{\partial \lambda}  &\left\{n^{-1} s^{- \lambda_0} \widehat{\mathbb{Y}}_n(\lambda_0)\right\} \\
			&- n^{-1}s_0^{- \lambda_0} \left({\mathbb{Y}}_n(\lambda_0) - \left(\boldsymbol{\widehat{\varepsilon}_{|z}\widehat{f}_z}\right)_n\right) ^T \;    {\mathbb{B}} _n s_0^{- \lambda_0}\frac{\partial }{\partial \lambda} \left\{ n^{-1}{\mathbb{Y}}_n(\lambda_0)\right\}  = o_{\mathbb{P} }\left(n^{-1/2}\right),
		\end{aligned}
	\end{equation}	
uniformly with respect to $s$, $\boldsymbol d$ and $h$. We start by showing that 
	\begin{multline*} 
		n^{-1}s^{- \lambda_0}\widehat{\mathbb{Y}}_n(\lambda_0) ^T \;  \widehat {\mathbb{B}} _n n^{-1} s^{- \lambda_0}\frac{\partial }{\partial \lambda}   		\widehat{\mathbb{Y}}_n(\lambda_0) \\
		- n^{-1}s_0^{- \lambda_0} \left({\mathbb{Y}}_n(\lambda_0) - \left(\boldsymbol{\widehat{\varepsilon}_{|z}\widehat{f}_z}\right)_n\right)^T \;    {\mathbb{B}} _n n^{-1}s_0^{- \lambda_0}\frac{\partial }{\partial \lambda}   {\mathbb{Y}}_n(\lambda_0)  = o_{\mathbb{P} }\left(n^{-1/2}\right),
	\end{multline*}
uniformly with respect to $s$, $\boldsymbol d$ and $h$. By the property $\widehat {\mathbb{X}} _n ^T\widehat {\mathbb{B}} _n = {\mathbb{X}} _n ^T {\mathbb{B}} _n = \boldsymbol 0_n ,$
we could equivalently prove that  
	\begin{multline}\label{neg11_bb}
		n^{-1}s^{- \lambda_0}\left( \widehat{\mathbb{Y}}_n(\lambda_0) - \widehat {\mathbb{X}} _n \boldsymbol \beta_0 \right)^T \;  \widehat {\mathbb{B}} _n n^{-1} s^{- \lambda_0}\frac{\partial }{\partial \lambda}   \widehat{\mathbb{Y}}_n(\lambda_0) \\
		- n^{-1}s_0^{- \lambda_0}\left( {\mathbb{Y}}_n(\lambda_0)  -  {\mathbb{X}} _n \boldsymbol \beta_0 - \left(\boldsymbol{\widehat{\varepsilon}_{|z}\widehat{f}_z}\right)_n\right)^T \;    {\mathbb{B}} _n n^{-1}s_0^{- \lambda_0}\frac{\partial }{\partial \lambda} {\mathbb{Y}}_n(\lambda_0)  = o_{\mathbb{P} }\left(n^{-1/2}\right),
	\end{multline}
uniformly with respect to $s \in S_n$, $\boldsymbol d\in\mathcal D$ and $h\in\mathcal H_{sc,n}$. Note that by definition $\mathbb{Y}_n(\lambda_0)  -  \mathbb{X} _n \boldsymbol \beta_0 = \left(\boldsymbol{ \varepsilon f_z}\right)_n$. To obtain \eqref{neg11_bb}, we  decompose the difference in a sum of the following four terms:
	\begin{align*}
		R_{n1} &=  n^{-1} s^{- \lambda_0}\left( \left[  \widehat{\mathbb{Y}}_n(\lambda_0) - \widehat {\mathbb{X}} _n \boldsymbol \beta_0\right] - \left[ {\mathbb{Y}}_n(\lambda_0) - {\mathbb{X}} _n \boldsymbol \beta_0 \right] + \left(\boldsymbol{\widehat{\varepsilon}_{|z}\widehat{f}_z}\right)_n \right)^T \; \widehat {\mathbb{B}} _n  n^{-1}s^{- \lambda_0}\frac{\partial }{\partial \lambda}   \widehat{\mathbb{Y}}_n(\lambda_0)  , \\ 
		R_{n2} &= n^{-1} \left(s^{- 2\lambda_0} - s_0^{- 2\lambda_0}\right)\left( {\mathbb{Y}}_n(\lambda_0)  -  {\mathbb{X}} _n \boldsymbol \beta_0  - \left(\boldsymbol{\widehat{\varepsilon}_{|z}\widehat{f}_z}\right)_n\right)^T \; \widehat{\mathbb{B}} _n  n^{-1}\frac{\partial }{\partial \lambda}   \widehat{\mathbb{Y}}_n(\lambda_0)\\
		R_{n3} &=  n^{-1}s_0^{- \lambda_0}\left( {\mathbb{Y}}_n(\lambda_0)  -  {\mathbb{X}} _n \boldsymbol \beta_0  - \left(\boldsymbol{\widehat{\varepsilon}_{|z}\widehat{f}_z}\right)_n\right) ^T \;  \left[ \widehat {\mathbb{B}} _n -{\mathbb{B}} _n\right]n^{-1}s_0^{- \lambda_0} \frac{\partial }{\partial \lambda}     \widehat{\mathbb{Y}}_n(\lambda_0), \\
		\text{ and} \qquad    
		R_{n4} &= n^{-1} s_0^{- \lambda_0} \left( {\mathbb{Y}}_n(\lambda_0)  -  {\mathbb{X}} _n \boldsymbol \beta_0 - \left(\boldsymbol{\widehat{\varepsilon}_{|z}\widehat{f}_z}\right)_n\right)^T \; {\mathbb{B}} _n  n^{-1}s_0^{- 
		\lambda_0}\left(  \frac{\partial }{\partial \lambda}   \widehat{\mathbb{Y}}_n(\lambda_0) - \frac{\partial }{\partial \lambda}   {\mathbb{Y}}_n(\lambda_0)\right).
	\end{align*}
Note that
$$
	\widehat {\mathbb{B}}_n =  \mathbb{S}_n^T   \left( \boldsymbol{I}_{n\times n}  -   P_{ \mathbb{S}_n\widehat{\mathbb{X}}_n  }   \right) \mathbb{S}_n \quad \text{ and } \quad   {\mathbb{B}}_n =  \mathbb{S}_n^T  \left( \boldsymbol{I}_{n\times n}  -   P_{ \mathbb{S}_n  {\mathbb{X}}_n  }   \right) \mathbb{S}_n,
$$
where $P_{ \mathbb{S}_n \widehat{\mathbb{X}}_n }$ and $P_{ \mathbb{S}_n  {\mathbb{X}}_n }$ are the orthogonal projectors on the  subspaces generated by $ \mathbb{S}_n\widehat{\mathbb{X}}_n$ and $ \mathbb{S}_n {\mathbb{X}}_n$, that is 
$$
	P_{ \mathbb{S}_n \widehat{\mathbb{X}}_n }=     \mathbb{S}_n \widehat{\mathbb{X}}_n   \left(\widehat{\mathbb{X}}_n ^T\mathbb{D}_n \widehat{\mathbb{X}}_n   \right)^{-1} \widehat{\mathbb{X}}_n ^T  \mathbb{S}_n^T 
	\quad \text{and} \quad 
	P_{ \mathbb{S}_n {\mathbb{X}}_n }=      \mathbb{S}_n {\mathbb{X}}_n   \left( {\mathbb{X}}_n ^T\mathbb{D}_n  {\mathbb{X}}_n   \right)^{-1} {\mathbb{X}}_n ^T \mathbb{S}_n^T
$$
with 
$$
	\mathbb{S}_n = \left( \boldsymbol{I}_{n\times n}  - P_{ \boldsymbol\Omega_n ^{1/2}{\boldsymbol{1}}_n   }\right) \boldsymbol\Omega_n^{1/2}.
$$
Here, $P_{ \boldsymbol\Omega_n^{1/2} {\boldsymbol{1}}_n   }$ is the projector on the subspace generated by the vector $\boldsymbol\Omega_n^{1/2} {\boldsymbol{1}}_n$, that is 
$$
	P_{ \boldsymbol\Omega_n^{1/2} {\boldsymbol{1}}_n } =   \frac{1}{\boldsymbol{1}_n^T \boldsymbol\Omega_n  \boldsymbol{1}_n}  \boldsymbol\Omega_n  ^{1/2} \boldsymbol{1}_n  \boldsymbol{1}_n^T  \boldsymbol\Omega_n  ^{1/2},
$$
and $\boldsymbol\Omega_n^{1/2} $ is the positive definite square root of $\boldsymbol\Omega_n $. 
Deduce that
\newpage
	\begin{multline*}
		\left| R_{n2} \right| \leq \big| s^{- 2\lambda_0} - s_0^{- 2\lambda_0} \big| \left\| \boldsymbol \Omega_n^{1/2} n^{-1}  \left[ {\mathbb{Y}}_n(\lambda_0) - {\mathbb{X}} _n \boldsymbol \beta_0 - \left(\boldsymbol{\widehat{\varepsilon}_{|z}\widehat{f}_z}\right)_n\right]  \right\|\\
		\times  \left\| \left( \boldsymbol{I}_{n\times n}  - P_{ \boldsymbol\Omega_n ^{1/2}{\boldsymbol{1}}_n   }\right)  \left( \boldsymbol{I}_{n\times n} - P_{ \mathbb{S}_n  \widehat {\mathbb{X}}_n}\right) \left( \boldsymbol{I}_{n\times n}  - P_{ \boldsymbol\Omega_n ^{1/2}{\boldsymbol{1}}_n   }\right)  \right\|_{\rm Sp} 
		\left\| \boldsymbol \Omega_n^{1/2}    n^{-1} \frac{\partial }{\partial \lambda}   \widehat{\mathbb{Y}}_n(\lambda_0)  \right\|_{\rm{Sp}},
	\end{multline*}
	\begin{multline*}
		\left| R_{n3}\right| \leq s_0^{- 2\lambda_0} \left\| \boldsymbol \Omega_n^{1/2} n^{-1}  \left[ {\mathbb{Y}}_n(\lambda_0) - {\mathbb{X}} _n \boldsymbol \beta_0  - \left(\boldsymbol{\widehat{\varepsilon}_{|z}\widehat{f}_z}\right)_n\right]  \right\| \\ 
		\times \left\| \left( \boldsymbol{I}_{n\times n}  - P_{ \boldsymbol\Omega_n ^{1/2}{\boldsymbol{1}}_n   }\right)  \left( P_{ \mathbb{S}_n  {\mathbb{X}}_n}  -   P_{ \mathbb{S}_n  \widehat{\mathbb{X}}_n  }   \right) \left( \boldsymbol{I}_{n\times n}  - P_{ \boldsymbol\Omega_n ^{1/2}{\boldsymbol{1}}_n   }\right)  \right\|_{\rm Sp}
		\left\| \boldsymbol \Omega_n^{1/2}    n^{-1}  \frac{\partial }{\partial \lambda}   \widehat{\mathbb{Y}}_n(\lambda_0) \right\|_{\rm{Sp}}
	\end{multline*}
and
	\begin{multline*}
		\left| R_{n4}\right| \leq s_0^{- 2\lambda_0} \left\| \boldsymbol \Omega_n^{1/2} n^{-1}  \left[ {\mathbb{Y}}_n(\lambda_0) - {\mathbb{X}} _n \boldsymbol \beta_0 - \left(\boldsymbol{\widehat{\varepsilon}_{|z}\widehat{f}_z}\right)_n\right]  \right\| \\ 
		\times \left\| \left( \boldsymbol{I}_{n\times n}  - P_{ \boldsymbol\Omega_n ^{1/2}{\boldsymbol{1}}_n   }\right)  \left( \boldsymbol{I}_{n\times n}  -   P_{ \mathbb{S}_n  {\mathbb{X}}_n  }   \right) \left( \boldsymbol{I}_{n\times n}  - P_{ \boldsymbol\Omega_n ^{1/2}{\boldsymbol{1}}_n   }\right)  \right\|_{\rm Sp}
		 \left\| \boldsymbol \Omega_n^{1/2}    n^{-1}  \left( \frac{\partial }{\partial \lambda}   \widehat{\mathbb{Y}}_n(\lambda_0) - \frac{\partial }{\partial \lambda}   {\mathbb{Y}}_n(\lambda_0)\right)  \right\|_{\rm{Sp}}.
	\end{multline*}
The uniform rate $o_{\mathbb{P} }\left(n^{-1/2}\right)$ as  in 
\eqref{neg11_bb} follows for $R_{n2}$, $R_{n3}$ and $R_{n4}$ from the fact that the spectral norm of a product of projectors  is at most equal to 1, the spectral norm of $P_{ \mathbb{S}_n  {\mathbb{X}}_n} - P_{ \mathbb{S}_n  \widehat {\mathbb{X}}_n} $ tends to zero, $\sup_{s\in S_n}\big| s^{- 2\lambda_0} - s_0^{- 2\lambda_0} \big| = o_{\mathbb{P} }(1)$ as well as $\sup_{s\in S_n} s^{- 2\lambda_0}  = O_{\mathbb{P} }(1)$,  and from Lemmas \ref{yn_hatt_5}, \ref{yn_der} and \ref{yn_hatt_7}. For the term $R_{n1}$, we could write
	\begin{multline*}
		\left| R_{n1} \right| \leq s^{- 2\lambda_0}\left\| \boldsymbol\Omega_n ^{1/2} n^{-1}  \left(\left[ {\mathbb{Y}}_n(\lambda_0) - {\mathbb{X}} _n \boldsymbol \beta_0 \right] - \left(\boldsymbol{\widehat{\varepsilon}_{|z}\widehat{f}_z}\right)_n -  \left[ \widehat{\mathbb{Y}}_n(\lambda_0) - \widehat{\mathbb{X}} _n \boldsymbol \beta_0 \right]\right) \right\| \\ 
		\times \left\| \left( \boldsymbol{I}_{n\times n}  - P_{ \boldsymbol\Omega_n ^{1/2}{\boldsymbol{1}}_n   }\right)  \left( \boldsymbol{I}_{n\times n}  - P_{ \mathbb{S}_n  \widehat {\mathbb{X}}_n}   \right) \left( \boldsymbol{I}_{n\times n}  - P_{ \boldsymbol\Omega_n ^{1/2}{\boldsymbol{1}}_n   }\right)  \right\|_{\rm Sp}
		\left\| \boldsymbol \Omega_n^{1/2}    n^{-1} \frac{\partial }{\partial \lambda}   \widehat{\mathbb{Y}}_n(\lambda_0)  \right\|_{\rm{Sp}},
	\end{multline*}
and use Lemmas \ref{yn_der} and \ref{yn_hatt_6} and again the facts that the spectral norm of a product of projectors  is at most equal to 1 and $\sup_{s\in S_n} s^{- 2\lambda_0}  = O_{\mathbb{P} }(1)$ to deduce that it is of rate $o_{\mathbb{P} }\left(n^{-1/2}\right) $, uniformly with respect to $s$,  $\boldsymbol d$ and $h$.
Now the proof of the property \eqref{neg11_bb} is complete. Next, to complete the arguments for \eqref{neg1}, we show that 
	\begin{equation*}
		n^{-1}s^{- \lambda_0}\widehat{\mathbb{Y}}_n(\lambda_0) ^T \;  \widehat {\mathbb{B}} _n n^{-1} \log(s)s^{- \lambda_0}  \widehat{\mathbb{Y}}_n(\lambda_0)   = o_{\mathbb{P} }\left(n^{-1/2}\right),
	\end{equation*}
uniformly with respect to $s$, $\boldsymbol d$ and $h$. By the property $\widehat {\mathbb{X}} _n ^T\widehat {\mathbb{B}} _n = \boldsymbol 0_n ,$ and since we could consider $S_n$ to the right of the origin, we could equivalently prove that  
	\begin{equation}\label{neg12}
		n^{-1}s^{- \lambda_0}\left(\widehat{\mathbb{Y}}_n(\lambda_0) - \widehat {\mathbb{X}} _n \boldsymbol \beta_0 \right) ^T \; \widehat {\mathbb{B}} _n n^{-1}  
		\left( \widehat{\mathbb{Y}}_n(\lambda_0) - \widehat {\mathbb{X}} _n \boldsymbol \beta_0 \right)   = o_{\mathbb{P} }\left(n^{-1/2}\right),
	\end{equation}
uniformly with respect to $s \in S_n$, $\boldsymbol d\in\mathcal D$ and $h\in\mathcal H_{sc,n}$. By a decomposition as in the proof of \eqref{neg11_bb} we get that 
	\begin{multline*}
		n^{-1}s^{- \lambda_0}\left( \widehat{\mathbb{Y}}_n(\lambda_0) - \widehat {\mathbb{X}} _n \boldsymbol \beta_0 \right)^T \;  \widehat {\mathbb{B}} _n n^{-1} 
		\left( \widehat{\mathbb{Y}}_n(\lambda_0) - \widehat {\mathbb{X}} _n \boldsymbol \beta_0 \right) \\
		-
		n^{-1}s^{- \lambda_0} \left( \mathbb{Y}_n(\lambda_0) -  {\mathbb{X}} _n \boldsymbol \beta_0 - \left(\boldsymbol{\widehat{\varepsilon}_{|z}\widehat{f}_z}\right)_n\right)^T \; {\mathbb{B}} _n n^{-1} 
		\left(
		{\mathbb{Y}}_n(\lambda_0) - {\mathbb{X}} _n \boldsymbol \beta_0 - \left(\boldsymbol{\widehat{\varepsilon}_{|z}\widehat{f}_z}\right)_n\right)
		= o_{\mathbb{P} }\left(n^{-1/2}\right).
	\end{multline*}
We obtain  
	\begin{align*}
		n^{-1}s^{- \lambda_0} \left( \mathbb{Y}_n(\lambda_0) -  {\mathbb{X}} _n \boldsymbol \beta_0 - \left(\boldsymbol{\widehat{\varepsilon}_{|z}\widehat{f}_z}\right)_n\right)^T \; {\mathbb{B}} _n n^{-1}  
		\left(
		{\mathbb{Y}}_n(\lambda_0) - {\mathbb{X}} _n \boldsymbol \beta_0 - \left(\boldsymbol{\widehat{\varepsilon}_{|z}\widehat{f}_z}\right)_n\right) = o_{\mathbb{P} }\left(n^{-1/2}\right),
	\end{align*}
from Lemmas \ref{yn_hatt_5} and \ref{yn_hatt_7} and the facts that the spectral norm of a product of projectors  is at most equal to 1 and $\sup_{s\in S_n} s^{- \lambda_0}  = O_{\mathbb{P} }(1)$ as well as $\sup_{s\in S_n} \log(s)s^{- \lambda_0}  = O_{\mathbb{P} }(1)$ such that \eqref{neg12} follows. 
The property \eqref{neg1} follows now from \eqref{neg11_bb} and \eqref{neg12}. 

Next, we show that
	\begin{equation}\label{neg2}
		\begin{aligned}
			n^{-1} \frac{\partial }{\partial \lambda} \left\{ s^{- \lambda_0}\widehat{\mathbb{Y}}_n(\lambda_0) \right\}^T  \widehat {\mathbb{B}} _n  n^{-1} \frac{\partial }{\partial \lambda} &\left\{ s^{- \lambda_0}\widehat{\mathbb{Y}}_n(\lambda_0) \right\}  \\
			&-   n^{-1}  s_0^{- \lambda_0}\frac{\partial }{\partial \lambda} {\mathbb{Y}}_n(\lambda_0)^T   {\mathbb{B}}_n n^{-1} s_0^{- \lambda_0}\frac{\partial }{\partial \lambda}  {\mathbb{Y}}_n(\lambda_0) 
			= o_{\mathbb{P} }\left(1\right),
		\end{aligned}
	\end{equation} 	
uniformly with respect to $s$, $\boldsymbol d$ and $h$. We start by showing that 
	\begin{equation}\label{neg21}
		\begin{aligned}
			n^{-1} s^{- \lambda_0}\frac{\partial }{\partial \lambda} \widehat{\mathbb{Y}}_n(\lambda_0)^T  \widehat {\mathbb{B}} _n  n^{-1} s^{- \lambda_0}&\frac{\partial }{\partial \lambda}  \widehat{\mathbb{Y}}_n(\lambda_0)  \\
			&-   n^{-1}s_0^{- \lambda_0}  \frac{\partial }{\partial \lambda} {\mathbb{Y}}_n(\lambda_0)^T   {\mathbb{B}}_n n^{-1}s_0^{- \lambda_0} \frac{\partial }{\partial \lambda} {\mathbb{Y}}_n(\lambda_0)  
			= o_{\mathbb{P} }\left(1\right),
		\end{aligned}
	\end{equation}	
uniformly with respect to $s \in S_n$, $\boldsymbol d\in\mathcal D$ and $h\in\mathcal H_{sc,n}$. To obtain \eqref{neg21}, we  decompose the difference in a sum of the following four terms:
	\begin{align*}
		\widetilde R_{n1} &=  n^{-1} s^{- \lambda_0}\left( \frac{\partial }{\partial \lambda} \widehat{\mathbb{Y}}_n(\lambda_0) - \frac{\partial }{\partial \lambda}  {\mathbb{Y}}_n(\lambda_0)  \right)^T \; \widehat {\mathbb{B}} _n  n^{-1}s^{- \lambda_0}\frac{\partial }{\partial \lambda}   \widehat{\mathbb{Y}}_n(\lambda_0)  , \\ 
		\widetilde R_{n2} &= n^{-1} \left(s^{- 2\lambda_0} - s_0^{- 2\lambda_0}\right)\frac{\partial }{\partial \lambda}  {\mathbb{Y}}_n(\lambda_0)^T \; \widehat{\mathbb{B}} _n  n^{-1}\frac{\partial }{\partial \lambda}   \widehat{\mathbb{Y}}_n(\lambda_0),\\
		\widetilde R_{n3} &=  n^{-1}s_0^{- \lambda_0}\frac{\partial }{\partial \lambda}  {\mathbb{Y}}_n(\lambda_0)^T \;  \left[ \widehat {\mathbb{B}} _n -{\mathbb{B}} _n\right]n^{-1}s_0^{- \lambda_0} \frac{\partial }{\partial \lambda}     \widehat{\mathbb{Y}}_n(\lambda_0), \\
		\text{ and} \qquad    
		\widetilde R_{n4} &= n^{-1} s_0^{- \lambda_0} \frac{\partial }{\partial \lambda}  {\mathbb{Y}}_n(\lambda_0)^T \; {\mathbb{B}} _n  n^{-1}s_0^{- 
		\lambda_0}\left(  \frac{\partial }{\partial \lambda}   \widehat{\mathbb{Y}}_n(\lambda_0) - \frac{\partial }{\partial \lambda}   {\mathbb{Y}}_n(\lambda_0)\right).
	\end{align*}
It follows now that 
	\begin{multline*}
		\left| \widetilde R_{n1} \right| \leq s^{- 2\lambda_0}\left\| \boldsymbol \Omega_n^{1/2} n^{-1}  \left(\frac{\partial }{\partial \lambda} \mathbb{Y}_n(\lambda_0) -  \frac{\partial }{\partial \lambda} \widehat{\mathbb{Y}}_n(\lambda_0)\right) \right\|_{\rm{Sp}} \\ \times 
		\left\| \left( \boldsymbol{I}_{n\times n}  - P_{ \boldsymbol\Omega_n ^{1/2}{\boldsymbol{1}}_n   }\right)  \left( \boldsymbol{I}_{n\times n}  - P_{ \mathbb{S}_n  \widehat {\mathbb{X}}_n}   \right) \left( \boldsymbol{I}_{n\times n}  - P_{ \boldsymbol\Omega_n ^{1/2}{\boldsymbol{1}}_n   }\right)  \right\|_{\rm Sp}
		\left\| \boldsymbol \Omega_n^{1/2}    n^{-1}\frac{\partial }{\partial \lambda}   \widehat{\mathbb{Y}}_n(\lambda_0) \right\|_{\rm{Sp}},
	\end{multline*}
	\begin{multline*}
		\left| \widetilde R_{n2}\right| \leq \big| s^{- 2\lambda_0} - s_0^{- 2\lambda_0} \big| \left\| \boldsymbol \Omega_n^{1/2} n^{-1}  \frac{\partial }{\partial \lambda}\mathbb{Y}_n(\lambda_0)  \right\|_{\rm{Sp}} \\ 
		\times \left\| \left( \boldsymbol{I}_{n\times n}  - P_{ \boldsymbol\Omega_n ^{1/2}{\boldsymbol{1}}_n   }\right)  \left( \boldsymbol{I}_{n\times n}  -   P_{ \mathbb{S}_n  \widehat{\mathbb{X}}_n  }   \right) \left( \boldsymbol{I}_{n\times n}  - P_{ \boldsymbol\Omega_n ^{1/2}{\boldsymbol{1}}_n   }\right)  \right\|_{\rm Sp}\\ 
		\times \left\| \boldsymbol \Omega_n^{1/2}    n^{-1}  \frac{\partial }{\partial \lambda}   \widehat{\mathbb{Y}}_n(\lambda_0) \right\|_{\rm{Sp}},
	\end{multline*}	
	\begin{multline*}
		\left| \widetilde R_{n3} \right| \leq s_0^{- 2\lambda_0} \left\| \boldsymbol \Omega_n^{1/2} n^{-1}  \frac{\partial }{\partial \lambda}\mathbb{Y}_n(\lambda_0) \right\|_{\rm{Sp}} \\ 
		\times \left\| \left( \boldsymbol{I}_{n\times n}  - P_{ \boldsymbol\Omega_n ^{1/2}{\boldsymbol{1}}_n   }\right)  \left( P_{ \mathbb{S}_n  {\mathbb{X}}_n  } - P_{ \mathbb{S}_n  \widehat {\mathbb{X}}_n}\right) \left( \boldsymbol{I}_{n\times n}  - P_{ \boldsymbol\Omega_n ^{1/2}{\boldsymbol{1}}_n   }\right)  \right\|_{\rm Sp}
		\left\| \boldsymbol \Omega_n^{1/2}    n^{-1} \frac{\partial }{\partial \lambda}   \widehat{\mathbb{Y}}_n(\lambda_0) \right\|_{\rm{Sp}},
	\end{multline*}
and \newpage
	\begin{multline*}
		\left| \widetilde R_{n4}\right| \leq s_0^{- 2\lambda_0} \left\| \boldsymbol \Omega_n^{1/2} n^{-1}  \frac{\partial }{\partial \lambda}\mathbb{Y}_n(\lambda_0)  \right\|_{\rm{Sp}} \\ 
		\times \left\| \left( \boldsymbol{I}_{n\times n}  - P_{ \boldsymbol\Omega_n ^{1/2}{\boldsymbol{1}}_n   }\right)  \left( \boldsymbol{I}_{n\times n}  -   P_{ \mathbb{S}_n  {\mathbb{X}}_n  }   \right) \left( \boldsymbol{I}_{n\times n}  - P_{ \boldsymbol\Omega_n ^{1/2}{\boldsymbol{1}}_n   }\right)  \right\|_{\rm Sp}
		 \left\| \boldsymbol \Omega_n^{1/2}    n^{-1}  \left(  \frac{\partial }{\partial \lambda}   \widehat{\mathbb{Y}}_n(\lambda_0) - \frac{\partial }{\partial \lambda}   {\mathbb{Y}}_n(\lambda_0)\right)\right\|_{\rm{Sp}}.
	\end{multline*}
The uniform rate $o_{\mathbb{P} }\left(1\right)$ as  in equation \eqref{neg21} follows for $\widetilde R_{n1}$, $\widetilde R_{n2}$, $\widetilde R_{n3}$ and $\widetilde R_{n4}$ from the fact that the spectral norm of a product of projectors  is at most equal to 1, the spectral norm of $P_{ \mathbb{S}_n  {\mathbb{X}}_n} - P_{ \mathbb{S}_n  \widehat {\mathbb{X}}_n} $ tends to zero, $\sup_{s\in S_n}\big| s^{- 2\lambda_0} - s_0^{- 2\lambda_0} \big| = o_{\mathbb{P} }(1)$ as well as $\sup_{s\in S_n} s^{- 2\lambda_0}  = O_{\mathbb{P} }(1)$  and from Lemma \ref{yn_der}. Now the proof of \eqref{neg21} is complete. In order to proof \eqref{neg2} it remains to show
	\begin{equation}\label{neg22}
		n^{-1} \log(s)s^{- \lambda_0} \widehat{\mathbb{Y}}_n(\lambda_0)^T  \widehat {\mathbb{B}} _n  n^{-1} s^{- \lambda_0}\frac{\partial }{\partial \lambda}  \widehat{\mathbb{Y}}_n(\lambda_0)   
	= o_{\mathbb{P} }\left(1\right),
	\end{equation}
and 
	\begin{equation}\label{neg23}
		n^{-1} \log(s)s^{- \lambda_0} \widehat{\mathbb{Y}}_n(\lambda_0)^T  \widehat {\mathbb{B}} _n  n^{-1} \log(s)s^{- \lambda_0}  \widehat{\mathbb{Y}}_n(\lambda_0)   
	= o_{\mathbb{P} }\left(1\right),
	\end{equation}
uniformly with respect to $s \in S_n$, $\boldsymbol d\in\mathcal D$ and $h\in\mathcal H_{sc,n}$. We obtain \eqref{neg22} and \eqref{neg23} by a similiar reasoning as in the proof of \eqref{neg12}. The details are omitted. Now the proof of \eqref{neg2} is complete.	

Next,  we show that 
	\begin{equation}\label{neg3}
		n^{-1} s^{- \lambda_0}\widehat{\mathbb{Y}}_n(\lambda_0)^T  \widehat {\mathbb{B}} _n \frac{\partial ^2}{\partial \lambda^2} \left\{ n^{-1} s^{- \lambda_0}\widehat{\mathbb{Y}}_n(\lambda_0) \right\} 
		= o_{\mathbb{P} }\left(1\right)
	\end{equation}
uniformly with respect to $s \in S_n$, $\boldsymbol d\in\mathcal D$ and $h\in\mathcal H_{sc,n}$. Note that $ \frac{\partial^2 }{\partial \lambda^2} \{n^{-1}s^{- \lambda_0} \widehat{\mathbb{Y}}_n(\lambda_0) \}  = s^{- \lambda_0} n^{-1} \frac{\partial^2 }{\partial \lambda^2} \widehat{\mathbb{Y}}_n(\lambda_0) - 2\log(s)s^{- \lambda_0} n^{-1}\frac{\partial }{\partial \lambda} \widehat{\mathbb{Y}}_n(\lambda_0) + \log(s)^2s^{- \lambda_0} \widehat{\mathbb{Y}}_n(\lambda_0)$. We start by showing that 
	\begin{equation}\label{neg31}
		n^{-1} s^{- \lambda_0}\widehat{\mathbb{Y}}_n(\lambda_0)^T \; \widehat {\mathbb{B}} _n s^{- \lambda_0}  n^{-1}\frac{\partial ^2}{\partial \lambda^2}  \widehat{\mathbb{Y}}_n(\lambda_0) = o_{\mathbb{P} }\left(1\right),
	\end{equation}
uniformly with respect to $s \in S_n$, $\boldsymbol d\in\mathcal D$ and $h\in\mathcal H_{sc,n}$. Once again we can equivalently consider 
	\begin{equation}\label{neg31_bb}
		n^{-1} s^{- \lambda_0}\left(\widehat{\mathbb{Y}}_n(\lambda_0)- \widehat {\mathbb{X}} _n \boldsymbol \beta_0 \right)^T  \widehat {\mathbb{B}} _n 
		 n^{-1}\frac{\partial ^2}{\partial \lambda^2}  \widehat{\mathbb{Y}}_n(\lambda_0) 
		= o_{\mathbb{P} }\left(1\right),
	\end{equation}
uniformly with respect to $s \in S_n$, $\boldsymbol d\in\mathcal D$ and $h\in\mathcal H_{sc,n}$. To obtain \eqref{neg31_bb}, we  consider
	\begin{align*}
		\Big|n^{-1}s^{- \lambda_0}&\left( \widehat{\mathbb{Y}}_n(\lambda_0) - \widehat {\mathbb{X}} _n \boldsymbol \beta_0 \right)^T \;  \widehat {\mathbb{B}} _n 
		n^{-1}\frac{\partial ^2}{\partial \lambda^2} \widehat{\mathbb{Y}}_n(\lambda_0) \Big|  
		\\&\leq 
		\Bigg(\left\| \boldsymbol\Omega_n ^{1/2} n^{-1} \left(\left[ \widehat{\mathbb{Y}}_n(\lambda_0) - \widehat{\mathbb{X}} _n \boldsymbol \beta_0 \right] - \left[ {\mathbb{Y}}_n(\lambda_0) - {\mathbb{X}} _n \boldsymbol \beta_0 \right] + \left(\boldsymbol{\widehat{\varepsilon}_{|z}\widehat{f}_z}\right)_n\right)\right\|
	   \\&\hskip 2cm 
	    +
	     \left\| \boldsymbol\Omega_n ^{1/2} n^{-1}  \left[ {\mathbb{Y}}_n(\lambda_0) - {\mathbb{X}} _n \boldsymbol \beta_0 - \left(\boldsymbol{\widehat{\varepsilon}_{|z}\widehat{f}_z}\right)_n\right]\right\|
	     \Bigg)
		\\&\qquad  
		\times \left\| \left( \boldsymbol{I}_{n\times n}  - P_{ \boldsymbol\Omega_n ^{1/2}{\boldsymbol{1}}_n   }\right)  \left( \boldsymbol{I}_{n\times n}  - P_{ \mathbb{S}_n  \widehat {\mathbb{X}}_n}   \right) \left( \boldsymbol{I}_{n\times n}  - P_{ \boldsymbol\Omega_n ^{1/2}{\boldsymbol{1}}_n   }\right)  \right\|_{\rm Sp} 
		\times \left\| n^{-1}\boldsymbol \Omega_n^{1/2}    \frac{\partial ^2}{\partial \lambda^2}  \widehat{\mathbb{Y}}_n(\lambda_0) \right\|_{\rm{Sp}}.
	\end{align*}
The uniform rate $o_{\mathbb{P} }\left(1\right)$ as  in equation \eqref{neg31_bb} follows from the fact that the spectral norm of a product of projectors  is at most equal to 1, $\sup_{s\in S_n} s^{- 2\lambda_0}  = O_{\mathbb{P} }(1)$  and from Lemmas \ref{yn_hatt_5}, \ref{yn_der}, \ref{yn_hatt_6} and \ref{yn_hatt_7}. Now the proof of property \eqref{neg31} is complete. The property \eqref{neg3} follows now together with \eqref{neg22} and \eqref{neg23}. Therefore, the proof of the asymptotic representation of 
$\widehat\lambda - \lambda_0$ in the first statement in the Proposition \ref{AN_prop} is complete. 

In addition we have that 
	\begin{align*}
		\underset{\boldsymbol d \in \mathcal{D}}{\sup}
		\left\|
		n^{-2}\frac{\partial }{\partial \lambda} {\mathbb{Y}}_n(\lambda_0) ^T   {\mathbb{B}} _n  \frac{\partial }{\partial \lambda}  {\mathbb{Y}}_n(\lambda_0)
		-
		E\left[n^{-2}\frac{\partial }{\partial \lambda} {\mathbb{Y}}_n(\lambda_0) ^T   {\mathbb{B}} _n  \frac{\partial }{\partial \lambda}  {\mathbb{Y}}_n(\lambda_0)\right]
		\right\|
		= 
		O_{\mathbb{P}}(n^{-1/2}),
	\end{align*}
by Lemma \ref{Dn_inverse}, where the expectation tends to a positive constant. Furthermore, it follows from the fact that the spectral norm of a product of projectors  is at most equal to 1 and from Lemmas \ref{yn_hatt_5}, \ref{yn_der} and \ref{yn_hatt_7} that 
	\begin{align*}
		\underset{\boldsymbol d \in \mathcal{D}}{\sup}\left|n^{-2}\frac{\partial }{\partial \lambda} {\mathbb{Y}}_n(  \lambda_0) ^T{\mathbb{B}} _n \left((\boldsymbol{\varepsilon f_z})_n - \left(\boldsymbol{\widehat{\varepsilon}_{|z}\widehat{f}_z}\right)_n \right)\right| = O_{\mathbb{P} }\left(n^{-1/2}\right)
	\end{align*}
such that $\widehat\lambda - \lambda_0 = O_{\mathbb{P} }\left(n^{-1/2}\right)$ uniformly with respect to $s \in S_n$, $\boldsymbol d\in\mathcal D$ and $h\in\mathcal H_{sc,n}$.

We consider now the representation of $\widehat{\boldsymbol \beta}  (\widehat\lambda)  - \boldsymbol \beta_0  $. We have \color{black}
	\begin{equation*}
		\widehat{\boldsymbol \beta}  (\widehat\lambda) = \left(\widehat{\mathbb{X}}_n ^T{\mathbb{D}} _n\widehat{\mathbb{X}}_n   \right)^{-1} \widehat{\mathbb{X}}_n ^T {\mathbb{D}} _n \widehat{\mathbb{Y}}_n(\widehat\lambda) 
		= \left(n^{-2}\widehat{\mathbb{X}}_n ^T{\mathbb{D}} _n\widehat{\mathbb{X}}_n   \right)^{-1} n^{-1}\widehat{\mathbb{X}}_n ^T {\mathbb{D}} _n  n^{-1}\widehat{\mathbb{Y}}_n(\widehat\lambda). 
	\end{equation*}
We can write 
	\begin{align*}
		n^{-1}\widehat{\mathbb{X}}_n ^T {\mathbb{D}} _n  n^{-1}\widehat{\mathbb{Y}}_n(\widehat\lambda) 
		= 
		n^{-1}\widehat{\mathbb{X}}_n ^T {\mathbb{D}} _n  n^{-1}\widehat{\mathbb{Y}}_n(\lambda_0)
		+
		\left(n^{-1}\widehat{\mathbb{X}}_n ^T {\mathbb{D}} _n  n^{-1}\frac{\partial}{\partial \lambda}\widehat{\mathbb{Y}}_n(\lambda_0)  + 		R_{2,n}(\widetilde\lambda,\lambda_0)\right)(\widehat\lambda - \lambda_0),
	\end{align*}
where $\widetilde \lambda = c \widehat \lambda + (1-c)\lambda_0$ for some $c\in(0,1)$. We have that $\sup_{h\in \mathcal H_{sc,n}}\sup_{\boldsymbol d \in \mathcal D} \sup_{s\in S_n}| R_{2,n}(\widetilde\lambda,\lambda_0)| = o_{\mathbb{P}}(1)$, see Lemma \ref{remainder_2}. In addition, we get that
	\begin{align*}
		\left(n^{-2}\widehat{\mathbb{X}}_n ^T{\mathbb{D}} _n\widehat{\mathbb{X}}_n   \right)^{-1}n^{-1}\widehat{\mathbb{X}}_n ^T {\mathbb{D}} _n  n^{-1}\widehat{\mathbb{Y}}_n(\lambda_0)
		&=
		\left(n^{-2}\widehat{\mathbb{X}}_n ^T{\mathbb{D}} _n\widehat{\mathbb{X}}_n   \right)^{-1}n^{-1}\widehat{\mathbb{X}}_n ^T {\mathbb{D}} _n  n^{-1}\left(\widehat{\mathbb{Y}}_n(\lambda_0) - \widehat{\mathbb{X}}_n\boldsymbol \beta_0  +  \widehat{\mathbb{X}}_n\boldsymbol \beta_0 \right)\\
		&=
		\boldsymbol \beta_0 +
		\left(n^{-2}\widehat{\mathbb{X}}_n ^T{\mathbb{D}} _n\widehat{\mathbb{X}}_n   \right)^{-1}n^{-1}\widehat{\mathbb{X}}_n ^T {\mathbb{D}} _n  n^{-1}\left(\widehat{\mathbb{Y}}_n(\lambda_0) - \widehat{\mathbb{X}}_n\boldsymbol \beta_0\right).
	\end{align*}
In the first step we show that 
	\begin{align}\label{nb1}
		n^{-1}\widehat{\mathbb{X}}_n ^T \; {\mathbb{D}} _n \; n^{-1}\left(\widehat{\mathbb{Y}}_n(\lambda_0) - \widehat{\mathbb{X}}_n\boldsymbol \beta_0\right)
		-
		n^{-1}\mathbb{X}_n ^T {\mathbb{D}} _n\; n^{-1}\left(\mathbb{Y}_n(\lambda_0) - \mathbb{X}_n\boldsymbol \beta_0 - \left(\boldsymbol{\widehat{\varepsilon}_{|z}\widehat{f}_z}\right)_n \right) = o_{\mathbb{P} }(n^{-1/2}),
	\end{align}
uniformly with respect to $\boldsymbol d\in\mathcal D$ and $h\in\mathcal H_{sc,n}$.	
We have that 
	\begin{align*}
		n^{-1}\widehat{\mathbb{X}}_n ^T& {\mathbb{D}} _n\;  n^{-1}\left(\widehat{\mathbb{Y}}_n(\lambda_0) - \widehat{\mathbb{X}}_n\boldsymbol \beta_0\right)
		-
		n^{-1}\mathbb{X}_n ^T {\mathbb{D}} _n\;  n^{-1}\left(\mathbb{Y}_n(\lambda_0) - \mathbb{X}_n\boldsymbol \beta_0 - \left(\boldsymbol{\widehat{\varepsilon}_{|z}\widehat{f}_z}\right)_n \right) \\
		=&\;
		n^{-1}\widehat{\mathbb{X}}_n ^T {\mathbb{D}} _n\;  n^{-1}\left(\left(\widehat{\mathbb{Y}}_n(\lambda_0) - \widehat{\mathbb{X}}_n\boldsymbol \beta_0\right) - n^{-1}\left(\mathbb{Y}_n(\lambda_0) - \mathbb{X}_n\boldsymbol \beta_0 - \left(\boldsymbol{\widehat{\varepsilon}_{|z}\widehat{f}_z}\right)_n \right)\right)\\
		&+n^{-1}\left(\widehat{\mathbb{X}}_n - \mathbb{X}_n\right)^T {\mathbb{D}} _n\;  n^{-1}\left(\mathbb{Y}_n(\lambda_0) - \mathbb{X}_n\boldsymbol \beta_0 - \left(\boldsymbol{\widehat{\varepsilon}_{|z}\widehat{f}_z}\right)_n \right).
	\end{align*}
It follows that 
	\begin{multline*}
		\left\|n^{-1}\widehat{\mathbb{X}}_n ^T {\mathbb{D}} _n\left(  n^{-1}\left(\widehat{\mathbb{Y}}_n(\lambda_0) - \widehat{\mathbb{X}}_n\boldsymbol \beta_0\right) - n^{-1}\left(\mathbb{Y}_n(\lambda_0) - \mathbb{X}_n\boldsymbol \beta_0- \left(\boldsymbol{\widehat{\varepsilon}_{|z}\widehat{f}_z}\right)_n \right)\right)\right\|_{\rm Sp}
		\\\leq 
		\left\|\boldsymbol\Omega_n^{1/2} n^{-1}\widehat{\mathbb{X}}_n\right\|_{\rm{Sp}} \times 
		\left\|\left( \boldsymbol{I}_{n\times n}  - P_{ \boldsymbol\Omega_n^{1/2} {\boldsymbol{1}}_n   }\right)\right\|_{\rm Sp} \\\times 
		\left\|\boldsymbol\Omega_n^{1/2} n^{-1}\left( \left(\widehat{\mathbb{Y}}_n(\lambda_0) - \widehat{\mathbb{X}}_n\boldsymbol \beta_0\right) - \left(\mathbb{Y}_n(\lambda_0) - \mathbb{X}_n\boldsymbol \beta_0- \left(\boldsymbol{\widehat{\varepsilon}_{|z}\widehat{f}_z}\right)_n \right)\right)\right\|,
		\end{multline*}
and \newpage
	\begin{multline*}
		\left\|n^{-1}\left(\widehat{\mathbb{X}}_n - \mathbb{X}_n\right)^T {\mathbb{D}} _n\; n^{-1}\left(\mathbb{Y}_n(\lambda_0) - \mathbb{X}_n\boldsymbol \beta_0- \left(\boldsymbol{\widehat{\varepsilon}_{|z}\widehat{f}_z}\right)_n \right)\right\|_{\rm Sp}
		\\\leq 
		\left\|\boldsymbol\Omega_n^{1/2}n^{-1}\left(\widehat{\mathbb{X}}_n - \mathbb{X}_n\right)\right\|_{\rm{Sp}} \times  
		\left\|\left( \boldsymbol{I}_{n\times n}  - P_{ \boldsymbol\Omega_n^{1/2} {\boldsymbol{1}}_n   }\right)\right\|_{\rm Sp} \times 
		\left\|\boldsymbol\Omega_n^{1/2}n^{-1}\left(\mathbb{Y}_n(\lambda_0) - \mathbb{X}_n\boldsymbol \beta_0- \left(\boldsymbol{\widehat{\varepsilon}_{|z}\widehat{f}_z}\right)_n \right)\right\|.
	\end{multline*}
The uniform rate $o_{\mathbb{P} }\left(n^{-1/2}\right)$ as  in equation \eqref{nb1} follows from the fact that the spectral norm of a product of projectors  is at most equal to 1 and from Lemmas \ref{Xn_hat}, \ref{yn_hatt_5}, \ref{yn_hatt_6} and \ref{yn_hatt_7}.

In the next step we show that
	\begin{align}\label{nb2}
		n^{-1}\widehat{\mathbb{X}}_n ^T {\mathbb{D}} _n  n^{-1}\frac{\partial}{\partial \lambda}\widehat{\mathbb{Y}}_n(\lambda_0) (\widehat\lambda - \lambda_0) 
		-
		n^{-1}\mathbb{X}_n ^T {\mathbb{D}} _n  n^{-1}\frac{\partial}{\partial \lambda}\mathbb{Y}_n(\lambda_0) (\widehat\lambda - \lambda_0) = o_{\mathbb{P} }(n^{-1/2}) ,
	\end{align}
uniformly with respect to $s \in S_n$, $\boldsymbol d\in\mathcal D$ and $h\in\mathcal H_{sc,n}$.	
We have that 
	\begin{align*}
		n^{-1}\widehat{\mathbb{X}}_n ^T {\mathbb{D}} _n  n^{-1}\frac{\partial}{\partial \lambda}\widehat{\mathbb{Y}}_n(\lambda_0) (\widehat\lambda - \lambda_0) 
		&-
		n^{-1}\mathbb{X}_n ^T {\mathbb{D}} _n  n^{-1}\frac{\partial}{\partial \lambda}\mathbb{Y}_n(\lambda_0) (\widehat\lambda - \lambda_0) \\
		&=
		n^{-1}\widehat{\mathbb{X}}_n ^T {\mathbb{D}} _n  n^{-1}
		\left(\frac{\partial}{\partial \lambda}\widehat{\mathbb{Y}}_n(\lambda_0)  - \frac{\partial}{\partial \lambda}\mathbb{Y}_n(\lambda_0)\right)(\widehat\lambda - \lambda_0)\\
		& \quad + 
		n^{-1}\left(\widehat{\mathbb{X}}_n - \mathbb{X}_n\right)^T {\mathbb{D}} _nn^{-1}\frac{\partial}{\partial \lambda}\mathbb{Y}_n(\lambda_0) (\widehat\lambda - \lambda_0).
	\end{align*}
It follows that 
	\begin{multline*}
		\left\|n^{-1}\widehat{\mathbb{X}}_n ^T {\mathbb{D}} _n \; n^{-1}\left(\frac{\partial}{\partial \lambda}\widehat{\mathbb{Y}}_n(\lambda_0)  - \frac{\partial}{\partial \lambda}\mathbb{Y}_n(\lambda_0)\right)\right\|_{\rm Sp}
		\\\leq 
		\left\|\boldsymbol\Omega_n^{1/2}n^{-1}\widehat{\mathbb{X}}_n\right\|_{\rm{Sp}} \times  
		\left\|\left( \boldsymbol{I}_{n\times n}  - P_{ \boldsymbol\Omega_n^{1/2} {\boldsymbol{1}}_n   }\right)\right\|_{\rm Sp} \times 
		\left\|\boldsymbol\Omega_n^{1/2} \; n^{-1}\left(\frac{\partial}{\partial \lambda}\widehat{\mathbb{Y}}_n(\lambda_0)  - \frac{\partial}{\partial \lambda}\mathbb{Y}_n(\lambda_0)\right)\right\|_{\rm{Sp}}
	\end{multline*}
and 
	\begin{multline*}
		\left\|n^{-1}\left(\widehat{\mathbb{X}}_n - \mathbb{X}_n\right)^T {\mathbb{D}} _n n^{-1}\frac{\partial}{\partial \lambda}\mathbb{Y}_n(\lambda_0) \right\|_{\rm Sp}
		\\\leq 
		\left\|\boldsymbol\Omega_n^{1/2}n^{-1}\left(\widehat{\mathbb{X}}_n - \mathbb{X}_n\right)\right\|_{\rm{Sp}} \times  
		\left\|\left( \boldsymbol{I}_{n\times n}  - P_{ \boldsymbol\Omega_n^{1/2} {\boldsymbol{1}}_n   }\right)\right\|_{\rm Sp} \times 
		\left\|\boldsymbol\Omega_n^{1/2}n^{-1}\frac{\partial}{\partial \lambda}\mathbb{Y}_n(\lambda_0) \right\|_{\rm{Sp}}.
	\end{multline*}
The uniform rate $o_{\mathbb{P} }\left(n^{-1/2}\right)$ as  in equation \eqref{nb2} follows from the fact that the spectral norm of a product of projectors  is at most equal to 1, from Lemma \ref{Xn_hat} and \ref{yn_der} and the fact that $\widehat\lambda - \lambda_0 = O_{\mathbb{P} }\left(n^{-1/2}\right)$ uniformly with respect to $s \in S_n$, $\boldsymbol d\in\mathcal D$ and $h\in\mathcal H_{sc,n}$. In addition, it follows from Lemmas \ref{Dn_inverse} and \ref{spectral_B_hat2} that 
	\begin{align*}
		\left\| \left(n^{-2}\widehat{\mathbb{X}}_n ^T{\mathbb{D}} _n\widehat{\mathbb{X}}_n   \right)^{-1}
		- 
		\left(n^{-2}\mathbb{X}_n ^T{\mathbb{D}} _n\mathbb{X}_n  \right)^{-1}
		\right\|
		=
		o_{\mathbb{P} }\left(1\right).
	\end{align*}
Furthermore, we get that 
	\begin{align*}
		\left\| n^{-1}\mathbb{X}_n ^T {\mathbb{D}} _n  n^{-1}\left(\mathbb{Y}_n(\lambda_0) - \mathbb{X}_n\boldsymbol \beta_0- \left(\boldsymbol{\widehat{\varepsilon}_{|z}\widehat{f}_z}\right)_n \right) \right \|_{\rm{Sp}} = O_{\mathbb{P} }\left(n^{-1/2}\right)
	\end{align*}
and
	\begin{align*}
		\left\| n^{-1}\mathbb{X}_n ^T {\mathbb{D}} _n  n^{-1}\frac{\partial}{\partial \lambda}\mathbb{Y}_n(\lambda_0) (\widehat\lambda - \lambda_0) \right \|_{\rm{Sp}} = O_{\mathbb{P} }\left(n^{-1/2}\right)
	\end{align*}	
from Lemmas \ref{Xn_hat}, \ref{yn_hatt_5}, \ref{yn_der}, \ref{yn_hatt_7} and $\widehat\lambda - \lambda_0 = O_{\mathbb{P} }\left(n^{-1/2}\right)$ uniformly with respect to $s \in S_n$, $\boldsymbol d\in\mathcal D$ and $h\in\mathcal H_{sc,n}$. Gathering facts, the second statement follows, and the proof of Proposition \ref{AN_prop} is complete. \\
\end{proofof}

\medskip

\begin{proofof}{Theorem \ref{AN}} ~~\\
Let 
	\begin{align*}
		\boldsymbol{V}_n(\boldsymbol{d}) &= 
		\begin{pmatrix}
		\frac{\partial}{\partial \lambda}\mathbb{Y}_n(\lambda_0)^T{\mathbb{D}}_n(\boldsymbol{d})\frac{\partial}{\partial \lambda}\mathbb{Y}_n(\lambda_0) & 
		- \frac{\partial}{\partial \lambda}\mathbb{Y}_n(\lambda_0)^T{\mathbb{D}}_n(\boldsymbol{d}) \mathbb{X}_n\\
		- \mathbb{X}_n^T{\mathbb{D}}_n(\boldsymbol{d})\frac{\partial}{\partial \lambda}\mathbb{Y}_n(\lambda_0) &\mathbb{X}_n^T{\mathbb{D}}_n(\boldsymbol{d})\mathbb{X}_n
		\end{pmatrix}
		\hskip 0.5cm \text{and}  \\	 
		\boldsymbol{A}_n &= 
		\begin{pmatrix}
		\frac{\partial}{\partial \lambda}\mathbb{Y}_n(\lambda_0)^T\\
		- \mathbb{X}_n^T
		\end{pmatrix}
		{\mathbb{D}}_n\left( (\boldsymbol{\varepsilon f_z})_n - \left(\boldsymbol{\widehat{\varepsilon}_{|z}\widehat{f}_z}\right)_n\right).
	\end{align*}
We get from Proposition \ref{AN_prop} that 	
	\begin{align*}
		\left((\widehat \lambda, \widehat {\boldsymbol \beta}(\widehat \lambda )^T)^T  - (\lambda_0, \boldsymbol {\beta}_0 ^T)^T\right) = - \boldsymbol{V}_n(\boldsymbol{d})^{-1}\boldsymbol{A}_n + 
		  o_{\mathbb{P} }\left(n^{-1/2}\right).
	\end{align*}
Furthermore, it follows from Lemma \ref{Dn_inverse} that 
	\begin{align*}
		\left((\widehat \lambda, \widehat {\boldsymbol \beta}(\widehat \lambda )^T)^T  - (\lambda_0, \boldsymbol {\beta}_0 ^T)^T\right) = - \boldsymbol{V}(\boldsymbol{d})^{-1} n^{-2}\boldsymbol{A}_n + 
		  o_{\mathbb{P} }\left(n^{-1/2}\right).
	\end{align*}
Both results hold uniformly with respect to $s \in S_n$, $\boldsymbol d\in\mathcal D$ and $h\in\mathcal H_{sc,n}$.	
Note that $\boldsymbol{V}(\boldsymbol{d})$ is invertible as $E\left[\mathbb{X}_n^T{\mathbb{D}}_n\mathbb{X}_n\right]$ tends to a positive definite matrix and $E\left[\frac{\partial}{\partial \lambda}\mathbb{Y}_n(\lambda_0)^T{\mathbb{B}}_n\frac{\partial}{\partial \lambda}\mathbb{Y}_n(\lambda_0) \right]$ to a positive constant, see Lemma \ref{Dn_inverse}. We consider now $\boldsymbol A_n$ and start with $\mathbb{X}_n^T{\mathbb{D}}_n\left((\boldsymbol{\varepsilon f_z})_n - \left(\boldsymbol{\widehat{\varepsilon}_{|z}\widehat{f}_z}\right)_n\right)$. Recall that 
	\begin{multline*}
		\mathbb{X}_n^T{\mathbb{D}}_n\left((\boldsymbol{\varepsilon f_z})_n - \left(\boldsymbol{\widehat{\varepsilon}_{|z}\widehat{f}_z}\right)_n\right)	\\
		=
	  \mathbb{X}_n^T\boldsymbol\Omega_n\left((\boldsymbol{\varepsilon f_z})_n - \left(\boldsymbol{\widehat{\varepsilon}_{|z}\widehat{f}_z}\right)_n\right)
	  - \frac{1}{\boldsymbol{1}_n^T \boldsymbol\Omega_n  \boldsymbol{1}_n} \mathbb{X}_n^T\boldsymbol\Omega_n  \boldsymbol{1}_n  \boldsymbol{1}_n^T\boldsymbol\Omega_n  \left((\boldsymbol{\varepsilon f_z})_n - \left(\boldsymbol{\widehat{\varepsilon}_{|z}\widehat{f}_z}\right)_n\right).
	\end{multline*}
It follows from the results of Lemma \ref{Dn_inverse} that 
	\begin{align*}
		\sup_{\boldsymbol d \in \mathcal D} \left\|\frac{1}{\boldsymbol{1}_n^T \boldsymbol\Omega_n  \boldsymbol{1}_n}{\mathbb{X}}_n ^T\boldsymbol\Omega_n  \boldsymbol{1}_n  - 
		\frac{1}{E\left[\boldsymbol{1}_n^T \boldsymbol\Omega_n  \boldsymbol{1}_n\right]}E\left[{\mathbb{X}}_n ^T\boldsymbol\Omega_n  \boldsymbol{1}_n \right] \right\|_{\rm{Sp}}
		= O_{\mathbb{P}}(n^{-1/2})
	\end{align*}
and together with the results of Lemmas \ref{spectral_B_hat}, \ref{yn_hatt_5} and \ref{yn_hatt_7} we get that 
	\begin{multline*}
		\sup_{\boldsymbol d \in \mathcal D} 
		\Bigg\|\frac{1}{\boldsymbol{1}_n^T \boldsymbol\Omega_n  \boldsymbol{1}_n}{\mathbb{X}}_n ^T\boldsymbol\Omega_n  \boldsymbol{1}_n  \frac{1}{n^2} \boldsymbol{1}_n^T  \boldsymbol\Omega_n\left((\boldsymbol{\varepsilon f_z})_n - \left(\boldsymbol{\widehat{\varepsilon}_{|z}\widehat{f}_z}\right)_n\right) 
		\\- 
		  \frac{1}{E\left[\boldsymbol{1}_n^T \boldsymbol\Omega_n  \boldsymbol{1}_n\right]}E\left[{\mathbb{X}}_n ^T\boldsymbol\Omega_n  \boldsymbol{1}_n \right] \frac{1}{n^2}\boldsymbol{1}_n^T  \boldsymbol\Omega_n\left((\boldsymbol{\varepsilon f_z})_n - \left(\boldsymbol{\widehat{\varepsilon}_{|z}\widehat{f}_z}\right)_n\right)
		\Bigg\|_{\rm{Sp}}
		= o_{\mathbb{P}}(n^{-1/2}).
	\end{multline*}
In the next step we consider 
	\begin{align*}
		\frac{1}{n^2}\mathbb{X}_n^T\boldsymbol\Omega_n(\boldsymbol{\varepsilon f_z})_n 
		= 
		  \frac{1}{2n^2} \sum_{1\leq i\neq j \leq n} \left(\mathbb{X}_{n,i}\varepsilon_j{f}_z(\boldsymbol Z_j)\boldsymbol\Omega_{n,ij} + \mathbb{X}_{n,j}\varepsilon_i{f}_z(\boldsymbol Z_i)\boldsymbol\Omega_{n,ji}\right) 
			+
		  \frac{1}{n^2} \sum_{i=1}^{n} \mathbb{X}_{n,i}\varepsilon_i{f}_z(\boldsymbol Z_i).
	\end{align*}
It's easy to check that 
	\begin{align*}
		\left\| \frac{1}{n^2} \sum_{i=1}^{n} \mathbb{X}_{n,i}\varepsilon_i{f}_z(\boldsymbol Z_i)\right\| = o_{\mathbb{P}}(n^{-1}).
	\end{align*}
In addition, we have that $E\left[\mathbb{X}_{n,i}\varepsilon_j{f}_z(\boldsymbol Z_j)\boldsymbol\Omega_{n,ij}	\right] = 0$ and $E\left[\mathbb{X}_{n,i}\varepsilon_j{f}_z(\boldsymbol Z_j)\boldsymbol\Omega_{n,ij}	\mid \boldsymbol X_i, \boldsymbol Z_i\right] = 0$ as well as
	\begin{align*}
		E\left[\mathbb{X}_{n,i}\varepsilon_j{f}_z(\boldsymbol Z_j)\boldsymbol\Omega_{n,ij}	\mid Y_j, \boldsymbol X_j, \boldsymbol Z_j\right]
		=		
		\varepsilon_j{f}_z(\boldsymbol Z_j)E\left[\mathbb{X}_{n,i}\boldsymbol\Omega_{n,ij}	\mid \boldsymbol X_j, \boldsymbol Z_j\right].
	\end{align*}
Therefore, we get by applying Hoeffding's decomposition that
	\begin{align*}
		\sup_{\boldsymbol d \in \mathcal D}\left\|\frac{1}{n^2}\mathbb{X}_n^T\boldsymbol\Omega_n(\boldsymbol{\varepsilon f_z})_n 
		-
		  \frac{1}{n}\sum_{j=1}^{n}\varepsilon_j{f}_z(\boldsymbol Z_j)E\left[\mathbb{X}_{n,i}\boldsymbol\Omega_{n,ij}	\mid \boldsymbol X_j, \boldsymbol Z_j\right]\right\|
		  = O_{\mathbb{P}}(n^{-1}).
	\end{align*} 
By Lemma \ref{yn_hatt_5} it suffices to consider $\boldsymbol d = \rm{diag}(d_U, \ldots, d_U)$ such that the uniform result in the last display follows.  	
By the same reasoning we get that 
	\begin{align*}
		\sup_{\boldsymbol d \in \mathcal D}\left\|\frac{1}{n^2}\boldsymbol{1}_n^T\boldsymbol\Omega_n(\boldsymbol{\varepsilon f_z})_n 
		-
		\frac{1}{n}\sum_{j=1}^{n}\varepsilon_j{f}_z(\boldsymbol Z_j)E\left[\boldsymbol\Omega_{n,ij}	\mid \boldsymbol X_j, \boldsymbol Z_j\right]\right\|
		= O_{\mathbb{P}}(n^{-1}).
	\end{align*}

In the next step we consider 
	\begin{align*}
		\frac{1}{n^2}\mathbb{X}_n^T\boldsymbol\Omega_n\left(\boldsymbol{\widehat{\varepsilon}_{|z}\widehat{f}_z}\right)_n 
		= 
		  \frac{1}{n^2} \sum_{1\leq i\neq j \leq n} \mathbb{X}_{n,i}\left(\boldsymbol{\widehat{\varepsilon}_{|z}\widehat{f}_z}\right)_{n,j}\boldsymbol\Omega_{n,ij} 
			+
		  \frac{1}{n^2} \sum_{i=1}^{n} \mathbb{X}_{n,i}\left(\boldsymbol{\widehat{\varepsilon}_{|z}\widehat{f}_z}\right)_{n,i}.
	\end{align*}
It's easy to check that 
	\begin{align*}
		\left\| \frac{1}{n^2} \sum_{i=1}^{n} \mathbb{X}_{n,i}\left(\boldsymbol{\widehat{\varepsilon}_{|z}\widehat{f}_z}\right)_{n,i}\right\| = o_{\mathbb{P}}(n^{-1/2}).
	\end{align*}	
In addition, we have that 
	\begin{align*}
		\frac{1}{n^2} \sum_{1\leq i\neq j \leq n} \mathbb{X}_{n,i}\left(\boldsymbol{\widehat{\varepsilon}_{|z}\widehat{f}_z}\right)_{n,j}\boldsymbol\Omega_{n,ij}
		&=
		\frac{1}{n^2} \sum_{1\leq i\neq j \leq n} \mathbb{X}_{n,i}\frac{1}{n} \sum_{k=1, k\neq j}^{n} \varepsilon_k K_{h,jk}\boldsymbol\Omega_{n,ij}		\\
		&=
		\frac{1}{n^3} \sum_{1\leq i\neq j \neq k \leq n} \mathbb{X}_{n,i} \varepsilon_k K_{h,jk}\boldsymbol\Omega_{n,ij}		\\		
		&\quad + 
		\frac{1}{n^3} \sum_{1\leq i\neq j \leq n} \mathbb{X}_{n,i} \varepsilon_i K_{h,ij}\boldsymbol\Omega_{n,ij}\\
		&=A_n(h) + B_n(h).	
	\end{align*}

In the following we compute the mean and use the Hoeffding decomposition for the $U-$process $A_n(h)$. The kernel of $A_{n}(h)$ is not symmetric in its arguments. However, we could apply the usual symmetrization idea. Thus, by abuse, we will proceed as if the kernel of the $U-$statistic we handle is symmetric. For instance, for a second order $U-$statistic defined by a kernel $h( \boldsymbol U_i, \boldsymbol U_j)$, we could replace it  by the symmetric kernel $\left[h(\boldsymbol U_i,\boldsymbol U_j) + h(\boldsymbol U_j,\boldsymbol U_i)\right]/2$ from which we get the same $U-$statistic. Here, $\boldsymbol U_i = \left(Y_i,\boldsymbol X_i^T, \boldsymbol Z_i^T\right)^T$. 

In addition, we have that the kernel of $A_{n}(h)$ is Euclidean for a squared integrable envelope. See Lemma 22 in \citet{nolan1987u} and Lemma 2.14 in \citet{pakes1989simulation}. Therefore, we can in the following repeatedly apply Corollary 7 and the Maximal Inequality of \citet{sherman1994maximal}. All remainder terms are controlled by Assumption \ref{ass_con}.2.

Recall that by assumption $E\left[\varepsilon_k\mid \boldsymbol X_k, \boldsymbol Z_k\right] = 0$. Therefore, we get that $E\left[A_{n}(h)\right] = 0$ as well as 
	$$
		E\left[\mathbb{X}_{n,i} \varepsilon_k K_{h,jk}\boldsymbol\Omega_{n,ij}\mid \boldsymbol U_p, p\in \{i,j\}\right] = 0.
	$$  
Furthermore we get that 
	\begin{align*}
		E\left[\mathbb{X}_{n,i} \varepsilon_k K_{h,jk}\boldsymbol\Omega_{n,ij}\mid \boldsymbol U_k\right] 
		&=
		\varepsilon_k E\left[\mathbb{X}_{n,i}  K_{h,jk}\boldsymbol\Omega_{n,ij}^X\boldsymbol\Omega_{n,ij}^Z\mid \boldsymbol Z_k\right] 	\\
		&=
		\varepsilon_k E\left[\mathbb{X}_{n,i}  E\left[K_{h,jk}\boldsymbol\Omega_{n,ij}^Z\mid \boldsymbol Z_k,\boldsymbol Z_i, \boldsymbol X_j\right] \boldsymbol\Omega_{n,ij}^X\mid \boldsymbol Z_k\right] 	\\	
		&=
		\varepsilon_k E\left[\mathbb{X}_{n,i}  \left(f_z(\boldsymbol Z_k)\boldsymbol\Omega_{n,ik}^Z + O_{\mathbb{P}}(h^2) \right)\boldsymbol\Omega_{n,ij}^X\mid \boldsymbol Z_k\right] 	\\
		&=
		\varepsilon_k f_z(\boldsymbol Z_k) E\left[\mathbb{X}_{n,i}  \boldsymbol\Omega_{n,ik}^Z \boldsymbol\Omega_{n,ij}^X\mid \boldsymbol Z_k\right] 	
		+\varepsilon_k E\left[\mathbb{X}_{n,i}  \boldsymbol\Omega_{n,ij}^X\right]  O_{\mathbb{P}}(h^2).								
	\end{align*}

It follows from the results that the first order $U$--process of the Hoeffding decomposition of $A_{n}(h)$ is of order $O_{\mathbb{P}}(n^{-1/2})$ uniformly with respect to $h$ and $\boldsymbol d$.

We consider now the three second order $U-$processes of the Hoeffding decomposition of $A_{n}(h)$. We get that 
	\begin{align*}
		E\left[\mathbb{X}_{n,i} \varepsilon_k K_{h,jk}\boldsymbol\Omega_{n,ij}\mid \boldsymbol U_i, \boldsymbol U_j\right] = 0. 
	\end{align*}
In addition, 
	\begin{align*}
		E\left[\mathbb{X}_{n,i} \varepsilon_k K_{h,jk}\boldsymbol\Omega_{n,ij}\mid \boldsymbol U_i, \boldsymbol U_k\right] 
		&= 
	   	\mathbb{X}_{n,i} \varepsilon_k E\left[ K_{h,jk}\boldsymbol\Omega_{n,ij}^X\boldsymbol\Omega_{n,ij}^Z\mid \boldsymbol U_i, \boldsymbol Z_k\right] \\
	   	&=
	   	\mathbb{X}_{n,i} \varepsilon_k E\left[ \left(f_z(\boldsymbol Z_k)\boldsymbol\Omega_{n,ik}^Z + O_{\mathbb{P}}(h^2)\right)\boldsymbol\Omega_{n,ij}^X\mid \boldsymbol U_i, \boldsymbol
	   	Z_k\right]\\
	   	&=
	   	\mathbb{X}_{n,i} \varepsilon_k f_z(\boldsymbol Z_k)\boldsymbol\Omega_{n,ik}^Z E\left[ \boldsymbol\Omega_{n,ij}^X\mid \boldsymbol X_i\right]	 
	   	+
	   	\mathbb{X}_{n,i} E\left[ \boldsymbol\Omega_{n,ij}^X\mid \boldsymbol X_i\right] O_{\mathbb{P}}(h^2).
	\end{align*}
The last conditional expectation that we need to consider is given by
	\begin{align*}
		E\left[\mathbb{X}_{n,i} \varepsilon_k K_{h,jk}\boldsymbol\Omega_{n,ij}\mid \boldsymbol U_j, \boldsymbol U_k\right] 
		&=
		\varepsilon_k K_{h,jk} E\left[\mathbb{X}_{n,i} \boldsymbol\Omega_{n,ij}\mid \boldsymbol U_j\right] 	
		= 		h^{-q} h^{q}K_{h,jk} \tau(\boldsymbol U_j, \boldsymbol U_k).
	\end{align*}
Now, we apply the Maximal Inequality of \citet{sherman1994maximal}, page 448,  for the degenerate $U-$process given by the kernel $h^q K_{h,jk}  \tau(\boldsymbol U_j, \boldsymbol U_k)$, indexed by $h\in\mathcal H_{sc,n}$, with envelope $\|K\|_\infty \tau(\cdot,\cdot)$. (Herein, $\|\cdot\|_\infty $ denotes the uniform norm.) We take $p=1$ and $\beta\in (0,1)$ arbitrarily close to 1 to stand for Sherman's quantity $\alpha$. Since $K(\cdot)$ is of bounded variation and symmetric, without loss of generality we could consider 
that $K(\cdot)$ is nonincreasing on $[0,\infty)$. In this case, $0\leq K(\cdot/h)\leq K(\cdot/\overline h)$ with $\overline h = \sup \mathcal H_{sc,n} = : c_{max} n^{-\alpha}$.
Hence, using Jensen's inequality, we could bound the right-hand side of the Maximal Inequality of  \citet{sherman1994maximal} by a universal constant times
$$
	\left(E \left[ K^{2}\left(\frac{\boldsymbol Z_j - \boldsymbol Z_k}{c_{max} n^{-\alpha}}\right)\tau^{2} (\boldsymbol U_j, \boldsymbol U_k)\right]\right)^{\beta/2}.
$$
By standard changes of variables and suitable integrability conditions, the power $\beta/2$ of the expectation in the last display is bounded by  a constant times $n^{-\alpha \beta q/2}$. Consequently, the uniform rate of the $U-$process obtained conditioning on $\boldsymbol U_j, \boldsymbol U_k$ is $n^{-1}\times O_{\mathbb P} (n^{\alpha q \{1 -\beta /2\} })$. As $1/2 - \alpha q (1 - \beta /2) > 0 $ under our assumptions we get that $n^{-1}\times O_{\mathbb P} (n^{\alpha q \{1 -\beta /2\} })= o_{\mathbb P} (n^{-1/2})$. 
From all the results it follows that the second order $U-$processes of the Hoeffding decomposition of $A_{n}(h)$ are of order $o_{\mathbb P} (n^{-1/2})$ uniformly with respect to $h$ and $\boldsymbol d$.

Finally, we need to consider the third order $U-$process. We get that 
	\begin{align*}
		\mathbb{X}_{n,i} \varepsilon_k K_{h,jk}\boldsymbol\Omega_{n,ij} = h^{-q} h^{q}K_{h,jk} \tau_1(\boldsymbol U_i, \boldsymbol U_j, \boldsymbol U_k).
	\end{align*}
We can again use the Maximal Inequality of \citet{sherman1994maximal} to argue that this process is of order $o_{\mathbb P} (n^{-1/2})$ uniformly with respect to $h$ and $\boldsymbol d$. The details are omitted.	

It remains to consider $B_n(h)$. One can argue in a similar way as for $A_n(h)$ to get that $B_n(h)$ is of order $o_{\mathbb P} (n^{-1/2})$ uniformly with respect to $h$ and $\boldsymbol d$. The details are omitted.

From all the results it follows now that
	\begin{align*}
		\sup_{h \in \mathcal H_{sc,n}}\sup_{\boldsymbol d \in \mathcal D}
		\left\|\frac{1}{n^2}\mathbb{X}_n^T\boldsymbol\Omega_n\left(\boldsymbol{\widehat{\varepsilon}_{|z}\widehat{f}_z}\right)_n 
		- 	\frac{1}{n}	\sum_{k=1}^{n}\varepsilon_k f_z(\boldsymbol Z_k) E\left[\mathbb{X}_{n,i}  \boldsymbol\Omega_{n,ik}^Z \boldsymbol\Omega_{n,ij}^X\mid \boldsymbol Z_k\right] 
		\right\| = o_{\mathbb P} (n^{-1/2}).
	\end{align*}
By the same reasoning we get that 
	\begin{align*}
		\sup_{h \in \mathcal H_{sc,n}}\sup_{\boldsymbol d \in \mathcal D}
		\left\|\frac{1}{n^2}\boldsymbol 1_n^T\boldsymbol\Omega_n\left(\boldsymbol{\widehat{\varepsilon}_{|z}\widehat{f}_z}\right)_n 
		- 	\frac{1}{n}	\sum_{k=1}^{n}\varepsilon_k f_z(\boldsymbol Z_k) E\left[  \boldsymbol\Omega_{n,ik}^Z \boldsymbol\Omega_{n,ij}^X\mid \boldsymbol Z_k\right] 
		\right\| = o_{\mathbb P} (n^{-1/2}).
	\end{align*}
	
Therefore, we get that 
	\begin{align*}
		\sup_{h \in \mathcal H_{sc,n}}\sup_{\boldsymbol d \in \mathcal D}
		\Bigg\|&\frac{1}{n^2}\mathbb{X}_n^T{\mathbb{D}}_n\left((\boldsymbol{\varepsilon f_z})_n - \left(\boldsymbol{\widehat{\varepsilon}_{|z}\widehat{f}_z}\right)_n\right)	\\
		&-
		\frac{1}{n}\sum_{j=1}^{n}\varepsilon_j{f}_z(\boldsymbol Z_j)E\left[\left(\mathbb{X}_{n,i} -\frac{1}{E\left[\boldsymbol{1}_n^T \boldsymbol\Omega_n  \boldsymbol{1}_n\right]}E\left[{\mathbb{X}}_n ^T\boldsymbol\Omega_n  \boldsymbol{1}_n \right]\right)	\boldsymbol\Omega_{n,ij}\mid \boldsymbol X_j, \boldsymbol Z_j\right]\\
		&+
		\frac{1}{n}	\sum_{k=1}^{n}\varepsilon_k f_z(\boldsymbol Z_k) E\left[ \left(\mathbb{X}_{n,i} -\frac{1}{E\left[\boldsymbol{1}_n^T \boldsymbol\Omega_n  \boldsymbol{1}_n\right]}E\left[{\mathbb{X}}_n ^T\boldsymbol\Omega_n  \boldsymbol{1}_n \right]\right) \boldsymbol\Omega_{n,ik}^Z \boldsymbol\Omega_{n,ij}^X\mid \boldsymbol Z_k\right] 
		\Bigg\|
	    = o_{\mathbb{P}}(n^{-1/2}).
	\end{align*}	
	
By the same arguments we get that 
	\begin{align*}
		\sup_{h \in \mathcal H_{sc,n}}&\sup_{\boldsymbol d \in \mathcal D}\Bigg\|
		\frac{1}{n^2}\frac{\partial}{\partial \lambda}\mathbb{Y}_n(\lambda_0)^T{\mathbb{D}}_n\left((\boldsymbol{\varepsilon f_z})_n - \left(\boldsymbol{\widehat{\varepsilon}_{|z}\widehat{f}_z}\right)_n\right)\\	
		&-
		 \frac{1}{n}\sum_{j=1}^{n}\varepsilon_j{f}_z(\boldsymbol Z_j)E\left[\left(\frac{\partial}{\partial \lambda}\mathbb{Y}_{n,i}(\lambda_0) -\frac{1}{E\left[\boldsymbol{1}_n^T \boldsymbol\Omega_n  \boldsymbol{1}_n\right]}E\left[\frac{\partial}{\partial \lambda}\mathbb{Y}_n(\lambda_0) ^T\boldsymbol\Omega_n  \boldsymbol{1}_n \right]\right)	\boldsymbol\Omega_{n,ij}\mid \boldsymbol X_j, \boldsymbol Z_j\right]\\
		 &+
		\frac{1}{n}	\sum_{k=1}^{n}\varepsilon_k f_z(\boldsymbol Z_k) E\left[ \left(\frac{\partial}{\partial \lambda}\mathbb{Y}_{n,i}(\lambda_0) -\frac{1}{E\left[\boldsymbol{1}_n^T \boldsymbol\Omega_n  \boldsymbol{1}_n\right]}E\left[\frac{\partial}{\partial \lambda}\mathbb{Y}_n(\lambda_0) ^T\boldsymbol\Omega_n  \boldsymbol{1}_n \right]\right) \boldsymbol\Omega_{n,ik}^Z \boldsymbol\Omega_{n,ij}^X\mid \boldsymbol Z_k\right] 		  
		  \Bigg\|
		  = o_{\mathbb{P}}(n^{-1/2}).
	\end{align*}
The details are omitted.

Therefore, we get that 	
	\begin{align*}
		\left((\widehat \lambda, \widehat {\boldsymbol \beta}(\widehat \lambda )^T)^T  - (\lambda_0, \boldsymbol {\beta}_0 ^T)^T\right) &= - \boldsymbol{V}(\boldsymbol{d})^{-1}\Bigg(\frac{1}{n}\sum_{j=1}^{n}\varepsilon_j{f}_z(\boldsymbol Z_j)  E\left[\boldsymbol{\tau}_i(\boldsymbol{d})\,\boldsymbol\Omega_{n,ij}(\boldsymbol{d})\mid \boldsymbol X_j, \boldsymbol Z_j\right]\\
		&\quad -
		\frac{1}{n}	\sum_{k=1}^{n}\varepsilon_k f_z(\boldsymbol Z_k) E\left[\boldsymbol{\tau}_i(\boldsymbol{d})  \boldsymbol\Omega_{n,ik}^Z(\boldsymbol{d}) \boldsymbol\Omega_{n,ij}^X(\boldsymbol{d})\mid \boldsymbol Z_k\right]
		\Bigg)  + 
		  o_{\mathbb{P} }\left(n^{-1/2}\right)\\
		= - \boldsymbol{V}(\boldsymbol{d})^{-1}\Bigg(&\frac{1}{n}\sum_{j=1}^{n}\varepsilon_j{f}_z(\boldsymbol Z_j)  E\left[\boldsymbol{\tau}_i(\boldsymbol{d})\,\boldsymbol\Omega_{n,ij}^Z(\boldsymbol{d}) \left(\boldsymbol\Omega_{n,ij}^X(\boldsymbol{d}) - \boldsymbol\Omega_{n,ik}^X(\boldsymbol{d}) \right)
		\mid \boldsymbol X_j, \boldsymbol Z_j\right]
			\Bigg)  \\
		&+ 
		o_{\mathbb{P} }\left(n^{-1/2}\right) \\ 
		= - \boldsymbol{V}(\boldsymbol{d})^{-1}\Bigg(&\frac{1}{n}\sum_{j=1}^{n}\varepsilon_j{f}_z(\boldsymbol Z_j)  E\left[\boldsymbol{\tau}_i(\boldsymbol{d})\,\boldsymbol\Omega_{n,ij}^Z(\boldsymbol{d}) \left(\boldsymbol\Omega_{n,ij}^X(\boldsymbol{d}) - E\left[\boldsymbol \Omega^X_{n,ik}(\boldsymbol d)\mid \boldsymbol X_i\right]\right)
		\mid \boldsymbol X_j, \boldsymbol Z_j\right]
			\Bigg)  \\
		&+ 
		o_{\mathbb{P} }\left(n^{-1/2}\right) \\
		= - \boldsymbol{V}(\boldsymbol{d})^{-1}\Bigg(&\frac{1}{n}\sum_{j=1}^{n}\varepsilon_j{f}_z(\boldsymbol Z_j)  E\left[\boldsymbol{\tau}_i(\boldsymbol{d})\,\boldsymbol\Omega_{n,ij}^Z(\boldsymbol{d}) \boldsymbol \Phi^X_{n,ij}(\boldsymbol d)
		\mid \boldsymbol X_j, \boldsymbol Z_j\right]
			\Bigg)  \\
		&+ 
		o_{\mathbb{P} }\left(n^{-1/2}\right)		
	\end{align*}
uniformly over $h \in \mathcal H_{sc,n}$ and $\boldsymbol d \in \mathcal D$.

Following the lines of \citet{lavergne2013smooth}, we study the convergence of  
$$\frac{1}{n}\sum\limits_{j=1}^{n}\varepsilon_j{f}_z(\boldsymbol Z_j)E\left[\boldsymbol{\tau}_i(\boldsymbol{d})\,\boldsymbol\Omega_{n,ij}^Z(\boldsymbol{d}) \boldsymbol \Phi^X_{n,ij}(\boldsymbol d)\mid \boldsymbol X_j, \boldsymbol Z_j\right],$$ 
as a process indexed by $\boldsymbol{d}\in \mathcal D$. For this purpose, we apply Theorem 19.28 of \citet{van2000asymptotic}. The needed Lindeberg condition follows from our assumptions. 
In the following we will show that 
	\begin{equation}
		\begin{aligned}\label{arg_asy_norm}
			\underset{\left\|\boldsymbol d_1 - \boldsymbol d_2\right\|< \delta}{\sup}
			&E\big[\big\|\varepsilon_j{f}_z(\boldsymbol Z_j)E\left[\boldsymbol{\tau}_i(\boldsymbol{d}_1)\,\boldsymbol\Omega_{n,ij}^Z(\boldsymbol{d}_1) \boldsymbol \Phi^X_{n,ij}(\boldsymbol d_1)\mid \boldsymbol X_j, \boldsymbol Z_j\right]\\
			  &\hskip 3cm   -
			 \varepsilon_j{f}_z(\boldsymbol Z_j)E\left[\boldsymbol{\tau}_i(\boldsymbol{d}_2)\,\boldsymbol\Omega_{n,ij}^Z(\boldsymbol{d}_2) \boldsymbol \Phi^X_{n,ij}(\boldsymbol d_2)\mid \boldsymbol X_j, \boldsymbol Z_j\right]	\big\|^2\big]
			\rightarrow 0
		\end{aligned}
	\end{equation}
whenever $\delta \rightarrow 0$. We get that 
	\begin{align*}
		E&\left[\left\|\varepsilon_j{f}_z(\boldsymbol Z_j)E\left[\boldsymbol{\tau}_i(\boldsymbol{d}_1)\,\boldsymbol\Omega_{n,ij}^Z(\boldsymbol{d}_1) \boldsymbol \Phi^X_{n,ij}(\boldsymbol d_1)\mid \boldsymbol X_j, \boldsymbol Z_j\right]
		         -
		           \varepsilon_j{f}_z(\boldsymbol Z_j)E\left[\boldsymbol{\tau}_i(\boldsymbol{d}_2)\,\boldsymbol\Omega_{n,ij}^Z(\boldsymbol{d}_2) \boldsymbol \Phi^X_{n,ij}(\boldsymbol d_2)\mid \boldsymbol X_j, \boldsymbol Z_j\right]	\right\|^2\right]\\
		&=E\Big[E\left[\varepsilon_j^2{f}_z(\boldsymbol Z_j)^2\mid \boldsymbol X_j, \boldsymbol Z_j\right]
		        \big(\boldsymbol \tau_i(\boldsymbol{d}_1)^T \,\boldsymbol \tau_k(\boldsymbol{d}_1)\boldsymbol\Omega_{n,ij}^Z(\boldsymbol{d}_1) \boldsymbol \Phi^X_{n,ij}(\boldsymbol d_1)\boldsymbol\Omega_{n,kj}^Z(\boldsymbol{d}_1) \boldsymbol \Phi^X_{n,kj}(\boldsymbol d_1) \\
		              &\hskip 5cm - 2\boldsymbol \tau_i(\boldsymbol{d}_1)^T \,\boldsymbol \tau_k(\boldsymbol{d}_2)\boldsymbol\Omega_{n,ij}^Z(\boldsymbol{d}_1) \boldsymbol \Phi^X_{n,ij}(\boldsymbol d_1)\boldsymbol\Omega_{n,kj}^Z(\boldsymbol{d}_2) \boldsymbol \Phi^X_{n,kj}(\boldsymbol d_2) \\
		              &\hskip 5cm + \boldsymbol \tau_i(\boldsymbol{d}_2)^T \,\boldsymbol \tau_k(\boldsymbol{d}_2)\boldsymbol\Omega_{n,ij}^Z(\boldsymbol{d}_2) \boldsymbol \Phi^X_{n,ij}(\boldsymbol d_2)\boldsymbol\Omega_{n,kj}^Z(\boldsymbol{d}_2) \boldsymbol \Phi^X_{n,kj}(\boldsymbol d_2)\big)\Big],\qquad \boldsymbol{d}\in \mathcal D.
	\end{align*}
By the same Fourier transformation arguments as in the proof of Lemma \ref{lem_ident} and the Dominated Convergence Theorem, the statement in \eqref{arg_asy_norm} follows.  Therefore, 
	\begin{align*}
		\boldsymbol{V}(\boldsymbol{d})^{-1}\left(\frac{1}{\sqrt{n}}\sum_{j=1}^{n}\varepsilon_j{f}_z(\boldsymbol Z_j)E\left[\boldsymbol{\tau}_i(\boldsymbol{d})\,\boldsymbol\Omega_{n,ij}^Z(\boldsymbol{d}) \boldsymbol \Phi^X_{n,ij}(\boldsymbol d)\mid \boldsymbol X_j, \boldsymbol Z_j\right]\right), 
	\end{align*}
converges in distribution to a tight random process whose marginal distribution is zero-mean normal with covariance function $\boldsymbol V(\boldsymbol{d})^{-1}\boldsymbol\Delta(\boldsymbol{d})\boldsymbol V(\boldsymbol{d})^{-1}$. Here
	$$
		\boldsymbol\Delta(\boldsymbol{d}) =  E\left\{Var\left[\varepsilon_j\mid \boldsymbol X_j, \boldsymbol Z_j\right]{f}^2_z(\boldsymbol Z_j)
		\left(E\left[\boldsymbol{\tau}_i(\boldsymbol{d})\,\boldsymbol\Omega_{n,ij}^Z(\boldsymbol{d}) \boldsymbol \Phi^X_{n,ij}(\boldsymbol d)\mid \boldsymbol X_j, \boldsymbol Z_j\right]\right)^2
		\right\}.
	$$
In particular, this proves Theorem \ref{AN}.

\end{proofof}

\medskip

\begin{proofof}{Proposition \ref{prop_test_lambda}}~~\\
We have that 
	\begin{align*}
		n^{-1} \widehat{\mathbb{Y}}_n(\widehat \lambda)^T \;  \widehat {\mathbb{B}} _n \;n^{-1} \widehat{\mathbb{Y}}_n(\widehat \lambda) 
		&=   
		n^{-1} \widehat{\mathbb{Y}}_n(\lambda_0)^T \;  \widehat {\mathbb{B}} _n \;n^{-1} \widehat{\mathbb{Y}}_n(\lambda_0) \\ 
		&\quad + 
		 2n^{-1}\frac{\partial }{\partial \lambda} \widehat{\mathbb{Y}}_n(\lambda_0)^T \;  \widehat {\mathbb{B}} _n \;n^{-1} \widehat{\mathbb{Y}}_n(\lambda_0) 
		\left(\widehat \lambda - \lambda_0\right)\\
		&\quad +  
		\Bigg[  n^{-1}\frac{\partial }{\partial \lambda} \widehat{\mathbb{Y}}_n(\lambda_0)^T  \; \widehat {\mathbb{B}} _n \;n^{-1} \frac{\partial }{\partial \lambda} \widehat{\mathbb{Y}}_n(\lambda_0)   \\ 
		&\quad +
		 n^{-1}\widehat{\mathbb{Y}}_n(\lambda_0)^T \; \widehat {\mathbb{B}} _n  \;n^{-1}\frac{\partial ^2}{\partial \lambda^2} \widehat{\mathbb{Y}}_n(\lambda_0) + R_{1,n}(\widetilde \lambda,\lambda_0)\Bigg] \left(\widehat \lambda - \lambda_0\right)^2,
	\end{align*}
where $\widetilde \lambda = c \widehat \lambda + (1-c)\lambda_0$ for some $c\in(0,1)$. By the same reasoning as in Proposition \ref{AN_prop} we get that 
	\begin{align*}
		n^{-1} \widehat{\mathbb{Y}}_n(\widehat \lambda)^T \;  \widehat {\mathbb{B}} _n \;n^{-1} \widehat{\mathbb{Y}}_n(\widehat \lambda) 
		&=   
		n^{-1} \widehat{\mathbb{Y}}_n(\lambda_0)^T \;  \widehat {\mathbb{B}} _n \; n^{-1} \widehat{\mathbb{Y}}_n(\lambda_0) \\ 
		&\quad + 
		 2n^{-1}\frac{\partial }{\partial \lambda} {\mathbb{Y}}_n(\lambda_0) \;   {\mathbb{B}}_n \;n^{-1}\left((\boldsymbol{\varepsilon f_z})_n - \left(\boldsymbol{\widehat{\varepsilon}_{|z}\widehat{f}_z}\right)_n \right)
		\left(\widehat \lambda - \lambda_0\right)\\
		&\quad +  
		n^{-1}\frac{\partial }{\partial \lambda} {\mathbb{Y}}_n(\lambda_0)^T\; {\mathbb{B}} _n \; n^{-1}\frac{\partial }{\partial \lambda}  {\mathbb{Y}}_n(\lambda_0)  \left(\widehat \lambda - \lambda_0\right)^2\\
		&\quad + o_{\mathbb{P} }\left(1/n\right)
	\end{align*}
uniformly with respect to $s \in S_n$, $\boldsymbol d\in\mathcal D$ and $h\in\mathcal H_{sc,n}$. Therefore, it follows that under $H_0$
	\begin{align*}
		n^{-1} \widehat{\mathbb{Y}}_n(\lambda_R)^T \;  \widehat {\mathbb{B}} _n &\;n^{-1} \widehat{\mathbb{Y}}_n(\lambda_R) 
		-   
		n^{-1} \widehat{\mathbb{Y}}_n(\widehat \lambda)^T \;  \widehat {\mathbb{B}} _n \;n^{-1} \widehat{\mathbb{Y}}_n(\widehat \lambda)  \\ 
		&= 
		 \frac{1}{n^2}
		 \left(\frac{\partial }{\partial \lambda}{\mathbb{Y}}_n(\lambda_0)^T \;  {\mathbb{B}} _n \left((\boldsymbol{\varepsilon f_z})_n - \left(\boldsymbol{\widehat{\varepsilon}_{|z}\widehat{f}_z}\right)_n \right)\right)^2
		 \left[  \frac{\partial }{\partial \lambda} {\mathbb{Y}}_n(\lambda_0) ^T   {\mathbb{B}} _n  \frac{\partial }{\partial \lambda}  {\mathbb{Y}}_n(\lambda_0)  \right]^{-1}\\
		&\quad + o_{\mathbb{P} }\left(1/n\right)\\
		&= (1, \boldsymbol{0}_p^T)\boldsymbol V(\boldsymbol{d})^{-1} n^{-2}\boldsymbol{A}_n n^{-2}\boldsymbol{A}_n^T \boldsymbol V(\boldsymbol{d})^{-1}(1, \boldsymbol{0}_p^T)^T 
		   \left[n^{-1}  \frac{\partial }{\partial \lambda} {\mathbb{Y}}_n(\lambda_0) ^T   {\mathbb{B}} _n  n^{-1} \frac{\partial }{\partial \lambda}  {\mathbb{Y}}_n(\lambda_0)  \right]\\
		&\quad + o_{\mathbb{P} }\left(1/n\right)\\
		&= (1, \boldsymbol{0}_p^T)\boldsymbol V(\boldsymbol{d})^{-1} n^{-2}\boldsymbol{A}_n n^{-2}\boldsymbol{A}_n^T \boldsymbol V(\boldsymbol{d})^{-1}(1, \boldsymbol{0}_p^T)^T 
		 E\left[\frac{\partial }{\partial \lambda} {\mathbb{Y}}_n(\lambda_0) ^T   {\mathbb{B}} _n \frac{\partial }{\partial \lambda}  {\mathbb{Y}}_n(\lambda_0)  \right]\\
		&\quad + o_{\mathbb{P} }\left(1/n\right)		.	
	\end{align*}

When $H_0$ does not hold it follows by the same arguments as in the proof of Proposition \ref{AN_prop} that $n^{-1}DM_{\lambda}$ converges in probability to a positive constant.\\
\end{proofof}

\medskip

\begin{proofof}{Proposition \ref{prop_test_beta}}~~\\

Under $H_0$ we get that 
	\begin{align*}
		&\left( \widehat{\mathbb{Y}}_n(\lambda)  - \widehat{\mathbb{X}}_n \widehat{\boldsymbol{\beta}}_R(\lambda)\right)^T\boldsymbol  {\mathbb{D}}_n  
		\left( \widehat{\mathbb{Y}}_n(\lambda)  - \widehat{\mathbb{X}}_n \widehat{\boldsymbol{\beta}}_R(\lambda)\right)\\
		&\hskip 2cm = 
		\widehat{\mathbb{Y}}_n(\lambda)^T\widehat{\mathbb{B}}_n\widehat{\mathbb{Y}}_n(\lambda)
		+ \left(\boldsymbol{R}\widehat{\boldsymbol{\beta}}(\lambda) - \boldsymbol c\right)^T 
		\left(\boldsymbol{R} \left(\widehat{\mathbb{X}}_n^T{\mathbb{D}}_n\widehat{\mathbb{X}}_n\right)^{-1} \boldsymbol{R}^T \right)^{-1}
		\left(\boldsymbol{R}\widehat{\boldsymbol{\beta}}(\lambda) - \boldsymbol c\right) \\
		&\hskip 2cm = 
		\widehat{\mathbb{Y}}_n(\lambda)^T\widehat{\mathbb{B}}_n\widehat{\mathbb{Y}}_n(\lambda)
		+ \left(\boldsymbol{R}\widehat{\boldsymbol{\beta}}(\lambda) - \boldsymbol{R}{\boldsymbol{\beta}}_0\right)^T 
		\left(\boldsymbol{R} \left(\widehat{\mathbb{X}}_n^T{\mathbb{D}}_n\widehat{\mathbb{X}}_n\right)^{-1} \boldsymbol{R}^T \right)^{-1}
		\left(\boldsymbol{R}\widehat{\boldsymbol{\beta}}(\lambda) - \boldsymbol{R}{\boldsymbol{\beta}}_0\right) \\
		&\hskip 2cm = 
		\left(\widehat{\mathbb{Y}}_n(\lambda) - \widehat{\mathbb{X}}_n \boldsymbol{\beta}_0\right)^T\\
		&\hskip 3cm
		\left(
			\widehat{\mathbb{B}}_n + 
			{\mathbb{D}}_n\widehat{\mathbb{X}}_n\left(\widehat{\mathbb{X}}_n^T{\mathbb{D}}_n\widehat{\mathbb{X}}_n\right)^{-1} \boldsymbol{R}^T 
			\left(\boldsymbol{R} \left(\widehat{\mathbb{X}}_n^T{\mathbb{D}}_n\widehat{\mathbb{X}}_n\right)^{-1} \boldsymbol{R}^T \right)^{-1}
			\boldsymbol{R} \left(\widehat{\mathbb{X}}_n^T{\mathbb{D}}_n\widehat{\mathbb{X}}_n\right)^{-1}\widehat{\mathbb{X}}_n^T{\mathbb{D}}_n	
		\right)\\
		&\hskip 13cm
		\left(\widehat{\mathbb{Y}}_n(\lambda) - \widehat{\mathbb{X}}_n \boldsymbol{\beta}_0\right)\\
		&\hskip 2cm = 
		\left(\widehat{\mathbb{Y}}_n(\lambda) - \widehat{\mathbb{X}}_n \boldsymbol{\beta}_0\right)^T\widehat{\mathbb{B}}_{n,R} 
		\left(\widehat{\mathbb{Y}}_n(\lambda) - \widehat{\mathbb{X}}_n \boldsymbol{\beta}_0\right),
	\end{align*}
where
$$
	\widehat{\mathbb{B}}_{n,R} = \widehat{\mathbb{B}}_n + 
	 {\mathbb{D}}_n\widehat{\mathbb{X}}_n\left(\widehat{\mathbb{X}}_n^T{\mathbb{D}}_n\widehat{\mathbb{X}}_n\right)^{-1} \boldsymbol{R}^T 
	 \left(\boldsymbol{R} \left(\widehat{\mathbb{X}}_n^T{\mathbb{D}}_n\widehat{\mathbb{X}}_n\right)^{-1} \boldsymbol{R}^T \right)^{-1}
	 \boldsymbol{R} \left(\widehat{\mathbb{X}}_n^T{\mathbb{D}}_n\widehat{\mathbb{X}}_n\right)^{-1}\widehat{\mathbb{X}}_n^T{\mathbb{D}}_n.
$$	
Therefore, we get by the same reasoning as in the proof of Proposition \ref{AN_prop} that
	\begin{align*}
		\widehat \lambda_R - \lambda_0  &= - \left[  \frac{\partial }{\partial \lambda} {\mathbb{Y}}_n(\lambda_0) ^T   {\mathbb{B}}_{n,R}  \frac{\partial }{\partial \lambda}  {\mathbb{Y}}_n(\lambda_0)  \right]^{-1}    \!  \frac{\partial }{\partial \lambda} {\mathbb{Y}}_n(  \lambda_0) ^T{\mathbb{B}}_{n,R} \left[(\boldsymbol{\varepsilon f_z})_n - \left(\boldsymbol{\widehat{\varepsilon}_{|z}\widehat{f}_z}\right)_n \right]+ o_{\mathbb{P}}(n^{-1/2})\\
		&= - \boldsymbol{V}_R(\boldsymbol{d})n^{-2}\boldsymbol{A}_n + o_{\mathbb{P}}(n^{-1/2}),
	\end{align*}
uniformly with respect to $h\in\mathcal{H}_n^{sc}$, $\boldsymbol d\in \mathcal D$ and $s \in S_n$.	
Furthermore, we get that	
	\begin{align*}
		\left( \widehat{\mathbb{Y}}_n(\widehat \lambda_R)  - \widehat{\mathbb{X}}_n \widehat{\boldsymbol{\beta}}_R(\widehat \lambda_R)\right)^T\boldsymbol  {\mathbb{D}}_n & 
		\left( \widehat{\mathbb{Y}}_n(\widehat \lambda_R)  - \widehat{\mathbb{X}}_n \widehat{\boldsymbol{\beta}}_R(\widehat \lambda_R)\right)
		=
		\left(\widehat{\mathbb{Y}}_n(\widehat \lambda_R) - \widehat{\mathbb{X}}_n \boldsymbol{\beta}_0\right)^T\widehat{\mathbb{B}}_{n,R} 
		\left(\widehat{\mathbb{Y}}_n(\widehat \lambda_R) - \widehat{\mathbb{X}}_n \boldsymbol{\beta}_0\right)		
		\\
		&=
		\left(\widehat{\mathbb{Y}}_n(\lambda_0) - \widehat{\mathbb{X}}_n \boldsymbol{\beta}_0\right)^T\widehat{\mathbb{B}}_{n,R} 
		\left(\widehat{\mathbb{Y}}_n(\lambda_0) - \widehat{\mathbb{X}}_n \boldsymbol{\beta}_0\right)		
		\\		 
		&\quad + 
		2\frac{\partial }{\partial \lambda}\widehat{\mathbb{Y}}_n(\lambda_0)^T \;  \widehat{\mathbb{B}}_{n,R} 
		\left(\widehat{\mathbb{Y}}_n(\lambda_0) - \widehat{\mathbb{X}}_n \boldsymbol{\beta}_0\right)\left(\widehat \lambda_R - \lambda_0\right)\\
		&\quad +  
		\Bigg[  \frac{\partial }{\partial \lambda} \widehat{\mathbb{Y}}_n(\lambda_0)^T \; \widehat{\mathbb{B}}_{n,R} \;  \frac{\partial }{\partial \lambda} \widehat{\mathbb{Y}}_n(\lambda_0)  \\ 
		&\quad +
		\left(\widehat{\mathbb{Y}}_n(\lambda_0) - \widehat{\mathbb{X}}_n \boldsymbol{\beta}_0\right)^T  \widehat{\mathbb{B}}_{n,R} \;  \frac{\partial ^2}{\partial \lambda^2} \widehat{\mathbb{Y}}_n(\lambda_0) + R_{1,n}(\widetilde \lambda,\lambda_0)\Bigg] \left(\widehat \lambda_R - \lambda_0\right)^2,
	\end{align*}
where $\widetilde \lambda = c \widehat \lambda_R + (1-c)\lambda_0$ for some $c\in(0,1)$. By the same reasoning as in Proposition \ref{AN_prop} and using the asymptotic representation of $\left(\widehat \lambda_R - \lambda_0\right)$ we get that 
	\begin{align*}
		n^{-1} \left( \widehat{\mathbb{Y}}_n(\widehat \lambda_R)  - \widehat{\mathbb{X}}_n \widehat{\boldsymbol{\beta}}_R(\widehat \lambda_R)\right)^T &{\mathbb{D}}_n 
		\; n^{-1}\left( \widehat{\mathbb{Y}}_n(\widehat \lambda_R)  - \widehat{\mathbb{X}}_n \widehat{\boldsymbol{\beta}}_R(\widehat \lambda_R)\right)\\
		&=
		n^{-1}\left(\widehat{\mathbb{Y}}_n(\widehat \lambda_R) - \widehat{\mathbb{X}}_n \boldsymbol{\beta}_0\right)^T\widehat{\mathbb{B}}_{n,R} 
		\;n^{-1}\left(\widehat{\mathbb{Y}}_n(\widehat \lambda_R) - \widehat{\mathbb{X}}_n \boldsymbol{\beta}_0\right)		
		\\
		&=
		n^{-1}\left((\boldsymbol{\varepsilon f_z})_n - \left(\boldsymbol{\widehat{\varepsilon}_{|z}\widehat{f}_z}\right)_n \right)^T{\mathbb{B}}_{n,R} 
		\;n^{-1}\left((\boldsymbol{\varepsilon f_z})_n - \left(\boldsymbol{\widehat{\varepsilon}_{|z}\widehat{f}_z}\right)_n \right)
		\\		 
		&\quad + 
		2n^{-1}\frac{\partial }{\partial \lambda} {\mathbb{Y}}_n(\lambda_0)^T \;  {\mathbb{B}}_{n,R} 
		\;n^{-1}\left((\boldsymbol{\varepsilon f_z})_n - \left(\boldsymbol{\widehat{\varepsilon}_{|z}\widehat{f}_z}\right)_n \right)
		\left(\widehat \lambda_R - \lambda_0\right)\\
		&\quad +  
		\Bigg[ n^{-1} \frac{\partial }{\partial \lambda} {\mathbb{Y}}_n(\lambda_0)^T \; {\mathbb{B}}_{n,R} \; n^{-1} \frac{\partial }{\partial \lambda} {\mathbb{Y}}_n(\lambda_0)  \Bigg] 
		\left(\widehat \lambda_R - \lambda_0\right)^2\\
		&\quad + o_{\mathbb{P} }\left(1/n\right)
		\\
		&=
		n^{-1}\left((\boldsymbol{\varepsilon f_z})_n - \left(\boldsymbol{\widehat{\varepsilon}_{|z}\widehat{f}_z}\right)_n \right)^T{\mathbb{B}}_{n,R} 
		\;n^{-1}\left((\boldsymbol{\varepsilon f_z})_n - \left(\boldsymbol{\widehat{\varepsilon}_{|z}\widehat{f}_z}\right)_n \right)
		\\		 
		&\quad -
		n^{-2}\boldsymbol{A}_n^T\boldsymbol{V}_R(\boldsymbol{d})^T\boldsymbol{V}_R(\boldsymbol{d})\boldsymbol{A}_n n^{-2}
		\left[ n^{-1} \frac{\partial }{\partial \lambda} {\mathbb{Y}}_n(\lambda_0) ^T   {\mathbb{B}}_{n,R} n^{-1} \frac{\partial }{\partial \lambda}  {\mathbb{Y}}_n(\lambda_0)  \right] \\
		&\quad + o_{\mathbb{P} }\left(1/n\right)
	\end{align*}
uniformly with respect to $s \in S_n$, $\boldsymbol d\in\mathcal D$ and $h\in\mathcal H_{sc,n}$. We know from the proof of Proposition \ref{prop_test_lambda} that 
	\begin{align*}
		n^{-1} \widehat{\mathbb{Y}}_n(\widehat\lambda)^T \;  \widehat{\mathbb{B}} _n \;n^{-1} \widehat{\mathbb{Y}}_n(\widehat\lambda) 
		&=   
		n^{-1}\left((\boldsymbol{\varepsilon f_z})_n - \left(\boldsymbol{\widehat{\varepsilon}_{|z}\widehat{f}_z}\right)_n \right)^T \;  
		{\mathbb{B}} _n 
		n^{-1}\left((\boldsymbol{\varepsilon f_z})_n - \left(\boldsymbol{\widehat{\varepsilon}_{|z}\widehat{f}_z}\right)_n \right) \\ 
		&\quad -  n^{-2}\boldsymbol{A}_n^T \boldsymbol V(\boldsymbol{d})^{-1}(1, \boldsymbol{0}_p^T)^T (1, \boldsymbol{0}_p^T)\boldsymbol V(\boldsymbol{d})^{-1} \boldsymbol{A}_nn^{-2}\\
		&\quad \hskip 5cm \left[n^{-1}  \frac{\partial }{\partial \lambda} {\mathbb{Y}}_n(\lambda_0)^T \;  {\mathbb{B}} _n \; n^{-1} \frac{\partial }{\partial \lambda}  {\mathbb{Y}}_n(\lambda_0)  \right]\\
		&\quad + o_{\mathbb{P} }\left(1/n\right).	
	\end{align*}

Therefore, we get that 
	\begin{align*}
		n^{-1}& \left( \widehat{\mathbb{Y}}_n(\widehat \lambda_R)  - \widehat{\mathbb{X}}_n \widehat{\boldsymbol{\beta}}_R(\widehat \lambda_R)\right)^T {\mathbb{D}}_n 
		\;n^{-1}\left( \widehat{\mathbb{Y}}_n(\widehat \lambda_R)  - \widehat{\mathbb{X}}_n \widehat{\boldsymbol{\beta}}_R(\widehat \lambda_R)\right)
		-
		n^{-1} \widehat{\mathbb{Y}}_n(\widehat\lambda)^T \;  \widehat{\mathbb{B}} _n \;n^{-1} \widehat{\mathbb{Y}}_n(\widehat\lambda) 	\\
		&=	
		n^{-1}\left((\boldsymbol{\varepsilon f_z})_n - \left(\boldsymbol{\widehat{\varepsilon}_{|z}\widehat{f}_z}\right)_n \right)^T
		\left({\mathbb{B}}_{n,R} -  {\mathbb{B}} _n\right) 
		\;n^{-1}\left((\boldsymbol{\varepsilon f_z})_n - \left(\boldsymbol{\widehat{\varepsilon}_{|z}\widehat{f}_z}\right)_n \right)\\
		&\quad -
		n^{-2}\boldsymbol{A}_n^T\boldsymbol{V}_R(\boldsymbol{d})^T\boldsymbol{V}_R(\boldsymbol{d})\boldsymbol{A}_nn^{-2}
		\left[ n^{-1} \frac{\partial }{\partial \lambda} {\mathbb{Y}}_n(\lambda_0) ^T \;  {\mathbb{B}}_{n,R} \;n^{-1} \frac{\partial }{\partial \lambda}  {\mathbb{Y}}_n(\lambda_0)  \right] \\
		&\quad +  n^{-2}\boldsymbol{A}_n^T \boldsymbol V(\boldsymbol{d})^{-1}(1, \boldsymbol{0}_p^T)^T (1, \boldsymbol{0}_p^T)\boldsymbol V(\boldsymbol{d})^{-1} \boldsymbol{A}_nn^{-2}
		\left[n^{-1}  \frac{\partial }{\partial \lambda} {\mathbb{Y}}_n(\lambda_0) ^T  \; {\mathbb{B}}_n \; n^{-1} \frac{\partial }{\partial \lambda}  {\mathbb{Y}}_n(\lambda_0)  \right]\\
		&\quad + o_{\mathbb{P} }\left(1/n\right)\\	
		&=	
		n^{-2}\boldsymbol{A}_n^T\left(\boldsymbol{0}_{p\times 1}, \boldsymbol{I}_{p\times p}\right)^T\left({\mathbb{X}}_n^T{\mathbb{D}}_n{\mathbb{X}}_n\right)^{-1} \boldsymbol{R}^T 
		\left(\boldsymbol{R} \left({\mathbb{X}}_n^T{\mathbb{D}}_n{\mathbb{X}}_n\right)^{-1} \boldsymbol{R}^T \right)^{-1}
		\boldsymbol{R} \left({\mathbb{X}}_n^T{\mathbb{D}}_n{\mathbb{X}}_n\right)^{-1}\left(\boldsymbol{0}_{p\times 1}, \boldsymbol{I}_{p\times p}\right)\boldsymbol{A}_n\\
		&\quad -
		n^{-2}\boldsymbol{A}_n^T\boldsymbol{V}_R(\boldsymbol{d})^T\boldsymbol{V}_R(\boldsymbol{d})\boldsymbol{A}_nn^{-2}
		\left[ n^{-1} \frac{\partial }{\partial \lambda} {\mathbb{Y}}_n(\lambda_0)^T  \; {\mathbb{B}}_{n,R} \; n^{-1} \frac{\partial }{\partial \lambda}  {\mathbb{Y}}_n(\lambda_0)  \right] \\
		&\quad +  n^{-2}\boldsymbol{A}_n^T \boldsymbol V(\boldsymbol{d})^{-1}(1, \boldsymbol{0}_p^T)^T (1, \boldsymbol{0}_p^T)\boldsymbol V(\boldsymbol{d})^{-1} \boldsymbol{A}_nn^{-2}
		\left[n^{-1}  \frac{\partial }{\partial \lambda} {\mathbb{Y}}_n(\lambda_0)^T \;  {\mathbb{B}}_n \; n^{-1} \frac{\partial }{\partial \lambda}  {\mathbb{Y}}_n(\lambda_0)  \right]\\
		&\quad + o_{\mathbb{P} }\left(1/n\right)	\\		
		&=	
		n^{-2}\boldsymbol{A}_n^T\left(\boldsymbol{0}_{p\times 1}, \boldsymbol{I}_{p\times p}\right)^T E\left[{\mathbb{X}}_n^T{\mathbb{D}}_n{\mathbb{X}}_n\right]^{-1} \boldsymbol{R}^T 
		\left(\boldsymbol{R} E\left[{\mathbb{X}}_n^T{\mathbb{D}}_n{\mathbb{X}}_n\right]^{-1} \boldsymbol{R}^T \right)^{-1}
		\boldsymbol{R} E\left[{\mathbb{X}}_n^T{\mathbb{D}}_n{\mathbb{X}}_n\right]^{-1}\\
		&\hskip 12cm \left(\boldsymbol{0}_{p\times 1}, \boldsymbol{I}_{p\times p}\right)\boldsymbol{A}_nn^{-2}\\
		&\quad -
		n^{-2}\boldsymbol{A}_n^T\boldsymbol{V}_R(\boldsymbol{d})^T\boldsymbol{V}_R(\boldsymbol{d})\boldsymbol{A}_nn^{-2}
		E\left[\frac{\partial }{\partial \lambda} {\mathbb{Y}}_n(\lambda_0)^T \;  {\mathbb{B}}_{n,R} \; \frac{\partial }{\partial \lambda}  {\mathbb{Y}}_n(\lambda_0)  \right] \\
		&\quad +  n^{-2}\boldsymbol{A}_n^T \boldsymbol V(\boldsymbol{d})^{-1}(1, \boldsymbol{0}_p^T)^T (1, \boldsymbol{0}_p^T)\boldsymbol V(\boldsymbol{d})^{-1} \boldsymbol{A}_nn^{-2}
		E\left[ \frac{\partial }{\partial \lambda} {\mathbb{Y}}_n(\lambda_0)^T \;  {\mathbb{B}}_n \;\frac{\partial }{\partial \lambda}  {\mathbb{Y}}_n(\lambda_0)  \right]\\
		&\quad + o_{\mathbb{P} }\left(1/n\right)						
	\end{align*}	
uniformly with respect to $h\in\mathcal{H}_n^{sc}$, $\boldsymbol d\in \mathcal D$ and $s \in S_n$.		
When $H_0$ does not hold it follows by the same arguments as in the proof of Proposition \ref{AN_prop} that $n^{-1}DM_{\boldsymbol{\beta}}$ converges in probability to a positive constant.\\
\end{proofof}

\medskip

\begin{proofof}{Proposition \ref{prop_test_beta_lambda}}~~\\
We can use the arguments as in the proof of Proposition \ref{prop_test_beta}. The only difference is that we do not need to taylor 
	$$		
		n^{-1}\left( \widehat{\mathbb{Y}}_n(\lambda_R)  - \widehat{\mathbb{X}}_n \widehat{\boldsymbol{\beta}}_R( \lambda_R)\right)^T {\mathbb{D}}_n 
		\;n^{-1}\left( \widehat{\mathbb{Y}}_n(\lambda_R)  - \widehat{\mathbb{X}}_n \widehat{\boldsymbol{\beta}}_R(\lambda_R)\right)
	$$
as $\lambda_R$ is fixed. Therefore, we get that under $H_0$
	\begin{align*}
		n^{-1}& \left( \widehat{\mathbb{Y}}_n( \lambda_R)  - \widehat{\mathbb{X}}_n \widehat{\boldsymbol{\beta}}_R(\lambda_R)\right)^T {\mathbb{D}}_n 
		\;n^{-1}\left( \widehat{\mathbb{Y}}_n( \lambda_R)  - \widehat{\mathbb{X}}_n \widehat{\boldsymbol{\beta}}_R( \lambda_R)\right)
		-
		n^{-1} \widehat{\mathbb{Y}}_n(\widehat\lambda)^T \;  \widehat{\mathbb{B}} _n \;n^{-1} \widehat{\mathbb{Y}}_n(\widehat\lambda) 	\\
		&=	
		n^{-1}\left((\boldsymbol{\varepsilon f_z})_n - \left(\boldsymbol{\widehat{\varepsilon}_{|z}\widehat{f}_z}\right)_n \right)^T
		\left({\mathbb{B}}_{n,R} -  {\mathbb{B}} _n\right) 
		\;n^{-1}\left((\boldsymbol{\varepsilon f_z})_n - \left(\boldsymbol{\widehat{\varepsilon}_{|z}\widehat{f}_z}\right)_n \right)\\
		&\quad +  n^{-2}\boldsymbol{A}_n^T \boldsymbol V(\boldsymbol{d})^{-1}(1, \boldsymbol{0}_p^T)^T (1, \boldsymbol{0}_p^T)\boldsymbol V(\boldsymbol{d})^{-1} \boldsymbol{A}_nn^{-2}
		\left[n^{-1}  \frac{\partial }{\partial \lambda} {\mathbb{Y}}_n(\lambda_0)^T \;  {\mathbb{B}}_n \; n^{-1} \frac{\partial }{\partial \lambda}  {\mathbb{Y}}_n(\lambda_0)  \right]\\
		&\quad + o_{\mathbb{P} }\left(1/n\right)\\	
		&=	
		n^{-2}\boldsymbol{A}_n^T\left(\boldsymbol{0}_{p\times 1}, \boldsymbol{I}_{p\times p}\right)^T\left({\mathbb{X}}_n^T{\mathbb{D}}_n{\mathbb{X}}_n\right)^{-1} \boldsymbol{R}^T 
		\left(\boldsymbol{R} \left({\mathbb{X}}_n^T{\mathbb{D}}_n{\mathbb{X}}_n\right)^{-1} \boldsymbol{R}^T \right)^{-1}
		\boldsymbol{R} \left({\mathbb{X}}_n^T{\mathbb{D}}_n{\mathbb{X}}_n\right)^{-1}\left(\boldsymbol{0}_{p\times 1}, \boldsymbol{I}_{p\times p}\right)\boldsymbol{A}_n\\
		&\quad +  n^{-2}\boldsymbol{A}_n^T \boldsymbol V(\boldsymbol{d})^{-1}(1, \boldsymbol{0}_p^T)^T (1, \boldsymbol{0}_p^T)\boldsymbol V(\boldsymbol{d})^{-1} \boldsymbol{A}_nn^{-2}
		\left[n^{-1}  \frac{\partial }{\partial \lambda} {\mathbb{Y}}_n(\lambda_0)^T \;  {\mathbb{B}}_n \; n^{-1} \frac{\partial }{\partial \lambda}  {\mathbb{Y}}_n(\lambda_0)  \right]\\
		&\quad + o_{\mathbb{P} }\left(1/n\right)	\\		
		&=	
		n^{-2}\boldsymbol{A}_n^T\left(\boldsymbol{0}_{p\times 1}, \boldsymbol{I}_{p\times p}\right)^T E\left[{\mathbb{X}}_n^T{\mathbb{D}}_n{\mathbb{X}}_n\right]^{-1} \boldsymbol{R}^T 
		\left(\boldsymbol{R} E\left[{\mathbb{X}}_n^T{\mathbb{D}}_n{\mathbb{X}}_n\right]^{-1} \boldsymbol{R}^T \right)^{-1}
		\boldsymbol{R} E\left[{\mathbb{X}}_n^T{\mathbb{D}}_n{\mathbb{X}}_n\right]^{-1}\left(\boldsymbol{0}_{p\times 1}, \boldsymbol{I}_{p\times p}\right)\boldsymbol{A}_nn^{-2}\\
		&\quad +  n^{-2}\boldsymbol{A}_n^T \boldsymbol V(\boldsymbol{d})^{-1}(1, \boldsymbol{0}_p^T)^T (1, \boldsymbol{0}_p^T)\boldsymbol V(\boldsymbol{d})^{-1} \boldsymbol{A}_nn^{-2}
		E\left[ \frac{\partial }{\partial \lambda} {\mathbb{Y}}_n(\lambda_0)^T \; {\mathbb{B}} _n \; \frac{\partial }{\partial \lambda}  {\mathbb{Y}}_n(\lambda_0)  \right]\\
		&\quad + o_{\mathbb{P} }\left(1/n\right)						
	\end{align*}
uniformly with respect to $h\in\mathcal{H}_n^{sc}$, $\boldsymbol d\in \mathcal D$ and $s \in S_n$.	
When $H_0$ does not hold it follows by the same arguments as in the proof of Proposition \ref{AN_prop} that $n^{-1}DM_{\boldsymbol{\beta}, \lambda}$ converges in probability to a positive constant.\\
\end{proofof}

\newpage
	
\setcounter{section}{2} 
\setcounter{equation}{0}
	
\subsection*{Appendix B: Preliminary results}	
\addcontentsline{toc}{subsection}{Appendix B: Preliminary results}
\begin{lem}\label{spectral_B_hat}
Let Assumptions \ref{ass_dgp}.1 and \ref{ass_dgp}.3 hold. Then
	$$
	  \sup_{\boldsymbol d \in \mathcal D} \left\| \mathbb{D}_n \right\|_{\rm{Sp}}  \leq \sup_{\boldsymbol d \in \mathcal D} \left\| \boldsymbol\Omega _n \right\|_{\rm{Sp}} \leq n .
	$$
  Moreover, 
	$$
	   \sup_{\boldsymbol d \in  \mathcal D} \left\| {\mathbb{B}} _n  \right\|_{\rm{Sp}}  \leq \sup_{\boldsymbol d \in \mathcal D} \left\| \mathbb{D}_n    
	    \right\|_{\rm{Sp}}\qquad \text{and} \qquad 
		\sup_{h>0} \sup_{\boldsymbol d \in \mathcal D} \left\| \widehat {\mathbb{B}} _n \right\|_{\rm{Sp}} \leq \sup_{\boldsymbol d \in \mathcal D} \left\| \mathbb{D}_n \right\|_{\rm{Sp}}.
	$$
 \end{lem}

\begin{proofof}{Lemma \ref{spectral_B_hat}} ~~\\

For all vectors $\boldsymbol d$, the matrix $\boldsymbol\Omega _n$ is positive definite, see Lemma \ref{omega_mat}. This implies that its spectral norm is equal to the largest eigenvalue. On the other hand, for all vectors $\boldsymbol d$, the trace of $\boldsymbol\Omega _n$ is equal to $n$. Necessarily, the spectral norm of $ \boldsymbol\Omega _n $ is at most equal to $n$, uniformly with respect to $\boldsymbol d \in \mathcal D$. Next, it is easy to see that $\| \boldsymbol A_1 \|_{\rm{Sp}} \leq \| \boldsymbol A_2 \|_{\rm{Sp}} $ whenever $\boldsymbol A_1$ and $\boldsymbol A_2 - \boldsymbol A_1$ are positive semi-definite real matrices. Using repeatedly this property and the fact that  ${\mathbb{D}}_n$,  ${\mathbb{B}}_n$ and $\widehat{\mathbb{B}}_n$ are positive semi-definite (cf. proof of   Lemma \ref{omega_mat}), we deduce the remaining inequalities, that clearly hold uniformly. \\
\end{proofof}

\quad

\begin{lem}\label{matrix_alg} ~\\
	\begin{enumerate}
		\item[1.] For any positive definite real matrices $\boldsymbol A_1$ and $\boldsymbol A_2$
				  $$
				  \|\boldsymbol  A_2^{-1/2} - \boldsymbol  A_1^{-1/2}\|_{\rm{Sp}} \leq \frac{1}{2}\left[ \max\{ \| \boldsymbol  A_1^{-1}\|_{\rm{Sp}}, \|\boldsymbol  A_2^{-1}\|_{\rm{Sp}}  \} \right]^{3/2} \|\boldsymbol  A_2 - \boldsymbol  A_1\|_{\rm{Sp}}.
				  $$
	  
		\item[2.] Let $\boldsymbol A_1$ and $\boldsymbol A_2$ be $n\times p-$matrices such that 
					$\boldsymbol A_1^T\boldsymbol A_1 = \boldsymbol A_2^T\boldsymbol A_2 = \boldsymbol{I}_{p \times p}$. Then 
					$$
				  \left\| \boldsymbol A_1 \boldsymbol A_1 ^T - \boldsymbol A_2 \boldsymbol A_2 ^T \right\|_{\rm{Sp}}\leq 
					2 \left\|\boldsymbol A_1  - \boldsymbol A_2 \right\|_{\rm{Sp}}.
					$$
	\end{enumerate}

\end{lem}

\begin{proofof}{Lemma \ref{matrix_alg}} ~~\\

	\begin{enumerate}
	\item[1.] For any positive definite real matrices $\boldsymbol A_1$ and $\boldsymbol A_2$
			  \[
			    \|\boldsymbol  A_2^{1/2} - \boldsymbol  A_1^{1/2}\|_{\rm{Sp}} \leq \frac{1}{2}\left[ \max\{ \|\boldsymbol  A_1^{-1}\|_{\rm{Sp}}, \|\boldsymbol  A_2^{-1}\|_{\rm{Sp}}  \} \right]^{1/2} \|\boldsymbol  A_2 - \boldsymbol  A_1\|_{\rm{Sp}}
			    \]
			(see for instance \citet{horn}, page 557). Moreover, for any invertible matrices $\boldsymbol A_1$ and $\boldsymbol A_2$ we have the identity 
			$\boldsymbol A_2^{-1} - \boldsymbol A_1^{-1} = \boldsymbol A_2^{-1}(\boldsymbol A_1- \boldsymbol A_2)\boldsymbol A_1^{-1}$. Apply this identity with $\boldsymbol  A_1^{1/2}$ and $\boldsymbol  A_2^{1/2}$ and, using the fact that the spectral norm of a product of two matrices is smaller or equal to the product of the matrices' spectral norms, we deduce the statement.
	
	\item[2.]  We could write
				\begin{multline*}
					\left\| ( \boldsymbol A_1 \boldsymbol A_1 ^T - \boldsymbol A_2 \boldsymbol A_2 ^T ) \boldsymbol u \right\| = 
					\left\|  \boldsymbol A_1 (\boldsymbol A_1- \boldsymbol A_2)^T \boldsymbol u  + (\boldsymbol A_1 - \boldsymbol A_2 ) \boldsymbol A_2 ^T  \boldsymbol u \right\| \\ \leq \left(\left\|\boldsymbol A_1  \right\|_{\rm{Sp}}+\left\|\boldsymbol A_2  \right\|_{\rm{Sp}} \right) \left\|\boldsymbol A_1  - \boldsymbol A_2\right\|_{\rm{Sp}}\left\| \boldsymbol u \right\|.
				\end{multline*}
	Moreover, $\left\| \boldsymbol A_1 \boldsymbol u \right\|^2 = \boldsymbol u^T  \boldsymbol A_1 ^T  \boldsymbol A_1  \boldsymbol u  =\left\|   \boldsymbol u \right\|^2  $, and thus $\left\| \boldsymbol A_1 \right\|_{\rm{Sp}} = \left\| \boldsymbol A_2 \right\|_{\rm{Sp}}=1$. Thus, $2 \left\|\boldsymbol A_1  - \boldsymbol A_2 \right\|_{\rm{Sp}}$ is a bound for the norm of the difference between the orthogonal projectors defined respectively by $\boldsymbol A_1$ and $\boldsymbol A_2$.	 
	\end{enumerate}
\end{proofof}

\begin{lem}\label{Dn_inverse} 
If the Assumptions \ref{ass_dgp}.1, \ref{ass_dgp}.3, \ref{ass_ident}.1 and \ref{ass_ident}.5  hold true, $E\left[n^{-2}{\mathbb{X}}_n ^T\mathbb{D}_n  {\mathbb{X}}_n\right]$ tends to a positive definite $p\times p-$matrix and 
	$$
		\sup_{\boldsymbol d \in \mathcal D} \left\|n^{-2}{\mathbb{X}}_n ^T\mathbb{D}_n  {\mathbb{X}}_n - E\left[n^{-2}{\mathbb{X}}_n ^T\mathbb{D}_n  {\mathbb{X}}_n\right] \right\|_{\rm{Sp}}= O_{\mathbb{P}}(n^{-1/2}).
	$$
If in addition Assumption \ref{ass_asy_norm}.1 holds true, $E\left[n^{-2}\frac{\partial }{\partial \lambda} {\mathbb{Y}}_n(\lambda_0) ^T   {\mathbb{B}} _n  \frac{\partial }{\partial \lambda}  {\mathbb{Y}}_n(\lambda_0)\right]$ tends to a positive constant and	
	\begin{align*}
		\underset{\boldsymbol d \in \mathcal{D}}{\sup}
		\left\|
		n^{-2}\frac{\partial }{\partial \lambda} {\mathbb{Y}}_n(\lambda_0) ^T   {\mathbb{D}} _n  \frac{\partial }{\partial \lambda}  {\mathbb{Y}}_n(\lambda_0)
		-
		E\left[n^{-2}\frac{\partial }{\partial \lambda} {\mathbb{Y}}_n(\lambda_0) ^T   {\mathbb{D}} _n  \frac{\partial }{\partial \lambda}  {\mathbb{Y}}_n(\lambda_0)\right]
		\right\|
		&= 
		O_{\mathbb{P}}(n^{-1/2})
		\text{\qquad and} \\
		\underset{\boldsymbol d \in \mathcal{D}}{\sup}
		\left\|
		n^{-2}{\mathbb{X}}_n ^T {\mathbb{D}} _n  \frac{\partial }{\partial \lambda}  {\mathbb{Y}}_n(\lambda_0)
		-
		E\left[n^{-2}{\mathbb{X}}_n ^T {\mathbb{D}} _n  \frac{\partial }{\partial \lambda}  {\mathbb{Y}}_n(\lambda_0)\right]
		\right\|
		&= 
		O_{\mathbb{P}}(n^{-1/2}).
			\end{align*}
\end{lem}

\begin{proofof}{Lemma \ref{Dn_inverse}}~~\\

First, we investigate the behavior of $n^{-2}{\mathbb{X}}_n ^T\boldsymbol{\Omega}_n  {\mathbb{X}}_n $ that we decompose 
$$
\frac{1}{n^2}{\mathbb{X}}_n ^T\boldsymbol{\Omega}_n  {\mathbb{X}}_n = \frac{n-1}{n} \frac{1}{n(n-1)}\sum_{1\leq i\neq j \leq n}\boldsymbol O_{ij}
+\frac{1}{n^2} \sum_{1\leq i \leq n}   (\boldsymbol X_i- {E}[\boldsymbol X_i\mid \boldsymbol Z_i]) (\boldsymbol X_i- {E}[\boldsymbol X_i\mid \boldsymbol Z_i]) ^T {f}_z ^2 (\boldsymbol Z_i),
$$
where $\boldsymbol O_{ij} = \boldsymbol O_{ij}(\boldsymbol d) =  (\boldsymbol X_i- {E}[\boldsymbol X_i\mid \boldsymbol Z_i]) (\boldsymbol X_j-{E}[\boldsymbol X_j\mid \boldsymbol Z_j])^T {f}_z(\boldsymbol Z_i) {f}_z(\boldsymbol Z_j) \boldsymbol \Omega_{n,ij}$. It is obvious that under our assumptions the second sum, corresponding to the diagonal terms of the quadratic form $ {\mathbb{X}}_n ^T\boldsymbol{\Omega}_n  {\mathbb{X}}_n$, has the rate $O_{\mathbb{P}}(n^{-1})$. On the other hand, for any $\boldsymbol d\in  \mathcal D$ and any $\boldsymbol u \in\mathbb{R}^p$, using the Fourier Transform and the monotonicity of the exponential function, we have
	\begin{align*}
		E[\boldsymbol u^T &\boldsymbol O_{ij}(\boldsymbol d)  \boldsymbol u] 
		= 
		E\left[ \boldsymbol u^T (\boldsymbol X_i- {E}[\boldsymbol X_i\mid \boldsymbol Z_i])  (\boldsymbol X_j-{E}[\boldsymbol X_j\mid \boldsymbol Z_j]) ^T  \boldsymbol u  {f}_z(\boldsymbol Z_i) {f}_z(\boldsymbol Z_j) \boldsymbol \Omega_{n,ij}\right]\\
		&=  
		\frac{\pi^{-(p+q)/2}}{\sqrt{d_1\cdots d_{p+q}}}  \int_{\mathbb{R}^{p+q}} \left|E\left[\boldsymbol u^T (\boldsymbol X - {E}[\boldsymbol X\mid \boldsymbol Z])  {f}_z(\boldsymbol Z) \exp\left\{2 i \boldsymbol{w}^T \left( \boldsymbol X^T,\boldsymbol Z^T \right)^T\right\} \right] \right|^2 \exp\left\{- \boldsymbol{w}^T \boldsymbol D^{-1}  \boldsymbol{w}\right\}d\boldsymbol{w}\\
		&\geq 
		\frac{\pi^{-(p+q)/2}}{d_U^{(p+q)/2}} \\
		&\quad \times 
		\int_{\mathbb{R}^{p+q}} \left|E\left[\boldsymbol u^T (\boldsymbol X - {E}[\boldsymbol X\mid \boldsymbol Z])  {f}_z(\boldsymbol Z) \exp\left\{2 i \boldsymbol{w}^T \left( \boldsymbol X^T,\boldsymbol Z^T \right)^T\right\} \right] \right|^2 \exp\left\{- \boldsymbol{w}^T {\rm diag}(d_L,\ldots,d_L)^{-1}  \boldsymbol{w}\right\}d\boldsymbol{w}\\
		&= 
		\frac{d_L^{(p+q)/2}}{d_U^{(p+q)/2}}E\left[\boldsymbol u^T \boldsymbol O_{ij}({\rm diag} (d_L,\ldots,d_L)^{-1})  \boldsymbol u\right],
	\end{align*}
where $d_U$ is the upper bound and $d_L$ the lower bound of the  values on the diagonal of $\boldsymbol D $. Since by Assumption \ref{ass_ident}.1 the variable  $\boldsymbol u^T (\boldsymbol X - {E}[\boldsymbol X\mid \boldsymbol Z]) $ could not be equal to zero almost surely, we necessarily have $E[\boldsymbol u^T \boldsymbol O_{ij}(\boldsymbol d ) \boldsymbol u] >0$ and thus, $E[\boldsymbol O_{ij}(\boldsymbol d )]$ is positive definite. Moreover, it is clear from the last display that there exists a constant $C >0$ 
such that  $E[\boldsymbol O_{ij}(\boldsymbol d )] - C \boldsymbol{I}_{p\times p} $ is positive definite  for each $\boldsymbol d \in \mathcal D$. By the uniform convergence results of \citet{sherman1994maximal}, 
$$
	\sup_{\boldsymbol d \in \mathcal D} \left\|\frac{1}{n^2}{\mathbb{X}}_n ^T\boldsymbol{\Omega}_n  {\mathbb{X}}_n - E[\boldsymbol O_{ij} (\boldsymbol d )]\right\|_{\rm{Sp}}= O_{\mathbb{P}}(n^{-1/2}).
$$

Next, we derive the convergence of $E\left[n^{-2}{\mathbb{X}}_n ^T\mathbb{D}_n  {\mathbb{X}}_n\right]$. Let us decompose 
$$
	\mathbb{D}_n = \boldsymbol\Omega_n^{1/2} \left( \boldsymbol{I}_{n\times n}  - P_{ \boldsymbol\Omega_n^{1/2} {\boldsymbol{1}}_n   }\right) \boldsymbol\Omega_n^{1/2},
$$
where 
$$
	P_{ \boldsymbol\Omega_n^{1/2} {\boldsymbol{1}}_n } =   \frac{1}{\boldsymbol{1}_n^T \boldsymbol\Omega_n  \boldsymbol{1}_n}  \boldsymbol\Omega_n  ^{1/2} \boldsymbol{1}_n  \boldsymbol{1}_n^T  \boldsymbol\Omega_n  ^{1/2}.
$$
(Here, $\boldsymbol\Omega_n^{1/2} $ is the positive definite square root of $\boldsymbol\Omega_n $.) 
Let us define
$$
	P^0_{ \boldsymbol\Omega_n^{1/2} {\boldsymbol{1}}_n } =   \frac{1}{n^{-2}E\left[\boldsymbol{1}_n^T \boldsymbol\Omega_n  \boldsymbol{1}_n\right]}  \boldsymbol\Omega_n  ^{1/2} n^{-1}\boldsymbol{1}_n  n^{-1} \boldsymbol{1}_n^T  \boldsymbol\Omega_n  ^{1/2}.
$$
It is clear from above that $n^{-2}E\left[\boldsymbol{1}_n^T \boldsymbol\Omega_n  \boldsymbol{1}_n\right]$ converges at the rate $O_{\mathbb{P}}(n^{-1})$ to a strictly positive limit and $n^{-2}\boldsymbol{1}_n^T \boldsymbol\Omega_n  \boldsymbol{1}_n- n^{-2}E\left[\boldsymbol{1}_n^T \boldsymbol\Omega_n  \boldsymbol{1}_n\right]= O_{\mathbb{P}}(n^{-1/2})$, uniformly with respect to $\boldsymbol d\in\mathcal D$. 
Thus 
$$
	\sup_{\boldsymbol d \in \mathcal D}\left\| \frac{1}{(n^{-2}E \left[\boldsymbol{1}_n^T \boldsymbol\Omega_n  \boldsymbol{1}_n \right] )^{1/2}} \boldsymbol\Omega_n  ^{1/2} n^{-1}\boldsymbol{1}_n - \frac{1}{\left(n^{-2}\boldsymbol{1}_n^T \boldsymbol\Omega_n  \boldsymbol{1}_n\right)^{1/2}} \boldsymbol\Omega_n  ^{1/2} n^{-1}\boldsymbol{1}_n \right\| = O_{\mathbb{P}}(n^{-1/2}).
$$
Then, it follows that 
$$
	\left\| P_{ \boldsymbol\Omega_n^{1/2} {\boldsymbol{1}}_n }  - P^0_{ \boldsymbol\Omega_n^{1/2} {\boldsymbol{1}}_n } \right\|_{\rm{Sp}} = O_{\mathbb{P}}(n^{-1/2}).
$$
Hence, in order to show that asymptotically the spectrum of  $E\left[n^{-2}{\mathbb{X}}_n ^T\mathbb{D}_n  {\mathbb{X}}_n\right]$ 
stays away from zero, it suffices to show that the spectrum of the $p\times p-$matrix $E\left[n^{-2}{\mathbb{X}}_n ^T \mathbb{D}^0_n {\mathbb{X}}_n\right] $ stays away from zero, where  
$$
	\mathbb{D}^0_n = \boldsymbol\Omega_n^{1/2} \left( \boldsymbol{I}_{n\times n}  - P^0_{ \boldsymbol\Omega_n^{1/2} {\boldsymbol{1}}_n   }\right) \boldsymbol\Omega_n^{1/2}.
$$
For any $\boldsymbol u \in\mathbb{R}^p$ we have $E\left[n^{-2} \boldsymbol u^T {\mathbb{X}}_n ^T\mathbb{D}_n^0  {\mathbb{X}}_n \boldsymbol u \right] = \Delta_n / E\left[n^{-2} \boldsymbol{1}_n^T \boldsymbol\Omega_n  \boldsymbol{1}_n\right]$ with
$$
	\Delta_n= \Delta_n (\boldsymbol u) =  E\left[ \left\|  n^{-1}  \boldsymbol\Omega_n^{1/2}  \boldsymbol{1}_n  \right\|^2\right] E\left[ \left\|  n^{-1} \boldsymbol\Omega_n^{1/2} {\mathbb{X}}_n \boldsymbol u   \right\|^2\right] 
	- E\left[ \left| \left\langle n^{-1} \boldsymbol\Omega_n^{1/2} {\mathbb{X}}_n \boldsymbol u  ,n^{-1} \boldsymbol\Omega_n^{1/2}  \boldsymbol{1}_n \right\rangle \right|^2\right] .
$$
We aim showing that, for any fixed $\boldsymbol u \in\mathbb{R}^p$, $\Delta_n / E\left[n^{-2} \boldsymbol{1}_n^T \boldsymbol\Omega_n  \boldsymbol{1}_n\right]$ stays away from zero, uniformly with respect to $\boldsymbol d$. This will imply that the limit of $E\left[n^{-2}{\mathbb{X}}_n ^T \mathbb{D}^0_n {\mathbb{X}}_n\right] $ is a positive $p\times p-$matrix. Consider the second order polynomial 
	\begin{multline*}
		P_n(t) = P_n(t; \boldsymbol u) =  E\left[ \left\| n^{-1} \boldsymbol\Omega_n^{1/2} {\mathbb{X}}_n \boldsymbol u  + t n^{-1} \boldsymbol\Omega_n^{1/2} {\boldsymbol{1}}_n   \right\|^2\right] =  E\left[ \left\| n^{-1} \boldsymbol\Omega_n^{1/2} {\mathbb{X}}_n \boldsymbol u    \right\|^2\right] \\ + 2 t E\left[  \left\langle n^{-1} \boldsymbol\Omega_n^{1/2} {\mathbb{X}}_n \boldsymbol u  ,n^{-1} \boldsymbol\Omega_n^{1/2}  \boldsymbol{1}_n \right\rangle \right]
		+ t^2 E\left[ \left\| n^{-1} \boldsymbol\Omega_n^{1/2} {\boldsymbol{1}}_n   \right\|^2\right]\geq 0. 
	\end{multline*}
By elementary properties of second order polynomials, the minimal value of $P_n(t)$ is $\Delta_n / E\left[n^{-2} \boldsymbol{1}_n^T \boldsymbol\Omega_n  \boldsymbol{1}_n\right]$. If the minimal value of $P_n(t)$ goes to zero, then necessarily 
$$
	\inf_{ t }E\left[ n^{-2}\left({\mathbb{X}}_n \boldsymbol u  + t  {\boldsymbol{1}}_n \right)^T \boldsymbol\Omega_n 
	\left({\mathbb{X}}_n \boldsymbol u  + t  {\boldsymbol{1}}_n \right) \right] \rightarrow 0,
$$
uniformly with respect to $\boldsymbol d \in \mathcal D$. From the first part of the proof we could deduce that this contradicts Assumption \ref{ass_ident}.1.
Thus, necessarily the spectrum of the $p\times p-$matrix $E\left[n^{-2}{\mathbb{X}}_n ^T \mathbb{D}^0_n {\mathbb{X}}_n\right] $ stays away from zero. Finally, to derive the rate of uniform convergence of $n^{-2}{\mathbb{X}}_n ^T \mathbb{D}_n {\mathbb{X}}_n$, we could use again the uniform convergence results of  \citet{sherman1994maximal} after removing the diagonal terms, and  next study the part given by the diagonal terms. The details are omitted. 

Next, we derive the convergence of $E\left[n^{-2}\frac{\partial }{\partial \lambda} {\mathbb{Y}}_n(\lambda_0) ^T\mathbb{B}_n  \frac{\partial }{\partial \lambda} {\mathbb{Y}}_n(\lambda_0)\right]$. We get that 
$$
	\mathbb{D}_n = \boldsymbol\Omega_n^{1/2} \left( \boldsymbol{I}_{n\times n}  - P_{ \boldsymbol\Omega_n^{1/2} {\boldsymbol{1}}_n   }\right) \boldsymbol\Omega_n^{1/2} = 
	\mathbb{S}_n^T \mathbb{S}_n,
$$
where 
$$
	\mathbb{S}_n = \left( \boldsymbol{I}_{n\times n}  - P_{ \boldsymbol\Omega_n ^{1/2}{\boldsymbol{1}}_n   }\right) \boldsymbol\Omega_n^{1/2}.
$$
In addition, let $\mathbb{W}_n = \mathbb{S}_n {\mathbb{X}}_n   \left( {\mathbb{X}}_n ^T\mathbb{D}_n  {\mathbb{X}}_n   \right)^{-1} {\mathbb{X}}_n ^T \mathbb{D}_n\frac{\partial }{\partial \lambda} {\mathbb{Y}}_n(\lambda_0)$. Therefore, it follows that 
	\begin{align*}
		E\left[n^{-2}\frac{\partial }{\partial \lambda} {\mathbb{Y}}_n(\lambda_0) ^T\mathbb{B}_n  \frac{\partial }{\partial \lambda} {\mathbb{Y}}_n(\lambda_0)\right]
		=
		E\left[\left\| n^{-1} \mathbb{S}_n \frac{\partial }{\partial \lambda} {\mathbb{Y}}_n(\lambda_0) \right\|^2\right] - E\left[n^{-2} \mathbb{W}_n^T\mathbb{W}_n\right].
	\end{align*}
Consider now the second order polynomial 
	\begin{multline*}
		P_n(t) =  E\left[ \left\| n^{-1} \mathbb{S}_n \frac{\partial }{\partial \lambda} {\mathbb{Y}}_n(\lambda_0) + t n^{-1} \mathbb{W}_n   \right\|^2\right] =  E\left[ \left\|n^{-1} \mathbb{S}_n \frac{\partial }{\partial \lambda} {\mathbb{Y}}_n(\lambda_0)   \right\|^2\right] \\ 
		+ 2 t E\left[  \left\langle n^{-1} \mathbb{W}_n  ,n^{-1} \mathbb{S}_n \frac{\partial }{\partial \lambda} {\mathbb{Y}}_n(\lambda_0) \right\rangle \right]
		+ t^2 E\left[ n^{-2} \mathbb{W}_n^T\mathbb{W}_n\right]\geq 0. 
	\end{multline*}
By elementary properties of second order polynomials, the minimal value of $P_n(t)$ is $E\left[ \left\| n^{-1} \mathbb{S}_n \frac{\partial }{\partial \lambda} {\mathbb{Y}}_n(\lambda_0) \right\|^2\right] - E\left[ n^{-2} \mathbb{W}_n^T\mathbb{W}_n\right]$. If the minimal value of $P_n(t)$ goes to zero, then necessarily 
$$
	\inf_{ t }E\left[ n^{-2}\frac{\partial }{\partial \lambda} {\mathbb{Y}}_n(\lambda_0)^T \left(\boldsymbol{I}_{n\times n} + t\mathbb{D}_n{\mathbb{X}}_n   \left( {\mathbb{X}}_n ^T\mathbb{D}_n  {\mathbb{X}}_n   \right)^{-1} {\mathbb{X}}_n ^T \right)\mathbb{D}_n
	\left(\boldsymbol{I}_{n\times n} + t{\mathbb{X}}_n   \left( {\mathbb{X}}_n ^T\mathbb{D}_n  {\mathbb{X}}_n   \right)^{-1} {\mathbb{X}}_n ^T \mathbb{D}_n\right) \frac{\partial }{\partial \lambda} {\mathbb{Y}}_n(\lambda_0)\right] \rightarrow 0,
$$
uniformly with respect to $\boldsymbol d \in \mathcal D$. Note that by the same reasoning as for $E\left[n^{-2}{\mathbb{X}}_n ^T\mathbb{D}_n  {\mathbb{X}}_n\right]$ we get that $E\left[n^{-2}\boldsymbol u^T\mathbb{D}_n \boldsymbol u\right] > 0$ for all $\boldsymbol u \in\mathbb{R}^p$ with $\boldsymbol u \neq \boldsymbol{0}$. Therefore, we could deduce that the upper statement contradicts Assumption \ref{ass_asy_norm}.1. Thus, necessarily $E\left[n^{-2}\frac{\partial }{\partial \lambda} {\mathbb{Y}}_n(\lambda_0) ^T\mathbb{B}_n  \frac{\partial }{\partial \lambda} {\mathbb{Y}}_n(\lambda_0)\right]$ stays away from zero.

Finally, to derive the rates of uniform convergence of $n^{-2}\frac{\partial }{\partial \lambda} {\mathbb{Y}}_n(\lambda_0) ^T   {\mathbb{D}} _n  \frac{\partial }{\partial \lambda}  {\mathbb{Y}}_n(\lambda_0)$ and $n^{-2}{\mathbb{X}}_n^T   {\mathbb{D}} _n  \frac{\partial }{\partial \lambda}  {\mathbb{Y}}_n(\lambda_0)$, we could use again the uniform convergence results of  \citet{sherman1994maximal} after removing the diagonal terms, and  next study the part given by the diagonal terms. The details are omitted. Now the proof is complete. \\
\end{proofof}

\begin{lem}\label{inner_smooth}
Under the conditions of Theorem \ref{consist},
	\begin{equation*}	
		\sup_{h\in\mathcal{H}^{c}_n} \frac{1}{\sqrt{n}}\left\| \widehat{\mathbb{X}}_n - {\mathbb{X}}_n  \right\|_{\rm{Sp}} =  o_{\mathbb{P}}(1).
	\end{equation*}
\end{lem}

\begin{proofof}{Lemma \ref{inner_smooth}}~~\\

In order to prove the statement we consider 
	\begin{equation*}	
		\frac{1}{n}\left\| \left(\widehat{\mathbb{X}}_n - {\mathbb{X}}_n\right) \boldsymbol{u}  \right\|^2,
	\end{equation*}
where $\boldsymbol{u} \in \mathbb R^p$ and $\left\| \boldsymbol{u}\right\|= 1$. In the remaining of the proof we set without loss of generality $p=1$ to keep the notation simple, i.e. we consider 
	\begin{equation*}	
		\frac{1}{n}\left\| \widehat{\mathbb{X}}_n - {\mathbb{X}}_n  \right\|^2.
	\end{equation*}
We have that, for $1 \leq i \leq n$,
	\begin{align*}
		\widehat{\mathbb{X}}_{n,i} - {\mathbb{X}}_{n,i} 
		&= 
		  ( X_i-\widehat{E}[ X_i\mid \boldsymbol Z_i]) \widehat{f}_z(\boldsymbol Z_i) 
		- ( X_i-E[ X_i\mid \boldsymbol Z_i]) f_z(\boldsymbol Z_i) \\
		&=
		  \frac{1}{n} \sum_{j=1}^{n} \left(X_i - X_j\right) K_{h,ij}
	    - ( X_i-E[ X_i\mid \boldsymbol Z_i]) f_z(\boldsymbol Z_i) \\	
		&=
		  X_i \frac{1}{n} \sum_{j=1, j\neq i}^{n} \left(K_{h,ij}  - f_z(\boldsymbol Z_i)\right)
		  - \frac{1}{n} X_i f_z(\boldsymbol Z_i)+ \\
		&\qquad \frac{1}{n} \sum_{j=1, j\neq i}^{n} \left(E[ X_i\mid \boldsymbol Z_i]f_z(\boldsymbol Z_i) - X_jK_{h,ij} \right)
		+ \frac{1}{n} E[ X_i\mid \boldsymbol Z_i] f_z(\boldsymbol Z_i). 		    
	\end{align*}
We start by considering
	\begin{align*}
		&\frac{1}{n}\left\|\left(X_1 \frac{1}{n} \sum_{j=1, j\neq 1}^{n} \left(K_{h,1j}  - f_z(\boldsymbol Z_1)\right), \ldots, X_n \frac{1}{n} \sum_{j=1, j\neq n}^{n}\left( K_{h,nj}- f_z(\boldsymbol Z_n)\right)\right)^T\right\|^2\\
		&\quad =
		\frac{1}{n^3} \sum_{1 \leq i\neq j \leq n}^{n} X_i^2 \left(K_{h,ij}  - f_z(\boldsymbol Z_i)\right)^2
		+
		\frac{1}{n^3} \sum_{1 \leq i\neq j \neq k\leq n}^{n} X_i^2 \left(K_{h,ij}  - f_z(\boldsymbol Z_i)\right)\left(K_{h,ik}  - f_z(\boldsymbol Z_i)\right)\\
		&\quad = A_n + B_n.
	\end{align*}
It is easy to check that $\sup_{h\in\mathcal{H}^{c}_n}| A_n | = o_{\mathbb{P}}(1)$. We show in the following that 
	\begin{align}\label{uni_rate_insmooth_1}
		\sup_{h\in\mathcal{H}^{c}_n}| B_n | = o_{\mathbb{P}}(1)
	\end{align}
as well. Note that $\frac{n^3}{(n)_3}B_n$ is a $U$--process of order 3, where $(n)_k = n(n-1) \ldots (n-k+1)$.

For this $U$--process we compute the mean and use the Hoeffding decomposition. The kernel of $B_{n}$ is not symmetric in its arguments. However, we could apply the usual symmetrization idea. For instance, for a second order $U-$statistic defined by a kernel $h( \boldsymbol U_i, \boldsymbol U_j)$, we could replace it  by the symmetric kernel $\frac{1}{2}\left[h(\boldsymbol U_i,\boldsymbol U_j) + h(\boldsymbol U_j,\boldsymbol U_i)\right]$ from which we get the same $U-$statistic. Here, $\boldsymbol U_i = \left(X_i, \boldsymbol Z_i^T\right)^T$. We can proceed in the same way by considering all $3!$ permutations of the variables for $B_{n}$ so that we can apply the Hoeffding decomposition. Thus, by abuse, we will proceed as if the kernel of the $U-$statistic we handle is symmetric. For simpler notation, we use $E_i$, $E_{i,j}, \ldots$ for the conditional expectations $E\left[\cdot \mid \boldsymbol U_i\right],$ $E\left[\cdot \mid \boldsymbol U_i,\boldsymbol U_j\right], \ldots$. 

In addition, we have that $\{(x_i, \boldsymbol{z}_i, \boldsymbol{z}_j, \boldsymbol{z}_k)\mapsto  x_i^2 \left(K_{h,ij}  - f_z(\boldsymbol z_i)\right)\left(K_{h,ik}  - f_z(\boldsymbol z_i)\right):  h\in\mathcal{H}^{c}_n \}$ is Euclidean for a squared integrable envelope. See Lemma 22 in \citet{nolan1987u} and Lemma 2.14 in \citet{pakes1989simulation}. Therefore, we can in the following repeatedly apply Corollary 7 and the Maximal Inequality of \citet{sherman1994maximal}. All remainder terms are controlled by Assumption \ref{ass_con}.2.

We start by considering the mean. We get that
	\begin{align*}
		E\left[X_i^2 \left(K_{h,ij}  - f_z(\boldsymbol Z_i)\right)\left(K_{h,ik} - f_z(\boldsymbol Z_i)\right)\right] 
		&= 
		E\left[X_i^2 \left(K_{h,ij}  - f_z(\boldsymbol Z_i)\right)
		E_{i,j}\left[\left(K_{h,ik}  - f_z(\boldsymbol Z_i)\right)\right]	\right]	\\	
		&= 
		E\left[X_i^2 \left(K_{h,ij} - f_z(\boldsymbol Z_i)\right)
		h^2\gamma_1(\boldsymbol Z_i )\right](1 + o(1))	\\
		&= 
		h^4E\left[X_i^2 \gamma_1(\boldsymbol Z_i )^2\right](1 + o(1)), 			
	\end{align*}
where 
$$
	\gamma_1(\boldsymbol Z ) = \mu(K) \cdot {\rm tr } \{\boldsymbol H_{z,z} f_z(\boldsymbol Z)\},
$$
with $\int _{\mathbb{R}^{q}}\boldsymbol u \boldsymbol u^TK(\boldsymbol u) d\boldsymbol u = \mu(K) \boldsymbol{I}_{q\times q}$. $\boldsymbol H_{z,z} f_z$ denotes the matrix of second derivative of $f_z(\cdot)$ with respect to the components of $\boldsymbol Z \in \mathbb{R}^{q}$ and ${\rm tr } \{\cdot\}$ denotes the trace operator. Therefore, it follows that the mean of $B_{n}$ is of order $ o_{\mathbb P} (1)$ uniformly with respect to $h$.

We consider now the three first order $U$--processes of the Hoeffding decomposition of $B_n$. We get that, by the same reasoning as for the mean,
	\begin{align*}
		E_i\left[X_i^2 \left(K_{h,ij}  - f_z(\boldsymbol Z_i)\right)\left(K_{h,ik} - f_z(\boldsymbol Z_i)\right)\right] 
		= 
		h^4X_i^2 \gamma_1(\boldsymbol Z_i )^2(1 + o_{\mathbb{P}}(1)).			
	\end{align*}	
\newpage 
In addition, we get that 	
	\begin{align*}
		E_j\left[X_i^2 \left(K_{h,ij}  - f_z(\boldsymbol Z_i)\right)\left(K_{h,ik}  - f_z(\boldsymbol Z_i)\right)\right] 
		&= h^2E_j\left[X_i^2 \left(K_{h,ij} - f_z(\boldsymbol Z_i)\right)\gamma_1(\boldsymbol Z_i)\right](1 + o_{\mathbb{P}}(1))\\
		&= \left(h^2E_j\left[X_i^2 K_{h,ij}\gamma_1(\boldsymbol Z_i)\right]  - h^2E\left[X_i^2 f_z(\boldsymbol Z_i)\gamma_1(\boldsymbol Z_i)\right]\right)(1 + o_{\mathbb{P}}(1))\\
		&= h^2E\left[X_j^2\mid \boldsymbol{Z}_j\right] \gamma_1(\boldsymbol Z_j)f_z(\boldsymbol Z_j) + O_{\mathbb{P}}(h^4) + O_{\mathbb{P}}(h^2)  + o_{\mathbb{P}}(h^2)
		.			
	\end{align*}
The reasoning when conditioning on $\boldsymbol{U}_k$ is the same. Therefore, it follows together with Corollary 4 of \citet{sherman1994maximal} that the first order $U-$processes of the Hoeffding decomposition of $B_{n}$ are of order $o_{\mathbb P} (1)$ uniformly with respect to $h$.
 		
We consider now the three second order $U-$processes of the Hoeffding decomposition of $B_{n}$. We start by conditioning on $\left(\boldsymbol{U}_i, \boldsymbol{U}_j\right)$ the reasoning for $\left(\boldsymbol{U}_i, \boldsymbol{U}_k\right)$ being similar. 
	\begin{align*}
		E_{i,j}\left[X_i^2 \left(K_{h,ij}  - f_z(\boldsymbol Z_i)\right)\left(K_{h,ik} - f_z(\boldsymbol Z_i)\right)\right] 
		&=X_i^2 \left(K_{h,ij}  - f_z(\boldsymbol Z_i)\right) 
		E_{i}\left[\left(K_{h,ik} - f_z(\boldsymbol Z_i)\right)\right] \\
		&=X_i^2 \left(K_{h,ij}  - f_z(\boldsymbol Z_i)\right) 
		h^2 \gamma_1(\boldsymbol Z_i)(1 + o_{\mathbb{P}}(1))\\
		&=X_i^2 K_{h,ij}h^2 \gamma_1(\boldsymbol Z_i)(1 + o_{\mathbb{P}}(1))  
		- X_i^2f_z(\boldsymbol Z_i)h^2 \gamma_1(\boldsymbol Z_i)(1 + o_{\mathbb{P}}(1))\\
		&=h^{2-q} h^qK_{h,ij} \tau(\boldsymbol{U}_i) 
		  + O_{\mathbb{P}}(h^2)							
	\end{align*}
Now, we apply the Maximal Inequality of \citet{sherman1994maximal}, page 448,  for the degenerate $U-$process given by the kernel $h^qK_{h,ij} \tau(\boldsymbol{U}_i) $, indexed by $h\in\mathcal H_n^c$, with envelope $\|K\|_\infty \tau(\cdot)$. (Herein, $\|\cdot\|_\infty $ denotes the uniform norm.) We take $p=1$ and $\beta\in (0,1)$ arbitrarily close to 1 to stand for Sherman's quantity $\alpha$. Since $K(\cdot)$ is of bounded variation and symmetric, without loss of generality we could consider 
that $K(\cdot)$ is nonincreasing on $[0,\infty)$. In this case, $0\leq K(\cdot/h)\leq K(\cdot/\overline h)$ with $\overline h = \sup \mathcal H^{c}_n = : c_{max} n^{-\alpha}$.
Hence, using Jensen's inequality, we could bound the right-hand side of the Maximal Inequality of  \citet{sherman1994maximal} by a universal constant times
$$
	\left(E \left[ K^{2}\left(\frac{\boldsymbol Z_i - \boldsymbol Z_j}{c_{max} n^{-\alpha}}\right)\tau^{2} (\boldsymbol U_i)\right]\right)^{\beta/2}.
$$
By standard changes of variables and suitable integrability conditions, the power $\beta/2$ of the expectation in the last display is bounded by  a constant times $n^{-\alpha \beta q/2}$. Consequently, the uniform rate of the second $U-$processes obtained conditioning by $\boldsymbol U_i, \boldsymbol U_j$ and  $\boldsymbol U_i, \boldsymbol U_k$, respectively is $n^{-1}\times O_{\mathbb P} (n^{-\alpha \{2-q +\beta q/2\} })$. As $1 + \alpha (2 - q + \beta q/2) > 0 $ under our assumptions we get that $n^{-1}\times O_{\mathbb P} (n^{-\alpha \{2-q +\beta q/2\}})= o_{\mathbb P} (1)$.	

In addition, we get that 
	\begin{align*}
		&E_{j,k}\left[X_i^2 \left(K_{h,ij}  - f_z(\boldsymbol Z_i)\right)\left(K_{h,ik}  - f_z(\boldsymbol Z_i)\right)\right] \\		
		&\hskip 3cm =h^{-2q}E_{j,k}\left[X_i^2 h^{2q}K_{h,ij}K_{h,ik}	\right]
		-
		2E_{j}\left[X_i^2K_{h,ij}f_z(\boldsymbol Z_i)	\right]
		+
		E\left[X_i^2f_z(\boldsymbol Z_i)^2	\right]	.
	\end{align*}
Now, we apply the Maximal Inequality of \citet{sherman1994maximal}, page 448,  for the degenerate $U-$process given by the kernel $E_{j,k}\left[X_i^2h^{2q}K_{h,ij}K_{h,ik}\right]$, indexed by $h\in\mathcal H_n^c$, with envelope $E_{j,k}\left[X_i^2\|K\|_\infty^2\right] $. We take $p=1$ and $\beta\in (0,1)$ arbitrarily close to 1 to stand for Sherman's quantity $\alpha$. Using Jensen's inequality, we could bound the right-hand side of the Maximal Inequality of  \citet{sherman1994maximal} by a universal constant times
$$
	\left(E \left[ E_{j,k}\left[X_i^2K\left(\frac{\boldsymbol{Z}_i - \boldsymbol{Z}_j}{c_{max} n^{-\alpha}}\right)	K\left(\frac{\boldsymbol{Z}_i - \boldsymbol{Z}_k}{c_{max} n^{-\alpha}}\right)	\right]^2\right]\right)^{\beta/2}.
$$
By standard changes of variables and suitable integrability conditions, the power $\beta/2$ of the expectation in the last display is bounded by  a constant times $n^{-\alpha \beta q}$. Consequently, the uniform rate of the second $U-$process obtained conditioning by $\boldsymbol U_j, \boldsymbol U_k$ is $n^{-1}\times O_{\mathbb P} (n^{\alpha q \{2 -\beta \} })$. As $1 - \alpha q(2 - \beta) > 0 $ under our assumptions we get that $n^{-1}\times O_{\mathbb P} (n^{\alpha q \{2 -\beta \} })= o_{\mathbb P} (1)$.	By similar reasoning we can control the remaining two parts. The details are omitted. Therefore, the second order $U-$processes of $B_n$ are of order $o_{\mathbb P} (1)$. 

In order to finish the proof of \eqref{uni_rate_insmooth_1} it remains to consider the rate for the third order $U-$process of $B_n$. As the reasoning for this part is the same as for the second order $U-$process we omit the details here. Therefore, the statement in \eqref{uni_rate_insmooth_1} follows.

In the next part we consider 
	\begin{align*}
		&\frac{1}{n}\Bigg\|\Bigg(\frac{1}{n} \sum_{j=1, j\neq 1}^{n} \left(E[ X_1\mid \boldsymbol Z_1]f_z(\boldsymbol Z_1) - X_jK_{h,1j} \right), 
		\ldots, 
		\frac{1}{n} \sum_{j=1, j\neq n}^{n} \left(E[ X_n\mid \boldsymbol Z_n]f_z(\boldsymbol Z_n) - X_jK_{h,nj} \right)\Bigg)^T\Bigg\|^2\\
		&\quad =
		\frac{1}{n^3} \sum_{1 \leq i\neq j \leq n}^{n}\left(E[ X_i\mid \boldsymbol Z_i]f_z(\boldsymbol Z_i) - X_jK_{h,ij} \right)^2\\
		&\qquad +
		\frac{1}{n^3} \sum_{1 \leq i\neq j \neq k\leq n}^{n} \left(E[ X_i\mid \boldsymbol Z_i]f_z(\boldsymbol Z_i) - X_jK_{h,ij} \right)
		\left(E[ X_i\mid \boldsymbol Z_i]f_z(\boldsymbol Z_i) - X_kK_{h,ik} \right)\\
		&\quad = \widetilde A_n + \widetilde B_n.
	\end{align*}
It is easy to check that $\sup_{h\in\mathcal{H}^{c}_n}| \widetilde A_n | = o_{\mathbb{P}}(1)$. We show in the following that 
	\begin{align}\label{uni_rate_insmooth_2}
		\sup_{h\in\mathcal{H}^{c}_n}| \widetilde B_n | = o_{\mathbb{P}}(1)
	\end{align}
as well. Note that $\frac{n^3}{(n)_3}\widetilde B_n$ is a $U$--process of order 3. For this $U$--process we compute the mean and use the Hoeffding decomposition. The kernel of $\widetilde B_{n}$ is not symmetric in its arguments. However, we apply again the usual symmetrization idea.	 

In addition, we have that $\{(x_j, x_k,\boldsymbol{z}_i, \boldsymbol{z}_j, \boldsymbol{z}_k)\mapsto  \left(E[ X_i\mid \boldsymbol z_i]f_z(\boldsymbol z_i) - x_jK_{h,ij} \right)\left(E[ X_i\mid \boldsymbol z_i]f_z(\boldsymbol z_i) - x_kK_{h,ik} \right):  h\in\mathcal{H}^{c}_n \}$ is Euclidean for a squared integrable envelope. See Lemma 22 in \citet{nolan1987u} and Lemma 2.14 in \citet{pakes1989simulation}. Therefore, we can in the following repeatedly apply Corollary 7 and the Maximal Inequality of \citet{sherman1994maximal}. All remainder terms are controlled by Assumption \ref{ass_con}.2.

We start by considering the mean of $\widetilde B_n$. We get that
	\begin{align*}
		&E\left[\left(E[ X_i\mid \boldsymbol Z_i]f_z(\boldsymbol Z_i) - X_jK_{h,ij} \right)
				\left(E[ X_i\mid \boldsymbol Z_i]f_z(\boldsymbol Z_i) - X_kK_{h,ik} \right)\right] \\
		&\hskip 3cm= 
		E\left[\left(E[ X_i\mid \boldsymbol Z_i]f_z(\boldsymbol Z_i) - X_jK_{h,ij} \right)
		E_{i,j}\left[\left(E[ X_i\mid \boldsymbol Z_i]f_z(\boldsymbol Z_i) - E[ X_k\mid \boldsymbol Z_k]K_{h,ik} \right)\right]	\right]	\\	
		&\hskip 3cm= 
		-E\left[\left(E[ X_i\mid \boldsymbol Z_i]f_z(\boldsymbol Z_i) - X_jK_{h,ij} \right)
		h^2\gamma_2(\boldsymbol Z_i )\right](1 + o(1))	\\
		&\hskip 3cm= 
		h^4E\left[\gamma_2(\boldsymbol Z_i )^2\right](1 + o(1)), 			
	\end{align*}
where 	
$$
  \gamma_2(\boldsymbol Z ) = \mu(K) \cdot {\rm tr } \{\boldsymbol H_{z,z} \left(E[X\mid \cdot]f_z\right)(\boldsymbol Z)\}.
$$
$\boldsymbol H_{z,z} \left(E[X\mid \cdot]f_z\right)$ denotes the matrix of second derivative of $E[X\mid \cdot]f_z(\cdot)$ with respect to the components of $\boldsymbol Z \in \mathbb{R}^{q}$. Therefore, it follows that the mean of $\widetilde B_{n}$ is of order $ o_{\mathbb P} (1)$ uniformly with respect to $h$.

We consider now the three first order $U$--processes of the Hoeffding decomposition of $\widetilde B_n$. We get that by the same reasoning as for the mean 
	\begin{align*}
		E_i\left[\left(E[ X_i\mid \boldsymbol Z_i]f_z(\boldsymbol Z_i) - X_jK_{h,ij} \right)
				  \left(E[ X_i\mid \boldsymbol Z_i]f_z(\boldsymbol Z_i) - X_kK_{h,ij} \right)\right]  
		= 
		h^4 \gamma_2(\boldsymbol Z_i )^2(1 + o_{\mathbb{P}}(1)).			
	\end{align*}
In addition, we get that 	
	\begin{align*}
		&E_j\left[\left(E[ X_i\mid \boldsymbol Z_i]f_z(\boldsymbol Z_i) - X_jK_{h,ij} \right)
				  \left(E[ X_i\mid \boldsymbol Z_i]f_z(\boldsymbol Z_i) - X_kK_{h,ik} \right)\right] \\
		&\hskip 3cm= -h^2E_j\left[\left(E[ X_i\mid \boldsymbol Z_i]f_z(\boldsymbol Z_i) - X_jK_{h,ij} \right)\gamma_2(\boldsymbol Z_i)\right](1 + o_{\mathbb{P}}(1))\\
		&\hskip 3cm= \left(h^2E_j\left[E\left[X_j \mid \boldsymbol{Z}_j\right] K_{h,ij}\gamma_2(\boldsymbol Z_i)\right]  - h^2E\left[X_i f_z(\boldsymbol Z_i)\gamma_2(\boldsymbol Z_i)\right]\right)(1 + o_{\mathbb{P}}(1))\\
		&\hskip 3cm = h^2E\left[X_j\mid \boldsymbol{Z}_j\right] \gamma_2(\boldsymbol Z_j)f_z(\boldsymbol Z_j) + O_{\mathbb{P}}(h^4) + O_{\mathbb{P}}(h^2)  + o_{\mathbb{P}}(h^2)
		.			
	\end{align*}
The reasoning when conditioning on $\boldsymbol{U}_k$ is the same. Therefore, it follows together with Corollary 4 of \citet{sherman1994maximal} that the first order $U-$processes of the Hoeffding decomposition of $\widetilde B_{n}$ are of order $o_{\mathbb P} (1)$ uniformly with respect to $h$.

We consider now the three second order $U-$processes of the Hoeffding decomposition of $\widetilde B_{n}$. We start by conditioning on $\left(\boldsymbol{U}_i, \boldsymbol{U}_j\right)$ the reasoning for $\left(\boldsymbol{U}_i, \boldsymbol{U}_k\right)$ being similar. 
	\begin{align*}
		&E_{i,j}\left[\left(E[ X_i\mid \boldsymbol Z_i]f_z(\boldsymbol Z_i) - X_jK_{h,ij} \right)
				\left(E[ X_i\mid \boldsymbol Z_i]f_z(\boldsymbol Z_i) - X_kK_{h,ik} \right)\right]\\
		&\hskip 3cm =-\left(E[ X_i\mid \boldsymbol Z_i]f_z(\boldsymbol Z_i) - X_jK_{h,ij} \right) 
		h^2 \gamma_2(\boldsymbol Z_i)(1 + o_{\mathbb{P}}(1))\\
		&\hskip 3cm =X_j K_{h,ij}h^2 \gamma_2(\boldsymbol Z_i)(1 + o_{\mathbb{P}}(1))  
		- E\left[X_i \mid \boldsymbol{Z}_i\right]f_z(\boldsymbol Z_i)h^2 \gamma_2(\boldsymbol Z_i)(1 + o_{\mathbb{P}}(1))\\
		&\hskip 3cm =h^{2-q} h^qK_{h,ij} \tau(\boldsymbol{U}_i, \boldsymbol{U}_j) 
		  + O_{\mathbb{P}}(h^2)							
	\end{align*}
Now, we apply the Maximal Inequality of \citet{sherman1994maximal}, page 448,  for the degenerate $U-$process given by the kernel $h^qK_{h,ij} \tau(\boldsymbol{U}_i, \boldsymbol{U}_j) $, indexed by $h\in\mathcal H_n^c$, with envelope $\|K\|_\infty \tau(\cdot,\cdot)$. We take $p=1$ and $\beta\in (0,1)$ arbitrarily close to 1 to stand for Sherman's quantity $\alpha$. Using Jensen's inequality, we could bound the right-hand side of the Maximal Inequality of  \citet{sherman1994maximal} by a universal constant times
$$
	\left(E \left[ K^{2}\left(\frac{\boldsymbol Z_i - \boldsymbol Z_j}{c_{max} n^{-\alpha}}\right)\tau^{2} (\boldsymbol U_i, \boldsymbol U_j)\right]\right)^{\beta/2}.
$$
By standard changes of variables and suitable integrability conditions, the power $\beta/2$ of the expectation in the last display is bounded by  a constant times $n^{-\alpha \beta q/2}$. Consequently, the uniform rate of the second $U-$processes obtained conditioning by $\boldsymbol U_i, \boldsymbol U_j$ and  $\boldsymbol U_i, \boldsymbol U_k$, respectively is $n^{-1}\times O_{\mathbb P} (n^{-\alpha \{2-q +\beta q/2\} })$. As $1 + \alpha (2 - q + \beta q/2) > 0 $ under our assumptions we get that $n^{-1}\times O_{\mathbb P} (n^{-\alpha \{2-q +\beta q/2\}})= o_{\mathbb P} (1)$.		
	
In addition we get that 
	\begin{align*}
		&E_{j,k}\left[\left(E[ X_i\mid \boldsymbol Z_i]f_z(\boldsymbol Z_i) - X_jK_{h,ij} \right)
		\left(E[ X_i\mid \boldsymbol Z_i]f_z(\boldsymbol Z_i) - X_kK_{h,ik} \right)\right] \\		
		&\hskip 3cm =h^{-2q}X_j X_kE_{j,k}\left[h^{2q}K_{h,ij}	K_{h,ik}	\right]
		-
		  X_jE_{j}\left[E[ X_i\mid \boldsymbol Z_i]f_z(\boldsymbol Z_i)K_{h,ij}\right]\\
		& \hskip 3cm \quad 
		-
		X_kE_{k}\left[E[ X_i\mid \boldsymbol Z_i]f_z(\boldsymbol Z_i)K_{h,ik}\right]
		+
		E[ X_i\mid \boldsymbol Z_i]^2f_z(\boldsymbol Z_i)^2	.
	\end{align*}
Now, we apply the Maximal Inequality of \citet{sherman1994maximal}, page 448,  for the degenerate $U-$process given by the kernel $X_j X_kE_{j,k}\left[h^{2q}K_{h,ij}	K_{h,ik}	\right]$, indexed by $h\in\mathcal H_n^c$, with envelope $X_jX_k\|K\|_\infty^2 $. We take $p=1$ and $\beta\in (0,1)$ arbitrarily close to 1 to stand for Sherman's quantity $\alpha$. Using Jensen's inequality, we could bound the right-hand side of the Maximal Inequality of  \citet{sherman1994maximal} by a universal constant times
$$
	\left(E \left[X_j^2 X_k^2E_{j,k}\left[K\left(\frac{\boldsymbol{Z}_i - \boldsymbol{Z}_j}{c_{max} n^{-\alpha}}\right)	K\left(\frac{\boldsymbol{Z}_i - \boldsymbol{Z}_k}{c_{max} n^{-\alpha}}\right)	\right]^2\right]\right)^{\beta/2}.
$$
By standard changes of variables and suitable integrability conditions, the power $\beta/2$ of the expectation in the last display is bounded by  a constant times $n^{-\alpha \beta q}$. Consequently, the uniform rate of the second $U-$process obtained conditioning by $\boldsymbol U_j, \boldsymbol U_k$ is $n^{-1}\times O_{\mathbb P} (n^{\alpha q \{2 -\beta \} })$. As $1 - \alpha q(2 - \beta) > 0 $ under our assumptions we get that $n^{-1}\times O_{\mathbb P} (n^{\alpha q \{2 -\beta \} })= o_{\mathbb P} (1)$.	By similar reasoning we can control the remaining three parts. The details are omitted. Therefore, the second order $U-$processes of $\widetilde B_n$ are of order $o_{\mathbb P} (1)$. 

In order to finish the proof of \eqref{uni_rate_insmooth_2} it remains to consider the rate for the third order $U-$process of $\widetilde B_n$. As the reasoning for this part is the same as for the second order $U-$process we omit the details here. Therefore, the statement in \eqref{uni_rate_insmooth_2} follows.
	
It is obvious that 	
	\begin{align*}
		\frac{1}{n}\left\|\left(n^{-1} E[ X_1\mid \boldsymbol Z_1] f_z(\boldsymbol Z_1), 
		\ldots, 
		n^{-1} E[ X_n\mid \boldsymbol Z_n] f_z(\boldsymbol Z_n)\right)^T\right\|^2 = o_{\mathbb P} (1) \\
		\text{and} \quad 
		\frac{1}{n}\left\|\left(n^{-1} X_1 f_z(\boldsymbol Z_1), 
		\ldots, 
		n^{-1} X_n f_z(\boldsymbol Z_n)\right)^T\right\|^2 = o_{\mathbb P} (1). 		
	\end{align*}	
Therefore, the statement follows.	\\
\end{proofof}

\begin{lem}\label{spectral_B_hat2}
Under the conditions of Theorem \ref{consist},
$$
	\sup_{h\in\mathcal{H}^{c}_n} \sup_{\boldsymbol d \in \mathcal D}  \left\| \widehat {\mathbb{B}} _n -{\mathbb{B}} _n  \right\|_{\rm{Sp}} = o_{\mathbb{P}} (n).
$$
\end{lem}

\begin{proofof}{Lemma \ref{spectral_B_hat2}}~~\\

We could once again write
$$
	\mathbb{D}_n = \boldsymbol\Omega_n^{1/2} \left( \boldsymbol{I}_{n\times n}  - P_{ \boldsymbol\Omega_n^{1/2} {\boldsymbol{1}}_n   }\right) \boldsymbol\Omega_n^{1/2} = \mathbb{S}_n^T \mathbb{S}_n,
$$
where 
$$
	\mathbb{S}_n = \left( \boldsymbol{I}_{n\times n}  - P_{ \boldsymbol\Omega_n ^{1/2}{\boldsymbol{1}}_n   }\right) \boldsymbol\Omega_n^{1/2}
$$
and $P_{ \boldsymbol\Omega_n^{1/2} {\boldsymbol{1}}_n }$ is the projector on the $1-$dimensional subspace generated by the vector $\boldsymbol\Omega_n^{1/2} {\boldsymbol{1}}_n$, that is 
$$
	P_{ \boldsymbol\Omega_n^{1/2} {\boldsymbol{1}}_n } =   \frac{1}{\boldsymbol{1}_n^T \boldsymbol\Omega_n  \boldsymbol{1}_n}  \boldsymbol\Omega_n  ^{1/2} \boldsymbol{1}_n  \boldsymbol{1}_n^T  \boldsymbol\Omega_n  ^{1/2}.
$$
Here, $\boldsymbol\Omega_n^{1/2} $ is the positive definite square root of $\boldsymbol\Omega_n $.
Next, we could rewrite $\widehat {\mathbb{B}}_n $ and $  {\mathbb{B}}_n $  under the form
	\begin{equation*} 
		\widehat {\mathbb{B}}_n =  \mathbb{S}_n^T   \left( \boldsymbol{I}_{n\times n}  -   P_{ \mathbb{S}_n\widehat{\mathbb{X}}_n  }   \right) \mathbb{S}_n \quad \text{ and } \quad   {\mathbb{B}}_n =  \mathbb{S}_n^T  \left( \boldsymbol{I}_{n\times n}  -   P_{ \mathbb{S}_n  {\mathbb{X}}_n  }   \right) \mathbb{S}_n,
	\end{equation*}
with $P_{ \mathbb{S}_n \widehat{\mathbb{X}}_n }$ and $P_{ \mathbb{S}_n  {\mathbb{X}}_n }$ the orthogonal projectors on the  subspaces generated by $ \mathbb{S}_n\widehat{\mathbb{X}}_n$ and $ \mathbb{S}_n {\mathbb{X}}_n$, that is 
$$
	P_{ \mathbb{S}_n \widehat{\mathbb{X}}_n }=     \mathbb{S}_n \widehat{\mathbb{X}}_n   \left(\widehat{\mathbb{X}}_n ^T\mathbb{D}_n \widehat{\mathbb{X}}_n   \right)^{-1} \widehat{\mathbb{X}}_n ^T  \mathbb{S}_n^T 
	\quad \text{and} \quad 
	P_{ \mathbb{S}_n {\mathbb{X}}_n }=      \mathbb{S}_n {\mathbb{X}}_n   \left( {\mathbb{X}}_n ^T\mathbb{D}_n  {\mathbb{X}}_n   \right)^{-1} {\mathbb{X}}_n ^T \mathbb{S}_n^T .
$$
Thus, 
	\begin{equation*} 
		\widehat {\mathbb{B}} _n -{\mathbb{B}} _n = \mathbb{S}_n   ^T\left( P_{ \mathbb{S}_n  {\mathbb{X}}_n }  -   P_{ \mathbb{S}_n \widehat{\mathbb{X}}_n  }   \right) \mathbb{S}_n.
	\end{equation*}
In view of this decomposition, it suffices to control uniformly the norm of the difference between the projectors $P_{ \mathbb{S}_n \widehat{\mathbb{X}}_n } $ and $   P_{ \mathbb{S}_n \mathbb{X}_n  }$. Whenever the inverses exist, we decompose  
	\begin{multline*}
		\mathbb{S}_n \widehat{\mathbb{X}}_n   \left(\widehat{\mathbb{X}}_n ^T\mathbb{D}_n \widehat{\mathbb{X}}_n   \right)^{-1/2} - \mathbb{S}_n {\mathbb{X}}_n   \left({\mathbb{X}}_n ^T\mathbb{D}_n {\mathbb{X}}_n   \right)^{-1/2} = \mathbb{S}_n \widehat{\mathbb{X}}_n   \left[\left(\widehat{\mathbb{X}}_n ^T\mathbb{D}_n \widehat{\mathbb{X}}_n   \right)^{-1/2} - \left({\mathbb{X}}_n ^T\mathbb{D}_n {\mathbb{X}}_n   \right)^{-1/2}\right]\\
		+ \left[ \mathbb{S}_n \widehat{\mathbb{X}}_n - \mathbb{S}_n {\mathbb{X}}_n \right] \left({\mathbb{X}}_n ^T\mathbb{D}_n {\mathbb{X}}_n   \right)^{-1/2}.
	\end{multline*}
Meanwhile, for any $\boldsymbol u \in\mathbb{R}^p$, 
$$
	\boldsymbol u ^T{\mathbb{X}}_n^T {\mathbb{X}}_n \boldsymbol u = n \boldsymbol u ^T Var(\boldsymbol{X} - E[\boldsymbol X \mid \boldsymbol Z]) \boldsymbol u +O_{\mathbb{P}}(n^{1/2}),
$$
which indicates that the spectral norm of $n^{-1/2}{\mathbb{X}}_n$ converges at the rate $O_{\mathbb{P}}(n^{-1/2})$ to the largest eigenvalue of the variance of $\boldsymbol{X} - E[\boldsymbol X \mid \boldsymbol Z]$. From Lemma \ref{inner_smooth} and the triangle inequality, we deduce that the spectral norm of $n^{-1/2}\widehat{\mathbb{X}}_n$ converges also to the largest eigenvalue of the variance of $\boldsymbol{X} - E[\boldsymbol X \mid \boldsymbol Z]$. Next, let us write
$$
	\widehat{\mathbb{X}}_n ^T\mathbb{D}_n  \widehat{\mathbb{X}}_n - {\mathbb{X}}_n ^T\mathbb{D}_n  {\mathbb{X}}_n = \left(\widehat{\mathbb{X}}_n - {\mathbb{X}}_n \right) ^T\mathbb{D}_n  {\mathbb{X}}_n + {\mathbb{X}}_n ^T\mathbb{D}_n  \left(  \widehat{\mathbb{X}}_n - {\mathbb{X}}_n \right) +  \left(\widehat{\mathbb{X}}_n - {\mathbb{X}}_n \right) ^T\mathbb{D}_n   \left(  \widehat{\mathbb{X}}_n - {\mathbb{X}}_n \right) .
$$
Taking spectral norm on both sides and using the bounds of the spectral norms for 
$\mathbb{D}_n$,  $n^{-1/2}{\mathbb{X}}_n$ and $n^{-1/2}\widehat{\mathbb{X}}_n$, as well as the uniform bound derived in Lemma \ref{inner_smooth}, we deduce that
	\begin{equation}\label{hatDn_inverse}
		\sup_{h\in\mathcal{H}^{c}_n}\sup_{\boldsymbol d \in \mathcal D} \left\|\frac{1}{n^2}\widehat{\mathbb{X}}_n ^T\mathbb{D}_n  \widehat{\mathbb{X}}_n - \frac{1}{n^2}{\mathbb{X}}_n ^T \mathbb{D}_n {\mathbb{X}}_n \right\|_{\rm{Sp}}= o_{\mathbb{P}}(1).
	\end{equation}
Let  $\delta>0$ and consider the event $\mathcal{A}_{n} = \mathcal{A}_{1n} \cap \mathcal{A}_{2n}$ where 
$$
	\mathcal{A}_{1n} = \{ n^{-2}{\mathbb{X}}_n ^T\mathbb{D}_n  {\mathbb{X}}_n - (\delta/2) \boldsymbol{I}_{p\times p} \text{ is positive semi-definite} \}
$$
and $\mathcal{A}_{2n}$ is defined in a similar way with $\mathbb{X}_n$ replaced by $\widehat{\mathbb{X}}_n$. 
From Lemma \ref{Dn_inverse}, we know that $E\left[n^{-2}{\mathbb{X}}_n ^T \mathbb{D}_n {\mathbb{X}}_n\right]$ tends to a positive definite matrix. 
From this and equation \eqref{hatDn_inverse}, we could fix $\delta>0$ such that the probability of the event 
$\mathcal{A}_{n}$ tends to 1. On the event $\mathcal{A}_{n}$, using Lemma \ref{matrix_alg} and \ref{Dn_inverse} and equation \eqref{hatDn_inverse}, we deduce that
	\begin{align*}
		\sup_{\boldsymbol d \in \mathcal D}\left\| n \left({\mathbb{X}}_n ^T\mathbb{D}_n {\mathbb{X}}_n   \right)^{-1/2} \right\|_{\rm{Sp}} 
		&\leq 
		\sup_{\boldsymbol d \in \mathcal D}\left\|  \left(n^{-2}{\mathbb{X}}_n ^T\mathbb{D}_n {\mathbb{X}}_n   \right)^{-1/2}  -   E[n^{-2}{\mathbb{X}}_n ^T\mathbb{D}_n {\mathbb{X}}_n]^{-1/2} \right\|_{\rm{Sp}} \\ 
		&\qquad +  \sup_{\boldsymbol d \in \mathcal D}\left\|E[n^{-2}{\mathbb{X}}_n ^T\mathbb{D}_n {\mathbb{X}}_n]^{-1/2} \right\|_{\rm{Sp}}   \\
		&\leq \sqrt{2} \delta^{-3/2} O_{\mathbb{P}}(n^{-1/2}) +\sqrt{2/\delta}
	\end{align*}
and 
$$
	\sup_{h\in \mathcal H^{c}_n}\sup_{\boldsymbol d \in \mathcal D}\left\| n \left(\widehat {\mathbb{X}}_n ^T\mathbb{D}_n \widehat {\mathbb{X}}_n   \right)^{-1/2}- n \left({\mathbb{X}}_n ^T\mathbb{D}_n {\mathbb{X}}_n   \right)^{-1/2} \right\|_{\rm{Sp}} =  o_{\mathbb{P}}(1).
$$
Finally, note that 
$$
	\left\| \mathbb{S}_n \right\|_{\rm{Sp}} = \left\| \mathbb{S}_n^T \right\|_{\rm{Sp}} \leq \left\|  \left( \boldsymbol{I}_{n\times n}  - P_{ \boldsymbol\Omega_n^{1/2} {\boldsymbol{1}}_n   }\right) \right\|_{\rm{Sp}}\left\| \boldsymbol\Omega_n^{1/2}\right\|_{\rm{Sp}} \leq n^{1/2}.
$$
Gathering facts and using repeatedly the property  $\left\|\boldsymbol A_1   \boldsymbol A_2 \right\|_{\rm{Sp}} \leq \left\|\boldsymbol A_1  \right\|_{\rm{Sp}} \left\| \boldsymbol A_2 \right\|_{\rm{Sp}}$, Lemma \ref{matrix_alg} and Lemma \ref{inner_smooth}, we deduce that
	\begin{align*}
		\left\|   P_{ \mathbb{S}_n \widehat{\mathbb{X}}_n }  -   P_{ \mathbb{S}_n {\mathbb{X}}_n  }    \right\|_{\rm{Sp}} 
		&\leq 
		2\left\| n^{-1/2}\mathbb{S}_n n^{-1/2} \widehat{\mathbb{X}}_n   \left[n \left(\widehat{\mathbb{X}}_n ^T\mathbb{D}_n \widehat{\mathbb{X}}_n   \right)^{-1/2} - n\left({\mathbb{X}}_n ^T\mathbb{D}_n {\mathbb{X}}_n   \right)^{-1/2}\right]\right\|_{\rm{Sp}}  \\
		&\quad + 2\left\| n^{-1/2}\mathbb{S}_n n^{-1/2}\left(  \widehat{\mathbb{X}}_n - {\mathbb{X}}_n \right) n \left({\mathbb{X}}_n ^T\mathbb{D}_n {\mathbb{X}}_n   \right)^{-1/2}\right\|_{\rm{Sp}} \\ 
		&= o_{\mathbb{P}}(1).
	\end{align*}
Finally, 
	\begin{equation*}
		\left\|\widehat {\mathbb{B}} _n -{\mathbb{B}} _n \right\|_{\rm{Sp}}  \leq  \left\| \mathbb{S}_n   ^T\right\|_{\rm{Sp}} \left\| P_{ \mathbb{S}_n \widehat{\mathbb{X}}_n }  -   P_{ \mathbb{S}_n \widehat{\mathbb{X}}_n  }   \right\|_{\rm{Sp}}  \left\| \mathbb{S}_n\right\|_{\rm{Sp}}= o_{\mathbb{P}}(n).
	\end{equation*}
Now, the proof is complete. \\
\end{proofof}

\begin{lem}\label{Xn_hat}
Assume the conditions of Theorem \ref{consist} hold true. Then,
	$$
		\sup_{\boldsymbol d \in\mathcal D }  \left\| \boldsymbol \Omega_n^{1/2}  (\boldsymbol d)n^{-1} \mathbb{X}_n  \right\|_{\rm{Sp}}  = O_{\mathbb{P}}(1),
		\qquad and \qquad 
		\sup_{h\in\mathcal{H}^{c}_n} \sup_{\boldsymbol d \in\mathcal D }  \left\| \boldsymbol \Omega_n^{1/2}  (\boldsymbol d)n^{-1} \left[\widehat{\mathbb{X}}_n - \mathbb{X}_n \right]  \right\|_{\rm{Sp}}  = o_{\mathbb{P}}(1).
	$$	
As a consequence
	$$
		\sup_{h\in\mathcal{H}^{c}_n} \sup_{\boldsymbol d \in\mathcal D }  \left\| \boldsymbol \Omega_n^{1/2}  (\boldsymbol d)n^{-1} \widehat{\mathbb{X}}_n  \right\|_{\rm{Sp}}  = O_{\mathbb{P}}(1).
	$$		  	
\end{lem}

\newpage
\begin{proofof}{Lemma \ref{Xn_hat}}~~\\

We have that 
	\begin{align*}
		\left\| \boldsymbol \Omega_n^{1/2}  (\boldsymbol d)n^{-1} \mathbb{X}_n  \right\|_{\rm{Sp}} \leq \left\| \boldsymbol \Omega_n^{1/2}  (\boldsymbol d)n^{-1/2} \right\|_{\rm{Sp}} \left\|n^{-1/2} \mathbb{X}_n  \right\|
	\end{align*}

The first rate follows now from Lemma \ref{spectral_B_hat} and the fact that by our assumptions the expectation of $\| n^{-1/2} {\mathbb{X}} _n \|^2$ is finite. The second rate follows from Lemma \ref{spectral_B_hat} and \ref{inner_smooth}. The third rate is a direct consequence of the first two rates. \\
\end{proofof}

\begin{lem} \label{T_prop_Daniel}
Under Assumption \ref{ass_dgp}.1, there exists a  constant $C$, depending on $\Lambda$ such that 
	\begin{align*} 
		0 \leq \frac{\partial }{\partial \lambda} T(Y,\lambda)\leq C  \max\left( (Y\vee 1)^{\lambda_{\min}}  \log^2\left(Y\vee e \right),(Y\vee 1)^{\lambda_{\max}}  \log^2\left(Y\vee e \right) \right).
	\end{align*}

\end{lem}

\begin{proofof}{Lemma \ref{T_prop_Daniel}}~~\\

The derivative of $T(Y,\lambda)$ with respect to $\lambda$ is given by
	\begin{align*}
		\frac{\partial}{\partial \lambda} T(Y,\lambda)=
		\begin{cases}
		\lambda^{-2} \left(Y^{\lambda}(\lambda \log(Y) -1) + 1 \right)&, \;\;\lambda \neq 0\\
		2^{-1} \log^2(Y)&, \;\;\lambda = 0.
		\end{cases}
	\end{align*}
The derivative is positive and continuous at $\lambda = 0$. Furthermore, the sign of the derivative for $\lambda \neq 0$ is determined by $Y^{\lambda}(\lambda \log(Y) -1) + 1$. We get that 
	\begin{align*}
		\frac{\partial }{\partial \lambda} \{Y^{\lambda}(\lambda \log(Y) -1) + 1\}= \lambda Y^{\lambda} \log^2\left(Y\right) ,
	\end{align*}	
which is positive for $\lambda > 0$ and negative for $\lambda < 0$. This implies that the first derivative is always positive. Next, the upper bound is obvious. \\	
\end{proofof}	

\begin{lem} \label{T_sec_der}
Under Assumption \ref{ass_dgp}.1, there exists a  constant $C$, depending on $\Lambda$ such that 
	\begin{align*} 
		\left|\frac{\partial^2 }{\partial \lambda^2} T(Y,\lambda)\right|\leq C  \max\left( (Y\vee 1)^{\lambda_{\min}}  |\log\left(Y\vee e \right)|^3,(Y\vee 1)^{\lambda_{\max}}  |\log\left(Y\vee e \right)|^3 \right).
	\end{align*}

\end{lem}

\begin{proofof}{Lemma \ref{T_sec_der}}~~\\

The second derivative of $T(Y,\lambda)$ with respect to $\lambda$ is given by
	\begin{align*}
		\frac{\partial^2 T(Y,\lambda)}{\partial \lambda^2}=
		\begin{cases}
		\lambda^{-3} \left(Y^{\lambda} \lambda^2\log(Y)^2 - 2\left(Y^{\lambda}(\lambda \log(Y) -1) + 1 \right)\right) &, \lambda \neq 0\\
		3^{-1} \log(Y)^3              																		  &, \lambda = 0.
		\end{cases}
	\end{align*}		
Once again we consider the derivative of the nominator for $\lambda \neq 0$. The derivative is given by
	\begin{align*}
		\frac{\partial}{\partial \lambda}\left\{Y^{\lambda} \lambda^2\log(Y)^2 - 2\left(Y^{\lambda}(\lambda \log(Y) -1) + 1 \right)\right\}
		=
		\log(Y)^3 \lambda^2 Y^{\lambda}.
	\end{align*}
This derivative is positive if $Y>1$ and negative if $Y<1$. The second derivative of $T(Y,\lambda)$ with respect to $\lambda$ is continuous at $\lambda = 0$ and positive if $Y>1$ and negative if $Y<1$. Therefore, it follows that 
	\begin{align*}
		\frac{\partial^2 T(Y,\lambda)}{\partial \lambda^2}
		=
		\begin{cases}
		> 0&, Y>1\\
		< 0&, Y<1
		\end{cases}
	\end{align*}
for all $\lambda$. Next, the upper bound is obvious. \\	
\end{proofof}

\begin{lem} \label{T_third_der}
Under Assumption \ref{ass_dgp}.1, there exists a  constant $C$, depending on $\Lambda$ such that 
	\begin{align*} 
		0\leq\frac{\partial^3 }{\partial \lambda^3} T(Y,\lambda)\leq C  \max\left( (Y\vee 1)^{\lambda_{\min}}  \log\left(Y\vee e \right)^4,(Y\vee 1)^{\lambda_{\max}}  \log\left(Y\vee e \right)^4 \right).
	\end{align*}

\end{lem}

\begin{proofof}{Lemma \ref{T_third_der}}~~\\

The third derivative of $T(Y,\lambda)$ with respect to $\lambda$ is given by
	\begin{align*}
		\frac{\partial^3 T(Y,\lambda)}{\partial \lambda^3}=
		\begin{cases}
		\lambda^{-4} \left(Y^{\lambda} \lambda^2\log(Y)^2  (\lambda \log(Y) -3) + 6\left(Y^{\lambda}(\lambda \log(Y) -1) + 1 \right)\right) &, \lambda \neq 0\\
		4^{-1} \log(Y)^4              																							 &, \lambda = 0.
		\end{cases}
	\end{align*}				
Once again we consider the derivative of the nominator for $\lambda \neq 0$. The derivative is given by
	\begin{align*}
		\frac{\partial}{\partial \lambda}\left\{Y^{\lambda} \lambda^2\log(Y)^2  (\lambda \log(Y) -3) + 6\left(Y^{\lambda}(\lambda \log(Y) -1) + 1 \right)\right\}
		=
		\lambda^3 \log(Y)^4  Y^{\lambda},
	\end{align*}
which is positive for $\lambda > 0$ and negative for $\lambda < 0$. This implies that the third derivative is always positive. Next, the upper bound is obvious. \\
\end{proofof}

\begin{lem}\label{yn_2}
Under the conditions of Theorem \ref{consist},
$$
	\sup_{\lambda \in\Lambda}  \left\| n^{-1}  {\mathbb{Y}}_n(\lambda) \right\| = O_{\mathbb{P}}(n^{-1/2}).
$$

\end{lem}

\begin{proofof}{Lemma \ref{yn_2}}~~\\

First, note that the functions $\{y\mapsto  \lambda^{-1}(y^\lambda -1): y\geq c>0,  \lambda \in \Lambda \}$, with $c$ a fixed lower bound of the support of $Y$, are Lipschitz in the index parameter $\lambda$. See Lemma \ref{T_prop_Daniel}. Deduce that this family of functions of $Y$ is Euclidean for a to the power of four integrable envelope. See Lemma 2.13 in \citet{pakes1989simulation}.  Since the  Euclidean property is preserved by multiplication with a fixed function, the family $\{(y,\boldsymbol z) \mapsto  \lambda^{-1}(y^\lambda -1)  f_{z}(\boldsymbol z) : y\geq c>0, \boldsymbol z\in\mathbb{R}^q, \lambda \in \Lambda \}$ is also Euclidean for a to the power of four integrable envelope. See Lemma 2.14 in \citet{pakes1989simulation}. The Euclidean property is also preserved if the functions of $Y$ and $\boldsymbol Z$ are centered by their conditional expectation given $\boldsymbol Z$. See Lemma 5 in  \citet{sherman1994maximal}. Next, it is also preserved by taking the square of the functions in the family. The envelope is now squared integrable. See Lemma 2.14 in \citet{pakes1989simulation}. Deduce from Corollary 7  in \citet{sherman1994maximal} that
$$
	\sup_{\lambda \in\Lambda}  \left| \frac{1}{n}\sum_{i=1}^n (T(Y_i,\lambda) -  {E}[T(Y_i,\lambda)\mid \boldsymbol Z_i] )^2  { f}^2 _z(\boldsymbol Z_i) \right| =   O_{\mathbb{P}}(1).
$$
Then the required rate follows. \\
\end{proofof}

\begin{lem}\label{yn_hat2}
Under the conditions of Theorem \ref{consist},
$$
	\sup_{h\in\mathcal{H}^{c}_n}   \sup_{\lambda \in\Lambda}  \left\| n^{-1} \left[ \widehat{\mathbb{Y}}_n(\lambda) -  {\mathbb{Y}}_n(\lambda) \right]  \right\|   = o_{\mathbb{P}}(n^{-1/2}).
$$

\end{lem}

\begin{proofof}{Lemma \ref{yn_hat2}}~~\\

It suffices to decompose
$$
	\widehat{\mathbb{Y}}_n(\lambda) - {\mathbb{Y}}_n(\lambda) = {\boldsymbol{R}}_{1n} +  {\boldsymbol{R}}_{2n} 
$$
with
$$
	{\boldsymbol{R}}_{1n} = \left( T(Y_1,\lambda) \left( \widehat{f} _z(\boldsymbol Z_1)- { f} _z(\boldsymbol Z_1)\right)  ,\ldots, T(Y_n,\lambda) \left( \widehat{f} _z(\boldsymbol Z_n)- { f} _z(\boldsymbol Z_n)\right)  \right)^T
$$
and 
	\begin{multline*}
		{\boldsymbol{R}}_{2n} = \left( \left( {E}[T(Y_1,\lambda)\mid \boldsymbol Z_1] { f} _z(\boldsymbol Z_1) - \widehat{E}[T(Y_1,\lambda)\mid \boldsymbol Z_1]  \widehat  { f} _z(\boldsymbol Z_1) \right),  \right. \\ \left.   \ldots, \left( {E}[T(Y_n,\lambda)\mid \boldsymbol Z_n]  { f} _z(\boldsymbol Z_n)  - \widehat{E}[T(Y_n,\lambda)\mid \boldsymbol Z_n]  \widehat { f} _z(\boldsymbol Z_n)\right)    \right)^T.
	\end{multline*}
We can now use the same arguments as in Lemma \ref{inner_smooth} to show that 
$$
	\sup_{h\in\mathcal{H}^{c}_n}   \sup_{\lambda \in\Lambda}  \left\| n^{-1} 	{\boldsymbol{R}}_{1n} \right\|   = o_{\mathbb{P}}(n^{-1/2}) \quad and \quad 
	\sup_{h\in\mathcal{H}^{c}_n}   \sup_{\lambda \in\Lambda}  \left\| n^{-1} 	{\boldsymbol{R}}_{2n} \right\|   = o_{\mathbb{P}}(n^{-1/2}).
$$
The Euclidean properties needed follow from a similar discussion as in Lemma  \ref{yn_2}.\\
\end{proofof}

\begin{lem}\label{yn_hatt_5}
Assume the conditions of Proposition \ref{AN_prop} hold true. 
Then 
$$
	\sup_{\boldsymbol d \in\mathcal D }  \left\| \boldsymbol \Omega_n^{1/2}  (\boldsymbol d) \; n^{-1} \left[ {\mathbb{Y}}_n(\lambda_0) - {\mathbb{X}} _n \boldsymbol \beta_0 \right]   \right\|   = O_{\mathbb{P}}(n^{-1/2}).
$$
\end{lem}

\begin{proofof}{Lemma \ref{yn_hatt_5}}~~\\

By definition ${\mathbb{Y}}_n(\lambda_0) - {\mathbb{X}} _n \boldsymbol \beta_0  = (\boldsymbol{\varepsilon f_z})_n  = (\varepsilon_1 {f}_z(\boldsymbol Z_1),\ldots,\varepsilon_n {f}_z(\boldsymbol Z_n))^T$. 
Next, for any $\boldsymbol d$,  using the Fourier Transform (see the last part of the  proof of Lemma \ref{lem_ident}), we can write
$$
	0\leq  n^{-2}  (\boldsymbol{\varepsilon f_z})_n  ^T \boldsymbol \Omega_n(\boldsymbol d)  (\boldsymbol{\varepsilon f_z})_n \leq  
	\frac{d_U^{(p+q)/2}}{d_L^{(p+q)/2}} n^{-2}  (\boldsymbol{\varepsilon f_z})_n  ^T \boldsymbol \Omega_n({\rm diag}(d_U,\ldots,d_U) )  (\boldsymbol{\varepsilon f_z})_n.
$$ 
Simply calculating the  expectation, the last quadratic form in the last display has the rate $ O_{\mathbb{P} }\left(n^{-1}\right)$. The uniform rate follows.\\  
\end{proofof}

\begin{lem}\label{yn_der} 	

Assume the conditions of Theorem \ref{consist} hold true and let $\Lambda_{0n}$ be an arbitrary $o_{\mathbb{P} }\left(1\right)$ neighborhood of $\lambda_0$. Then, for $s \in\{1,2,3\}$, 
	$$
		\sup_{\lambda \in \Lambda_{0n}}\sup_{\boldsymbol d \in\mathcal D }   \left\|  \boldsymbol \Omega_n^{1/2}  (\boldsymbol d)n^{-1} \frac{\partial ^s}{\partial \lambda^s}{\mathbb{Y}}_n(\lambda)  \right\|_{\rm{Sp}}  = O_{\mathbb{P}}(1)
	$$ 
and	 
	$$
		\sup_{ \lambda \in \Lambda_{0n}}\sup_{h\in\mathcal{H}^{c}_n} \sup_{\boldsymbol d \in\mathcal D }  \left\| \boldsymbol \Omega_n^{1/2}  (\boldsymbol d)n^{-1} \left[ \frac{\partial ^s}{\partial \lambda^s}\widehat{\mathbb{Y}}_n(\lambda) -  \frac{\partial ^s}{\partial \lambda ^s} {\mathbb{Y}}_n(\lambda) \right]  \right\|_{\rm{Sp}}  = o_{\mathbb{P}}(1).
	$$
As a consequence
	$$
		\sup_{\lambda \in \Lambda_{0n}}\sup_{h\in\mathcal{H}^{c}_n} \sup_{\boldsymbol d \in\mathcal D }   \left\|  \boldsymbol \Omega_n^{1/2} (\boldsymbol d) n^{-1} \frac{\partial^s }{\partial \lambda^s}\widehat{\mathbb{Y}}_n(\lambda)  \right\|_{\rm{Sp}}  = O_{\mathbb{P}}(1).
	$$

\end{lem}

\begin{proofof}{Lemma \ref{yn_der}}~~\\

We have that 
	\begin{align*}
		\left\|  \boldsymbol \Omega_n^{1/2}  (\boldsymbol d)n^{-1} \frac{\partial ^s}{\partial \lambda^s}{\mathbb{Y}}_n(\lambda)  \right\|_{\rm{Sp}}
		\leq
		\left\|  \boldsymbol \Omega_n^{1/2}  (\boldsymbol d)n^{-1/2} \right\|_{\rm{Sp}}	\left\| n^{-1/2} \frac{\partial ^s}{\partial \lambda^s}{\mathbb{Y}}_n(\lambda)  \right\|	
	\end{align*}

The first rate follows now from Lemma \ref{spectral_B_hat} and the fact that, by our assumptions, the expectation of \\
$\sup\limits_{\lambda \in \Lambda_{0n}}\| n^{-1/2}(\partial^s/ \partial \lambda^s) {\mathbb{Y}}_n(\lambda)  \|^2$ is finite.

The second rate follows again from Lemma \ref{spectral_B_hat} and the same arguments as in Lemma \ref{inner_smooth} and \ref{yn_hat2}. The Euclidean properties needed follow from a similar discussion as in Lemma  \ref{yn_2}. The third rate is a direct consequence of the first two rates.\\
\end{proofof}

\begin{lem}\label{remainder} Assume the conditions of Theorem \ref{consist} hold true and let $\Lambda_{0n}$ be an arbitrary $o_{\mathbb{P} }\left(1\right)$ neighborhood of $\lambda_0$. Then, 
	\begin{multline*}
		\sup_{\lambda \in \Lambda_{0n}}\sup_{h\in\mathcal{H}^{c}_n} \sup_{s\in S_n} \sup_{\boldsymbol d \in \mathcal{D}} \Big|
		\frac{\partial }{\partial \lambda} \left\{ n^{-1}s^{-\lambda}  \widehat{\mathbb{Y}}_n(\lambda) \right\}^T  \widehat {\mathbb{B}} _n  \frac{\partial }{\partial \lambda} \left\{ n^{-1} s^{-\lambda}  \widehat{\mathbb{Y}}_n(\lambda) \right\}  \\
		-
		\frac{\partial }{\partial \lambda} \left\{ n^{-1}s^{-\lambda_0}  \widehat{\mathbb{Y}}_n(\lambda_0) \right\}^T  \widehat {\mathbb{B}} _n  \frac{\partial }{\partial \lambda} \left\{ n^{-1} s^{-\lambda_0}  \widehat{\mathbb{Y}}_n(\lambda_0) \right\}\Big| = o_{\mathbb{P}}(1) 
	\end{multline*}
and 
	\begin{multline*}
		\sup_{\lambda \in \Lambda_{0n}}\sup_{h\in\mathcal{H}^{c}_n} \sup_{s\in S_n} \sup_{\boldsymbol d \in \mathcal{D}} \Big|
		 n^{-1}s^{-\lambda}  \widehat{\mathbb{Y}}_n(\lambda)^T \; \widehat {\mathbb{B}} _n  \frac{\partial ^2}{\partial \lambda^2} \left\{ n^{-1}s^{-\lambda}  \widehat{\mathbb{Y}}_n(\lambda) \right\}\\
		-	
		\left\{ n^{-1}s^{-\lambda_0}  \widehat{\mathbb{Y}}_n(\lambda_0)  \right\}^T  \widehat {\mathbb{B}} _n  \frac{\partial ^2}{\partial \lambda^2} \left\{ n^{-1}s^{-\lambda_0}  \widehat{\mathbb{Y}}_n(\lambda_0) \right\}\Big| = o_{\mathbb{P}}(1).
	\end{multline*} 
\end{lem}

\begin{proofof}{Lemma \ref{remainder}} ~~\\

Note that $ \frac{\partial }{\partial \lambda} \{n^{-1}s^{- \lambda} \widehat{\mathbb{Y}}_n(\lambda) \}  = s^{- \lambda}\frac{\partial }{\partial \lambda} \{n^{-1} \widehat{\mathbb{Y}}_n(\lambda) \} - \log(s)s^{- \lambda} n^{-1} \widehat{\mathbb{Y}}_n(\lambda)$.

We have that 
	\begin{align*}
		&s^{-2\lambda} n^{-1}\frac{\partial }{\partial \lambda} \widehat{\mathbb{Y}}_n(\lambda)^T \; \widehat {\mathbb{B}} _n \;n^{-1} \frac{\partial }{\partial \lambda}    \widehat{\mathbb{Y}}_n(\lambda)
		-
		s^{-2\lambda_0}n^{-1}\frac{\partial }{\partial \lambda}  \widehat{\mathbb{Y}}_n(\lambda_0)^T  \;\widehat {\mathbb{B}} _n \; n^{-1}\frac{\partial }{\partial \lambda}  \widehat{\mathbb{Y}}_n(\lambda_0) \\
		&=
		s^{-2\lambda} n^{-1}\left[\frac{\partial }{\partial \lambda}\widehat{\mathbb{Y}}_n(\lambda)  -\frac{\partial }{\partial \lambda}  \widehat{\mathbb{Y}}_n(\lambda_0)\right]^T  \widehat {\mathbb{B}} _n \; n^{-1}\frac{\partial }{\partial \lambda}    \widehat{\mathbb{Y}}_n(\lambda) \\
		&\quad-
		s^{-2\lambda} n^{-1}\frac{\partial }{\partial \lambda}    \widehat{\mathbb{Y}}_n(\lambda_0)^T \; \widehat {\mathbb{B}} _n \; n^{-1} \left[ \frac{\partial }{\partial \lambda}  \widehat{\mathbb{Y}}_n(\lambda_0)  -\frac{\partial }{\partial \lambda} \widehat{\mathbb{Y}}_n(\lambda) \right]\\		
		&\quad+
		\left(s^{-2\lambda} - s^{-2\lambda_0}\right)n^{-1} \frac{\partial }{\partial \lambda}  \widehat{\mathbb{Y}}_n(\lambda_0)^T \; \widehat {\mathbb{B}} _n  \;n^{-1}\frac{\partial }{\partial \lambda}  \widehat{\mathbb{Y}}_n(\lambda_0)\\
		&=
		s^{-2\lambda} n^{-1}\frac{\partial^2 }{\partial \lambda^2}  \widehat{\mathbb{Y}}_n(\widetilde\lambda)^T  \; \widehat {\mathbb{B}} _n  \; n^{-1}\frac{\partial }{\partial \lambda}  \widehat{\mathbb{Y}}_n(\lambda)(\lambda - \lambda_0) \\
		&\quad+	
		s^{-2\lambda} n^{-1}\frac{\partial }{\partial \lambda}  \widehat{\mathbb{Y}}_n(\lambda_0)^T \; \widehat {\mathbb{B}} _n  \;n^{-1}
		\frac{\partial^2 }{\partial \lambda^2}\widehat{\mathbb{Y}}_n(\widetilde\lambda) (\lambda - \lambda_0)\\		
		&\quad+
		\left(s^{-2\lambda} - s^{-2\lambda_0}\right)n^{-1}\frac{\partial }{\partial \lambda}  \widehat{\mathbb{Y}}_n(\lambda_0)^T \; \widehat {\mathbb{B}} _n \; n^{-1} \frac{\partial }{\partial \lambda}   \widehat{\mathbb{Y}}_n(\lambda_0) ,		
	\end{align*}
where $\widetilde \lambda = c  \lambda + (1-c)\lambda_0$ for some $c\in(0,1)$. Recall that
$$
	\widehat {\mathbb{B}}_n =  \mathbb{S}_n^T   \left( \boldsymbol{I}_{n\times n}  -   P_{ \mathbb{S}_n\widehat{\mathbb{X}}_n  }   \right) \mathbb{S}_n,
$$
where $P_{ \mathbb{S}_n \widehat{\mathbb{X}}_n }$ is the orthogonal projector on the subspace generated by $ \mathbb{S}_n\widehat{\mathbb{X}}_n$ with 
$$
	\mathbb{S}_n = \left( \boldsymbol{I}_{n\times n}  - P_{ \boldsymbol\Omega_n ^{1/2}{\boldsymbol{1}}_n   }\right) \boldsymbol\Omega_n^{1/2}
$$
and $P_{ \boldsymbol\Omega_n^{1/2} {\boldsymbol{1}}_n   }$ is the projector on the subspace generated by the vector $\boldsymbol\Omega_n^{1/2} {\boldsymbol{1}}_n$. Deduce that
\newpage
	\begin{align*}
		\Big|s^{-2\lambda} n^{-1} \frac{\partial^2 }{\partial \lambda^2}  \widehat{\mathbb{Y}}_n(\widetilde\lambda)^T \; &\widehat {\mathbb{B}} _n  \; n^{-1}\frac{\partial }{\partial \lambda}    \widehat{\mathbb{Y}}_n(\lambda) (\lambda - \lambda_0)\Big|\\
		\leq
		s^{-2\lambda} &|(\lambda - \lambda_0)| \left\| \boldsymbol\Omega_n ^{1/2}n^{-1}\frac{\partial^2 }{\partial \lambda^2}  \widehat{\mathbb{Y}}_n(\widetilde\lambda) \right\|_{\rm Sp} \\
		&\times \left\| \left( \boldsymbol{I}_{n\times n}  - P_{ \boldsymbol\Omega_n ^{1/2}{\boldsymbol{1}}_n   }\right)  \left( \boldsymbol{I}_{n\times n}  -   P_{ \mathbb{S}_n  {\mathbb{X}}_n  }   \right) \left( \boldsymbol{I}_{n\times n}  - P_{ \boldsymbol\Omega_n ^{1/2}{\boldsymbol{1}}_n   }\right)  \right\|_{\rm Sp}\\
		&\hskip 7cm \times \left\| \boldsymbol\Omega_n ^{1/2}n^{-1}\frac{\partial}{\partial \lambda}  \widehat{\mathbb{Y}}_n(\lambda) \right\|_{\rm Sp}.
	\end{align*}
By the same reasoning we get that
	\begin{align*}
		\Big|s^{-2\lambda} n^{-1}\frac{\partial^2 }{\partial \lambda^2}  \widehat{\mathbb{Y}}_n(\widetilde\lambda)^T  \;&\widehat {\mathbb{B}} _n \; n^{-1}\frac{\partial }{\partial \lambda}  \widehat{\mathbb{Y}}_n(\lambda_0)(\lambda - \lambda_0)\Big|\\
		\leq
		s^{-2\lambda} &|(\lambda - \lambda_0)| \left\| \boldsymbol\Omega_n ^{1/2}n^{-1}\frac{\partial^2 }{\partial \lambda^2}  \widehat{\mathbb{Y}}_n(\widetilde\lambda) \right\|_{\rm Sp} \\
		&\times \left\| \left( \boldsymbol{I}_{n\times n}  - P_{ \boldsymbol\Omega_n ^{1/2}{\boldsymbol{1}}_n   }\right)  \left( \boldsymbol{I}_{n\times n}  -   P_{ \mathbb{S}_n  {\mathbb{X}}_n  }   \right) \left( \boldsymbol{I}_{n\times n}  - P_{ \boldsymbol\Omega_n ^{1/2}{\boldsymbol{1}}_n   }\right)  \right\|_{\rm Sp}\\
		&\hskip 7cm \times \left\| \boldsymbol\Omega_n ^{1/2}n^{-1}\frac{\partial}{\partial \lambda}  \widehat{\mathbb{Y}}_n(\lambda_0) \right\|_{\rm Sp}.
	\end{align*}
and
	\begin{align*}
		\Big|\left(s^{-2\lambda} - s^{-2\lambda_0}\right)n^{-1}\frac{\partial}{\partial \lambda} \widehat{\mathbb{Y}}_n(\lambda_0)^T \; &\widehat {\mathbb{B}} _n \;n^{-1} \frac{\partial }{\partial \lambda} \widehat{\mathbb{Y}}_n(\lambda_0) \Big|\\
		\leq
		&|s^{-2\lambda} - s^{-2\lambda_0}| \left\| \boldsymbol\Omega_n ^{1/2}n^{-1}\frac{\partial }{\partial \lambda}  \widehat{\mathbb{Y}}_n(\lambda_0) \right\|_{\rm Sp} \\
		&\times \left\| \left( \boldsymbol{I}_{n\times n}  - P_{ \boldsymbol\Omega_n ^{1/2}{\boldsymbol{1}}_n   }\right)  \left( \boldsymbol{I}_{n\times n}  -   P_{ \mathbb{S}_n  {\mathbb{X}}_n  }   \right) \left( \boldsymbol{I}_{n\times n}  - P_{ \boldsymbol\Omega_n ^{1/2}{\boldsymbol{1}}_n   }\right)  \right\|_{\rm Sp}\\
		&\hskip 7cm \times \left\| \boldsymbol\Omega_n ^{1/2}n^{-1}\frac{\partial}{\partial \lambda}  \widehat{\mathbb{Y}}_n(\lambda_0) \right\|_{\rm Sp}.
	\end{align*}
It follows now from the fact that the spectral norm of a product of projectors  is at most equal to 1, 
$\lambda \in \Lambda_{0n}$, $\sup_{\lambda\in \Lambda_{0n}}\sup_{s\in S_n}\big| s^{- 2\lambda} - s^{- 2\lambda_0} \big| = o_{\mathbb{P} }(1)$ as well as $\sup_{\lambda\in \Lambda_{0n}}\sup_{s\in S_n} s^{- 2\lambda}  = O_{\mathbb{P} }(1)$  and from Lemma \ref{yn_der} that 
	\begin{align*}
		&\sup_{\lambda \in \Lambda_{0n}}\sup_{h\in\mathcal{H}^{c}_n} \sup_{s\in S_n} \sup_{\boldsymbol d \in \mathcal{D}} \Big|
		s^{-2\lambda} n^{-1}\frac{\partial }{\partial \lambda} \widehat{\mathbb{Y}}_n(\lambda)^T \; \widehat {\mathbb{B}} _n \;n^{-1} \frac{\partial }{\partial \lambda}    \widehat{\mathbb{Y}}_n(\lambda)\\
		&\hskip 7cm-
		s^{-2\lambda_0}n^{-1}\frac{\partial }{\partial \lambda}  \widehat{\mathbb{Y}}_n(\lambda_0)^T  \;\widehat {\mathbb{B}} _n \; n^{-1}\frac{\partial }{\partial \lambda}  \widehat{\mathbb{Y}}_n(\lambda_0)\Big|  = o_{\mathbb{P}}(1). 
	\end{align*}
By similar reasoning, we get that 
	\begin{align*}
		&\sup_{\lambda \in \Lambda_{0n}}\sup_{h\in\mathcal{H}^{c}_n} \sup_{s\in S_n} \sup_{\boldsymbol d \in \mathcal{D}} \Big|
		\log(s)s^{- \lambda} n^{-1} \widehat{\mathbb{Y}}_n(\lambda)^T \; \widehat {\mathbb{B}} _n \;n^{-1} \frac{\partial }{\partial \lambda}    \widehat{\mathbb{Y}}_n(\lambda)\\
		&\hskip 7cm-
		\log(s)s^{- \lambda_0} n^{-1} \widehat{\mathbb{Y}}_n(\lambda_0)^T  \;\widehat {\mathbb{B}} _n \; n^{-1}\frac{\partial }{\partial \lambda}  \widehat{\mathbb{Y}}_n(\lambda_0)\Big|  = o_{\mathbb{P}}(1). 
	\end{align*}
and \newpage
	\begin{align*}
		&\sup_{\lambda \in \Lambda_{0n}}\sup_{h\in\mathcal{H}^{c}_n} \sup_{s\in S_n} \sup_{\boldsymbol d \in \mathcal{D}} \Big|
		\log(s)s^{- \lambda} n^{-1} \widehat{\mathbb{Y}}_n(\lambda)^T \; \widehat {\mathbb{B}} _n \;\log(s)s^{- \lambda} n^{-1} \widehat{\mathbb{Y}}_n(\lambda)\\
		&\hskip 6cm-
		\log(s)s^{- \lambda_0} n^{-1} \widehat{\mathbb{Y}}_n(\lambda_0)^T  \;\widehat {\mathbb{B}} _n \; \log(s)s^{- \lambda_0} n^{-1} \widehat{\mathbb{Y}}_n(\lambda_0)\Big|  = o_{\mathbb{P}}(1). 
	\end{align*}	
Therefore, the first statement follows. The second statement follows by the same reasoning. The details are omitted.\\
\end{proofof}

\begin{lem}\label{remainder_2} Assume the conditions of Theorem \ref{consist} hold true and let $\Lambda_{0n}$ be an arbitrary $o_{\mathbb{P} }\left(1\right)$ neighborhood of $\lambda_0$. Then, 
	\begin{align*}
		\sup_{\lambda \in \Lambda_{0n}}\sup_{h\in\mathcal{H}^{c}_n} \sup_{s\in S_n} \sup_{\boldsymbol d \in \mathcal{D}} \Big|
		n^{-1}\widehat{\mathbb{X}}_n ^T {\mathbb{D}} _n  n^{-1}
		\left(\frac{\partial}{\partial \lambda}\widehat{\mathbb{Y}}_n(\lambda) - \frac{\partial}{\partial \lambda}\widehat{\mathbb{Y}}_n(\lambda_0)\right)
		\Big| = o_{\mathbb{P}}(1).
	\end{align*}
\end{lem}

\begin{proofof}{Lemma \ref{remainder_2}} ~~\\

We have that 
	\begin{align*}
		n^{-1}\widehat{\mathbb{X}}_n ^T {\mathbb{D}} _n  n^{-1}\left(\frac{\partial}{\partial \lambda}\widehat{\mathbb{Y}}_n(\lambda) - \frac{\partial}{\partial \lambda}\widehat{\mathbb{Y}}_n(\lambda_0)\right)
		=
		n^{-1}\widehat{\mathbb{X}}_n ^T {\mathbb{D}} _n  n^{-1}\frac{\partial^2}{\partial \lambda^2}\widehat{\mathbb{Y}}_n(\widetilde\lambda) (\lambda - \lambda_0),		
	\end{align*}
where $\widetilde \lambda = c  \lambda + (1-c)\lambda_0$ for some $c\in(0,1)$. We get that 
	\begin{multline*}
		\Big|n^{-1}\widehat{\mathbb{X}}_n ^T {\mathbb{D}} _n  n^{-1}\frac{\partial^2}{\partial \lambda^2}\widehat{\mathbb{Y}}_n(\widetilde\lambda)\Big|\\
		\leq
		\left\| \boldsymbol\Omega_n ^{1/2}n^{-1}\widehat{\mathbb{X}}_n \right\|_{\rm Sp}
		\times 
		\left\| \left( \boldsymbol{I}_{n\times n}  - P_{ \boldsymbol\Omega_n ^{1/2}{\boldsymbol{1}}_n   }\right) \left( \boldsymbol{I}_{n\times n}  - P_{ \boldsymbol\Omega_n ^{1/2}{\boldsymbol{1}}_n   }\right)  \right\|_{\rm Sp}
		\times
		\left\| \boldsymbol\Omega_n ^{1/2}n^{-1}\frac{\partial^2 }{\partial \lambda^2}  \widehat{\mathbb{Y}}_n(\widetilde\lambda) \right\|_{\rm Sp}.
	\end{multline*}
The statement follows now from the fact that the spectral norm of a product of projectors  is at most equal to 1, 
$\lambda \in \Lambda_{0n}$ and  Lemma \ref{Xn_hat} and \ref{yn_der}.\\	
\end{proofof}

\begin{lem}\label{yn_hatt_6}
Assume the conditions of Proposition \ref{AN_prop} hold true. Then,
	\begin{align*}
		\sup_{h\in\mathcal{H}^{sc}_n} \sup_{\boldsymbol d \in\mathcal D }  \left\| \boldsymbol\Omega_n ^{1/2}(\boldsymbol d) n^{-1}  \left(\left[ {\mathbb{Y}}_n(\lambda_0) - {\mathbb{X}} _n \boldsymbol \beta_0 \right] 
		- \left(\boldsymbol{\widehat{\varepsilon}_{|z}\widehat{f}_z}\right)_n
		-  \left[ \widehat{\mathbb{Y}}_n(\lambda_0) - \widehat{\mathbb{X}} _n \boldsymbol \beta_0 \right]\right) \right\|  = o_{\mathbb{P}}(n^{-1/2}).
	\end{align*}
\end{lem}

\begin{proofof}{Lemma \ref{yn_hatt_6}} ~~\\

By the arguments used for Lemma \ref{yn_hatt_5}, it suffices to consider  $\boldsymbol d = {\rm diag}(d_U,\ldots,d_U)$. Moreover, for simpler notation,  we omit the argument $\boldsymbol d$ in $\boldsymbol \Omega_n(\boldsymbol d) $. 
We get that, for $1\leq i \leq n$,
	\begin{align*}
		&\left[  \widehat{\mathbb{Y}}_{n,i}(\lambda_0) - \widehat {\mathbb{X}} _{n,i} \boldsymbol \beta_0\right] - [ {\mathbb{Y}}_{n,i}(\lambda_0) - {\mathbb{X}}_{n,i} \boldsymbol \beta_0 ]  + \left(\boldsymbol{\widehat{\varepsilon}_{|z}\widehat{f}_z}\right)_{n,i}\\
		&=
		\left[\frac{1}{n}\sum_{k=1}^n \left( m(\boldsymbol Z_i) -  m(\boldsymbol Z_k) \right) K_{h,ik}	+ \frac{1}{n}\sum_{k=1}^n   \left(\varepsilon_i - \varepsilon_k \right) K_{h,ik}\right]
		- \varepsilon_i f_z(\boldsymbol Z_i) + \frac{1}{n}\sum_{k=1, k\neq i}^n \varepsilon_k  K_{h,ik}\\
		&=\frac{1}{n}\sum_{k=1}^n \left( m(\boldsymbol Z_i) -  m(\boldsymbol Z_k) \right) K_{h,ik}
		+ \frac{1}{n}\sum_{k=1, k\neq i}^n \varepsilon_i  \left(K_{h,ik} - f_z(\boldsymbol Z_i)\right) - \frac{1}{n}\varepsilon_i f_z(\boldsymbol Z_i).
	\end{align*}
\newpage
Let \\
$\left(\boldsymbol{\varepsilon} \left(\boldsymbol{f_z} - \boldsymbol{\widehat{f}_z}\right)\right)_n = \left(\frac{1}{n}\sum\limits_{k=1, k\neq 1}^n \varepsilon_1  \left(f_z(\boldsymbol Z_1) - K_{h,1k} \right),\ldots,\frac{1}{n}\sum\limits_{k=1, k\neq n}^n \varepsilon_n  \left(f_z(\boldsymbol Z_n) - K_{h,nk} \right)\right)^T$ and \\ $\left(\boldsymbol{m \widehat{f}_z} - \boldsymbol{\widehat{mf}_z}\right)_n = \left(\frac{1}{n}\sum\limits_{k=1}^n \left( m(\boldsymbol Z_1) -  m(\boldsymbol Z_k) \right) K_{h,1k},\ldots,\frac{1}{n}\sum\limits_{k=1}^n \left( m(\boldsymbol Z_n) -  m(\boldsymbol Z_k) \right) K_{h,nk}\right)^T$.

We start by showing that 
	\begin{align}\label{first_part}
		\sup_{h\in\mathcal{H}^{sc}_n} \sup_{\boldsymbol d \in\mathcal D } \left\|\boldsymbol\Omega_n ^{1/2} n^{-1}  \left(\boldsymbol{\varepsilon} \left(\boldsymbol{f_z} 	- \boldsymbol{\widehat{f}_z}\right)\right)_n\right\| = o_{\mathbb{P}}(n^{-1/2}).
	\end{align}

We get that 
	\begin{align*}
		\Big\|\boldsymbol\Omega_n ^{1/2} n^{-1}  &\left(\boldsymbol{\varepsilon} \left(\boldsymbol{f_z} - \boldsymbol{\widehat{f}_z}\right)\right)_n\Big\|^2\\
		&=
		\frac{1}{n^2} \sum_{1\leq i\neq j \leq n} \left(\boldsymbol{\varepsilon} \left(\boldsymbol{f_z} - \boldsymbol{\widehat{f}_z}\right)\right)_{n,i}\left(\boldsymbol{\varepsilon} \left(\boldsymbol{f_z} - \boldsymbol{\widehat{f}_z}\right)\right)_{n,j}\boldsymbol\Omega_{n,ij}
		+
		\frac{1}{n^2} \sum_{i = 1}^{n} \left(\boldsymbol{\varepsilon} \left(\boldsymbol{f_z} - \boldsymbol{\widehat{f}_z}\right)\right)_{n,i}^2 	\\
		&=
		A_n + B_n.
	\end{align*}
It is easy to check that $\sup_{h\in\mathcal{H}^{sc}_n}  B_n =  o_{\mathbb{P}}(n^{-1})$. Furthermore, we get that 
	\begin{align*}
		A_n = 
		\frac{1}{n^2} \sum_{1\leq i\neq j \leq n} \left[\frac{1}{n}\sum\limits_{1 \leq k \leq n, k\neq i} \varepsilon_i  \left(f_z(\boldsymbol Z_i) - K_{h,ik} \right) \right]
		\left[\frac{1}{n}\sum\limits_{1 \leq l \leq n, l\neq j} \varepsilon_j \left(f_z(\boldsymbol Z_j) - K_{h,jl} \right)\right]\boldsymbol\Omega_{n,ij}.
	\end{align*}
We show in the following that 
	\begin{align}\label{rate_star}
		\sup_{h\in\mathcal H^{sc}_n} \sup_{\boldsymbol d \in\mathcal D }| A_n| = \sup_{h\in\mathcal H^{sc}_n} \sup_{\boldsymbol d \in\mathcal D }| A_n(h)| = o_{\mathbb P}(n^{-1}).
	\end{align}
For this purpose, we define $(n)_k = n(n-1) \ldots (n-k+1)$ and decompose $A_n(h)$ into a sum of four $U$-- processes, i.e. 
	\begin{align*}
		A_n(h) = \frac{(n-1)_3}{n^3}A_{1,n}(h) + \frac{(n-1)_2}{n^2}A_{2,n}(h) + 2\frac{(n-1)_2}{n^2}A_{3,n}(h) + \frac{n-1}{n}A_{4,n}(h),
	\end{align*}
where
	\begin{align*}
		A_{1,n} = A_{1,n}(h) &= \frac{1}{(n)_4} \sum_{1 \leq i\neq j \neq k\neq l \leq n} \varepsilon_i  \left(f_z(\boldsymbol Z_i) - K_{h,ik} \right) \varepsilon_j \left(f_z(\boldsymbol Z_j) - 
		K_{h,jl} \right)\boldsymbol \Omega _{n,ij}\\ 
		A_{2,n} = A_{2,n}(h) &= \frac{1}{n(n)_3}\sum_{1 \leq i\neq j \neq k\leq n} \varepsilon_i  \left(f_z(\boldsymbol Z_i) - K_{h,ik} \right) \varepsilon_j \left(f_z(\boldsymbol Z_j) - 
		K_{h,jk} \right)\boldsymbol \Omega _{n,ij}\\ 	 
		A_{3,n} = A_{3,n}(h) &= \frac{1}{n(n)_3}\sum_{1 \leq i\neq j \neq l\leq n} \varepsilon_i  \left(f_z(\boldsymbol Z_i) - K_{h,ij} \right) \varepsilon_j \left(f_z(\boldsymbol Z_j) - 
		K_{h,jl} \right)\boldsymbol \Omega _{n,ij}\\ 	 
		\text{and} \quad 
		A_{4,n} = A_{4,n}(h) &= \frac{1}{n^2(n)_2}\sum_{1 \leq i\neq j \leq n} \varepsilon_i  \left(f_z(\boldsymbol Z_i) - K_{h,ij} \right) \varepsilon_j \left(f_z(\boldsymbol Z_j) - 
		K_{h,ij} \right)\boldsymbol \Omega _{n,ij}.
	\end{align*}

For each of these $U$--processes we compute the mean and use the Hoeffding decomposition. The kernels of $A_{1,n}$, $A_{2,n}$ and $A_{3,n}$ are not symmetric in their arguments. However, we could apply the usual symmetrization idea. For instance, for a second order $U-$statistic defined by a kernel $h( \boldsymbol U_i, \boldsymbol U_j)$, we could replace it  by the symmetric kernel $\frac{1}{2}\left[h(\boldsymbol U_i,\boldsymbol U_j) + h(\boldsymbol U_j,\boldsymbol U_i)\right]$ from which we get the same $U-$statistic. Here, $\boldsymbol U_i = \left(Y_i,\boldsymbol X_i^T, \boldsymbol Z_i^T\right)^T$. We can proceed in the same way by considering all $4!$ permutations of the variables for $A_{1,n}$ and $3!$ permutations for $A_{2,n}$ and $A_{3,n}$ so that we can apply the Hoeffding decomposition. Thus, by abuse, we will proceed as if the kernels of the $U-$statistics we handle are symmetric.  

In addition, we have that the kernels of $A_{1,n}$, $A_{2,n}$, $A_{3,n}$ and $A_{4,n}$ are Euclidean for a squared integrable envelope. See Lemma 22 in \citet{nolan1987u} and Lemma 2.14 in \citet{pakes1989simulation}. Therefore, we can in the following repeatedly apply Corollary 7 and the Maximal Inequality of \citet{sherman1994maximal}. All remainder terms are controlled by Assumption \ref{ass_con}.2.

We start by considering $A_{1,n}$. Recall that by assumption $E\left[\varepsilon_i\mid \boldsymbol X_i, \boldsymbol Z_i\right] = E\left[\varepsilon_j\mid \boldsymbol X_j, \boldsymbol Z_j\right] = 0$. Therefore, we get that $E\left[A_{1,n}\right] = 0$ as well as 
$$
	E\left[\varepsilon_i  \left(f_z(\boldsymbol Z_i) - K_{h,ik} \right) \varepsilon_j \left(f_z(\boldsymbol Z_j) - K_{h,jl} \right)\boldsymbol \Omega _{n,ij}\mid \boldsymbol U_p, p\in \{i,j,k,l\}\right] = 0.
$$  
Note that we need to consider the conditional expectations with respect to all four variables for the first order $U$--process of the Hoeffding decomposition of $A_{1,n}$ as we symmetrized the kernel. It follows from the results that the first order $U$--process of the Hoeffding decomposition of $A_{1,n}$ is $0$.

We consider now the six second order $U-$processes of the Hoeffding decomposition of $A_{1,n}$. There are two types of such processes. First, the ones that are $0$. This is the case when conditioning by the pairs $(\boldsymbol U_i, \boldsymbol U_l)$, $(\boldsymbol U_i, \boldsymbol U_k)$, $(\boldsymbol U_j, \boldsymbol U_l)$, $(\boldsymbol U_j, \boldsymbol U_k)$ and $(\boldsymbol U_l, \boldsymbol U_k)$. The second case occurs when conditioning on $(\boldsymbol U_i, \boldsymbol U_j)$. We get that 
	\begin{align*}
		E[\varepsilon_i  \left(f_z(\boldsymbol Z_i) - K_{h,ik} \right) \varepsilon_j &\left(f_z(\boldsymbol Z_j) - K_{h,jl} \right)\boldsymbol \Omega _{n,ij}\mid \boldsymbol U_i, \boldsymbol U_j] \\
		&= 
		\varepsilon_i \varepsilon_j E\left[ \left(f_z(\boldsymbol Z_i) - K_{h,ik} \right)  \left(f_z(\boldsymbol Z_j) - K_{h,jl} \right)\mid \boldsymbol Z_i, \boldsymbol Z_j\right]  \boldsymbol \Omega _{n,ij}	\\	
		&= 
		\varepsilon_i \varepsilon_j E\left[ \left(f_z(\boldsymbol Z_i) - K_{h,ik} \right)\mid \boldsymbol Z_i\right]    E\left[\left(f_z(\boldsymbol Z_j) - K_{h,jl} \right)\mid \boldsymbol Z_j\right]  \boldsymbol \Omega _{n,ij}	\\
		&= 
		\varepsilon_i \varepsilon_j h^4 \gamma_1(\boldsymbol Z_i)\gamma_1(\boldsymbol Z_j)(1 + o_{\mathbb P}(1)) \boldsymbol \Omega _{n,ij}, 			
	\end{align*}
where 
$$
	\gamma_1(\boldsymbol Z ) = \mu(K) \cdot {\rm tr } \{\boldsymbol H_{z,z} f_z(\boldsymbol Z)\},
$$
with $\int _{\mathbb{R}^{q}}\boldsymbol u \boldsymbol u^TK(\boldsymbol u) d\boldsymbol u = \mu(K) \boldsymbol{I}_{q\times q}$. $\boldsymbol H_{z,z} f_z$ denotes the matrix of second derivative of $f_z(\cdot)$ with respect to the components of $\boldsymbol Z \in \mathbb{R}^{q}$ and ${\rm tr } \{\cdot\}$ denotes the trace operator. Therefore, it follows together with Corollary 4 of \citet{sherman1994maximal} that the second order $U-$processes of the Hoeffding decomposition of $A_{1,n}$ are of order $O_{\mathbb{P}}(n^{-1}n^{-4\alpha})= o_{\mathbb P} (n^{-1})$ uniformly with respect to $h$ and $\boldsymbol d$.

We consider now the four $U-$processes of order three obtained by conditioning on any subset of three of the four vectors $\boldsymbol U_i$, $\boldsymbol U_k$,  $\boldsymbol U_j$ and $ \boldsymbol U_l$. There are two types of such processes. First, the ones that are $0$. This is the case when conditioning by $(\boldsymbol U_i, \boldsymbol U_l, \boldsymbol U_k)$ or $(\boldsymbol U_j, \boldsymbol U_l, \boldsymbol U_k)$. The second case occurs when conditioning on $(\boldsymbol U_i, \boldsymbol U_j, \boldsymbol U_l)$ or $(\boldsymbol U_i, \boldsymbol U_j, \boldsymbol U_k)$, the other one being similar. We get that
	\begin{align*}
		E[\varepsilon_i  \left(f_z(\boldsymbol Z_i) - K_{h,ik} \right) \varepsilon_j (f_z(\boldsymbol Z_j) &- K_{h,jl} )\boldsymbol \Omega _{n,ij}\mid \boldsymbol U_i, \boldsymbol U_j, \boldsymbol U_k] \\
		&= 
		\varepsilon_i \left(f_z(\boldsymbol Z_i) - K_{h,ik} \right) \varepsilon_j E\left[ \left(f_z(\boldsymbol Z_j) - K_{h,jl} \right)\mid \boldsymbol Z_j\right]  \boldsymbol \Omega _{n,ij}	\\	
		&= 
		\varepsilon_i \left(f_z(\boldsymbol Z_i) - K_{h,ik} \right) \varepsilon_j h^2 \gamma_1(\boldsymbol Z_j)(1 + o_{\mathbb P}(1)) \boldsymbol \Omega _{n,ij}\\
		&=
		\varepsilon_i f_z(\boldsymbol Z_i) \varepsilon_j h^2 \gamma_1(\boldsymbol Z_j)(1 + o_{\mathbb P}(1)) \boldsymbol \Omega _{n,ij}\\	
		& \quad - \varepsilon_i K_{h,ik} \varepsilon_j h^2 \gamma_1(\boldsymbol Z_j)(1 + o_{\mathbb P}(1)) \boldsymbol \Omega _{n,ij}\\
		&=
		\varepsilon_i f_z(\boldsymbol Z_i) \varepsilon_j h^2 \gamma_1(\boldsymbol Z_j)(1 + o_{\mathbb P}(1)) \boldsymbol \Omega _{n,ij}\\	
		& \quad - h^{2-q} h^qK_{h,ik} \tau(\boldsymbol U_i, \boldsymbol U_j)(1 + o_{\mathbb P}(1)) 		
		.		
	\end{align*}
Now, we apply the Maximal Inequality of \citet{sherman1994maximal}, page 448,  for the degenerate $U-$process given by the kernel $h^q K_{h,ik}  \tau (\boldsymbol U_i, \boldsymbol U_j)$, indexed by $h\in\mathcal H_n^{sc}$, with envelope $\|K\|_\infty \tau(\cdot,\cdot)$. (Herein, $\|\cdot\|_\infty $ denotes the uniform norm.) We take $p=1$ and $\beta\in (0,1)$ arbitrarily close to 1 to stand for Sherman's quantity $\alpha$. Since $K(\cdot)$ is of bounded variation and symmetric, without loss of generality we could consider 
that $K(\cdot)$ is nonincreasing on $[0,\infty)$. In this case, $0\leq K(\cdot/h)\leq K(\cdot/\overline h)$ with $\overline h = \sup \mathcal H^{sc}_n = : c_{max} n^{-\alpha}$.
Hence, using Jensen's inequality, we could bound the right-hand side of the Maximal Inequality of  \citet{sherman1994maximal} by a universal constant times
$$
	\left(E \left[ K^{2}\left(\frac{\boldsymbol Z_i - \boldsymbol Z_k}{c_{max} n^{-\alpha}}\right)\tau^{2} (\boldsymbol U_i, \boldsymbol U_j)\right]\right)^{\beta/2}.
$$
By standard changes of variables and suitable integrability conditions, the power $\beta/2$ of the expectation in the last display is bounded by  a constant times $n^{-\alpha \beta q/2}$. Consequently, the uniform rate of the second $U-$processes obtained conditioning by $\boldsymbol U_i, \boldsymbol U_j, \boldsymbol U_k$ and  $\boldsymbol U_i,\boldsymbol U_j, \boldsymbol U_l$, respectively is $n^{-3/2}\times O_{\mathbb P} (n^{-\alpha \{2-q +\beta q/2\} })$. As $1/2 + \alpha (2 - q + \beta q/2) > 0 $ under our assumptions we get that $n^{-3/2}\times O_{\mathbb P} (n^{-\alpha \{2-q +\beta q/2\}})= o_{\mathbb P} (n^{-1})$ such that the third order $U-$processes of the Hoeffding decomposition of $A_{1,n}$ are of order $o_{\mathbb P} (n^{-1})$.

Finally, we consider the remaining $U-$process of order four. This process is given by 
	\begin{align*}
		\varepsilon_i  \left(f_z(\boldsymbol Z_i) - K_{h,ik} \right) \varepsilon_j &(f_z(\boldsymbol Z_j) - K_{h,jl} )\boldsymbol \Omega _{n,ij}\\
		&=
		\varepsilon_i \varepsilon_j \left(f_z(\boldsymbol Z_i)f_z(\boldsymbol Z_j) - f_z(\boldsymbol Z_j)K_{h,ik}  - f_z(\boldsymbol Z_i)K_{h,jl} + K_{h,ik}K_{h,jl}\right) \boldsymbol \Omega _{n,ij}\\
		&=\varepsilon_i \varepsilon_j f_z(\boldsymbol Z_i)f_z(\boldsymbol Z_j)\boldsymbol \Omega _{n,ij}
		- h^{-q}\tau_1(\boldsymbol U_i, \boldsymbol U_j) h^{q}K_{h,jl}
		- h^{-q}\tau_2(\boldsymbol U_i, \boldsymbol U_j) h^{q}K_{h,ik}\\
		& \quad + h^{-2q}\tau_3(\boldsymbol U_i, \boldsymbol U_j) h^{2q}K_{h,ik}K_{h,jl}.
	\end{align*}

Now, we apply again the Maximal Inequality of \citet{sherman1994maximal}, page 448,  for the degenerate $U-$process given by the kernel $\tau_3(\boldsymbol U_i, \boldsymbol U_j)h^qK_{h,ik}h^qK_{h,jl}$, indexed by $h\in\mathcal H_n^{sc}$, with envelope $\|K\|_\infty^2 \tau(\cdot,\cdot)$. We take again $p=1$ and $\beta\in (0,1)$ arbitrarily close to 1 to stand for Sherman's quantity $\alpha$. 
Hence, using Jensen's inequality, we could bound the right-hand side of the Maximal Inequality of  \citet{sherman1994maximal} by a universal constant times
$$
	\left(E \left[ K^{2}\left(\frac{\boldsymbol Z_i - \boldsymbol Z_k}{c_{max} n^{-\alpha}}\right) K^{2}\left(\frac{\boldsymbol Z_j - \boldsymbol Z_l}{c_{max} n^{-\alpha}}\right)\tau_3^{2}(\boldsymbol U_i,\boldsymbol U_j)\right]\right)^{\beta/2}.
$$
By standard changes of variables and suitable integrability conditions, the power $\beta/2$ of the expectation in the last display is bounded by  a constant times $n^{-\alpha \beta q}$. Consequently, the uniform rate of the fourth order $U-$process is $n^{-2}\times O_{\mathbb P} (n^{\alpha q \{2 - \beta\} })$. Since $1> \alpha q$ under our assumptions we get that $n^{-2}\times O_{\mathbb P} (n^{\alpha q \{2 - \beta\} }) = o_{\mathbb P}(n^{-1})$. By the same reasoning we can control $h^{-q}\tau_1(\boldsymbol U_i, \boldsymbol U_j) h^{q}K_{h,jl}$ and $h^{-q}\tau_2(\boldsymbol U_i, \boldsymbol U_j) h^{q}K_{h,ik}$. The details are omitted.

From all the results it follows that 
	\begin{align*}
		\sup_{h\in\mathcal H^{sc}_n} \sup_{\boldsymbol d \in\mathcal D } | A_{1,n}| = \sup_{h\in\mathcal H^{sc}_n} \sup_{\boldsymbol d \in\mathcal D }| A_{1,n}(h)| = o_{\mathbb P}(n^{-1}).
	\end{align*}

In the next step we consider $A_{2,n}$. We get that $E\left[A_{2,n}\right] = 0$ as well as 
$$
	E\left[\varepsilon_i  \left(f_z(\boldsymbol Z_i) - K_{h,ik} \right) \varepsilon_j \left(f_z(\boldsymbol Z_j) - K_{h,jk} \right)\boldsymbol \Omega _{n,ij}\mid \boldsymbol U_p, p\in \{i,j,k\}\right] = 0.
$$  
In addition, it is easy to see that the second and third order $U$--processes of the Hoeffding decomposition of $A_{2,n}$ are of order $o_{\mathbb P}(n^{-1})$ if we apply the Maximal Inequality of \citet{sherman1994maximal}. From all the results it follows that 
	\begin{align*}
		\sup_{h\in\mathcal H^{sc}_n}\sup_{\boldsymbol d \in\mathcal D } | A_{2,n}| = \sup_{h\in\mathcal H^{sc}_n}\sup_{\boldsymbol d \in\mathcal D } | A_{2,n}(h)| = o_{\mathbb P}(n^{-1}).
	\end{align*} 
As it follows by the same reasoning as for $A_{2,n}$ that
	\begin{align*}
		\sup_{h\in\mathcal H^{sc}_n}\sup_{\boldsymbol d \in\mathcal D } | A_{3,n}| = \sup_{h\in\mathcal H^{sc}_n} \sup_{\boldsymbol d \in\mathcal D }| A_{3,n}(h)| = o_{\mathbb P}(n^{-1}).
	\end{align*}	
we omit the details here.	

Finally, we get that $E\left[A_{4,n}\right] = 0$ as well as 
$$
	E\left[\varepsilon_i  \left(f_z(\boldsymbol Z_i) - K_{h,ij} \right) \varepsilon_j \left(f_z(\boldsymbol Z_j) - K_{h,ij} \right)\boldsymbol \Omega _{n,ij}\mid \boldsymbol U_p, p\in \{i,j\}\right] = 0.
$$   
In addition, it is easy to see that the second order $U$--process of the Hoeffding decomposition of $A_{4,n}$ is of order $o_{\mathbb P}(n^{-1})$ if we apply the Maximal Inequality of \citet{sherman1994maximal}. 
Deduce that 
	\begin{align*}
		\sup_{h\in\mathcal H^{sc}_n}\sup_{\boldsymbol d \in\mathcal D }|A_{4,n}| = o_{\mathbb{P}}(n^{-1}). 
	\end{align*}
With all these results \eqref{rate_star} and, in particular, \eqref{first_part} follow.

In the next part we show that  	
	\begin{align} \label{second_part}
		\sup_{h\in\mathcal{H}^{sc}_n} \sup_{\boldsymbol d \in\mathcal D } \left\|\boldsymbol\Omega_n ^{1/2} n^{-1}  \left(\boldsymbol{m \widehat{f}_z} - \boldsymbol{\widehat{mf}_z}\right)_n\right\| = o_{\mathbb{P}}(n^{-1/2}).
	\end{align}
We get that 
	\begin{align*}
		\Big\|\boldsymbol\Omega_n ^{1/2} n^{-1}  &\left(\boldsymbol{m \widehat{f}_z} - \boldsymbol{\widehat{mf}_z}\right)_n\Big\|^2\\
		&=
		  \frac{1}{n^2} \sum_{1\leq i\neq j \leq n} \left(\boldsymbol{m \widehat{f}_z} - \boldsymbol{\widehat{mf}_z}\right)_{n,i}\left(\boldsymbol{m \widehat{f}_z} - \boldsymbol{\widehat{mf}_z}\right)_{n,j}\boldsymbol\Omega_{n,ij}
		+
		  \frac{1}{n^2} \sum_{i = 1}^{n} \left(\boldsymbol{m \widehat{f}_z} - \boldsymbol{\widehat{mf}_z}\right)_{n,i}^2 	\\
		&=
		  \widetilde A_n + \widetilde B_n.
	\end{align*}
It is easy to check that $\sup_{h\in\mathcal{H}^{sc}_n}  \widetilde B_n =  o_{\mathbb{P}}(n^{-1})$. Furthermore, we get that 
	\begin{align*}
		\widetilde A_n = 
		  \frac{1}{n^2} \sum_{1\leq i\neq j \leq n} \left[\frac{1}{n}\sum\limits_{1 \leq k \leq n, k\neq i} \left(m(\boldsymbol Z_i) -  m(\boldsymbol Z_k) \right) K_{h,ik} \right]
		\left[\frac{1}{n}\sum\limits_{1 \leq l \leq n, l\neq j} \left( m(\boldsymbol Z_j) -  m(\boldsymbol Z_l) \right)K_{h,jl} \right]\boldsymbol\Omega_{n,ij}.
	\end{align*}
We show in the following that 
	\begin{align}\label{rate_star2}
		\sup_{h\in\mathcal H^{sc}_n} \sup_{\boldsymbol d \in\mathcal D }| \widetilde A_n| = \sup_{h\in\mathcal H^{sc}_n} \sup_{\boldsymbol d \in\mathcal D }| \widetilde A_n(h)| = o_{\mathbb P}(n^{-1}).
	\end{align}
We decompose $\widetilde A_n(h)$ into a sum of four $U$-- processes, i.e. 
	\begin{align*}
		\widetilde A_n(h) = \frac{(n-1)_3}{n^3}\widetilde A_{1,n}(h) + \frac{(n-1)_2}{n^2}\widetilde A_{2,n}(h) + 2\frac{(n-1)_2}{n^2}\widetilde A_{3,n}(h) - \frac{n-1}{n}\widetilde A_{4,n}(h),
	\end{align*}
where
	\begin{align*}
		\widetilde A_{1,n} = \widetilde A_{1,n}(h) &= \frac{1}{(n)_4} \sum_{1 \leq i\neq j \neq k\neq l \leq n} \left( m(\boldsymbol Z_i) - m(\boldsymbol Z_k)\right)K_{h,ik}\left( m(\boldsymbol Z_j) - m(\boldsymbol Z_l)\right)K_{h,jl}
		\boldsymbol \Omega _{n,ij}\\ 
		\widetilde A_{2,n} = \widetilde A_{2,n}(h) &= \frac{1}{n(n)_3}\sum_{1 \leq i\neq j \neq k\leq n} \left( m(\boldsymbol Z_i) - m(\boldsymbol Z_k)\right)K_{h,ik}\left( m(\boldsymbol Z_j) - m(\boldsymbol Z_k)\right)K_{h,jk}
		\boldsymbol \Omega _{n,ij}\\ 	 
		\widetilde A_{3,n} = \widetilde A_{3,n}(h) &= \frac{1}{n(n)_3}\sum_{1 \leq i\neq j \neq l\leq n} \left( m(\boldsymbol Z_i) - m(\boldsymbol Z_j)\right)K_{h,ij}\left( m(\boldsymbol Z_j) - m(\boldsymbol Z_l)\right)K_{h,jl}
		\boldsymbol \Omega _{n,ij}\\ 	 
		\text{and} \quad 
		\widetilde A_{4,n} = \widetilde A_{4,n}(h) &= \frac{1}{n^2(n)_2}\sum_{1 \leq i\neq j \leq n} \left( m(\boldsymbol Z_i) - m(\boldsymbol Z_j)\right)^2K_{h,ij}^2\boldsymbol \Omega _{n,ij}.
	\end{align*}
For each of these $U$--processes we compute the mean and use the Hoeffding decomposition. The kernels of $\widetilde A_{1,n}$, $\widetilde A_{2,n}$ and $\widetilde A_{3,n}$ are not symmetric in their arguments. However, we could apply the usual symmetrization idea. Here, $\widetilde{\boldsymbol U}_i = \left(\boldsymbol X_i^T, \boldsymbol Z_i^T\right)^T$. Thus, by abuse, we will proceed as if the kernels of the $U-$statistics we handle are symmetric. For simpler formulae, we use the short notation $m_i, m_k,\ldots$ instead of $m(\boldsymbol Z_i), m(\boldsymbol Z_k),\ldots$.  

In addition, we have that the kernels of $\widetilde A_{1,n}$, $\widetilde A_{2,n}$, $\widetilde A_{3,n}$ and $\widetilde A_{4,n}$ are Euclidean for a squared integrable envelope. See Lemma 22 in \citet{nolan1987u} and Lemma 2.14 in \citet{pakes1989simulation}. Therefore, we can in the following repeatedly apply Corollary 7 and the Maximal Inequality of \citet{sherman1994maximal}. All remainder terms are controlled by Assumption \ref{ass_con}.2.

We start by considering the expectation of $\widetilde A_{1,n}$. We get that  \newpage
	\begin{align*}
		E[\widetilde A_{1,n}] &= E\left[\left( m_i - m_k\right)K_{h,ik}\left( m_j - m_l\right)K_{h,jl}\boldsymbol \Omega _{n,ij}\right]\\
		&= E\left[ E\left[ \left(m_i -  m_k \right) K_{h,ik} \boldsymbol \Omega^Z _{n,ij}\mid \boldsymbol X_i,\widetilde{\boldsymbol U}_j,\boldsymbol Z_l\right] \left( m_j - m_l\right)K_{h,jl} \boldsymbol \Omega^X_{n,ij} \right].
	\end{align*}
Next, by Taylor expansion and Dominated convergence 
	\begin{align*}
		E\left[m_i  K_{h,ik} \boldsymbol \Omega^Z_{n,ij} \mid \boldsymbol X_i,\widetilde{\boldsymbol U}_j,\boldsymbol Z_l \right] &= E \left[m_i E\left[ K_{h,ik}  \mid  \boldsymbol Z_i  \right]  \boldsymbol  \Omega^Z _{n,ij}  \mid \boldsymbol X_i, \boldsymbol Z_j \right]\\
		&=  E \left[ m_i (f_z(\boldsymbol Z_i) + h^2 \gamma_1(\boldsymbol Z_i) (1+o_{\mathbb P }(1)))  \boldsymbol  \Omega^Z _{n,ij} \mid \boldsymbol X_i,\boldsymbol Z_j \right]\\
		&=  E \left[ m_i  f_z(\boldsymbol Z_i)  \boldsymbol  \Omega^Z_{n,ij}  \mid \boldsymbol X_i, \boldsymbol Z_j \right]\\ 
		&\quad  +h^2 E \left[  m_i  \gamma_1(\boldsymbol Z_i)  \boldsymbol  \Omega^Z _{n,ij} \mid \boldsymbol X_i,\boldsymbol Z_j \right] (1+o_{\mathbb P }(1)).
	\end{align*}
Similarly,  
	\begin{align*}
		E\left[  m_k  K_{h,ik} \boldsymbol \Omega^Z _{n,ij} \mid \boldsymbol X_i,\widetilde{\boldsymbol U}_j,{\boldsymbol Z}_l\right] &= E \left[ E\left[ m_k K_{h,ik} \mid \boldsymbol Z_i \right]  \boldsymbol  \Omega^Z _{n,ij}  \mid \boldsymbol X_i,\boldsymbol Z_j \right]\\
		&= E \left[ ( m_i f_z(\boldsymbol Z_i)    +  h^2 \gamma_2(\boldsymbol Z_i)(1+o_{\mathbb P }(1)) \boldsymbol \Omega ^Z_{n,ij}\mid   \boldsymbol X_i, \boldsymbol Z_j \right]   \\
		&= E \left[ m_i f_z(\boldsymbol Z_i)  \boldsymbol \Omega^Z _{n,ij}  \mid\boldsymbol X_i, \boldsymbol Z_j \right] \\ 
		& \quad + h^2 E \left[\gamma_2(\boldsymbol Z_i) \boldsymbol \Omega ^Z_{n,ij}\mid   \boldsymbol X_i, \boldsymbol Z_j \right] (1+o_{\mathbb P }(1)),
	\end{align*}
where 
$$
  \gamma_2(\boldsymbol Z ) = \mu(K) \cdot {\rm tr } \{\boldsymbol H_{z,z} \left(mf_z\right)(\boldsymbol Z)\}.
$$
$\boldsymbol H_{z,z} \left(mf_z\right)$ denotes the matrix of second derivative of $mf_z(\cdot)$ with respect to the components of $\boldsymbol Z \in \mathbb{R}^{q}$. Thus,
	\begin{align*}
		E[\widetilde A_{1,n}] 
		&= E\left[ E\left[ (m_i -  m_k ) K_{h,ik} \boldsymbol \Omega^Z _{n,ij}\mid \boldsymbol X_i,\widetilde{\boldsymbol U}_j,\boldsymbol Z_l\right] ( m_j - m_l)K_{h,jl} \boldsymbol \Omega^X_{n,ij} \right]\\
		&=h^2 
		E\left[ \gamma_3( \boldsymbol X_i, \boldsymbol Z_j) ( m_j - m_l)K_{h,jl} \boldsymbol \Omega^X_{n,ij} \right] (1+o(1))
		\\ 
		&=h^2 E\left[ \gamma_3( \boldsymbol X_i, \boldsymbol Z_j) E\left[  ( m_j - m_l)K_{h,jl} \mid \boldsymbol X_i,\boldsymbol X_j, \boldsymbol Z_j \right] \boldsymbol \Omega^X_{n,ij} \right] (1+o(1))\\
		&=h^2 E\left[ \gamma_3( \boldsymbol X_i	, \boldsymbol Z_j) E\left[  ( m_j - m_l)K_{h,jl} \mid \boldsymbol Z_j \right] \boldsymbol \Omega^X_{n,ij} \right] (1+o(1)),
	\end{align*}
where $\gamma_3( \boldsymbol X_i, \boldsymbol Z_j) = E \left[ m_i \gamma_1(\boldsymbol Z_i)  \boldsymbol  \Omega^Z _{n,ij} \mid \boldsymbol X_i,\boldsymbol Z_j \right] - E \left[\gamma_2(\boldsymbol Z_i) \boldsymbol \Omega ^Z_{n,ij}\mid \boldsymbol X_i,\boldsymbol Z_j \right]$. In the next step we consider \\
$E\left[  ( m_j - m_l)K_{h,jl} \mid\boldsymbol Z_j \right]$. We get that 
	\begin{align*}
		E\left[m_j K_{h,jl} \mid \boldsymbol Z_j \right] 
		&= m_jE\left[ K_{h,jl} \mid \boldsymbol Z_j \right]\\
		&= m_j\left(f_z(\boldsymbol Z_j) + h^2 \gamma_1(\boldsymbol Z_j) (1+o_{\mathbb P }(1))\right)\\		
		&= m_jf_z(\boldsymbol Z_j) + h^2 m_j\gamma_1(\boldsymbol Z_j) (1+o_{\mathbb P }(1)).		
	\end{align*}
In addition, we get that 	
	\begin{align*}
		E\left[m_l K_{h,jl} \mid \boldsymbol Z_j \right] 			
		&= E\left[ m_j f_z(\boldsymbol Z_j)  +  h^2 \gamma_2(\boldsymbol Z_j)(1+o_{\mathbb P }(1))\mid\boldsymbol Z_j \right]\\
		&=  m_j f_z(\boldsymbol Z_j)   +  h^2 \gamma_2(\boldsymbol Z_j)(1+o_{\mathbb P }(1)).								
	\end{align*}	
Therefore, we get that 
	\begin{align*}
		E[\widetilde A_{1,n}] &= h^2 E\left[ \gamma_3( \boldsymbol X_i	, \boldsymbol Z_j) E\left[  ( m_j - m_l)K_{h,jl} \mid \boldsymbol Z_j \right] \boldsymbol \Omega^X_{n,ij} \right] (1+o(1)) \\
		&= h^4 E\left[\gamma_3( \boldsymbol X_i, \boldsymbol Z_j)\left(m_j \gamma_1(\boldsymbol Z_j) - \gamma_2(\boldsymbol Z_j)\right)  
		   \boldsymbol\Omega^X_{n,ij}\right](1+o(1)).   
	\end{align*}
This implies that $E[\widetilde A_{1,n}] = o_{\mathbb P} (n^{-1})$ uniformly with respect to $h\in\mathcal H_n^{sc}$.

We consider now the first order $U$--process of the Hoeffding decomposition for $\widetilde A_{1,n}$. As we symmetrized the kernel we need to consider the conditional expectations with respect to all four variables. By the same reasoning as for $E[\widetilde A_{1,n}]$ we get that 
	\begin{align*}
		E\big[(m_i -   m_k)K_{h,ik}( m_j - m_l)&K_{h,jl}\boldsymbol \Omega _{n,ij}\mid \widetilde{\boldsymbol U}_i\big] 
		= h^2 E\left[\gamma_3(\boldsymbol X_j, \boldsymbol Z_i) (m_i-m_k)K_{h,ik}\boldsymbol \Omega_{n,ij}^X\mid \widetilde{\boldsymbol U}_i\right](1 + o_{\mathbb P }(1)) \\
		&= h^2 E\left[\gamma_3(\boldsymbol X_j, \boldsymbol Z_i) E\left[(m_i-m_k)K_{h,ik}\mid\boldsymbol Z_i\right]\boldsymbol \Omega_{n,ij}^X\mid \widetilde{\boldsymbol U}_i\right](1 + o_{\mathbb P }(1))\\
		&= h^2 E\left[(m_i-m_k)K_{h,ik}\mid\boldsymbol Z_i\right]E\left[\gamma_3(\boldsymbol X_j, \boldsymbol Z_i) \boldsymbol \Omega_{n,ij}^X\mid \widetilde{\boldsymbol U}_i\right](1 + o_{\mathbb P }(1))\\
		&=h^4 \left(m_i\gamma_1(\boldsymbol Z_i) - \gamma_2(\boldsymbol Z_i)\right)E\left[\gamma_3(\boldsymbol X_j, \boldsymbol Z_i) \boldsymbol \Omega_{n,ij}^X \mid \widetilde{\boldsymbol U}_i\right](1 + o_{\mathbb P }(1)).			      
	\end{align*}  
Note that the reasoning for $E\left[(m_i -   m_k)K_{h,ik}( m_j - m_l)K_{h,jl}\boldsymbol \Omega _{n,ij} \mid \widetilde{\boldsymbol U}_j\right]$ is exactly the same. In addition, we have that 
	\begin{align*}
		E\left[(m_i -   m_k)K_{h,ik}(m_j - m_l)K_{h,jl}\boldsymbol \Omega _{n,ij}\mid \widetilde{\boldsymbol U}_k\right] 
		=
		h^2 E\left[\gamma_3(\boldsymbol X_j,\boldsymbol Z_i)(m_i -   m_k)K_{h,ik}\boldsymbol \Omega^X _{n,ij}\mid \boldsymbol Z_k\right]. 							  			      
	\end{align*} 
We get that 
	\begin{align*}
		E\big[\gamma_3(\boldsymbol X_j,\boldsymbol Z_i)m_i K_{h,ik}\boldsymbol \Omega^X _{n,ij}\mid \boldsymbol Z_k\big] 
		&=
		  E\big[K_{h,ik} E\big[\gamma_3(\boldsymbol X_j,\boldsymbol Z_i)m_i \boldsymbol \Omega^X _{n,ij}\mid \boldsymbol Z_i\big] \mid \boldsymbol Z_k\big] \\
		&=
		  m_k f_z(\boldsymbol Z_k) E\big[\gamma_3(\boldsymbol X_j,\boldsymbol Z_k)\boldsymbol \Omega^X _{n,ij}\mid \boldsymbol Z_k\big] + O_{\mathbb P} (h^2)	
	\end{align*}
and 
	\begin{align*}
		E\big[\gamma_3(\boldsymbol X_j,\boldsymbol Z_i)m_k K_{h,ik}\boldsymbol \Omega^X _{n,ij}\mid \boldsymbol Z_k\big] 
		&=
		  m_kE\big[K_{h,ik} E\big[\gamma_3(\boldsymbol X_j,\boldsymbol Z_i) \boldsymbol \Omega^X _{n,ij}\mid \boldsymbol Z_i\big] \mid \boldsymbol Z_k\big] \\
		&=
		  m_k f_z(\boldsymbol Z_k) E\big[\gamma_3(\boldsymbol X_j,\boldsymbol Z_k) \boldsymbol \Omega^X _{n,ij}\mid \boldsymbol Z_k\big] + O_{\mathbb P} (h^2).						
	\end{align*}	
Note that the reasoning for $E\left[(m_i -   m_k)K_{h,ik}( m_j - m_l)K_{h,jl}\boldsymbol \Omega _{n,ij} \mid \widetilde{\boldsymbol U}_l\right]$ is exactly the same.	
Therefore, it follows together with Corollary 4 of \citet{sherman1994maximal} that the first order $U-$processes of the Hoeffding decomposition for $\widetilde A_{1,n}$ are of order $o_{\mathbb P} (n^{-1})$ uniformly with respect to $h$. 

We consider now the six second order $U-$processes of the Hoeffding decomposition for $\widetilde A_{1,n}$. There are two types of such processes. First, the ones where the two kernels $K_{h,ik}$ and $K_{h,jl}$ are both integrated with respect to one of the variables they contain. This is the case when conditioning by the pairs $(\widetilde{\boldsymbol U}_i, \widetilde{\boldsymbol U}_j)$, $(\widetilde{\boldsymbol U}_i, \widetilde{\boldsymbol U}_l)$, $(\widetilde{\boldsymbol U}_k, \widetilde{\boldsymbol U}_j)$ and $(\widetilde{\boldsymbol U}_k, \widetilde{\boldsymbol U}_l)$. We get that
	\begin{align*}
		&E\left[(m_i -   m_k)K_{h,ik}(m_j - m_l)K_{h,jl}\boldsymbol \Omega _{n,ij} \mid \widetilde{\boldsymbol U}_i, \widetilde{\boldsymbol U}_j\right] \\
		&\quad=
		  E\left[E\left[(m_i -   m_k)K_{h,ik}\mid \boldsymbol Z_i\right]( m_j - m_l)K_{h,jl}\mid \boldsymbol Z_i, \boldsymbol Z_j\right]\boldsymbol \Omega _{n,ij}\\
		&\quad=
		  \left[m_i\left(f_z(\boldsymbol Z_i) + h^2\gamma_1(\boldsymbol Z_i)(1 + o_{\mathbb P }(1))\right) - \left(m_if_z(\boldsymbol Z_i) + h^2\gamma_2(\boldsymbol Z_i)(1 + o_{\mathbb P }(1))\right)\right] \\
		&\quad~~~~\left[m_j\left(f_z(\boldsymbol Z_j) + h^2\gamma_1(\boldsymbol Z_j)(1 + o_{\mathbb P }(1))\right) - \left(m_jf_z(\boldsymbol Z_j) + h^2\gamma_2(\boldsymbol Z_j)(1 + o_{\mathbb P }(1))\right)\right]\boldsymbol \Omega _{n,ij}\\
		&\quad=
		h^4\left[m_i\gamma_1(\boldsymbol Z_i) - \gamma_2(\boldsymbol Z_i)\right] \left[m_j\gamma_1(\boldsymbol Z_j) - \gamma_2(\boldsymbol Z_j)\right]
		(1 + o_{\mathbb P }(1)) \boldsymbol \Omega _{n,ij}.  	      
	\end{align*}  
In addition, we get that	
	\begin{align*}
		&E\left[(m_i -   m_k)K_{h,ik}( m_j - m_l)K_{h,jl}\boldsymbol \Omega _{n,ij}\mid \widetilde{\boldsymbol U}_i, \widetilde{\boldsymbol U}_l\right] \\
		&\quad=E\left[E\left[(m_i -   m_k)K_{h,ik}\mid \boldsymbol Z_i\right](m_j - m_l)K_{h,jl}\boldsymbol \Omega _{n,ij}\mid \widetilde{\boldsymbol U}_i, \widetilde{\boldsymbol U}_l\right]\\
		&\quad= \left[m_i\left(f_z(\boldsymbol Z_i) + h^2\gamma_1(\boldsymbol Z_i)(1 + o_{\mathbb P }(1))\right) - \left(m_if_z(\boldsymbol Z_i) + h^2\gamma_2(\boldsymbol Z_i)(1 + o_{\mathbb P }(1))\right)\right]\\
		&\qquad~ E\left[(m_j - m_l)K_{h,jl}\boldsymbol \Omega _{n,ij}\mid \widetilde{\boldsymbol U}_i, \widetilde{\boldsymbol U}_l\right]\\
		&\quad=  h^2\left[m_i\gamma_1(\boldsymbol Z_i)(1 + o_{\mathbb P }(1)) - \gamma_2(\boldsymbol Z_i)(1 + o_{\mathbb P }(1))\right]\\			
		&\qquad~ \left[m_l f_z(\boldsymbol Z_l)  \boldsymbol \Omega _{n,il}^Z E\left[ \boldsymbol \Omega _{n,ij}^X\mid {\boldsymbol X}_i\right] - 
		                 m_lf_z(\boldsymbol Z_l)  \boldsymbol \Omega _{n,il}^Z E\left[ \boldsymbol \Omega _{n,ij}^X\mid {\boldsymbol X}_i\right] + O_{\mathbb P} (h^2)\right] =  O_{\mathbb P} (h^4).			
	\end{align*}	
The reasoning when conditioning on $(\widetilde{\boldsymbol U}_k, \widetilde{\boldsymbol U}_j)$ is the same. For the fourth part we get that
	\begin{align*}
		&E\left[(m_i -   m_k)K_{h,ik}( m_j - m_l)K_{h,jl}\boldsymbol \Omega _{n,ij}\mid \widetilde{\boldsymbol U}_k, \widetilde{\boldsymbol U}_l\right]\\ 
		&=E\left[( m_i - m_k)K_{h,ik}E\left[( m_j - m_l)K_{h,jl}\boldsymbol \Omega _{n,ij}\mid \boldsymbol Z_l, \widetilde{\boldsymbol U}_i\right]\mid \widetilde{\boldsymbol U}_k, \widetilde{\boldsymbol U}_l\right]	\\
		&=E\left[(m_i - m_k)K_{h,ik}
		         \left(m_l f_z(\boldsymbol Z_l) \boldsymbol \Omega _{n,il}^Z E\left[ \boldsymbol \Omega _{n,ij}^X\mid {\boldsymbol X}_i\right] - 
		               m_l f_z(\boldsymbol Z_l) \boldsymbol \Omega _{n,il}^Z E\left[ \boldsymbol \Omega _{n,ij}^X\mid {\boldsymbol X}_i\right]
		               + O_{\mathbb P} (h^2)
		         \right)\mid \widetilde{\boldsymbol U}_k, \widetilde{\boldsymbol U}_l\right]\\
		&= \left(m_k f_z(\boldsymbol Z_k)  - m_k f_z(\boldsymbol Z_k) + O_{\mathbb P}(h^2)\right) O_{\mathbb P}(h^2) = O_{\mathbb P}(h^4).
	\end{align*}
Applying the results of \citet{sherman1994maximal}, the four $U-$processes for which the two kernels $K_{h,ik}$ and $K_{h,jl}$ are both integrated with respect to one of their variables have the uniform rate $o_{\mathbb P } (n^{-1})$. 

Next, we investigate one of the two $U-$processes of the Hoeffding decomposition obtained by conditioning on the pairs $(\widetilde{\boldsymbol U}_i, \widetilde{\boldsymbol U}_k)$ and  $(\widetilde{\boldsymbol U}_j, \widetilde{\boldsymbol U}_l)$, the other one being similar. We have 
	\begin{align*}
		E\big[ ( m_i -   m_k)K_{h,ik}( m_j - m_l)&K_{h,jl}\boldsymbol \Omega _{n,ij}\mid \widetilde{\boldsymbol U}_j, \widetilde{\boldsymbol U}_l\big] 
		= ( m_j -   m_l)K_{h,jl}E\left[( m_i - m_k)K_{h,ik}\boldsymbol \Omega _{n,ij}\mid \widetilde{\boldsymbol U}_j\right]\\
		&= ( m_j -   m_l)K_{h,jl}E\left[ E\left[ ( m_i - m_k)K_{h,ik}\boldsymbol \Omega^Z _{n,ij}\mid \boldsymbol X_i, \boldsymbol Z_j \right] \boldsymbol \Omega^X _{n,ij}\mid \widetilde{\boldsymbol U}_j\right] \\
		&= h^2 ( m_j -   m_l)K_{h,jl} E \left[ \gamma_3( \boldsymbol X_i, \boldsymbol Z_j)  \boldsymbol \Omega^X_{n,i,j} \mid \widetilde{\boldsymbol U}_j\right] (1+o_{\mathbb P }(1))\\
		&=: h^{2-q} (1+o_{\mathbb P }(1)) \times h^q K_{h,jl}  \tau (\widetilde{\boldsymbol U}_j, \widetilde{\boldsymbol U}_l).
	\end{align*}
Now, we apply the Maximal Inequality of \citet{sherman1994maximal}, page 448,  for the degenerate $U-$process given by the kernel $h^q K_{h,jl}  \tau (\widetilde{\boldsymbol U}_j, \widetilde{\boldsymbol U}_l)$, indexed by $h\in\mathcal H_n^{sc}$, with envelope $\|K\|_\infty \tau(\cdot,\cdot)$. We take $p=1$ and $\beta\in (0,1)$ arbitrarily close to 1 to stand for Sherman's quantity $\alpha$. Using Jensen's inequality, we could bound the right-hand side of the Maximal Inequality of  \citet{sherman1994maximal} by a universal constant times
$$
  \left(E \left[ K^{2}\left(\frac{\boldsymbol Z_j - \boldsymbol Z_l}{c_{max} n^{-\alpha}}\right)\tau^{2} (\widetilde{\boldsymbol U}_j, \widetilde{\boldsymbol U}_l)\right]\right)^{\beta/2}.
$$
By standard changes of variables and suitable integrability conditions, the power $\beta/2$ of the expectation in the last display is bounded by  a constant times $n^{-\alpha \beta q/2}$. Consequently, the uniform rate of the second $U-$processes obtained conditioning by $\widetilde{\boldsymbol U}_i, \widetilde{\boldsymbol U}_k$ and  $\widetilde{\boldsymbol U}_j, \widetilde{\boldsymbol U}_l$, respectively is $n^{-1}\times O_{\mathbb P} (n^{-\alpha \{2-q +\beta q/2\} })$. Since $\beta<1$ could be arbitrarily close to 1, we have $2-q +\beta q/2 > 0$, and, thus,  $n^{-1}\times O_{\mathbb P} (n^{-\alpha \{2-q +\beta q/2\}})= o_{\mathbb P} (n^{-1})$.

We consider now the four $U-$processes of order three obtained by conditioning on any subset of three of the four vectors $\widetilde{\boldsymbol U}_i$, $\widetilde{\boldsymbol U}_k$,  $\widetilde{\boldsymbol U}_j$ and $ \widetilde{\boldsymbol U}_l$. We start by conditioning on $(\widetilde{\boldsymbol U}_i,\widetilde{\boldsymbol U}_j, \widetilde{\boldsymbol U}_l)$ and  $(\widetilde{\boldsymbol U}_i,\widetilde{\boldsymbol U}_j, \widetilde{\boldsymbol U}_k)$, the other one being similar.
	\begin{align*}
		E[&( m_i - m_k)K_{h,ik}( m_j - m_l)K_{h,jl} \boldsymbol \Omega _{n,ij} \mid \widetilde{\boldsymbol U}_i,\widetilde{\boldsymbol U}_j, \widetilde{\boldsymbol U}_l] 	 \\
		&=   
		  E\left[( m_i - m_k)K_{h,ik}\mid \widetilde{\boldsymbol U}_i\right]( m_j - m_l)K_{h,jl} \boldsymbol \Omega _{n,ij} \\	    
		&= \left[m_i\left(f_z(\boldsymbol Z_i) + h^2\gamma_1(\boldsymbol Z_i)(1 + o_{\mathbb P }(1))\right) - \left(m_i f_z(\boldsymbol Z_i) + h^2\gamma_2(\boldsymbol Z_i)(1 + o_{\mathbb P }(1))\right)\right]( m_j - m_l)K_{h,jl} \boldsymbol \Omega _{n,ij}\\
		&= h^{2-q}\tau(\widetilde{\boldsymbol U}_i, \widetilde{\boldsymbol U}_j, \widetilde{\boldsymbol U}_l)h^qK_{h,jl}(1 + o_{\mathbb P }(1)). 	   
	\end{align*}
Now, we apply again the Maximal Inequality of \citet{sherman1994maximal}, page 448,  for the degenerate $U-$process given by the kernel $h^q K_{h,jl}  \tau (\widetilde{\boldsymbol U}_i,\widetilde{\boldsymbol U}_j, \widetilde{\boldsymbol U}_l)$, indexed by $h\in\mathcal H_n^{sc}$, with envelope $\|K\|_\infty \tau(\cdot,\cdot,\cdot)$. We take again $p=1$ and $\beta\in (0,1)$ arbitrarily close to 1 to stand for Sherman's quantity $\alpha$. Hence, using Jensen's inequality, we could bound the right-hand side of the Maximal Inequality of  \citet{sherman1994maximal} by a universal constant times
$$
  \left(E \left[ K^{2}\left(\frac{\boldsymbol Z_j - \boldsymbol Z_l}{c_{max} n^{-\alpha}}\right)\tau^{2} (\widetilde{\boldsymbol U}_i, \widetilde{\boldsymbol U}_j, \widetilde{\boldsymbol U}_l)\right]\right)^{\beta/2}.
$$
By standard changes of variables and suitable integrability conditions, the power $\beta/2$ of the expectation in the last display is bounded by  a constant times $n^{-\alpha \beta q/2}$. Consequently, the uniform rate of the $U-$processes obtained conditioning by $\widetilde{\boldsymbol U}_i, \widetilde{\boldsymbol U}_j,\widetilde{\boldsymbol U}_l$ and  $\widetilde{\boldsymbol U}_i,\widetilde{\boldsymbol U}_j, \widetilde{\boldsymbol U}_k$, respectively is $n^{-3/2}\times O_{\mathbb P} (n^{-\alpha \{2-q +\beta q/2\} })$. Since $1/2 + \alpha \{2-q +\beta q/2\} > 0$ under our assumptions $q<4$ and $\alpha \in (1/4, 1/q)$ we get that $n^{-3/2}\times O_{\mathbb P} (n^{-\alpha \{2-q +\beta q/2\} }) = o_{\mathbb P}(n^{-1})$.

In addition, we get by conditioning on $\widetilde{\boldsymbol U}_i,\widetilde{\boldsymbol U}_k, \widetilde{\boldsymbol U}_l$ and  $\widetilde{\boldsymbol U}_j,\widetilde{\boldsymbol U}_k, \widetilde{\boldsymbol U}_l$, the other one being similar, that	
	\begin{align*}
		E[( m_i - m_k)&K_{h,ik}( m_j - m_l)K_{h,jl} \boldsymbol \Omega _{n,ij}\mid \widetilde{\boldsymbol U}_i,\widetilde{\boldsymbol U}_k, \widetilde{\boldsymbol U}_l] 	 \\
		&=   	
		  ( m_i - m_k)K_{h,ik}E[( m_j - m_l)K_{h,jl} \boldsymbol \Omega _{n,ij}\mid \widetilde{\boldsymbol U}_i, \widetilde{\boldsymbol U}_l] 	 \\	
		&= 
		( m_i - m_k)K_{h,ik}
		\left(
		m_lf_z(\boldsymbol Z_l) \boldsymbol \Omega _{n,il}^Z E\left[ \boldsymbol \Omega _{n,ij}^X\mid {\boldsymbol X}_i\right] - 
	    m_lf_z(\boldsymbol Z_l) \boldsymbol \Omega _{n,il}^Z E\left[ \boldsymbol \Omega _{n,ij}^X\mid {\boldsymbol X}_i\right] + O_{\mathbb P} (h^2)
	    \right)	\\
		&= 
		h^{2-q}\tau(\widetilde{\boldsymbol U}_i, \widetilde{\boldsymbol U}_k, \widetilde{\boldsymbol U}_l)h^qK_{h,ik} (1 + O_{\mathbb P }(h^2))	    		   	     
	\end{align*}

Now, we apply again the Maximal Inequality of \citet{sherman1994maximal}, page 448,  for the degenerate $U-$process given by the kernel $h^q K_{h,ik}  \tau (\widetilde{\boldsymbol U}_i,\widetilde{\boldsymbol U}_k, \widetilde{\boldsymbol U}_l)$, indexed by $h\in\mathcal H_n^{sc}$, with envelope $\|K\|_\infty \tau(\cdot,\cdot,\cdot)$. We take again $p=1$ and $\beta\in (0,1)$ arbitrarily close to 1 to stand for Sherman's quantity $\alpha$. Hence, using Jensen's inequality, we could bound the right-hand side of the Maximal Inequality of  \citet{sherman1994maximal} by an universal constant times
$$
  \left(E \left[ K^{2}\left(\frac{\boldsymbol Z_i - \boldsymbol Z_k}{c_{max} n^{-\alpha}}\right)\tau^{2} (\widetilde{\boldsymbol U}_i, \widetilde{\boldsymbol U}_k, \widetilde{\boldsymbol U}_l)\right]\right)^{\beta/2}.
$$
By standard changes of variables and suitable integrability conditions, the power $\beta/2$ of the expectation in the last display is bounded by  a constant times $n^{-\alpha \beta q/2}$. Consequently, the uniform rate of the $U-$processes obtained conditioning by $\widetilde{\boldsymbol U}_i, \widetilde{\boldsymbol U}_k,\widetilde{\boldsymbol U}_l$ and  $\widetilde{\boldsymbol U}_j,\widetilde{\boldsymbol U}_k, \widetilde{\boldsymbol U}_l$, respectively is $n^{-3/2}\times O_{\mathbb P} (n^{-\alpha \{2-q +\beta q/2\} })$. Since $1/2 +\alpha \{2-q +\beta q/2\}> 0$ under our assumptions $q<4$ and $\alpha \in (1/4, 1/q)$ we get that $n^{-3/2}\times O_{\mathbb P} (n^{\alpha q \{1 - \beta/2\} }) = o_{\mathbb P}(n^{-1})$.	

Finally, we consider the remaining $U-$process of order four. This process is given by 
	\begin{align*}
		( m(\boldsymbol Z_i) - m(\boldsymbol Z_k))K_{h,ik}( m(\boldsymbol Z_j) - m(\boldsymbol Z_l))K_{h,jl}\boldsymbol \Omega _{n,ij}
		=h^{-2q}\tau(\widetilde{\boldsymbol U}_i, \widetilde{\boldsymbol U}_j, \widetilde{\boldsymbol U}_k, \widetilde{\boldsymbol U}_l)h^qK_{h,ik}h^qK_{h,jl}.
	\end{align*}

Now, we apply again the Maximal Inequality of \citet{sherman1994maximal}, page 448,  for the degenerate $U-$process given by the kernel $\tau(\widetilde{\boldsymbol U}_i, \widetilde{\boldsymbol U}_j, \widetilde{\boldsymbol U}_k, \widetilde{\boldsymbol U}_l)h^qK_{h,ik}h^qK_{h,jl}$, indexed by $h\in\mathcal H_n^{sc}$, with envelope $\|K\|_\infty^2 \tau(\cdot,\cdot,\cdot,\cdot)$. We take again $p=1$ and $\beta\in (0,1)$ arbitrarily close to 1 to stand for Sherman's quantity $\alpha$. 
Hence, using Jensen's inequality, we could bound the right-hand side of the Maximal Inequality of  \citet{sherman1994maximal} by an universal constant times
$$
  \left(E \left[ K^{2}\left(\frac{\boldsymbol Z_i - \boldsymbol Z_k}{c_{max} n^{-\alpha}}\right) K^{2}\left(\frac{\boldsymbol Z_j - \boldsymbol Z_l}{c_{max} n^{-\alpha}}\right)\tau^{2}(\widetilde{\boldsymbol U}_i,\widetilde{\boldsymbol U}_j, \widetilde{\boldsymbol U}_k, \widetilde{\boldsymbol U}_l)\right]\right)^{\beta/2}.
$$
By standard changes of variables and suitable integrability conditions, the power $\beta/2$ of the expectation in the last display is bounded by  a constant times $n^{-\alpha \beta q}$. Consequently, the uniform rate of the fourth order $U-$process is $n^{-2}\times O_{\mathbb P} (n^{\alpha q \{2 - \beta\} })$. Since $1> \alpha q$ under our assumptions $q<4$ and $\alpha \in (1/4, 1/q)$ we get that $n^{-2}\times O_{\mathbb P} (n^{\alpha q \{2 - \beta\} }) = o_{\mathbb P}(n^{-1})$.

From all the results it follows that 
	\begin{align*}
		\sup_{h\in\mathcal H^{sc}_n} \sup_{\boldsymbol d \in\mathcal D } | \widetilde A_{1,n}| = \sup_{h\in\mathcal H^{sc}_n} \sup_{\boldsymbol d \in\mathcal D }| \widetilde A_{1,n}(h)| = o_{\mathbb P}(n^{-1}).
	\end{align*}

In the next step we consider $\widetilde A_{2,n}$. We get that
	\begin{align*}
		&E[( m_i - m_k)K_{h,ik}( m_j - m_k)K_{h,jk}\boldsymbol \Omega _{n,ij}]
		=
		  E[( m_i - m_k)K_{h,ik}E[( m_j - m_k)K_{h,jk}\boldsymbol \Omega _{n,ij}\mid \widetilde{\boldsymbol U}_i, \widetilde{\boldsymbol U}_k]]\\
		&=
		  E\left[( m_i - m_k)K_{h,ik}
		  \left(
		  f_z(\boldsymbol Z_k)m_k\boldsymbol \Omega _{n,ik}^Z E\left[ \boldsymbol \Omega _{n,ij}^X\mid {\boldsymbol X}_i\right] - 
		  f_z(\boldsymbol Z_k)m_k\boldsymbol \Omega _{n,ik}^Z E\left[ \boldsymbol \Omega _{n,ij}^X\mid {\boldsymbol X}_i\right] + O_{\mathbb P}(h^2)\right)\right]	\\
		&=
		  E[( m_i - m_k)K_{h,ik}] O(h^2) = O(h^4).
	\end{align*}
This implies that $E[\widetilde A_{2,n}] = o_{\mathbb P} (n^{-1})$ uniformly with respect to $h\in\mathcal H_n^{sc}$. 	

We consider now the first order $U$--process of the Hoeffding decomposition for $\widetilde A_{2,n}$. As we symmetrized the kernel we need to consider the conditional expectations with respect to all four variables. By the same reasoning as for $E[\widetilde A_{2,n}]$ we get that 
	\begin{align*}
		&E[( m_i - m_k)K_{h,ik}( m_j - m_k)K_{h,jk}\boldsymbol \Omega _{n,ij} \mid \widetilde{\boldsymbol U}_i]\\
		&=
		  E[( m_i - m_k)K_{h,ik}E[( m_j - m_k)K_{h,jk}\boldsymbol \Omega _{n,ij}\mid \boldsymbol Z_k, \widetilde{\boldsymbol U}_i]\mid \widetilde{\boldsymbol U}_i]\\
		&=
		  E\left[
		  ( m_i - m_k)K_{h,ik}
		  \left(
		  m_k f_z(\boldsymbol Z_k) \boldsymbol \Omega _{n,ik}^Z E\left[ \boldsymbol \Omega _{n,ij}^X\mid {\boldsymbol X}_i\right] - 
		  m_k f_z(\boldsymbol Z_k) \boldsymbol \Omega _{n,ik}^Z E\left[ \boldsymbol \Omega _{n,ij}^X\mid {\boldsymbol X}_i\right] + O_{\mathbb P}(h^2)\right)\mid \widetilde{\boldsymbol U}_i\right]	\\
		&=
		  \left(m_if_z(\boldsymbol Z_i) - m_if_z(\boldsymbol Z_i) + O_{\mathbb P}(h^2)\right)O_{\mathbb P}(h^2) = O_{\mathbb P}(h^4).
	\end{align*}
Note that the reasoning when conditioning on $\widetilde{\boldsymbol U}_j$ and $\widetilde{\boldsymbol U}_k$ is the same. Therefore, we get that the first order $U$--processes of the Hoeffding decompositions for $\widetilde A_{2,n}$ are of order $o_{\mathbb P}(n^{-1})$.

It is easy to see that the second and third order $U$--processes of the Hoeffding decomposition for $\widetilde A_{2,n}$ are of order $o_{\mathbb P}(n^{-1})$ if we apply the Maximal Inequality of \citet{sherman1994maximal}. From all the results it follows that 
	\begin{align*}
		\sup_{h\in\mathcal H^{sc}_n}\sup_{\boldsymbol d \in\mathcal D } | \widetilde A_{2,n}| = \sup_{h\in\mathcal H^{sc}_n}\sup_{\boldsymbol d \in\mathcal D } | \widetilde A_{2,n}(h)| = o_{\mathbb P}(n^{-1}). 
	\end{align*} 
As it follows by the same reasoning as for $\widetilde A_{2,n}$ that
	\begin{align*}
		\sup_{h\in\mathcal H^{sc}_n}\sup_{\boldsymbol d \in\mathcal D } | \widetilde A_{3,n}| = \sup_{h\in\mathcal H^{sc}_n} \sup_{\boldsymbol d \in\mathcal D }| \widetilde A_{3,n}(h)| = o_{\mathbb P}(n^{-1})
	\end{align*}	
we omit the details here. Finally, we get by standard change of variables that
	\begin{align*}
		E[\widetilde A_{4,n}] = n^{-2} E[(m_i - m_j )^2K_{h,ij}^2\boldsymbol \Omega _{n,ij}] = O(n^{-2}n^{\alpha q})  = o(n^{-1})
	\end{align*}
as well as
	\begin{align*}
		 n^{-2} E[(m_i - m_j )^2K_{h,ij}^2\boldsymbol \Omega _{n,ij} \mid \widetilde{\boldsymbol U}_i] = O_{\mathbb{P}}(n^{-2}n^{\alpha q})  
	\end{align*}	
and
	\begin{align*}
		n^{-2} E[(m_i - m_j )^2K_{h,ij}^2\boldsymbol \Omega _{n,ij} \mid \widetilde{\boldsymbol U}_j] = O_{\mathbb{P}}(n^{-2}n^{\alpha q}). 
	\end{align*}
Using the Hoeffding decomposition and applying Corollary 4 of \citet{sherman1994maximal}, we deduce that
	\begin{align*}
		n^2h^{2q} (\widetilde A_{4,n} - E[\widetilde A_{4,n}]) = O_{\mathbb{P}}(n^{-1}) + O_{\mathbb{P}}(n^{-1/2}n^{-\alpha q}) 
	\end{align*}
uniformly with respect to $h$. Deduce that 

	\begin{align*}
		\sup_{h\in\mathcal H^{sc}_n}\sup_{\boldsymbol d \in\mathcal D }|\widetilde A_{4,n}| = O_{\mathbb{P}}(n^{-3}n^{2\alpha q}) + O_{\mathbb{P}}(n^{-5/2}n^{\alpha q}) + O_{\mathbb{P}}(n^{-2}n^{\alpha q}) = o_{\mathbb{P}}(n^{-1}). 
	\end{align*}
With all these results \eqref{rate_star2} and, in particular, \eqref{second_part} follow.	

We know from Lemma \ref{yn_hatt_5} that 	
	\begin{align*}
		\Big\|\boldsymbol\Omega_n ^{1/2} n^{-2}  \left(\boldsymbol{\varepsilon} \boldsymbol{f_z}\right)_n\Big\| = O_{\mathbb{P}}(n^{-3/2}) \\
	\end{align*}	
such that from all these results the statement follows.\\
\end{proofof}

\begin{lem}\label{yn_hatt_7}
Assume the conditions of Proposition \ref{AN_prop} hold true. Then,
	\begin{align*}
		\sup_{h\in\mathcal{H}^{sc}_n} \sup_{\boldsymbol d \in\mathcal D }  
		\left\| \boldsymbol\Omega_n ^{1/2} n^{-1}  \left(\boldsymbol{\widehat{\varepsilon}_{|z}\widehat{f}_z}\right)_n \right\|  = O_{\mathbb{P}}(n^{-1/2}).
	\end{align*}
\end{lem}

\begin{proofof}{Lemma \ref{yn_hatt_7}} ~~\\

By the arguments used for Lemma \ref{yn_hatt_5}, it suffices to consider  $\boldsymbol d = {\rm diag}(d_U,\ldots,d_U)$. Moreover, for simpler notation,  we omit the argument $\boldsymbol d$ in $\boldsymbol \Omega_n(\boldsymbol d) $. 
We get that 
	\begin{align*}
		\Big\|\boldsymbol\Omega_n ^{1/2} n^{-1}  &\left(\boldsymbol{\widehat{\varepsilon}_{|z}\widehat{f}_z}\right)_n \Big\|^2\\
		&=
		\frac{1}{n^2} \sum_{1\leq i\neq j \leq n} \left(\boldsymbol{\widehat{\varepsilon}_{|z}\widehat{f}_z}\right)_{n,i}\left(\boldsymbol{\widehat{\varepsilon}_{|z}\widehat{f}_z}\right)_{n,j}\boldsymbol\Omega_{n,ij}
		+
		\frac{1}{n^2} \sum_{i = 1}^{n} \left(\boldsymbol{\widehat{\varepsilon}_{|z}\widehat{f}_z}\right)_{n,i}^2 	\\
		&=
		A_n + B_n.
	\end{align*}
It is easy to check that $\sup_{h\in\mathcal{H}^{sc}_n}  B_n =  o_{\mathbb{P}}(n^{-1})$. Furthermore, we get that 
	\begin{align*}
		A_n = 
		\frac{1}{n^2} \sum_{1\leq i\neq j \leq n} \left[\frac{1}{n}\sum\limits_{1 \leq k \leq n, k\neq i} \varepsilon_k K_{h,ik} \right]
		\left[\frac{1}{n}\sum\limits_{1 \leq l \leq n, l\neq j} \varepsilon_l K_{h,jl} \right]\boldsymbol\Omega_{n,ij}.
	\end{align*}
We show in the following that 
	\begin{align}\label{rate_star3}
		\sup_{h\in\mathcal H^{sc}_n} \sup_{\boldsymbol d \in\mathcal D }| A_n| = \sup_{h\in\mathcal H^{sc}_n} \sup_{\boldsymbol d \in\mathcal D }| A_n(h)| = O_{\mathbb P}(n^{-1}).
	\end{align}
For this purpose, we define $(n)_k = n(n-1) \ldots (n-k+1)$ and decompose $A_n(h)$ into a sum of four $U$-- processes, i.e. 
	\begin{align*}
		A_n(h) = \frac{(n-1)_3}{n^3}A_{1,n}(h) + \frac{(n-1)_2}{n^2}A_{2,n}(h) + 2\frac{(n-1)_2}{n^2}A_{3,n}(h) + \frac{n-1}{n}A_{4,n}(h),
	\end{align*}
where
	\begin{align*}
		A_{1,n} = A_{1,n}(h) &= \frac{1}{(n)_4} \sum_{1 \leq i\neq j \neq k\neq l \leq n} \varepsilon_k K_{h,ik}  \varepsilon_l K_{h,jl} \boldsymbol \Omega _{n,ij}\\ 
		A_{2,n} = A_{2,n}(h) &= \frac{1}{n(n)_3}\sum_{1 \leq i\neq j \neq k\leq n} \varepsilon_k^2 K_{h,ik} K_{h,jk} \boldsymbol \Omega _{n,ij}\\ 	 
		A_{3,n} = A_{3,n}(h) &= \frac{1}{n(n)_3}\sum_{1 \leq i\neq j \neq l\leq n} \varepsilon_j K_{h,ij} \varepsilon_l K_{h,jl} \boldsymbol \Omega _{n,ij}\\ 	 
		\text{and} \quad 
		A_{4,n} = A_{4,n}(h) &= \frac{1}{n^2(n)_2}\sum_{1 \leq i\neq j \leq n} \varepsilon_i  \varepsilon_j K_{h,ij}^2 \boldsymbol \Omega _{n,ij}.
	\end{align*}

For each of these $U$--processes we compute the mean and use the Hoeffding decomposition. The kernels of $A_{1,n}$, $A_{2,n}$ and $A_{3,n}$ are not symmetric in their arguments. However, we could apply the usual symmetrization idea. Thus, by abuse, we will proceed as if the kernels of the $U-$statistics we handle are symmetric. Here, $\boldsymbol U_i = \left(Y_i,\boldsymbol X_i^T, \boldsymbol Z_i^T\right)^T$. 

In addition, we have that the kernels of $A_{1,n}$, $A_{2,n}$, $A_{3,n}$ and $A_{4,n}$ are Euclidean for a squared integrable envelope. See Lemma 22 in \citet{nolan1987u} and Lemma 2.14 in \citet{pakes1989simulation}. Therefore, we can in the following repeatedly apply Corollary 7 and the Maximal Inequality of \citet{sherman1994maximal}. All remainder terms are controlled by Assumption \ref{ass_con}.2.

We start by considering $A_{1,n}$. Recall that by assumption $E\left[\varepsilon_k\mid \boldsymbol X_k, \boldsymbol Z_k\right] = E\left[\varepsilon_l\mid \boldsymbol X_l, \boldsymbol Z_l\right] = 0$. Therefore, we get that $E\left[A_{1,n}\right] = 0$ as well as 
$$
	E\left[\varepsilon_k K_{h,ik}  \varepsilon_l K_{h,jl} \boldsymbol \Omega _{n,ij}\mid \boldsymbol U_p, p\in \{i,j,k,l\}\right] = 0.
$$  
Note that we need to consider the conditional expectations with respect to all four variables for the first order $U$--process of the Hoeffding decomposition of $A_{1,n}$ as we symmetrized the kernel. It follows from the results that the first order $U$--process of the Hoeffding decomposition of $A_{1,n}$ is $0$.

We consider now the six second order $U-$processes of the Hoeffding decomposition of $A_{1,n}$. There are two types of such processes. First, the ones that are $0$. This is the case when conditioning by the pairs $(\boldsymbol U_i, \boldsymbol U_l)$, $(\boldsymbol U_i, \boldsymbol U_k)$, $(\boldsymbol U_j, \boldsymbol U_l)$, $(\boldsymbol U_j, \boldsymbol U_k)$ and $(\boldsymbol U_i, \boldsymbol U_j)$. The second case occurs when conditioning on $(\boldsymbol U_k, \boldsymbol U_l)$. We get that 
	\begin{align*}
		E[\varepsilon_k K_{h,ik}  \varepsilon_l K_{h,jl} &\boldsymbol \Omega _{n,ij}\mid \boldsymbol U_k, \boldsymbol U_l] \\
		&= 
		\varepsilon_k \varepsilon_l E[ K_{h,ik} K_{h,jl} \boldsymbol \Omega _{n,ij}^X \boldsymbol \Omega _{n,ij}^Z \mid \boldsymbol Z_k, \boldsymbol Z_l]	\\	
		&= 
		\varepsilon_k \varepsilon_l E[ E[K_{h,ik} \boldsymbol \Omega _{n,ij}^X \boldsymbol \Omega _{n,ij}^Z\mid \boldsymbol Z_k, \boldsymbol Z_j] K_{h,jl}  \mid \boldsymbol Z_k, \boldsymbol Z_l]	\\
		&= 
		\varepsilon_k \varepsilon_l E[ \left(f_z(\boldsymbol Z_k)E[ \boldsymbol \Omega _{n,ij}^X \mid  \boldsymbol Z_j] \boldsymbol \Omega _{n,kj}^Z + O_{\mathbb{P}}(h^2)\right) K_{h,jl}  \mid \boldsymbol Z_k, \boldsymbol Z_l]\\
		&= 
		\varepsilon_k \varepsilon_l f_z(\boldsymbol Z_k) E[ f_z(\boldsymbol Z_l) E[ \boldsymbol \Omega _{n,ij}^X] \boldsymbol \Omega _{n,kl}^Z + O_{\mathbb{P}}(h^2)\mid \boldsymbol Z_k, \boldsymbol Z_l]\\
		&\quad + \varepsilon_k \varepsilon_l O_{\mathbb{P}}(h^2) \left(f_z(\boldsymbol Z_l)  + O_{\mathbb{P}}(h^2)\right) \\
		&= 
		\varepsilon_k \varepsilon_l f_z(\boldsymbol Z_k) f_z(\boldsymbol Z_l) E[ \boldsymbol \Omega _{n,ij}^X ] \boldsymbol \Omega _{n,kl}^Z 
		+ \varepsilon_k \varepsilon_l f_z(\boldsymbol Z_k)  O_{\mathbb{P}}(h^2) 	\\
		& \quad + \varepsilon_k \varepsilon_l f_z(\boldsymbol Z_l)  O_{\mathbb{P}}(h^2)		
		+ \varepsilon_k \varepsilon_l O_{\mathbb{P}}(h^4).
	\end{align*}
Therefore, it follows together with Corollary 4 of \citet{sherman1994maximal} that the second order $U-$processes of the Hoeffding decomposition of $A_{1,n}$ are of order $O_{\mathbb{P}}(n^{-1})$ uniformly with respect to $h$ and $\boldsymbol d$.
	
We consider now the four $U-$processes of order three obtained by conditioning on any subset of three of the four vectors $\boldsymbol U_i$, $\boldsymbol U_k$,  $\boldsymbol U_j$ and $ \boldsymbol U_l$. There are two types of such processes. First, the ones that are $0$. This is the case when conditioning by $(\boldsymbol U_i, \boldsymbol U_j, \boldsymbol U_k)$ or $(\boldsymbol U_i, \boldsymbol U_j, \boldsymbol U_l)$. The second case occurs when conditioning on $(\boldsymbol U_i, \boldsymbol U_k, \boldsymbol U_l)$ or $(\boldsymbol U_j, \boldsymbol U_k, \boldsymbol U_l)$, the other one being similar. We get that
	\begin{align*}
		E[\varepsilon_k K_{h,ik}  \varepsilon_l K_{h,jl} &\boldsymbol \Omega _{n,ij}\mid \boldsymbol U_i, \boldsymbol U_k, \boldsymbol U_l] \\
		&= 
		\varepsilon_k \varepsilon_l  K_{h,ik}  E[ K_{h,jl} 	\boldsymbol \Omega _{n,ij}^X \boldsymbol \Omega _{n,ij}^Z\mid \boldsymbol U_i, \boldsymbol Z_l] \\	
		&= 
		\varepsilon_k \varepsilon_l  K_{h,ik} f_z(\boldsymbol Z_l) \boldsymbol \Omega _{n,il}^Z E[ \boldsymbol \Omega _{n,ij}^X \mid \boldsymbol X_i] \\
		&\quad + \varepsilon_k \varepsilon_l  K_{h,ik} 	O_{\mathbb{P}}(h^2)	\\	
		&= 
		h^{-q} h^q K_{h,ik}  \tau_1(\boldsymbol U_i, \boldsymbol U_k, \boldsymbol U_l) \\
		&\quad + \tau_2(\boldsymbol U_k, \boldsymbol U_l) h^{-q} h^qK_{h,ik} 	O_{\mathbb{P}}(h^2).		
	\end{align*}
Now, we apply the Maximal Inequality of \citet{sherman1994maximal}, page 448,  for the degenerate $U-$process given by the kernel $h^q K_{h,ik}  \tau_1(\boldsymbol U_i, \boldsymbol U_k, \boldsymbol U_l)$, indexed by $h\in\mathcal H_n^{sc}$, with envelope $\|K\|_\infty \tau_1(\cdot,\cdot,\cdot)$. The reasoning for $h^q K_{h,ik}  \tau_2(\boldsymbol U_k, \boldsymbol U_l)$ is the same.  (Herein, $\|\cdot\|_\infty $ denotes the uniform norm.) We take $p=1$ and $\beta\in (0,1)$ arbitrarily close to 1 to stand for Sherman's quantity $\alpha$. Since $K(\cdot)$ is of bounded variation and symmetric, without loss of generality we could consider 
that $K(\cdot)$ is nonincreasing on $[0,\infty)$. In this case, $0\leq K(\cdot/h)\leq K(\cdot/\overline h)$ with $\overline h = \sup \mathcal H^{sc}_n = : c_{max} n^{-\alpha}$.
Hence, using Jensen's inequality, we could bound the right-hand side of the Maximal Inequality of  \citet{sherman1994maximal} by a universal constant times
$$
	\left(E \left[ K^{2}\left(\frac{\boldsymbol Z_i - \boldsymbol Z_k}{c_{max} n^{-\alpha}}\right)\tau_1^{2} (\boldsymbol U_i, \boldsymbol U_k, \boldsymbol U_l)\right]\right)^{\beta/2}.
$$
By standard changes of variables and suitable integrability conditions, the power $\beta/2$ of the expectation in the last display is bounded by  a constant times $n^{-\alpha \beta q/2}$. Consequently, the uniform rate of the second $U-$processes obtained conditioning by $\boldsymbol U_i, \boldsymbol U_k, \boldsymbol U_l$ and  $\boldsymbol U_j,\boldsymbol U_k, \boldsymbol U_l$, respectively is $n^{-3/2}\times O_{\mathbb P} (n^{\alpha q\{1-\beta /2\} })$. As $1/2 - \alpha q (1- \beta /2) > 0 $ under our assumptions we get that $n^{-3/2}\times O_{\mathbb P} (n^{\alpha q\{1 - \beta /2\}})= o_{\mathbb P} (n^{-1})$ such that the third order $U-$processes of the Hoeffding decomposition of $A_{1,n}$ are of order $o_{\mathbb P} (n^{-1})$.

Finally, we consider the remaining $U-$process of order four. This process is given by 
	\begin{align*}
		\varepsilon_k K_{h,ik}  \varepsilon_l K_{h,jl} \boldsymbol \Omega _{n,ij}=h^{-2q}\tau_3(\boldsymbol U_i, \boldsymbol U_j,\boldsymbol U_k, \boldsymbol U_l) h^{2q}K_{h,ik}K_{h,jl}.
	\end{align*}

Now, we apply again the Maximal Inequality of \citet{sherman1994maximal}, page 448,  for the degenerate $U-$process given by the kernel $\tau_3(\boldsymbol U_i, \boldsymbol U_j,\boldsymbol U_k, \boldsymbol U_l)h^qK_{h,ik}h^qK_{h,jl}$, indexed by $h\in\mathcal H_n^{sc}$, with envelope $\|K\|_\infty^2 \tau(\cdot,\cdot)$. We take again $p=1$ and $\beta\in (0,1)$ arbitrarily close to 1 to stand for Sherman's quantity $\alpha$. 
Hence, using Jensen's inequality, we could bound the right-hand side of the Maximal Inequality of  \citet{sherman1994maximal} by an universal constant times
$$
	\left(E \left[ K^{2}\left(\frac{\boldsymbol Z_i - \boldsymbol Z_k}{c_{max} n^{-\alpha}}\right) K^{2}\left(\frac{\boldsymbol Z_j - \boldsymbol Z_l}{c_{max} n^{-\alpha}}\right)\tau_3^{2}(\boldsymbol U_i,\boldsymbol U_j,\boldsymbol U_k,\boldsymbol U_l)\right]\right)^{\beta/2}.
$$
By standard changes of variables and suitable integrability conditions, the power $\beta/2$ of the expectation in the last display is bounded by  a constant times $n^{-\alpha \beta q}$. Consequently, the uniform rate of the fourth order $U-$process is $n^{-2}\times O_{\mathbb P} (n^{\alpha q \{2 - \beta\} })$. Since $1> \alpha q$ under our assumptions we get that $n^{-2}\times O_{\mathbb P} (n^{\alpha q \{2 - \beta\} }) = o_{\mathbb P}(n^{-1})$. 

From all the results it follows that 
	\begin{align*}
		\sup_{h\in\mathcal H^{sc}_n} \sup_{\boldsymbol d \in\mathcal D } | A_{1,n}| = \sup_{h\in\mathcal H^{sc}_n} \sup_{\boldsymbol d \in\mathcal D }| A_{1,n}(h)| = O_{\mathbb P}(n^{-1}).
	\end{align*}

In the next step we consider $A_{2,n}$. We get that 
	\begin{align*}
		n E\left[A_{2,n}\right] 
		&= E\left[\varepsilon_k^2 K_{h,ik} K_{h,jk} \boldsymbol \Omega _{n,ij}\right]\\
		&= E\left[E\left[\varepsilon_k^2\mid \boldsymbol X_k, \boldsymbol Z_k\right] K_{h,ik} K_{h,jk} \boldsymbol \Omega _{n,ij}\right] \\		
		&= E\left[\sigma^2 \left(\boldsymbol X_k, \boldsymbol Z_k\right) E\left[K_{h,ik} \boldsymbol \Omega _{n,ij}^X\boldsymbol \Omega _{n,ij}^Z \mid \boldsymbol Z_j, \boldsymbol Z_k \right]K_{h,jk}\right] \\			
		&= E\left[\sigma^2 \left(\boldsymbol X_k, \boldsymbol Z_k\right) f_z(\boldsymbol Z_k)\boldsymbol \Omega _{n,kj}^Z E\left[ \boldsymbol \Omega _{n,ij}^X \mid \boldsymbol Z_j \right]K_{h,jk}\right] \\		
		&\quad +  E\left[\sigma^2 \left(\boldsymbol X_k, \boldsymbol Z_k\right) K_{h,jk}\right] O(h^2)\\
		&= E\left[\sigma^2 \left(\boldsymbol X_k, \boldsymbol Z_k\right) f_z(\boldsymbol Z_k) E\left[\boldsymbol \Omega _{n,kj}^Z E\left[ \boldsymbol \Omega _{n,ij}^X \mid \boldsymbol Z_j \right]K_{h,jk}\mid \boldsymbol Z_k\right]\right] \\		
		&\quad +  E\left[\sigma^2 \left(\boldsymbol X_k, \boldsymbol Z_k\right) E\left[K_{h,jk}\mid \boldsymbol Z_k\right]\right] O(h^2)\\
		&= E\left[\sigma^2 \left(\boldsymbol X_k, \boldsymbol Z_k\right) f_z(\boldsymbol Z_k)^2  E\left[ \boldsymbol \Omega _{n,ij}^X \right]\right] \\		
		&\quad + 2 E\left[\sigma^2 \left(\boldsymbol X_k, \boldsymbol Z_k\right) f_z(\boldsymbol Z_k)\right] O(h^2)		\\		
		&\quad +  E\left[\sigma^2 \left(\boldsymbol X_k, \boldsymbol Z_k\right)\right] O(h^4).				
	\end{align*}
Therefore, $E\left[A_{2,n}\right] = O(n^{-1})$.

In addition, we get by a similar reasoning that the first, second and third order $U$--processes of the Hoeffding decomposition of $A_{2,n}$ are of order $o_{\mathbb P}(n^{-1})$ if we apply the Maximal Inequality of \citet{sherman1994maximal}. The details are omitted. From all the results it follows that 
	\begin{align*}
		\sup_{h\in\mathcal H^{sc}_n}\sup_{\boldsymbol d \in\mathcal D } | A_{2,n}| = \sup_{h\in\mathcal H^{sc}_n}\sup_{\boldsymbol d \in\mathcal D } | A_{2,n}(h)| = O_{\mathbb P}(n^{-1}).
	\end{align*}

In the next step we consider $A_{3,n}$. We get that $E\left[A_{3,n}\right] = 0$ as well as 
	$$
		E\left[\varepsilon_j K_{h,ij} \varepsilon_l K_{h,jl} \boldsymbol \Omega _{n,ij}\mid \boldsymbol U_p, p\in \{i,j,l\}\right] = 0.
	$$  
In addition, it is easy to see that the second and third order $U$--processes of the Hoeffding decomposition of $A_{3,n}$ are of order $o_{\mathbb P}(n^{-1})$ if we apply the Maximal Inequality of \citet{sherman1994maximal}. From all the results it follows that 
		\begin{align*}
			\sup_{h\in\mathcal H^{sc}_n}\sup_{\boldsymbol d \in\mathcal D } | A_{3,n}| = \sup_{h\in\mathcal H^{sc}_n}\sup_{\boldsymbol d \in\mathcal D } | A_{3,n}(h)| = o_{\mathbb P}(n^{-1}).
		\end{align*}

Finally, we get that $E\left[A_{4,n}\right] = 0$ as well as 
$$
	E\left[\varepsilon_i  \varepsilon_j K_{h,ij}^2 \boldsymbol \Omega _{n,ij}\mid \boldsymbol U_p, p\in \{i,j\}\right] = 0.
$$   
In addition, it is easy to see that the second order $U$--process of the Hoeffding decomposition of $A_{4,n}$ is of order $o_{\mathbb P}(n^{-1})$ if we apply the Maximal Inequality of \citet{sherman1994maximal}. 
Deduce that 
	\begin{align*}
		\sup_{h\in\mathcal H^{sc}_n}\sup_{\boldsymbol d \in\mathcal D }|A_{4,n}| = o_{\mathbb{P}}(n^{-1}). 
	\end{align*}
From all the results \eqref{rate_star3} follows and, therefore, the statement.	\\
\end{proofof}	

\clearpage
\setcounter{section}{3} 
\subsection*{Appendix C: Additional simulation results}	
\addcontentsline{toc}{subsection}{Appendix C: Additional simulation results}

	\begin{table}[H]
		\caption{\textit{Bias and Standard Deviation of the estimators for $\lambda$ and $\beta$ in Model 2. } }
		\begin{tabular}{@{}lcd{3.5}d{3.5}d{3.5}d{3.5}d{3.5}d{3.5}}
    \toprule\midrule
    & $s$   & \multicolumn{3}{c}{Bias} &\multicolumn{3}{c}{St. dev.}\\\cmidrule(lr){3-5} \cmidrule(lr){6-8} $n$ &  & \mc{250} & \mc{500} & \mc{1000} & \mc{250} & \mc{500} & \mc{1000} \\ 
   \midrule
$\lambda$ estimator &  &  &  &  &  &  &  \\ 
   \midrule
  SmoothMD with $\gamma$ & $G_n$    & 0.006   & 0.004 & 0.002   & 0.094 & 0.066  & 0.047 \\ 
  SmoothMD without $\gamma$ & $G_n$ & 0.006   & 0.004 & 0.002   & 0.09  & 0.063  & 0.045 \\ 
  NL2SLS                    & $G_n$ & -0.0003 & 0.041 & -0.001  & 0.059 & 0.041  & 0.028 \\ 
   \midrule
$\beta$ estimator &  &  &  &  &  &  &  \\ 
   \midrule
  SmoothMD with $\gamma$ & $G_n$    & 0.025 & 0.014  & 0.007  & 0.182 & 0.125 & 0.089\\ 
  SmoothMD without $\gamma$ & $G_n$ & 0.024 & 0.013  & 0.007  & 0.173 & 0.119 & 0.084 \\ 
  NL2SLS                    & $G_n$ & 0.005 & 0.003  & 0.0003 & 0.109 & 0.074 & 0.051 \\  
  \midrule\bottomrule
\end{tabular}
		\vskip 0.3cm
		\textit{Notes: For the SmoothMD estimators, $h \propto n^{-1/3.5}$. The components of $\boldsymbol d$ are set equal to the componentwise standard deviations for all variables. The grid for $\lambda$ is $[\lambda_0-0.8,\lambda_0+0.8]$. 2000 Monte Carlo samples were used for all simulations. }
		\label{Bias_Std_Model2}
	\end{table}

	\begin{table}[H]
		\caption{\textit{Bias and Standard Deviation of the estimators for $\lambda$ and $\beta$ in Model 3. } }
		\begin{tabular}{@{}lcd{3.5}d{3.5}d{3.5}d{3.5}d{3.5}d{3.5}}
    \toprule\midrule
    & $s$   & \multicolumn{3}{c}{Bias} &\multicolumn{3}{c}{St. dev.}\\\cmidrule(lr){3-5} \cmidrule(lr){6-8} $n$ &  & \mc{250} & \mc{500} & \mc{1000} & \mc{250} & \mc{500} & \mc{1000} \\ 
   \midrule
$\lambda$ estimator &  &  &  &  &  &  &  \\ 
   \midrule
  SmoothMD with $\gamma$ & $G_n$    & -0.002  & 0.002 &  -0.003  & 0.146 & 0.102  & 0.073 \\ 
  SmoothMD without $\gamma$ & $G_n$ & -0.003  & 0.002 &  -0.003  & 0.145 & 0.101  & 0.072 \\ 
  NL2SLS                    & $G_n$ & -0.003  & 0.002 &  -0.002  & 0.123 & 0.086  & 0.06 \\ 
   \midrule
$\beta$ estimator &  &  &  &  &  &  &  \\ 
   \midrule
  SmoothMD with $\gamma$ & $G_n$    & 0.016 & 0.004  & 0.007  & 0.171 & 0.118 & 0.084\\ 
  SmoothMD without $\gamma$ & $G_n$ & 0.017 & 0.005  & 0.007  & 0.17  & 0.117 & 0.084 \\ 
  NL2SLS                    & $G_n$ & 0.013 & 0.002  & 0.004  & 0.145 & 0.099 & 0.07\\  
   \midrule\bottomrule
\end{tabular}
		\vskip 0.3cm
		\textit{Notes: For the SmoothMD estimators, $h \propto n^{-1/3.5}$. The components of $\boldsymbol d$ are set equal to the componentwise standard deviations for all variables. The grid for $\lambda$ is $[\lambda_0-0.8,\lambda_0+0.8]$. 2000 Monte Carlo samples were used for all simulations. }
		\label{Bias_Std_Model3}
	\end{table}

	\begin{table}[H]
		\caption{\textit{Empirical Level for Z-Tests of the estimators for $\lambda$ and $\beta$ in Model 1. } }
\begin{tabular}{@{}lcd{3.5}d{3.5}d{3.5}d{3.5}d{3.5}d{3.5}}
    \toprule\midrule& $s$   & \multicolumn{3}{c}{5\% level} &\multicolumn{3}{c}{10\% level}\\\cmidrule(lr){3-5} \cmidrule(lr){6-8} $n$ &  & \mc{250} & \mc{500} & \mc{1000} & \mc{250} & \mc{500} & \mc{1000} \\ 
   \midrule
Test for $\lambda$  &  &  &  &  &  &  &  \\ 
   \midrule
SmoothMD with $\gamma$     & $G_n$ & 5.75 & 6.0  & 4.55 & 10.2  & 11.15 & 10.75 \\ 
SmoothMD* with $\gamma$    & $G_n$ & 5.95 & 6.35 & 4.65 & 10.75 & 11.35 & 11.1  \\ 
SmoothMD without $\gamma$  & $G_n$ & 6.45 & 6.6  & 5.4  & 11.1  & 12.1  & 10.9  \\ 
  NL2SLS                   & $G_n$ & 9.25 & 7.75 & 5.95 & 15.1  & 13.9  & 11.55 \\ 
   \midrule
Test for $\beta$ &  &  &  &  &  &  &  \\ 
   \midrule
SmoothMD with $\gamma$     & $G_n$ & 5.85 & 4.5  & 3.85 & 10.25 & 9.15  & 7.9  \\ 
SmoothMD* with $\gamma$    & $G_n$ & 6.3  & 5.15 & 4.3  & 11.55 & 10.65 & 8.55 \\ 
SmoothMD without $\gamma$  & $G_n$ & 6.2  & 4.7  & 3.95 & 10.65 & 9.75  & 7.9  \\ 
  NL2SLS                   & $G_n$ & 7.65 & 6.1  & 4.45 & 12.85 & 12.15 &  8.6 \\ 
\midrule\bottomrule
\end{tabular}

		\vskip 0.3cm
		\textit{Notes: For the SmoothMD estimators, $h \propto n^{-1/3.5}$. The components of $\boldsymbol d$ are set equal to the componentwise standard deviations for all variables. The variances are estimated by the Eiker-White variance estimator. For SmoothMD* the additional variance part due to the estimation of $\boldsymbol{\eta}$ is not taken into account. For SmoothMD the additional variance part is taken into account. 2000 Monte Carlo samples were used for all simulations.}
		\label{Emp_LevZ_Model1}
	\end{table}		
	
	\begin{table}[H]
		\caption{\textit{Empirical Level for Z-Tests of the estimators for $\lambda$ and $\beta$ in Model 2. } }
\begin{tabular}{@{}lcd{3.5}d{3.5}d{3.5}d{3.5}d{3.5}d{3.5}}
    \toprule\midrule& $s$   & \multicolumn{3}{c}{5\% level} &\multicolumn{3}{c}{10\% level}\\\cmidrule(lr){3-5} \cmidrule(lr){6-8} $n$ &  & \mc{250} & \mc{500} & \mc{1000} & \mc{250} & \mc{500} & \mc{1000} \\ 
   \midrule
Test for $\lambda$  &  &  &  &  &  &  &  \\ 
   \midrule
SmoothMD with $\gamma$     & $G_n$ & 7.3 & 5.85 & 5.5  & 12.7 & 10.5  & 10.25 \\ 
SmoothMD* with $\gamma$    & $G_n$ & 6.7 & 5.85 & 5.3  & 12.5 & 10.15 & 10.1  \\ 
SmoothMD without $\gamma$  & $G_n$ & 6.7 & 5.65 & 5.1  & 12.1 & 10.25 & 9.95  \\ 
  NL2SLS                   & $G_n$ & 6.2 & 5.45 & 4.8  & 12.3 & 10.25 & 10    \\ 
   \midrule
Test for $\beta$ &  &  &  &  &  &  &  \\ 
   \midrule
SmoothMD with $\gamma$     & $G_n$ & 7.15 & 5.25 & 5.55 & 11.75 & 9.95  & 10.35 \\ 
SmoothMD* with $\gamma$    & $G_n$ & 6.95 & 5.15 & 5.5  & 11.45 & 9.75  & 10.2  \\ 
SmoothMD without $\gamma$  & $G_n$ & 6.3  & 5.25 & 4.9  & 11.25 & 9.9   & 10.1  \\ 
  NL2SLS                   & $G_n$ & 6.55 & 4.95 & 5.1  & 12.25 & 10.95 & 9.75  \\ 
\midrule\bottomrule
\end{tabular}

		\vskip 0.3cm
		\textit{Notes: For the SmoothMD estimators, $h \propto n^{-1/3.5}$. The components of $\boldsymbol d$ are set equal to the componentwise standard deviations for all variables. The variances are estimated by the Eiker-White variance estimator. For SmoothMD* the additional variance part due to the estimation of $\boldsymbol{\eta}$ is not taken into account. For SmoothMD the additional variance part is taken into account. 2000 Monte Carlo samples were used for all simulations.}
		\label{Emp_LevZ_Model2}
	\end{table}		

	\begin{table}[H]
		\caption{\textit{Empirical Level for Z-Tests of the estimators for $\lambda$ and $\beta$ in Model 3. } }
\begin{tabular}{@{}lcd{3.5}d{3.5}d{3.5}d{3.5}d{3.5}d{3.5}}
    \toprule\midrule& $s$   & \multicolumn{3}{c}{5\% level} &\multicolumn{3}{c}{10\% level}\\\cmidrule(lr){3-5} \cmidrule(lr){6-8} $n$ &  & \mc{250} & \mc{500} & \mc{1000} & \mc{250} & \mc{500} & \mc{1000} \\ 
   \midrule
Test for $\lambda$  &  &  &  &  &  &  &  \\ 
   \midrule
SmoothMD with $\gamma$     & $G_n$ & 6.45 & 5.55 & 4.85 & 11.2  & 9.65  & 9.5   \\ 
SmoothMD* with $\gamma$    & $G_n$ & 6.95 & 6.2  & 5.55 & 12.05 & 10.5  & 10.55 \\ 
SmoothMD without $\gamma$  & $G_n$ & 6.4  & 5.55 & 5.0  & 11.5  & 10.15 & 9.7   \\ 
  NL2SLS                   & $G_n$ & 5.75 & 5.6  & 5.15 & 11.8  & 10.75 & 10.05 \\ 
   \midrule
Test for $\beta$ &  &  &  &  &  &  &  \\ 
   \midrule
SmoothMD with $\gamma$     & $G_n$ & 5.9  & 5.15 & 4.95 & 11.1  & 9.8   & 9.25 \\ 
SmoothMD* with $\gamma$    & $G_n$ & 6.4  & 5.5  & 5.5  & 12.15 & 10.95 & 10.3 \\ 
SmoothMD without $\gamma$  & $G_n$ & 6.05 & 5.35 & 4.9  & 10.9  & 9.8   & 9.1  \\ 
  NL2SLS                   & $G_n$ & 6.5  & 5.75 & 5.45 & 11.4  & 10.95 & 9.15 \\ 
\midrule\bottomrule
\end{tabular}

		\vskip 0.3cm
		\textit{Notes: For the SmoothMD estimators, $h \propto n^{-1/3.5}$. The components of $\boldsymbol d$ are set equal to the componentwise standard deviations for all variables. The variances are estimated by the Eiker-White variance estimator. For SmoothMD* the additional variance part due to the estimation of $\boldsymbol{\eta}$ is not taken into account. For SmoothMD the additional variance part is taken into account. 2000 Monte Carlo samples were used for all simulations.}
		\label{Emp_LevZ_Model3}
	\end{table}

	\begin{table}[H]
		\caption{\textit{Empirical Level for distance metric statistics of the estimators for $\lambda$ and $\beta$ in Model 1. } }
\begin{tabular}{@{}lcd{3.5}d{3.5}d{3.5}d{3.5}d{3.5}d{3.5}}
    \toprule\midrule& $s$   & \multicolumn{3}{c}{5\% level} &\multicolumn{3}{c}{10\% level}\\\cmidrule(lr){3-5} \cmidrule(lr){6-8} $n$ &  & \mc{250} & \mc{500} & \mc{1000} & \mc{250} & \mc{500} & \mc{1000} \\ 
   \midrule
Test for $\lambda$  &  &  &  &  &  &  &  \\ 
   \midrule
SmoothMD with $\gamma$     & $G_n$ & 7.0  & 6.8  & 5.35 & 11.9  & 12.7  & 11.2  \\ 
SmoothMD* with $\gamma$    & $G_n$ & 6.95 & 7.45 & 5.35 & 12.1  & 13.25 & 11.65 \\ 
SmoothMD without $\gamma$  & $G_n$ & 6.7  & 7.1  & 5.4  & 12.05 & 12.8  & 11.25 \\ 
   \midrule
Test for $\beta$ &  &  &  &  &  &  &  \\ 
   \midrule
SmoothMD with $\gamma$     & $G_n$ & 5.9  & 4.5  & 3.9   & 10.55 & 9.45  & 7.9   \\ 
SmoothMD* with $\gamma$    & $G_n$ & 6.5  & 5.0  & 4.35  & 11.75 & 10.65 & 8.5   \\ 
SmoothMD without $\gamma$  & $G_n$ & 6.35 & 4.75 & 4.0   & 10.8  & 9.75  & 7.85  \\ 
\midrule\bottomrule
\end{tabular}

		\vskip 0.3cm
		\textit{Notes: For the SmoothMD estimators, $h \propto n^{-1/3.5}$. The components of $\boldsymbol d$ are set equal to the componentwise standard deviations for all variables. The variances are estimated by the Eiker-White variance estimator. For SmoothMD* the additional variance part due to the estimation of $\boldsymbol{\eta}$ is not taken into account. For SmoothMD the additional variance part is taken into account. 2000 Monte Carlo samples were used for all simulations.}
		\label{Emp_DistM_Model1}
	\end{table}

	\begin{table}[H]
		\caption{\textit{Empirical Level for distance metric statistics of the estimators for $\lambda$ and $\beta$ in Model 3. } }
\begin{tabular}{@{}lcd{3.5}d{3.5}d{3.5}d{3.5}d{3.5}d{3.5}}
    \toprule\midrule& $s$   & \multicolumn{3}{c}{5\% level} &\multicolumn{3}{c}{10\% level}\\\cmidrule(lr){3-5} \cmidrule(lr){6-8} $n$ &  & \mc{250} & \mc{500} & \mc{1000} & \mc{250} & \mc{500} & \mc{1000} \\ 
   \midrule
Test for $\lambda$  &  &  &  &  &  &  &  \\ 
   \midrule
SmoothMD with $\gamma$     & $G_n$ & 12.15 & 9.55 & 7.05 & 14.1  & 12.6  & 10.65 \\ 
SmoothMD* with $\gamma$    & $G_n$ & 12.85 & 10.1 & 7.85 & 14.65 & 12.65 & 11.0  \\ 
SmoothMD without $\gamma$  & $G_n$ & 12.3  & 9.55 & 6.95 & 14.4  & 12.7  & 10.3  \\ 
   \midrule
Test for $\beta$ &  &  &  &  &  &  &  \\ 
   \midrule
SmoothMD with $\gamma$     & $G_n$ & 12.15 & 9.25  & 6.55 & 14.4  & 12.05 & 10.3  \\ 
SmoothMD* with $\gamma$    & $G_n$ & 12.6  & 10.05 & 7.2  & 14.6  & 12.4  & 10.95 \\ 
SmoothMD without $\gamma$  & $G_n$ & 11.85 & 8.95  & 6.7  & 14.15 & 11.75 & 10.7  \\ 
\midrule\bottomrule
\end{tabular}

		\vskip 0.3cm
		\textit{Notes: For the SmoothMD estimators, $h \propto n^{-1/3.5}$. The components of $\boldsymbol d$ are set equal to the componentwise standard deviations for all variables. The variances are estimated by the Eiker-White variance estimator. For SmoothMD* the additional variance part due to the estimation of $\boldsymbol{\eta}$ is not taken into account. For SmoothMD the additional variance part is taken into account. 2000 Monte Carlo samples were used for all simulations.}
		\label{Emp_DistM_Model3}
	\end{table}			
	
		\begin{figure}[h]
			\begin{minipage}{0.4\textwidth} 
				\caption{\textit{Power function of the distance metric \\ statistic
				for $\lambda$ of Model 1 with $n=500$.}}
				\includegraphics[scale=0.4]{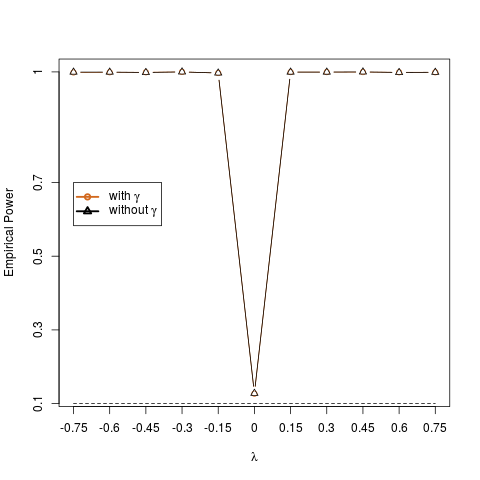}
				\label{fig_power_1.1}
			\end{minipage}
			\hfill
			\begin{minipage}{0.4\textwidth}
				\caption{\textit{Power function of the distance metric \\ statistic
				for $\beta$ of Model 1 with $n=250$.}}
				\includegraphics[scale=0.4]{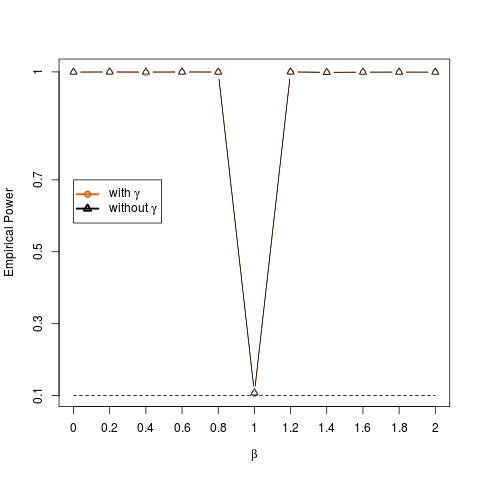}
				\label{fig_power_1.2}
			\end{minipage}
			\begin{spacing}{0.8}
				\flushleft{\textit{\footnotesize Notes: For the SmoothMD estimators, $h \propto n^{-1/3.5}$. The components of $\boldsymbol d$ are set equal to the componentwise standard deviations for all variables. The variances are estimated by the Eiker-White variance estimator. Only the SmoothMD estimators that take the additional variance part due to the estimation of $\boldsymbol{\eta}$ into account are considered. 2000 Monte Carlo samples were used for all simulations. The nominal level is $ 10\%$.}}		
			\end{spacing}	
		\end{figure}

		\begin{figure}[h]
			\begin{minipage}{0.4\textwidth} 
				\caption{\textit{Estimated $m(Z)$ for Model 1 with $n=500$.}}
				\includegraphics[scale=0.4]{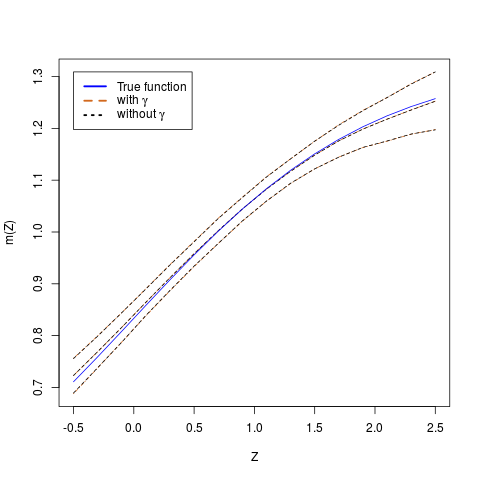}
			\end{minipage}
			\hfill
			\begin{minipage}{0.4\textwidth}
				\caption{\textit{Estimated $m(Z)$ for Model 2 with $n=500$.}}
				\includegraphics[scale=0.4]{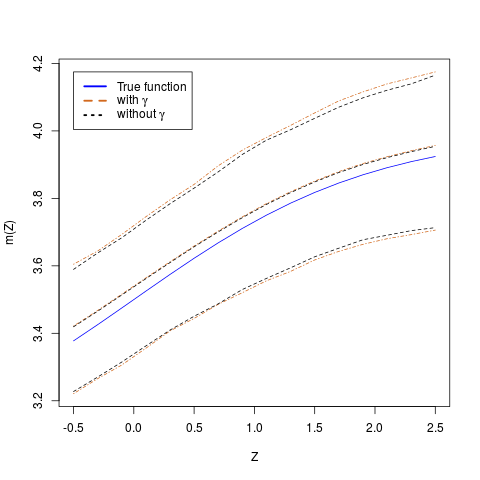}			
				\label{fig_m_2.1}
			\end{minipage}
		\begin{spacing}{0.8}
			\flushleft{\textit{\footnotesize Notes: For the estimation the NW estimator with normal kernel and $h \propto n^{-1/3.5}$ is employed. The $25\%$ and $75\%$ quantiles as well as the mean are reported. 2000 Monte Carlo samples were used for all simulations.}}		
		\end{spacing}		
		\end{figure}

\vskip 4cm
~~\\
\newpage

\end{appendix}
\clearpage

\bibliographystyle{cas-model2-names}

\bibliography{bibliography}

\end{document}